\documentclass[aps,prd,twocolumn,nofootinbib,superscriptaddress,preprintnumbers,floatfix]{revtex4-2}
\usepackage{float,dcolumn}
\usepackage{amsmath,amssymb}
\usepackage{mathtools,slashed}
\usepackage{bm}
\usepackage{xcolor}
\usepackage[caption=false]{subfig}
\usepackage{verbatim}
\usepackage{array}
\usepackage{multirow}
\usepackage{booktabs}
\usepackage{url}
\usepackage{color}
\usepackage[T1]{fontenc}
\usepackage[latin9]{inputenc}
\usepackage{graphicx}
\usepackage{esint}
\usepackage[hypertexnames=false]{hyperref}
\usepackage{cleveref}
\usepackage{atbegshi}
\usepackage{placeins}

\usepackage[normalem]{ulem}

\Crefname{equation}{Eq.}{Eqs.}
\Crefname{section}{Sec.}{Secs.}
\Crefname{figure}{Fig.}{Figs.}
\Crefname{appendix}{Appendix}{Appendices}

\newcommand{\cD}[0]{\mathcal D}

\newcommand{\cK}[0]{\mathcal K}
\newcommand{\cL}[0]{\mathcal L}
\newcommand{\cM}[0]{\mathcal M}
\newcommand{\cO}[0]{\mathcal O}
\newcommand{\cR}[0]{\mathcal R}

\newcommand{\cT}[0]{\mathcal T}

\newcommand{\cZ}[0]{\mathcal Z}

\newcommand{\kdf}{\mathcal{K}_{\text{df},3} }
\newcommand\Kdf{\mathcal{K}_{\rm df,3}}

\hypersetup{ 
pdfnewwindow=true,
colorlinks=true,    
linkcolor=blue,
citecolor=blue,
filecolor=blue,
urlcolor=blue
} 

\begin{document}

\preprint{\vbox{\hbox{MIT-CTP/5846} }}
\title{Two- and three-meson scattering amplitudes  with physical quark masses \\ from lattice QCD } 

\author{Sebastian M. Dawid}
\affiliation{Physics Department, University of Washington, Seattle, WA 98195-1560, USA}

\author{Zachary T. Draper}
\affiliation{Physics Department, University of Washington, Seattle, WA 98195-1560, USA}

\author{Andrew D. Hanlon}
\affiliation{Department of Physics, Kent State University, Kent, OH 44242, USA}

\author{Ben H\"{o}rz}
\affiliation{Intel Deutschland GmbH, Dornacher Str. 1, 85622 Feldkirchen, Germany}

\author{Colin Morningstar}
\affiliation{Department of Physics, Carnegie Mellon University, Pittsburgh, Pennsylvania 15213, USA}

\author{Fernando Romero-L\'{o}pez}
\affiliation{Albert Einstein Center, Institute for Theoretical Physics, University of Bern, 3012 Bern, Switzerland}
\affiliation{Center for Theoretical Physics, Massachusetts Institute of Technology, Cambridge, MA 02139, USA}

\author{Stephen R. Sharpe}
\affiliation{Physics Department, University of Washington, Seattle, WA 98195-1560, USA}

\author{Sarah Skinner}
\affiliation{Department of Physics, Carnegie Mellon University, Pittsburgh, Pennsylvania 15213, USA}

\date{\today}

\begin{abstract}
We study systems of two and three mesons composed of pions and kaons at maximal isospin using four CLS ensembles with $a\approx 0.063\;$fm, including one with approximately physical quark masses. Using the stochastic Laplacian-Heaviside method, we determine the energy spectrum of these systems including many levels in different momentum frames and irreducible representations. Using the relativistic two- and three-body finite-volume formalism, we constrain the two and three-meson K matrices, including not only the leading $s$ wave, but also $p$ and $d$ waves. By solving the three-body integral equations, we determine, for the first time, the physical-point scattering amplitudes for $3\pi^+$, $3K^+$, $\pi^+\pi^+ K^+$  and $K^+ K^+ \pi^+$ systems. 
These are determined for total angular momentum $J^P=0^-$, $1^+$, and $2^-$.
We also obtain accurate results for $2\pi^+$, $\pi^+ K^+$, and $2K^+$ phase shifts. We compare our results to Chiral Perturbation Theory, and to phenomenological fits.

\end{abstract}

\keywords{lattice QCD, scattering amplitudes}
\maketitle

\section{Introduction}

Three-hadron systems represent an important frontier in the first-principles understanding of the hadron spectrum from Quantum Chromodynamics (QCD).  Many hadronic resonances exhibit decay modes involving at least three hadrons, such as the doubly-charmed tetraquark, $T_{cc}^+(3875)$, the Roper resonance, and three-pion resonances such as $\omega(782)$, $h_1(1170)$, and $a_1(1260)$. Theoretical and numerical advances in Lattice QCD (LQCD) spectroscopy are making it feasible to study such systems from first principles~\cite{Detmold:2008gh,Beane:2007qr,Briceno:2012rv,Polejaeva:2012ut,Hansen:2014eka,Hansen:2015zga,Briceno:2017tce,Hammer:2017uqm,Hammer:2017kms,Mai:2017bge,Briceno:2018aml,Briceno:2018mlh,Pang:2019dfe,Jackura:2019bmu,Blanton:2019igq,Briceno:2019muc,
Romero-Lopez:2019qrt,Pang:2020pkl,Blanton:2020gha,Blanton:2020jnm,Romero-Lopez:2020rdq,Blanton:2020gmf,Muller:2020vtt,Blanton:2021mih,Muller:2021uur,Muller:2022oyw,Pang:2023jri,Bubna:2023oxo,Briceno:2024txg,Xiao:2024dyw,Hansen:2024ffk,Draper:2023xvu,Feng:2024wyg,Jackura:2023qtp}; see also recent reviews~\cite{Hansen:2019nir,Rusetsky:2019gyk,Mai:2021lwb,Romero-Lopez:2022usb}. Indeed, there are already pioneering applications to three-body resonances~\cite{Mai:2021nul,Garofalo:2022pux,Yan:2024gwp}.

Significant efforts have been devoted to studying weakly interacting multihadron systems using LQCD~\cite{Beane:2007es,Detmold:2008fn,Detmold:2008yn,Detmold:2011kw,Mai:2018djl,Horz:2019rrn,Blanton:2019vdk,Culver:2019vvu,Mai:2019fba,Fischer:2020jzp,Hansen:2020otl,Alexandru:2020xqf,Brett:2021wyd,Blanton:2021llb,NPLQCD:2020ozd,Abbott:2023coj,Abbott:2024vhj},
and we study here the example of three light mesons (pions and/or kaons) at maximal isospin.
These systems have simple dynamics, lacking resonances both in the full system and in two-particle subchannels.
This makes them ideal for benchmarking first-principles multiparticle methods.
Furthermore, since we work with pseudo-Goldstone bosons, we can compare results with predictions from chiral perturbation theory (ChPT),
offering valuable insights into multi-hadron interactions. 

In this work, we use the stochastic Laplacian-Heaviside method to compute finite-volume multimeson energy spectra from LQCD. We then use the relativistic two- and three-particle finite-volume formalism of Refs.~\cite{Kim:2005gf,Hansen:2014eka,Hansen:2015zga} (the so-called RFT formalism), which allows one to constrain two- and three-particle K matrices from LQCD energies. Physical scattering amplitudes are then obtained from the K matrices by solving appropriate integral equations, which incorporate $s$-channel unitarity~\cite{Hansen:2015zga}. 
This workflow allows an indirect extraction of infinite-volume three-particle scattering amplitudes from the energies of three particles in a finite box.

Specifically, we analyze the following systems composed of $\pi^+$ and $K^+$ mesons: $2\pi^+$, $\pi^+ K^+$, $2K^+$, $3\pi^+$, $\pi^+\pi^+K^+$, $K^+K^+\pi^+$, and $3K^+$. Extending our previous studies~\cite{Blanton:2021llb,Blanton:2021eyf}, this work includes an additional lattice ensemble with quark masses very close to the physical point. With the exception of Ref.~\cite{Fischer:2020jzp}, which studied the $2\pi^+$ and $3\pi^+$ systems at the physical point, no comparable studies have been performed. 
While our main focus is on the application of three-particle methods at the physical point, we can study the chiral dependence of the amplitudes by combining all available ensembles (four in all).

An important byproduct of studies of three-particle systems is that one obtains results  for the interactions of the two-meson subsystems, here $\pi^+\pi^+$, $K^+K^+$, and $\pi^+ K^+$. For all of these systems, and in particular, for $\ell>0$ partial waves, we constrain the two-meson phase shifts directly at the physical point for the first time. In the cases in which fits to experimental data are available, we find $1-2\sigma$ agreement with our LQCD results.

A novel aspect of this work is the determination of three-meson scattering amplitudes. These are obtained by solving integral equations whose inputs are the two- and three-meson K matrices obtained from fits to the LQCD energies. Three-meson amplitudes have only been explored before for the $3\pi^+$ system with total angular momentum $J^P=0^-$ and at heavier than physical quark masses~\cite{Hansen:2020otl}. This work provides, for the first time, direct results for amplitudes at the physical point, and not only for $3\pi^+$ but also for systems involving kaons. Another new feature is that we determine amplitudes with higher angular momentum, $J^P=1^+$ and $2^-$. 

The intermediate two- and three-hadron K matrices, while scheme-dependent and unphysical, can be compared to ChPT predictions. These are available both at leading order (LO), and, in some cases, next-to-leading order (NLO). These comparisons provide an important cross-check on our computations, and, in some cases, allow for the extraction of low-energy effective couplings.  

This paper is released in parallel with a letter with the highlights of this work~\cite{Dawid:2025zxc}.

This paper is organized as follows. Section~\ref{sec:lattice} summarizes the LQCD setup and the determination of LQCD energies. Section~\ref{sec:formalism} describes the finite-volume formalism,  the integral equations, the parametrizations of the K matrices, and the fitting procedure. Fits to the spectra and constraints on threshold parameters are shown in \Cref{sec:results}. Comparisons to ChPT are discussed in \Cref{sec:comp_ChPT}, and results for the three-meson scattering amplitudes are provided in \Cref{sec:amplitudes}. We conclude in \Cref{sec:concl}. We include four appendices: \Cref{app:operators} lists the interpolating operators,\Cref{app:E2502meson} collects a few tables with two-meson results, \Cref{app:reananlysis} presents a reanalysis of previously published data, and \Cref{app:pwproj} discusses the partial-wave projection of all quantities in the integral equations.

\section{Lattice QCD setup and results}
\label{sec:lattice}

In this section, we describe the LQCD calculations of this work, including the ensembles, the interpolating operators, and the extraction of energy levels. The methods used here are largely the same as in our previous work~\cite{Blanton:2021llb,Draper:2023boj}, and we only repeat the important details. What is new here is the use of an additional ensemble with close to physical quark masses, and the inclusion of the geometric fit model for the single-meson correlators.

\subsection{Ensembles}

This work uses the $N_f=2+1$ CLS LQCD ensembles~\cite{Bruno:2014jqa} at a fixed lattice spacing, generated with nonperturbatively $\mathcal{O}(a)$-improved Wilson fermions and the tree-level $\mathcal{O}(a^2)$-improved L\"uscher-Weisz gauge action. There have been several determinations of the lattice spacing for the ensembles used in this work. It was first determined to be $a \approx 0.06426(76)$ fm using a linear combination of decay constants~\cite{Bruno:2016plf}. More recently, it has been updated to $a \approx 0.0633(4)(6)$ fm~\cite{Strassberger:2021tsu} also using decay constants. Using baryon masses, $a \approx 0.06379(37)$ fm was reported~\cite{RQCD:2022xux}. For the purpose of this paper, it suffices to consider the lattice spacing to be $a \simeq 0.063$ fm.

The main results of this work are obtained using the E250 ensemble, with almost physical pion and kaon masses. The mass dependence of scattering quantities is explored by combining the results with those from three additional ensembles at heavier-than-physical pion masses. These ensembles follow a chiral trajectory along which the trace of the quark mass matrix is kept approximately constant, ${\rm tr\ } M_q = 2m_{l} + m_s \simeq {\rm const}$ ($m_l$ is the up/down quark mass, and $m_s$ is the strange quark mass). Open temporal boundary conditions (BCs)~\cite{Luscher:2012av} are used on all ensembles, except for E250 where periodic BCs are used. A summary of all ensembles is given in \Cref{tab:ensems}. The details of the N203, N200, and D200 ensembles have already been reported in our previous work~\cite{Blanton:2021llb,Draper:2023boj}.
Single meson masses and decay constants for these ensembles are listed in \Cref{tab:masses_decay_constants}. These will be required for the analysis of LQCD energies and for the investigation of the chiral dependence of scattering observables.

\begin{table*}
  \centering
  \tabcolsep=0.15cm
  \begin{tabular}{c c c c c c c c c c}
    \toprule
    & $(L/a)^3 \times (T/a)\phantom{^3}$ & BCs & $M_\pi \, [\mathrm{MeV}]$ & $M_K \, [\mathrm{MeV}]$ & $N_\mathrm{cfg}$ & $t_\mathrm{src}/a$ & $N_\mathrm{ev}$ & dilution & $N_r(\ell/s)$ \\
    \midrule
    N203 & $48^3 \times 128$& open & 340 & 440 & 771 & 32, 52 & 192 & (LI12,SF) & 6/3 \\
    N200 & $48^3 \times 128$& open & 280 & 460 & 1712 & 32, 52 & 192 & (LI12,SF) & 6/3 \\
    D200 & $64^3 \times 128$& open & 200 & 480 & 2000 & 35, 92 & 448 & (LI16,SF) & 6/3 \\
    E250 & $96^3 \times 192$& periodic & 130 & 500 & 505  & 4 random & 1536 & (LI16,SF) & 6/3 \\
    \bottomrule
  \end{tabular}
  \caption{Details on the ensembles used in this work:  name, geometry, boundary conditions,
  approximate meson masses, number of configurations $N_{\rm cfg}$, source positions $t_{\rm src}$
  used, number of eigenvectors $N_{\rm ev}$ of the covariant Laplacian retained, dilution scheme
  (see Ref.~\cite{Morningstar:2011ka} for details), and
  number of noises $N_{\rm r}$ used for the light ($l$) and strange ($s$) quark sources.
  The lattice spacing is $a \approx 0.063$ fm.
  For the E250 ensemble, both forward and backward correlators were evaluated for the 4 evenly spaced source times, resulting in 8 effective sources.}
  \label{tab:ensems}
\end{table*}

\begin{table*}
   \centering
   \begin{tabular}{c c c c c c c c}
     \toprule
     & $a M_\pi$ & $a M_K$ & $M_\pi L$ & $M_K L$ & $ M_\pi / F_\pi$  & $M_K/F_K$ & $M_K/M_\pi$ \\
     \midrule
     N203 & 0.11261(20) & 0.14392(15) & 5.4053(96) & 6.9082(72) & 3.4330(89) & 4.1530(72) &  1.2780(13) \\
     N200 & 0.09208(22)  &  0.15052(14)  & 4.420(11)   & 7.2250(67)   &  2.964(10) & 4.348(11) & 1.6347(29)   \\
     D200 & 0.06562(19) & 0.15616(12) & 4.200(12)  & 9.9942(77) & 2.2078(67) & 4.5132(93) &  2.3798(59) \\
     E250 &  0.04217(20) & 0.159170(93)    &   4.049(19) & 15.2803(90)   &  1.4927(62)  &  4.6415(56) &  3.774(17) \\
     \bottomrule
   \end{tabular}
   \caption{ Masses and decay constants for the ensembles used in this work.
   For the N203. N200 and D200 ensembles, the masses were determined in our previous publications~\cite{Blanton:2021llb,Draper:2023boj}. The decay constants are taken from Ref.~\cite{Ce:2022eix}. }
   \label{tab:masses_decay_constants}
\end{table*}

\subsection{Correlation functions}

The low-lying finite-volume spectrum can be determined from two-point correlation functions of interpolating operators that overlap with the sought-after states. To this end, we compute Hermitian correlator matrices, where the elements of this matrix are of the form
\begin{align}
\begin{split}
C_{i j}\left(t_{\text {sep }}\right) &\equiv\left\langle\mathcal{O}_i\left(t_{\rm sep}+t_0\right) \overline{{\mathcal{O}}}_j\left(t_0\right)\right\rangle \\ &=\sum_{n=0}^{\infty}\left\langle\Omega\left|\mathcal{O}_i\right| n\right\rangle\left\langle\Omega\left|\mathcal{O}_j\right| n\right\rangle^* e^{-E_n t_{\text {sep }}},
\label{eq:spec_decomp}
\end{split}
\end{align}
where $\mathcal{O}_i$ and $\overline{\mathcal{O}}_j$ are annihilation and creation operators, and $\left\langle\Omega\left|\mathcal{O}_i\right| n\right\rangle$ are the matrix elements of the operators  between the vacuum $|\Omega\rangle$ and the $n$-th energy eigenstate $| n \rangle$ with energy $E_n$. In the second equality above, we give the spectral decomposition of the matrix elements, which shows how these correlation functions depend on the finite-volume spectrum. For the E250 ensemble, due to the temporal PBCs, the source timeslice at $t_0$ can be shifted arbitrarily which helps to reduce autocorrelations. The use of a matrix of correlators, rather than a single correlator, is due to the practical difficulties of reliably extracting several multi-meson excited states. By solving a generalized eigenvalue problem (GEVP) for an $N \times N$ correlator matrix, the leading behavior of each of the resulting eigenvalues is dominated by a different energy $E_n$ from the set of the lowest $N$ energy eigenstates that have non-zero overlap with the states created by at least one operator entering the correlator matrix~\cite{Luscher:1990ck}. This allows for a determination of the $N$ lowest finite-volume energies with a given set of quantum numbers from simple fits to each of the eigenvalues. Additionally, under certain circumstances, the NLO behavior of the eigenvalues can be shown to be controlled by $E_N$, which leads to less excited-state contamination~\cite{Blossier:2009kd}.

The scattering analysis done in later sections requires both single-meson ground state energies at rest and several multi-meson energies. For the determination of the single-meson ground state energies, a single symmetric correlator (i.e. a one-by-one correlator matrix) is sufficient. The single-meson interpolators used in this work are given by
\begin{align}
  H_{\pi^+} (\textbf{p}, t) &= \sum_{\textbf{x}} e^{-i \textbf{p} \cdot \textbf{x}} \; \overline{d}(x) \gamma_5 u(x) , \\
  H_{K^+} (\textbf{p}, t) &= \sum_{\textbf{x}} e^{-i \textbf{p} \cdot \textbf{x}} \; \overline{s}(x) \gamma_5 u(x) ,
\end{align}
where $u,d,s$ label the up, down, and strange quark fields, respectively.
In principle, only the single-meson ground state energies at rest are needed for the scattering analysis, since the moving energies are obtained through the continuum dispersion relation.
However, in our extraction of multi-meson energies later on, we utilize ratios of correlators which necessarily involve single-meson correlators with non-zero momentum.

Multi-meson energies are more difficult to determine reliably, in part due to the larger number of multi-meson energies sought-after, but also due to an exponentially decreasing signal-to-noise ratio along with smaller energy gaps between the states.
Fortunately, these latter two issues are typically not very severe when dealing with systems of mesons, but to ameliorate the issue we utilize correlator matrices where the number of interpolating operators corresponds to the number of states we would like to extract.
The correlator matrix is then amenable to the use of the GEVP method mentioned above and described in more detail below.
The two- and three-hadron interpolators used in this work are formed from the single-meson interpolators given above
\begin{align}
    \begin{split}
        & [H_{f_1} H_{f_2}]_{\Lambda} (\textbf{P},t) = \sum_{\textbf{p}_1, \textbf{p}_2} c^{\textbf{P}, \Lambda}_{\textbf{p}_1,f_1;\textbf{p}_2,f_2} H_{f_1}(\textbf{p}_1, t) H_{f_2}(\textbf{p}_2, t) , \\
  &[H_{f_1} H_{f_2} H_{f_3}]_{\Lambda} (\textbf{P},t) = \\ &\sum_{\textbf{p}_1, \textbf{p}_2, \textbf{p}_3} c^{\textbf{P}, \Lambda}_{\textbf{p}_1,f_1;\textbf{p}_2,f_2;\textbf{p}_3,f_3} H_{f_1}(\textbf{p}_1, t) H_{f_2}(\textbf{p}_2, t) H_{f_3}(\textbf{p}_3, t),
    \end{split}
\end{align}
where the total momentum is $\textbf{P} = \sum_{i} \textbf{p}_i$, the irreducible representation (irrep) of the little group of $\textbf{P}$ is denoted by $\Lambda$, $f_i \in \{\pi^+, K^+\}$ is the flavor of each meson, and $c^{\textbf{P},\Lambda}$ are  Clebsch-Gordan coefficients that ensure the operator transforms according to the irrep $\Lambda$.
The determination of the Clebsch-Gordan coefficients follows the same procedure as in our previous work~\cite{Blanton:2021llb,Draper:2023boj}, which follows Ref.~\cite{Morningstar:2013bda}\footnote{The coefficients can be shared upon request.}.
The momentum sums run over all momenta related via rotations within the little group of $\textbf{P}$, such that $\sum_{i} \textbf{p}_i = \textbf{P}$. Tables of the multi-meson interpolating operators used for the E250 calculation are given in \Cref{app:operators}.

To evaluate the correlation functions, quark propagators from all spatial points in the source timeslice to all points in the sink timeslice are needed. These propagators are computed using the stochastic Laplacian-Heaviside (sLapH) method~\cite{Morningstar:2011ka}, a stochastic variant of the distillation method~\cite{Peardon:2009gh}. Both sLapH and distillation employ a quark smearing procedure that projects the propagators onto the space spanned by the $N_{\rm ev}$ lowest modes of $-\widetilde{\Delta}$ (on each timeslice), where $\widetilde{\Delta}$ is the gauge-covariant 3-dimensional Laplacian expressed in terms of the stout-smeared gauge field. Since the number of eigenvectors needed for a constant smearing radius grows in proportion to the spatial volume $L^3$, the advantage of the stochastic variant, sLapH, lies in a better cost scaling with the volume. Variance reduction is accomplished in sLapH by diluting the noises in a judicious manner (see Refs.~\cite{Blanton:2021llb,Draper:2023boj}). Details about $N_{\rm ev}$, the dilution scheme, and the number of noises are given in \Cref{tab:ensems}.

\begin{figure}[h!]
    \includegraphics[width=8cm]{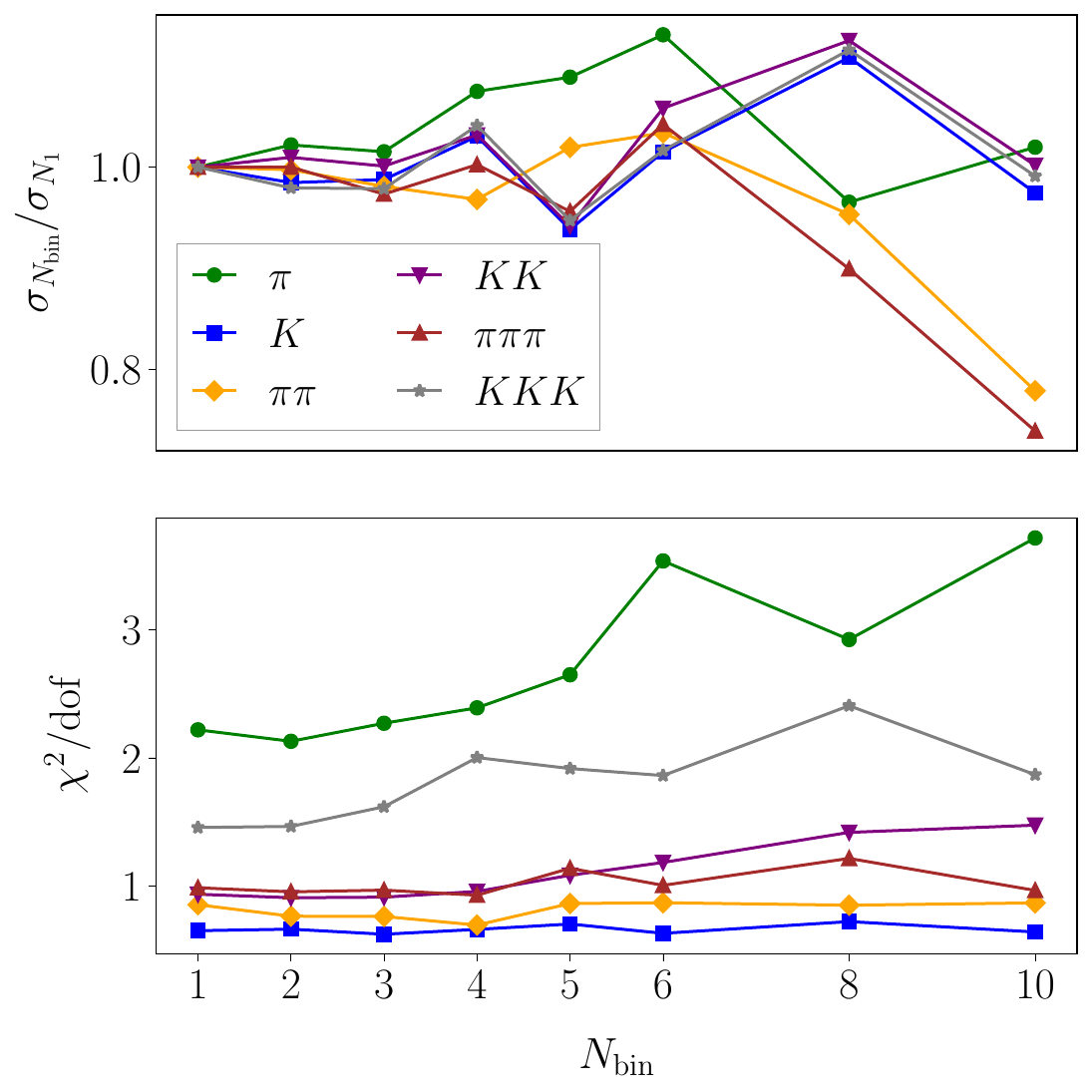}
    \caption{Results of the rebinning analysis to quantify autocorrelations in single- and multi-meson correlators. For the $\pi\pi$ system, we choose the $A_1^+(8)$ level 0; for $KK$, $E(9)$ level 0; for $\pi\pi\pi$, $B_1^-$ level 0 ; for $KKK$, $E_u(0)$ level 0. The choice of irrep and momentum for the multi-meson correlators was random, in order to not bias our results.}
    \label{fig:rebin}
\end{figure}

Before proceeding to the analysis of the correlators, the question of autocorrelation is addressed. It was mentioned earlier that the use of randomly shifted source times on each configuration can help to alleviate autocorrelations, so we expect autocorrelations to be small on E250.
If this is true, we should find that the errors on the energies do not increase and the $\chi^2$ for the fits do not decrease as the amount of rebinning $N_{\rm bin}$ is increased, as rebinning tends to reduce autocorrelations.
By rebinning, we mean that we take the results for a correlator on the first $N_{\rm bin}$ configurations and average them to create the first bin, then the average of the next $N_{\rm bin}$ configurations to create the second bin, and so on.
Then we analyze the correlator with this set of binned results.
The dependence of the errors and $\chi^2$ on the bin size for various energies are shown in \Cref{fig:rebin}. As can be seen, no significant increase of errors occurs when going from a bin size of 1 to 10. This indicates, that the effects of autocorrelations are negligible for these observables. This is further supported by the $\chi^2$ value shown in the lower panel, which does not decrease for increasing bin size. Rather, it fluctuates or even increases, potentially due to worse estimates of the covariance matrix as the number of total bins decreases with increasing bin size. Therefore, we use $N_{\rm bin} = 1$ for the analysis on E250.

\subsection{Extraction of energies}

\begin{figure*}[th!]
     \centering
     \subfloat[Pion \label{fig:singlepi}]{\includegraphics[width=0.49\textwidth]{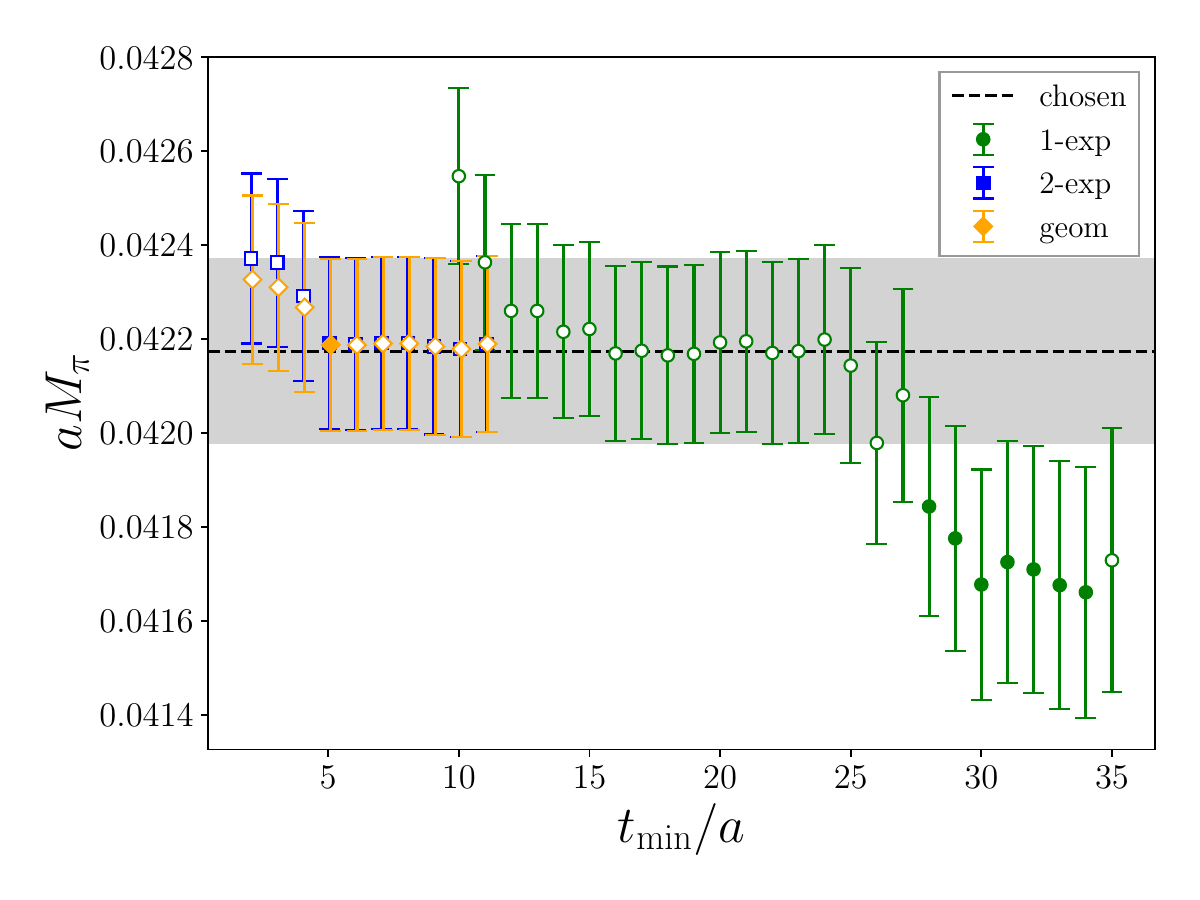}
    }
    \hfill
    \subfloat[Kaon \label{fig:singlek}]{%
     \includegraphics[width=0.49\textwidth]{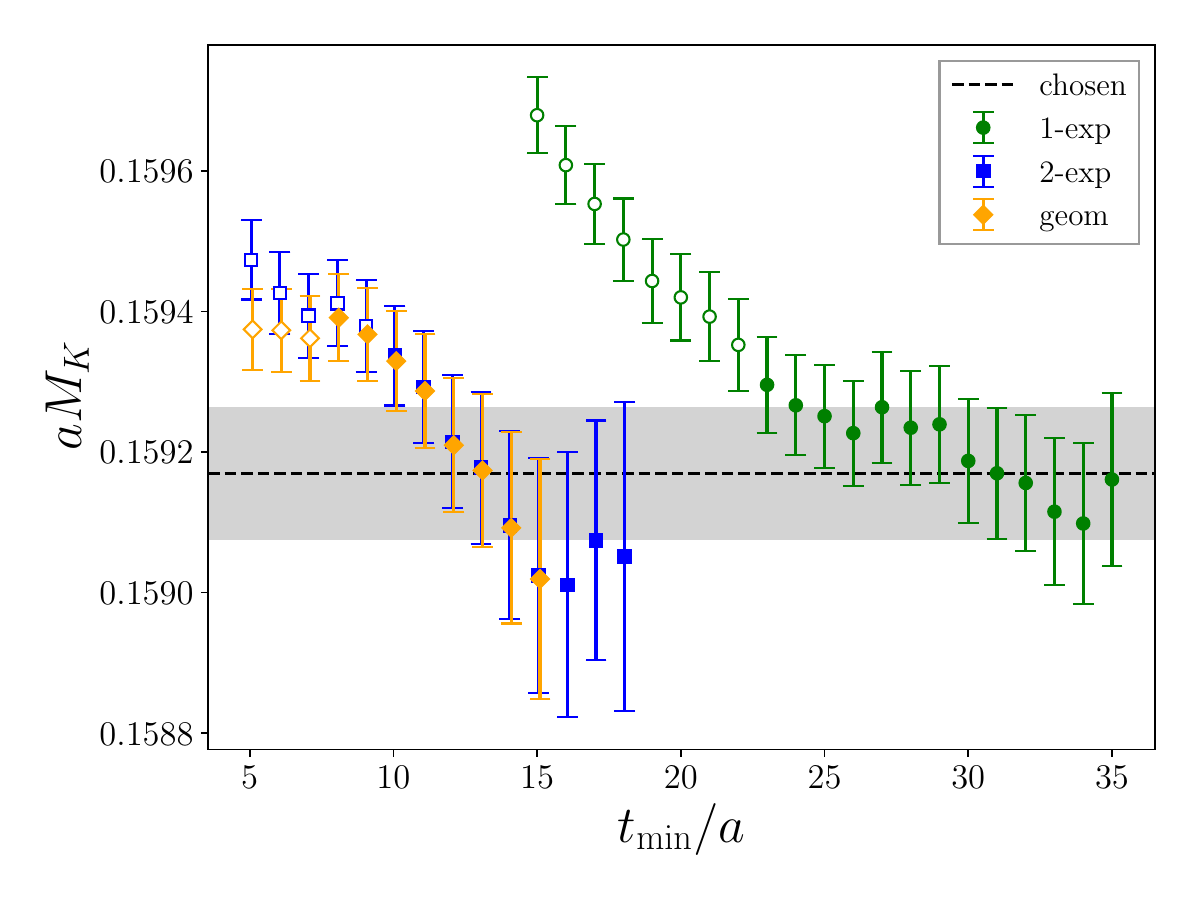}
    }
    \caption{Dependence of the single-meson energy extractions on the first time separation included in the fit, $t_{\rm min}$, and the choice of model. Three different models, described in the text, are considered: single-exponential (``1-exp''), two-exponential (``2-exp''), and geometric (``geom''). The grey bands indicate the fit result based on the chosen $t_{\rm min}$ and model. Here the chosen model is the single-exponential. Solid markers indicate a $p$-value $\ge 0.1$, whereas hollow markers indicate a $p$-value $< 0.1$.}
    \label{fig:singlehads}
\end{figure*}

The spectral decomposition in \Cref{eq:spec_decomp} shows that it is in principle possible to extract the spectrum from two-point correlators if the interpolating operators included have nonzero overlap with the sought-after states. In practice, however, statistical noise complicates this extraction. In this section, we detail the strategies we follow to deal with this difficulty. Two independent analyses were performed in order to verify correctness and to reduce the influence of human bias.

\subsubsection{Single-meson masses}

In order to extract the single-meson energies in this work, it is sufficient to use a single, symmetric correlator at large-time separations, such that there is ground-state dominance. We fit a single exponential in the range $t \in [t_{\rm min}, t_{\rm max}]$, with $t_{\rm min}$ at least as large as the onset of the observed ground-state saturation:
\begin{equation}
    C(t) = A_0 \exp(-E_0 t),
\end{equation}
where $E_0$ and $A_0$ are fit parameters. Other fit models are used for consistency, e.g.~two-exponential fits:
\begin{equation}
     C(t) = \sum_{i=0,1} A_i \exp(-E_i t),
\end{equation}
as well as the so-called ``geometric fits''~\cite{Bulava:2022vpq}:
\begin{equation}
     C(t) = A \frac{\exp(-E_0 t)}{1 - B \exp(-\Delta E t) },
\end{equation}
which model a tower of equally spaced inelastic excited states. In \Cref{fig:singlehads}, we show the results for the mass of the pion and kaon as a function of the lower end of the fit range, $t_{\rm min}$, for all three models. All uncertainties in the energy extractions are estimated with jackknife resampling. The single pion shows a long plateau at early time separations. However, single exponential fits show a decrease in the central value for fit ranges starting around $t_{\rm min}/a \approx 27$. Given the consistency across all fit models at earlier values of $t_{\rm min}$, we consider this downward trend to be a fluctuation, and opt for the earlier plateau. For the single kaon, single exponential fits at large $t_{\rm min}$ show consistency with two-exponential and geometric fits. The extracted energies show little to no dependence on the precise value of $t_{\rm max}$ so long as it is large enough to give the fit enough freedom. Therefore, we opt for $t_{\rm max}/a = 40$, which is the largest time separation computed.

\begin{figure*}[th!]
    \centering
    \includegraphics[width=0.32\textwidth]{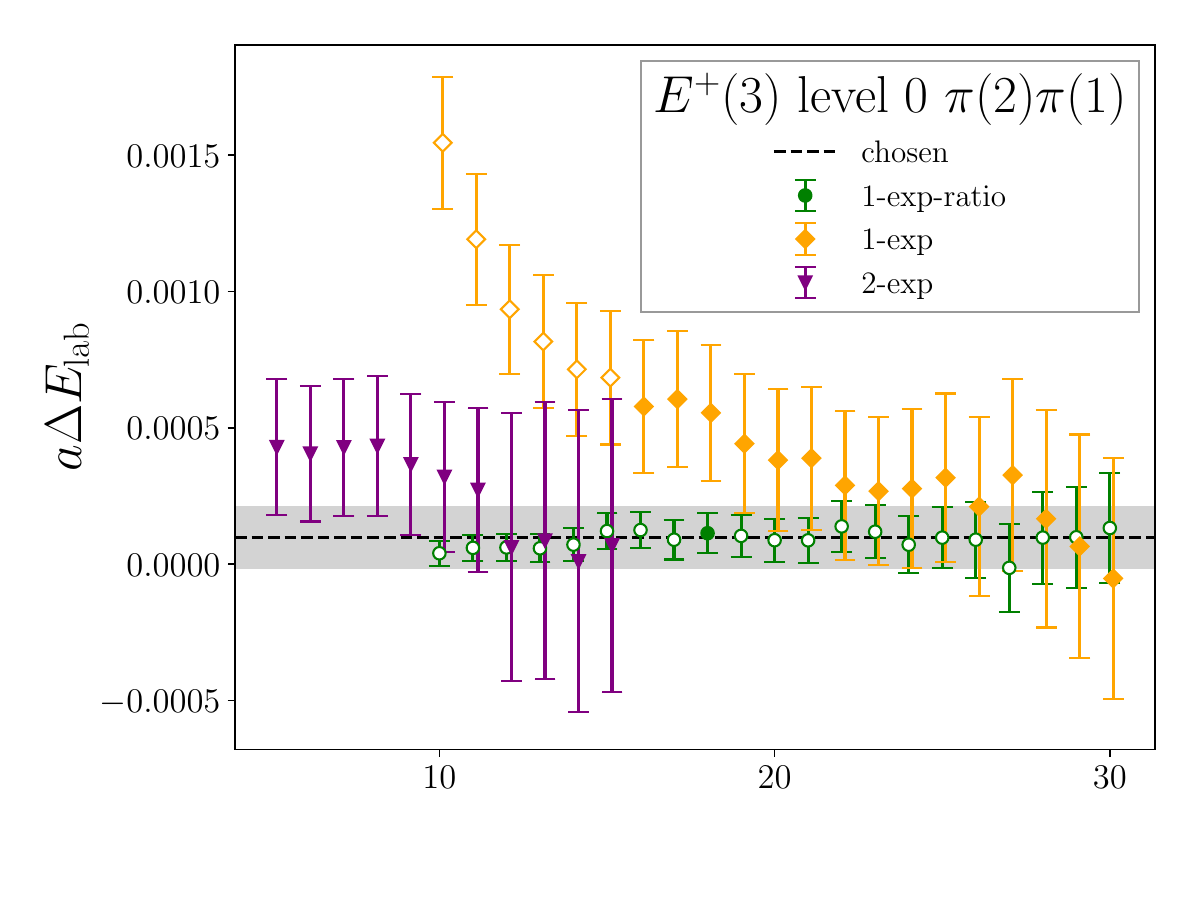}
    \includegraphics[width=0.32\textwidth]{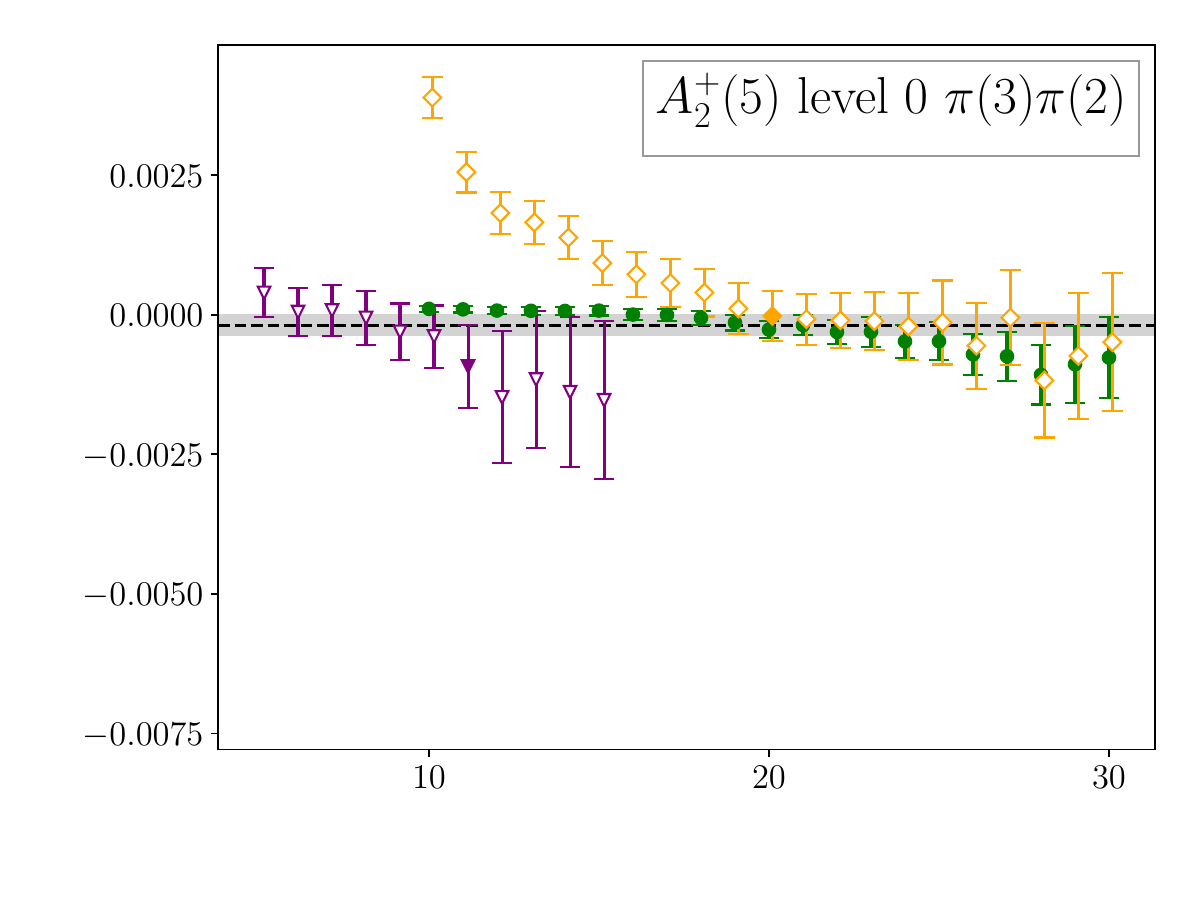}
    \includegraphics[width=0.32\textwidth]{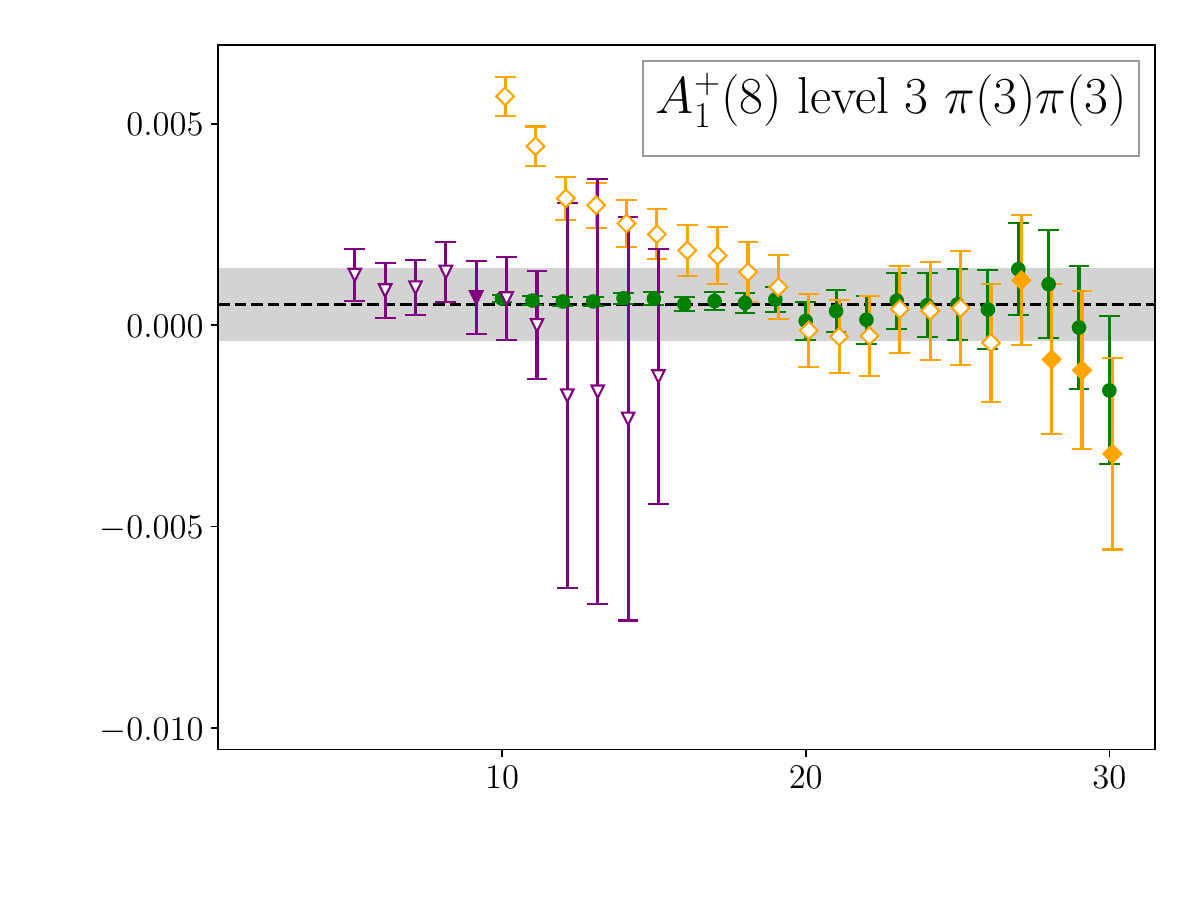}
    \includegraphics[width=0.32\textwidth]{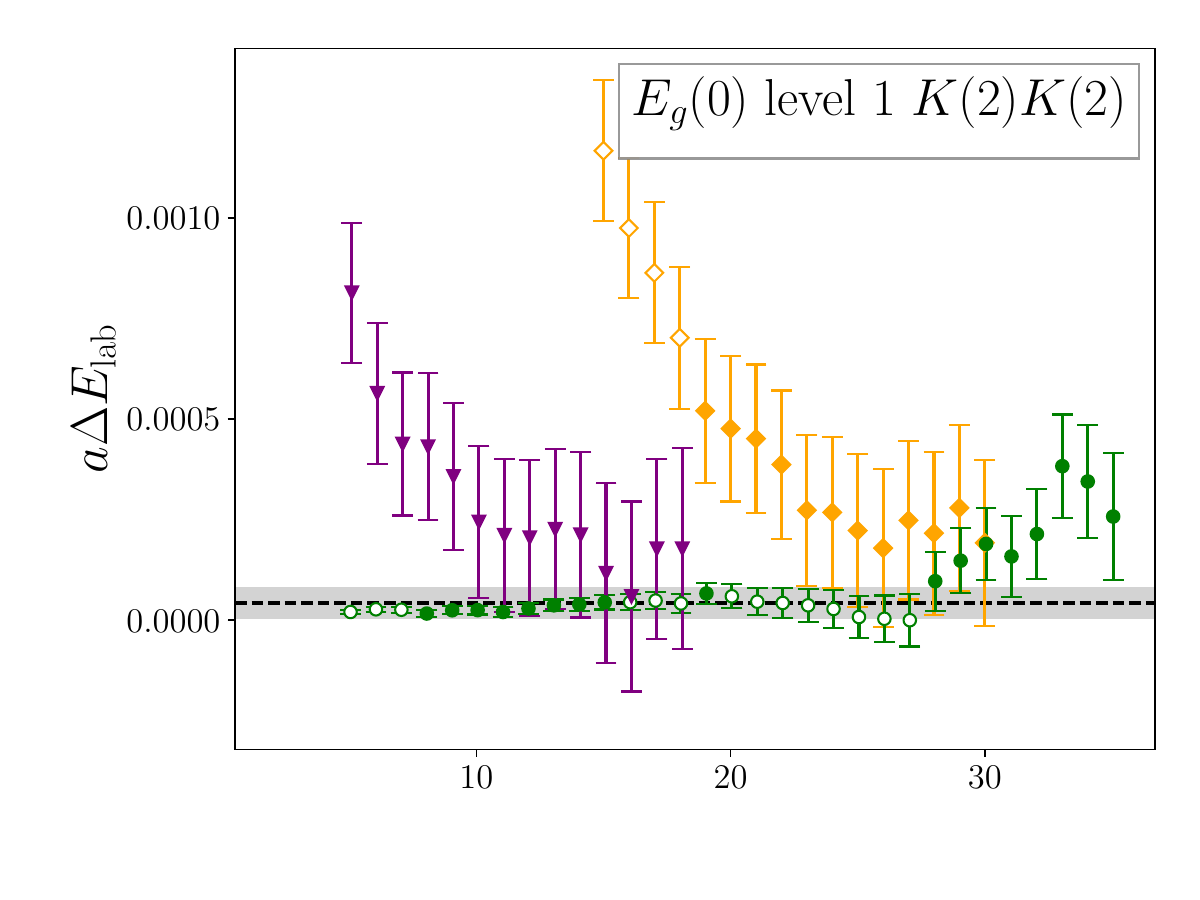}
    \includegraphics[width=0.32\textwidth]{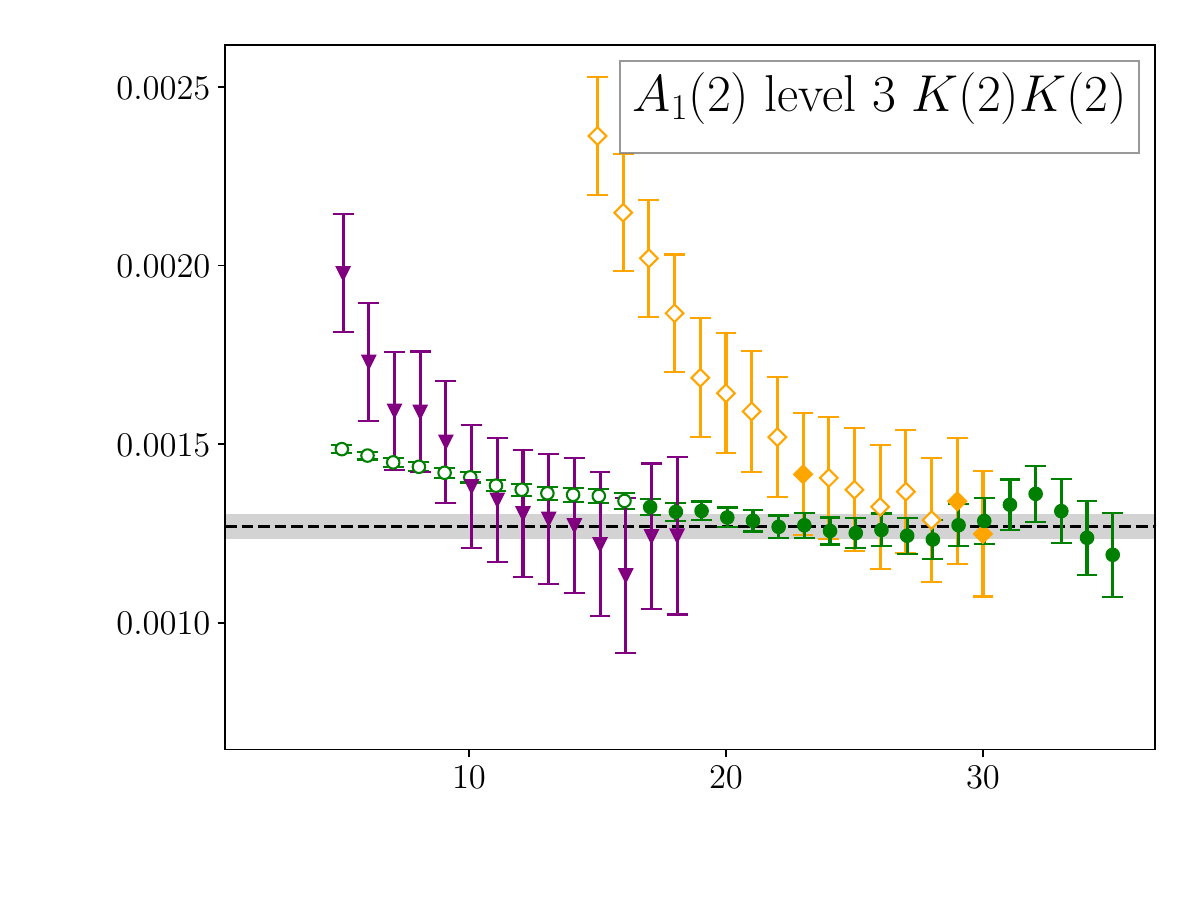}
    \includegraphics[width=0.32\textwidth]{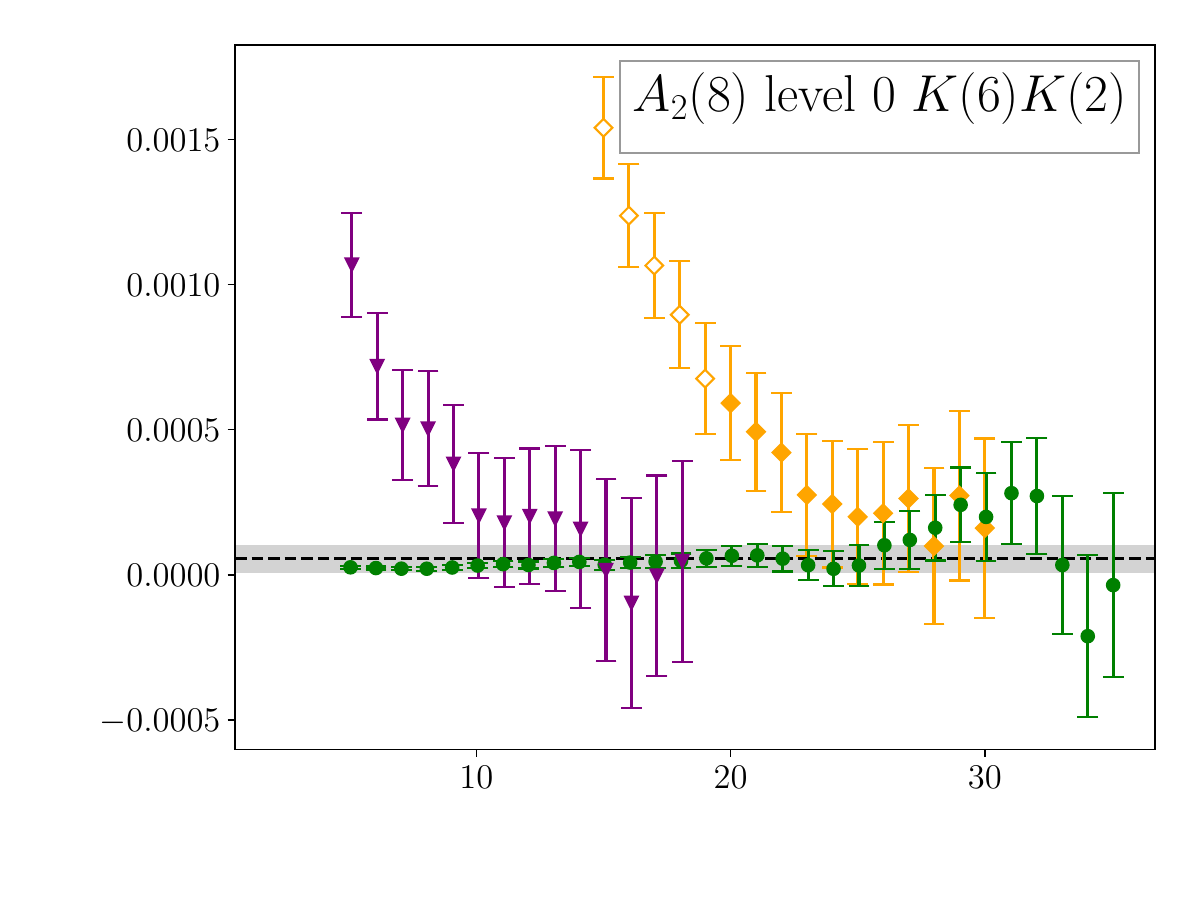}
    \includegraphics[width=0.32\textwidth]{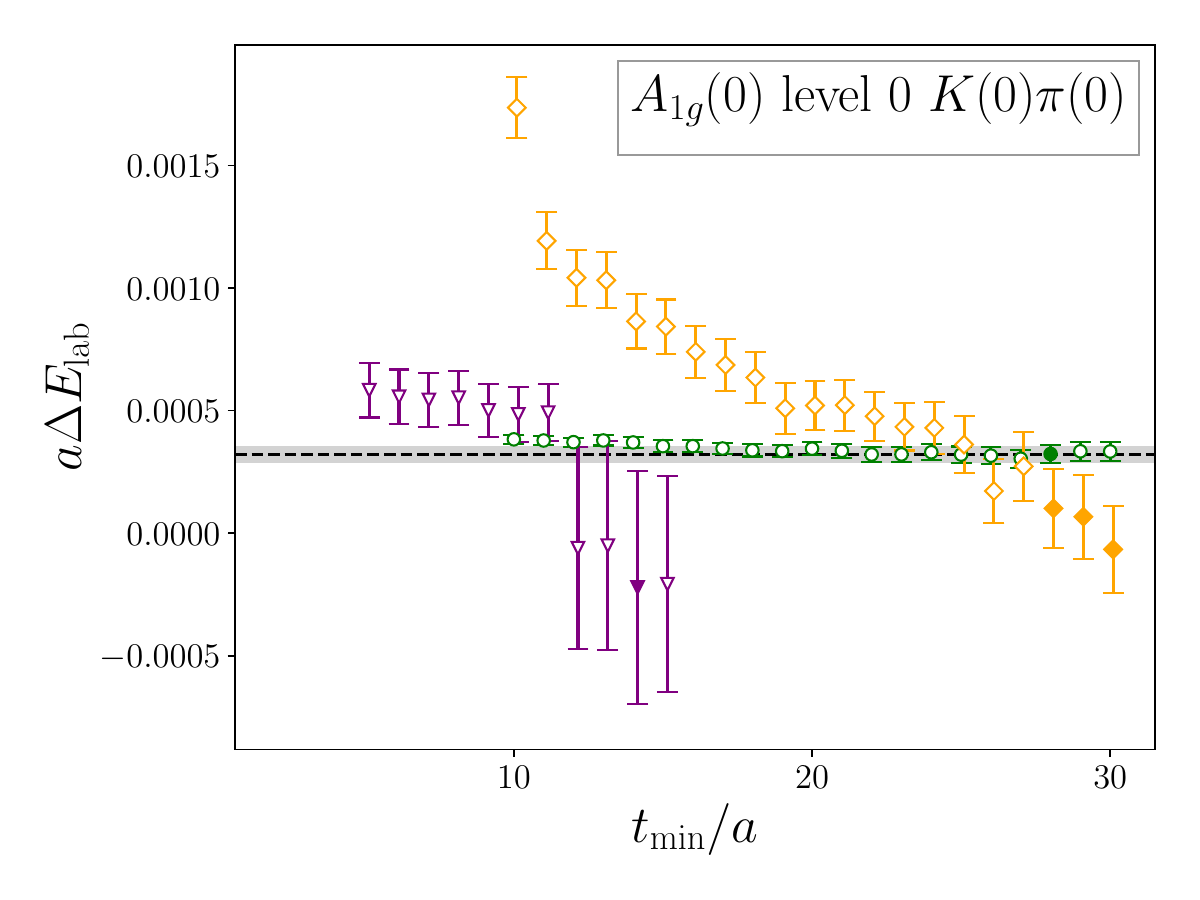}
    \includegraphics[width=0.32\textwidth]{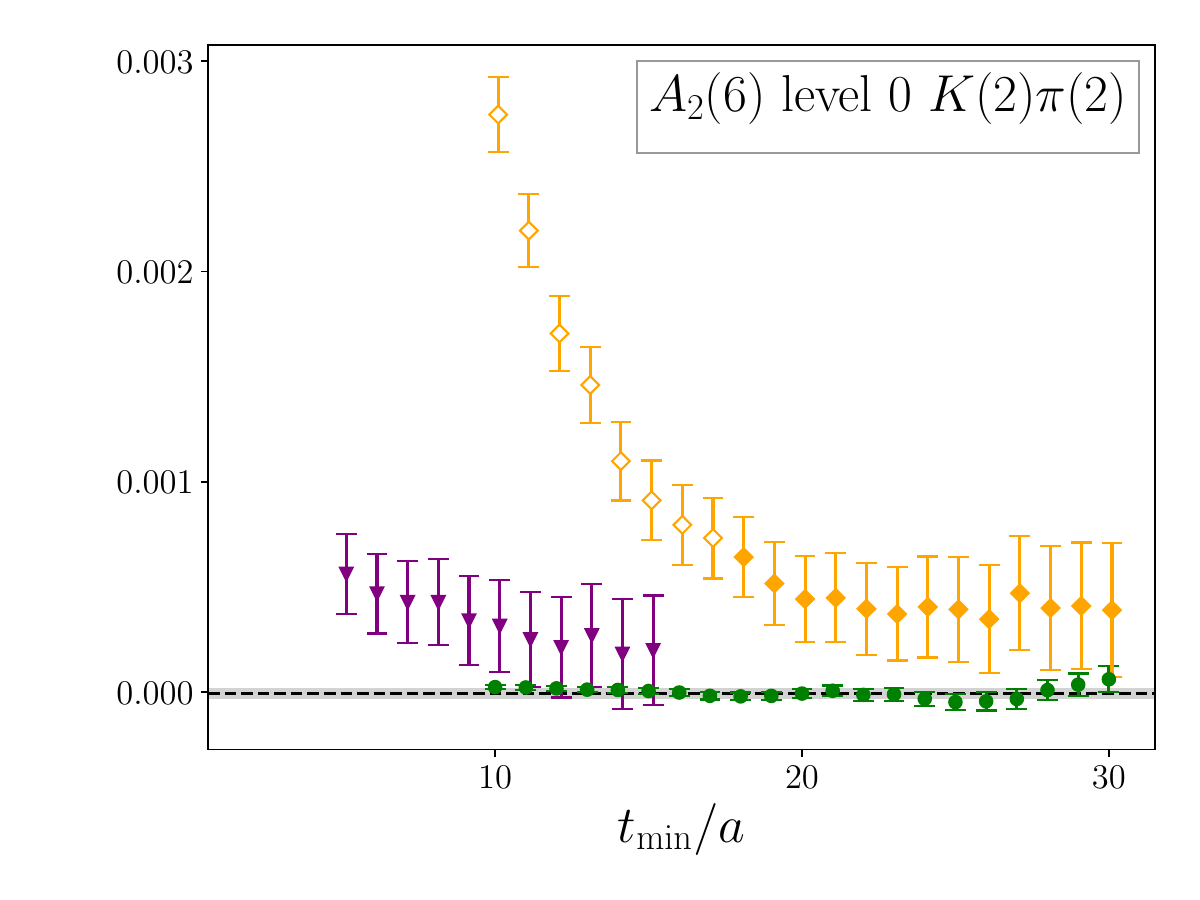}
    \includegraphics[width=0.32\textwidth]{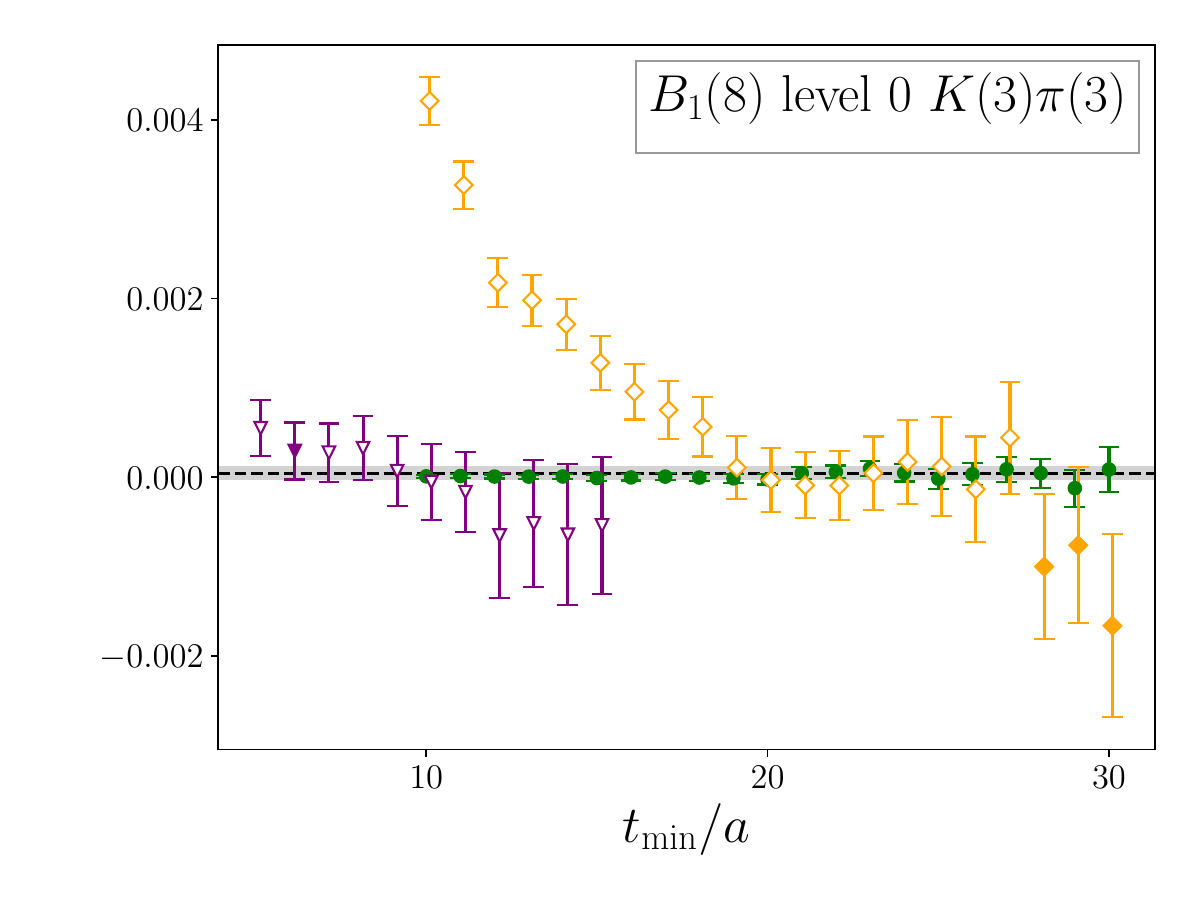}
    \caption{The dependence of a selected set of two-meson energies on the lower end of the fit range $t_{\rm min}$ and the fit model. We consider single-exponential fits to the ratio correlator (green), and single- (yellow) and two-exponential (purple) fits directly to the rotated correlator.
    The labels in each panel indicate the irrep, total momentum squared (in parentheses after the irrep label), the level number within the irrep (starting at $0$), and the chosen free level that enters the denominator of the ratio. Also included is the chosen energy extraction, which is indicated by the black dashed line and grey error band. Solid markers indicate a $p$-value $\ge 0.1$, whereas hollow markers indicate a $p$-value $< 0.1$.}
    \label{fig:twohadsE}
\end{figure*}
\begin{figure*}[th!]
    \centering
    \includegraphics[width=0.32\textwidth]{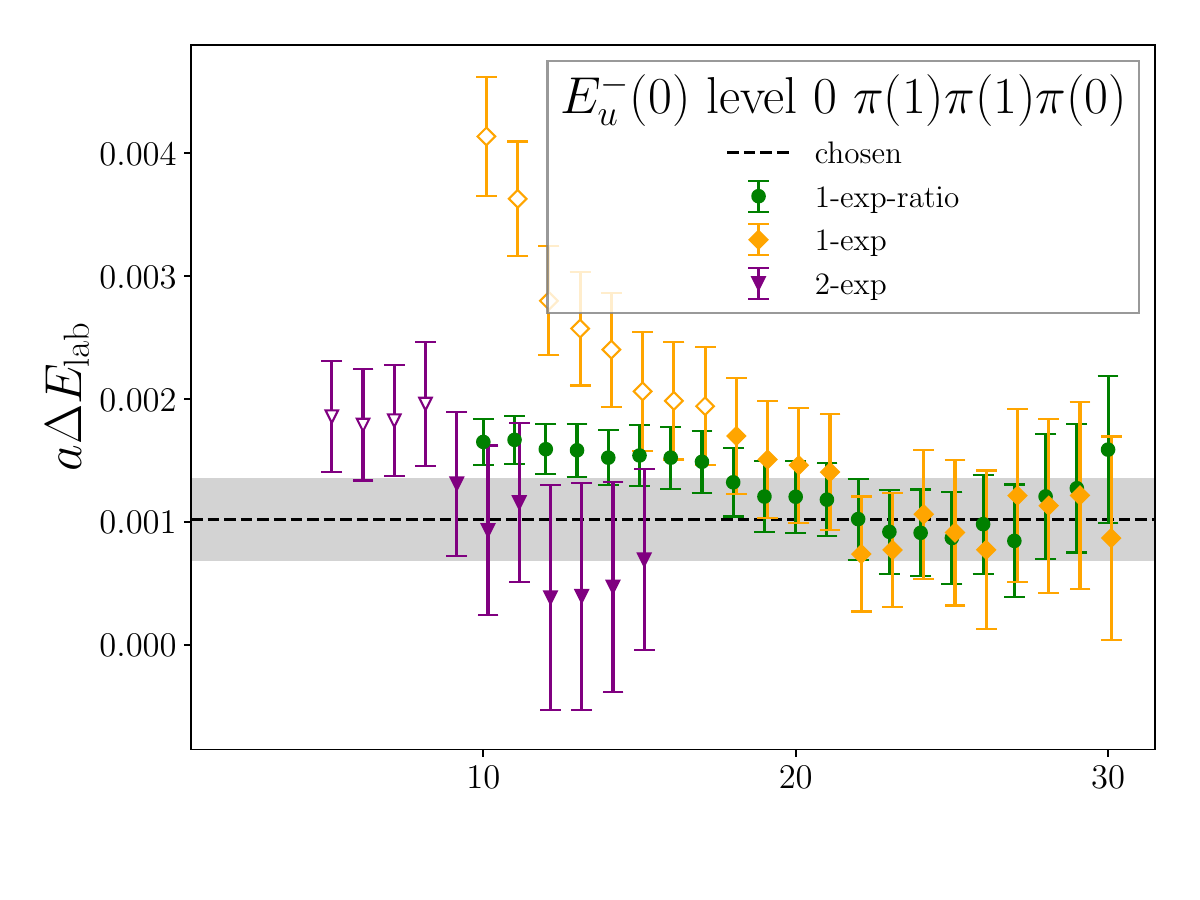}
    \includegraphics[width=0.32\textwidth]{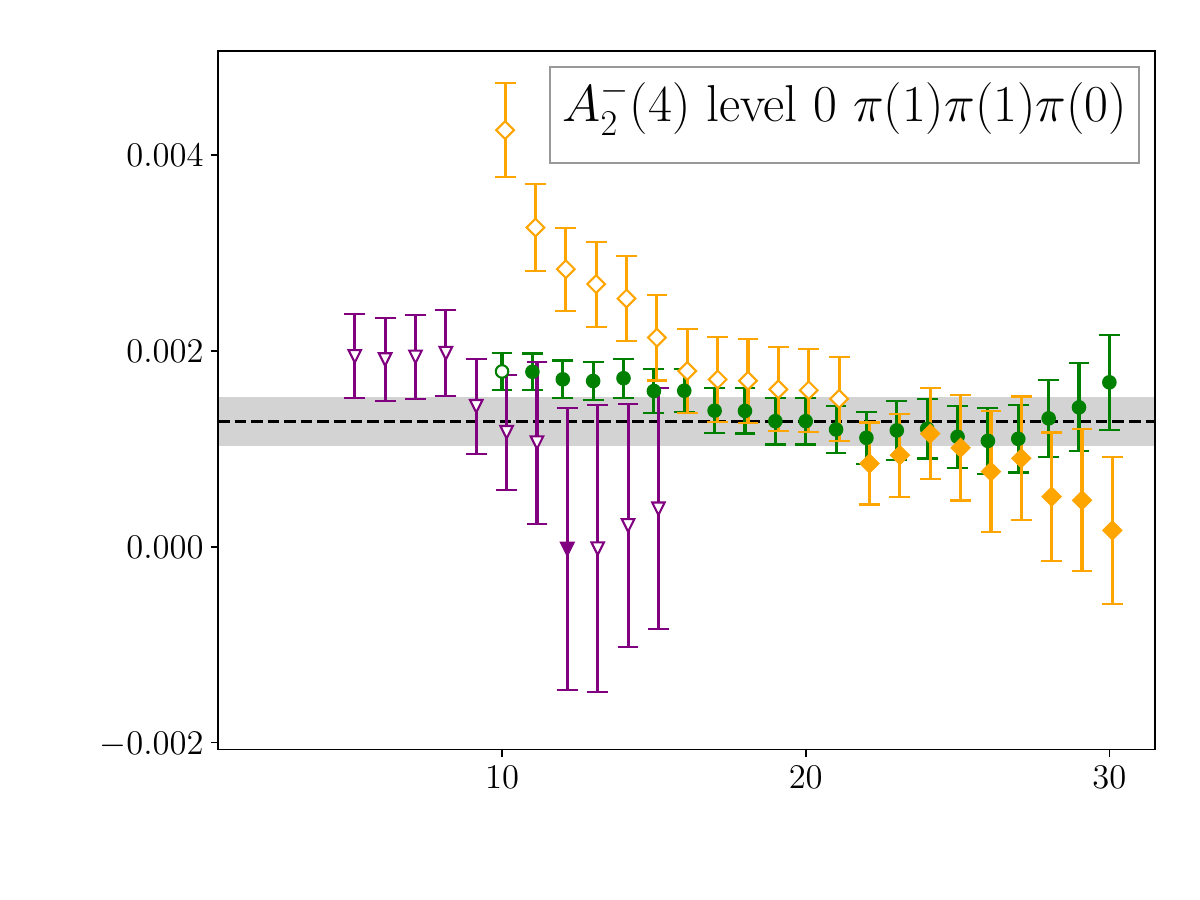}
    \includegraphics[width=0.32\textwidth]{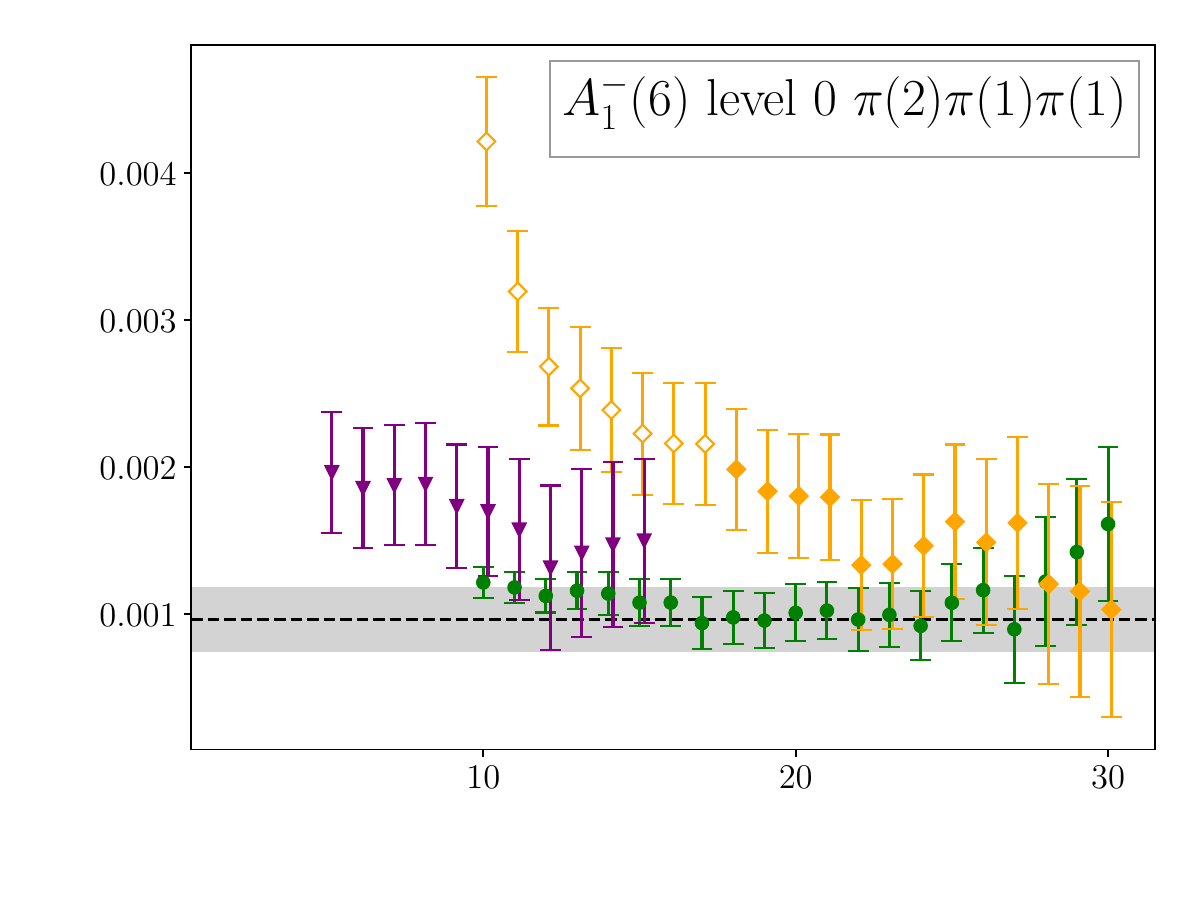}
    \includegraphics[width=0.32\textwidth]{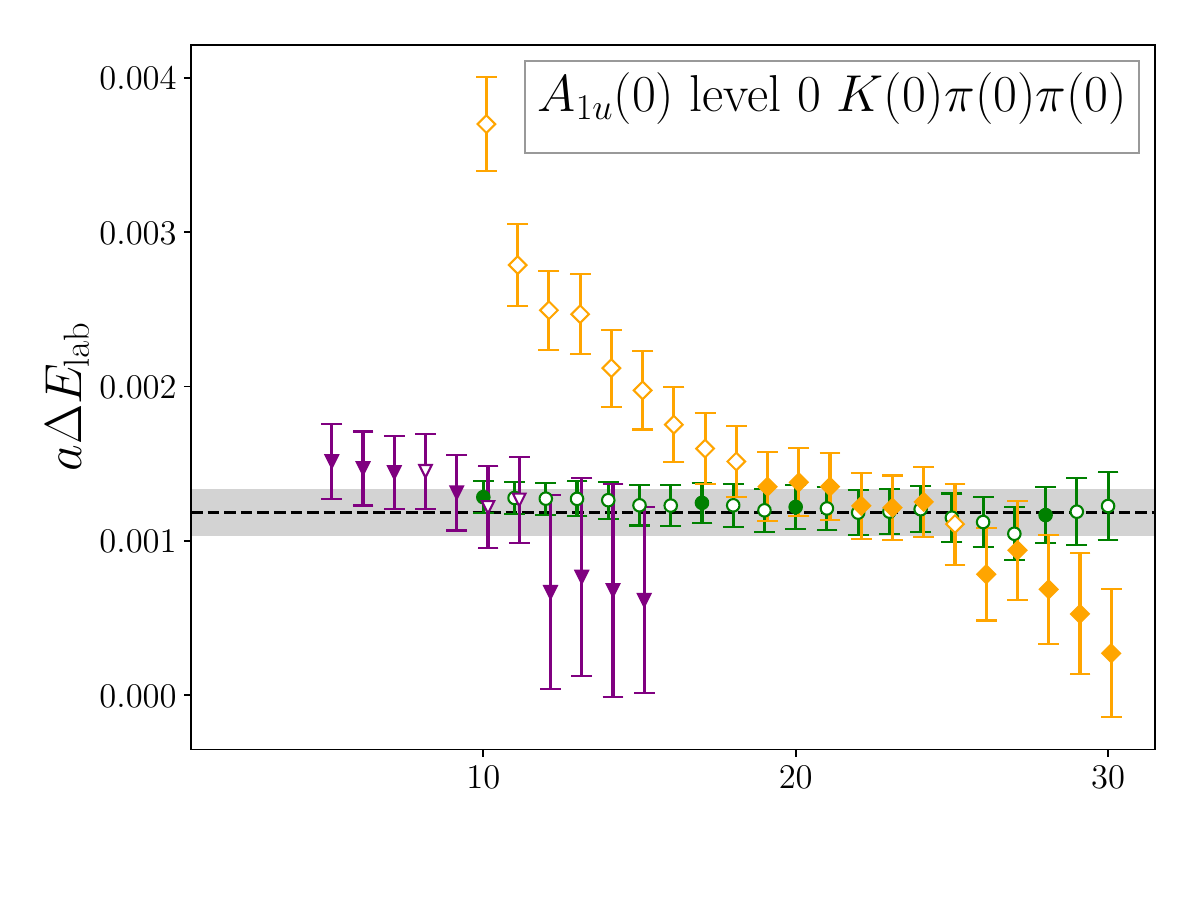}
    \includegraphics[width=0.32\textwidth]{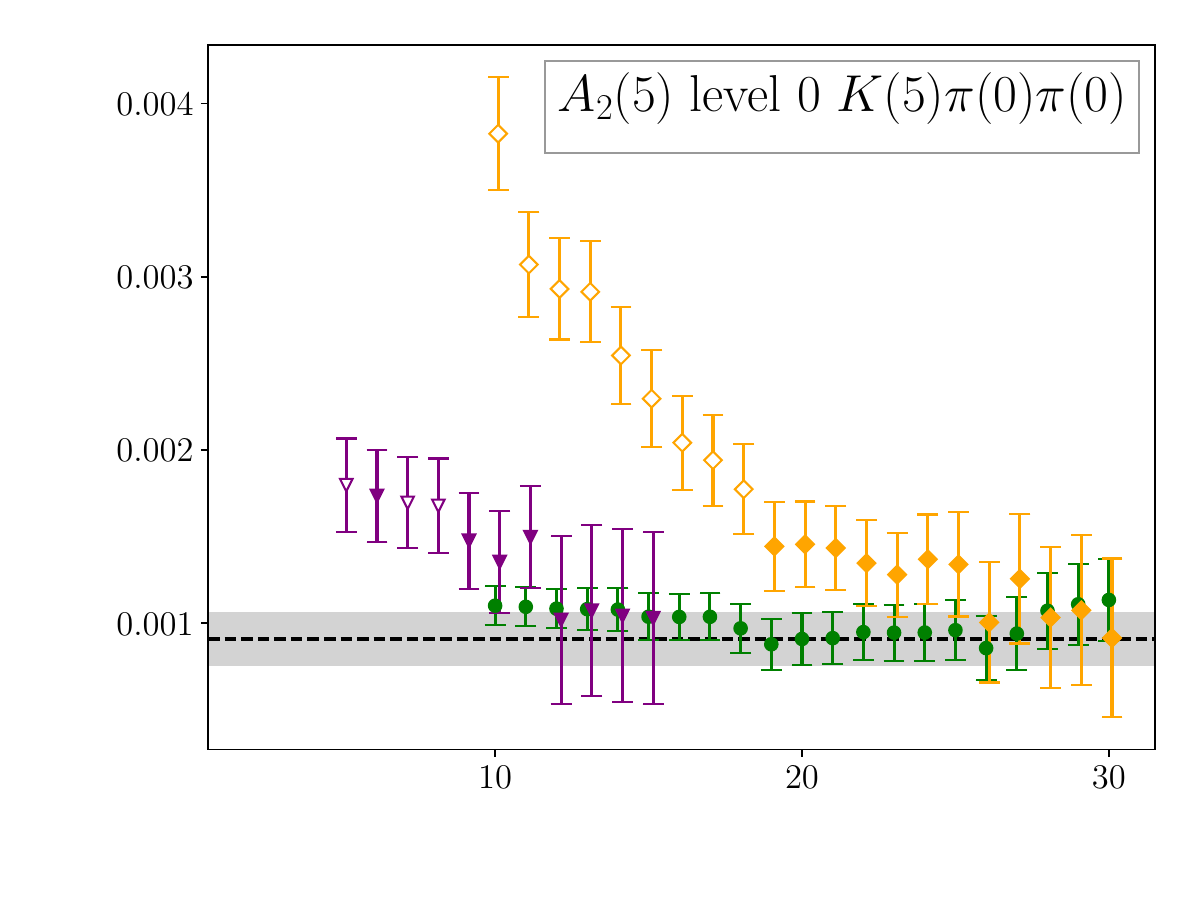}
    \includegraphics[width=0.32\textwidth]{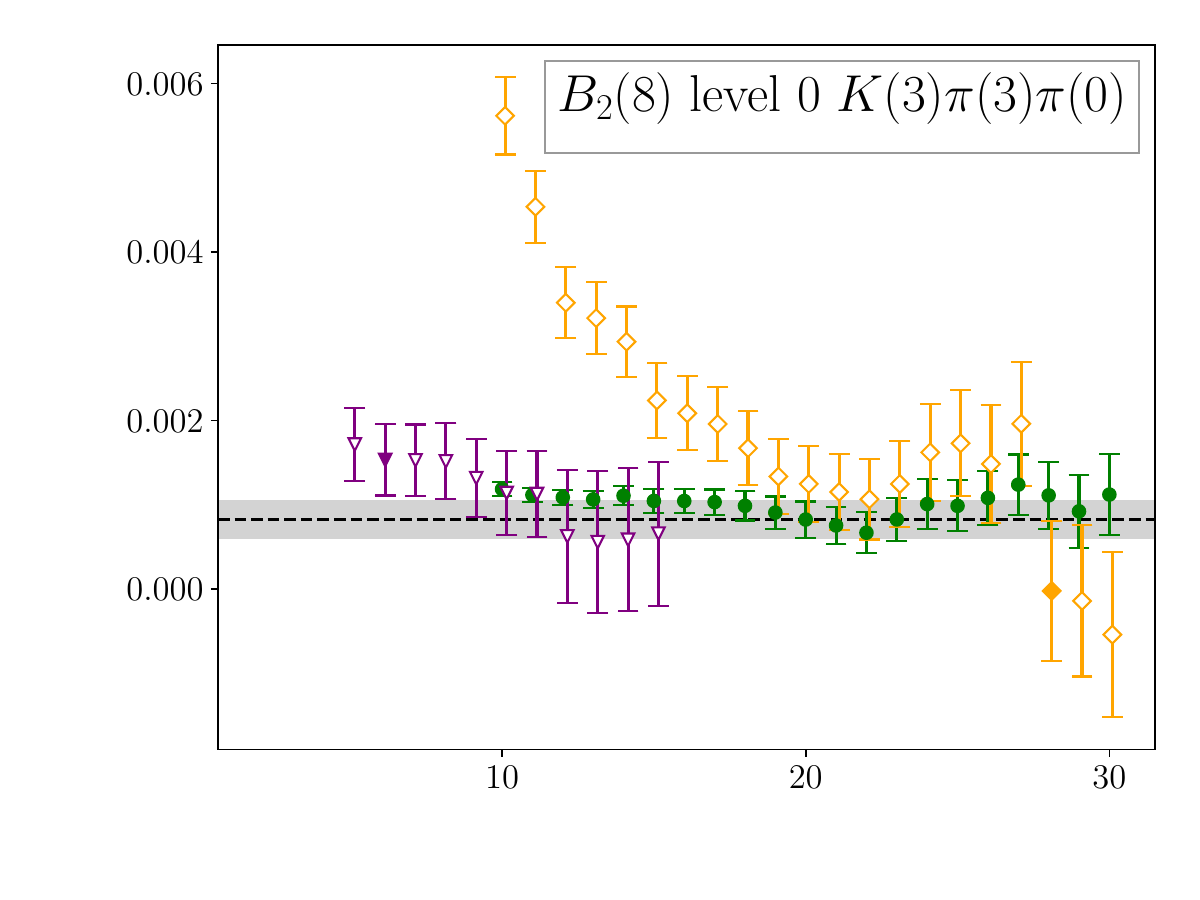}
    \includegraphics[width=0.32\textwidth]{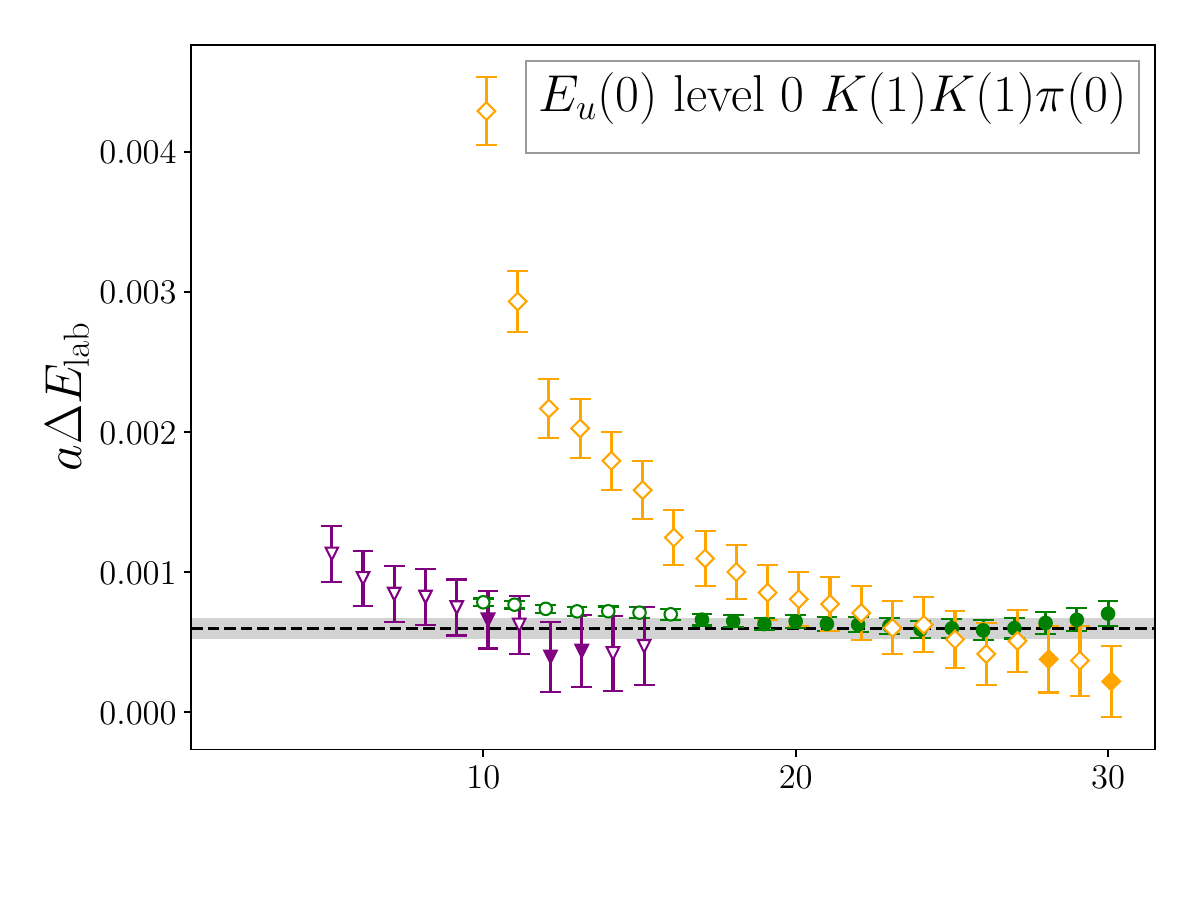}
    \includegraphics[width=0.32\textwidth]{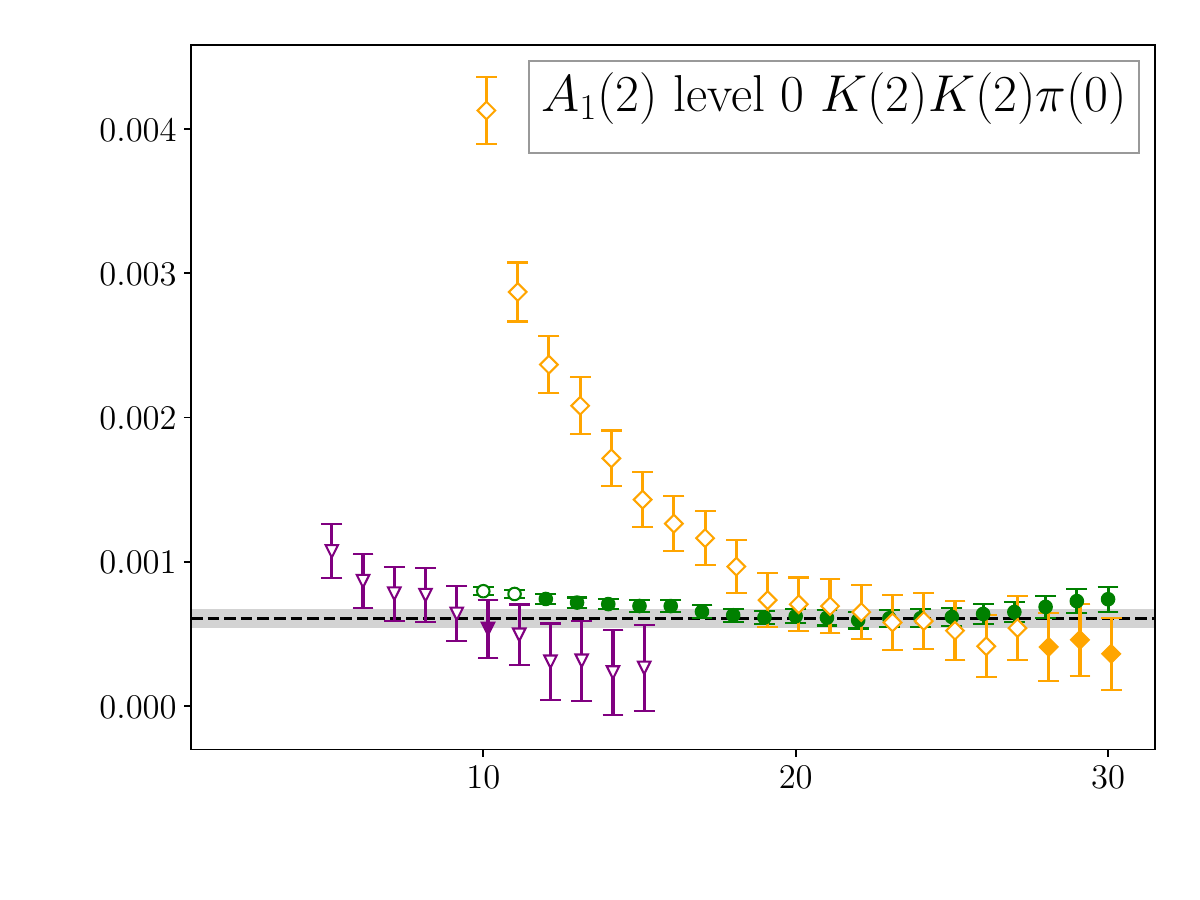}
    \includegraphics[width=0.32\textwidth]{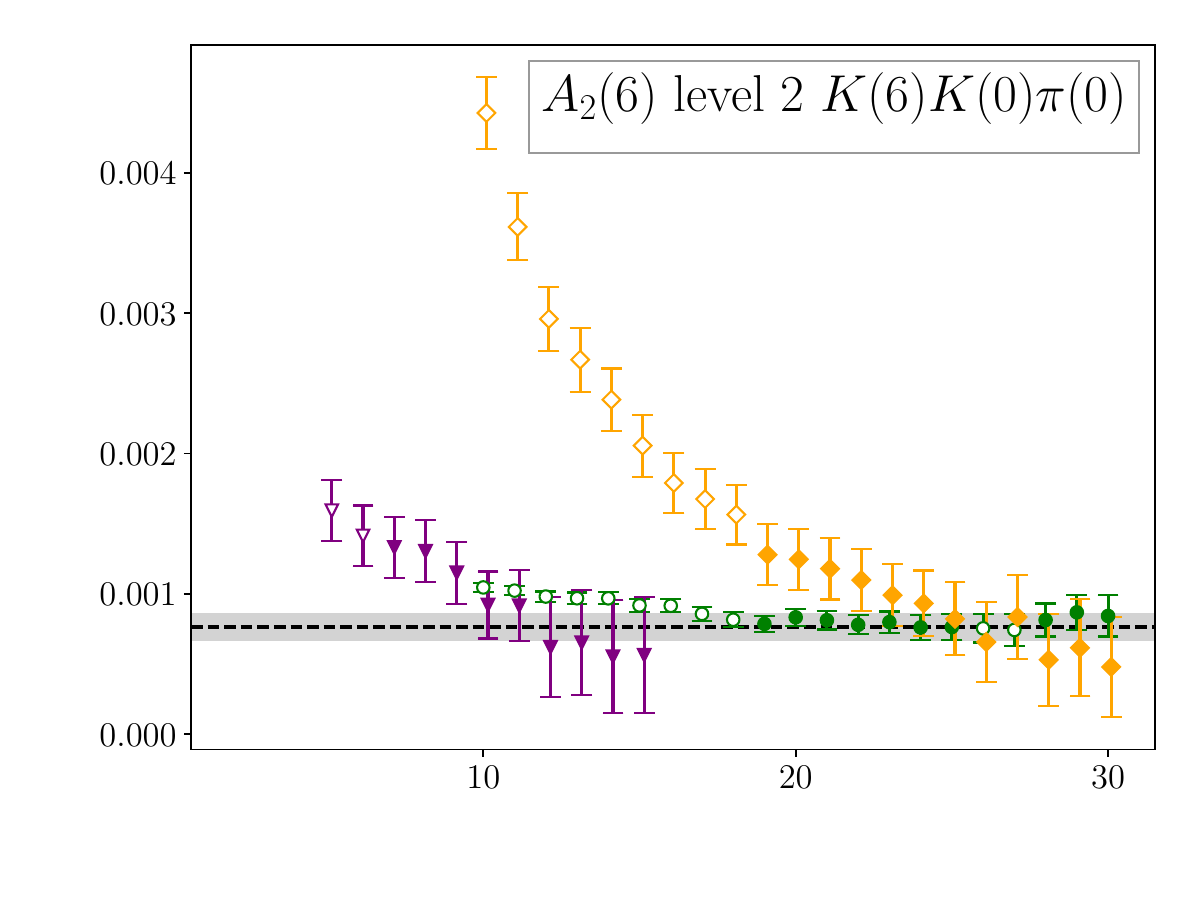}
     \includegraphics[width=0.32\textwidth]{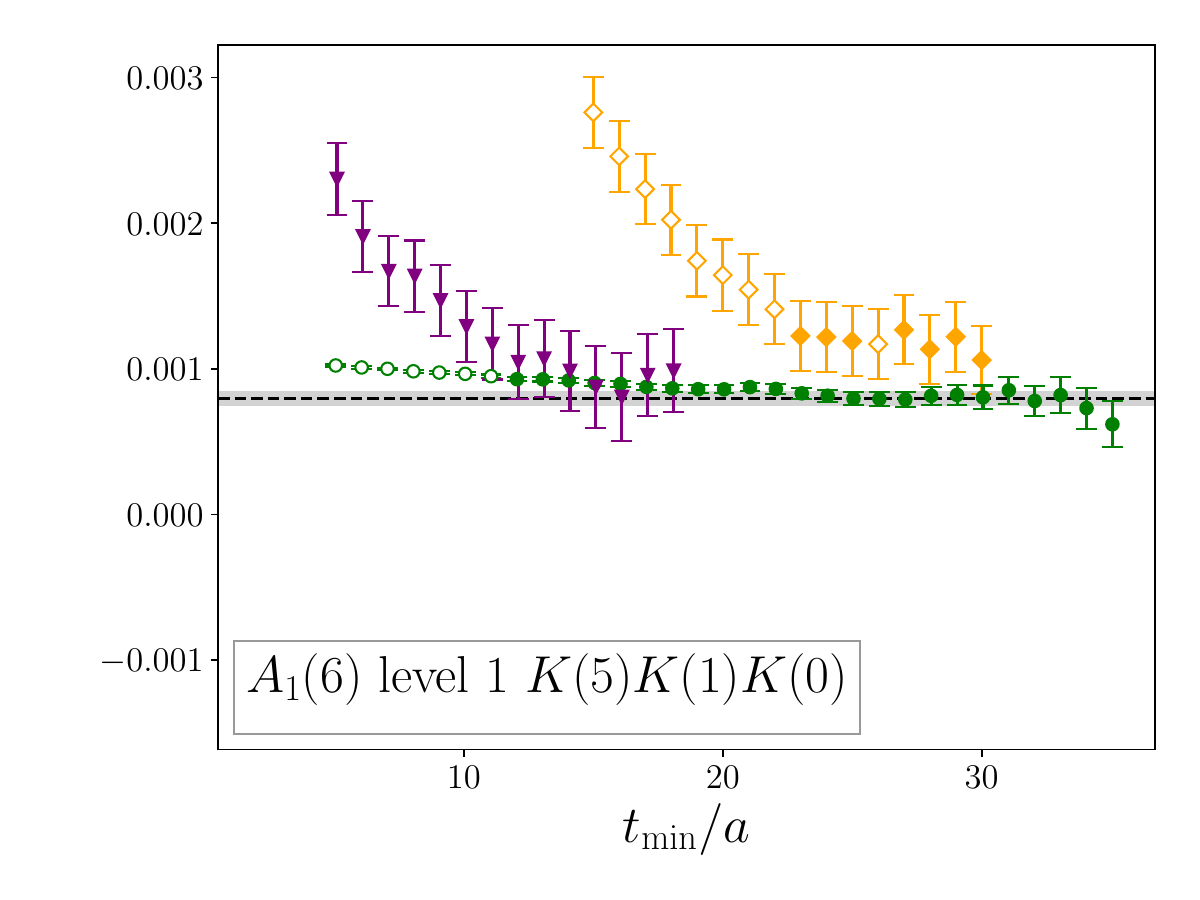}
    \includegraphics[width=0.32\textwidth]{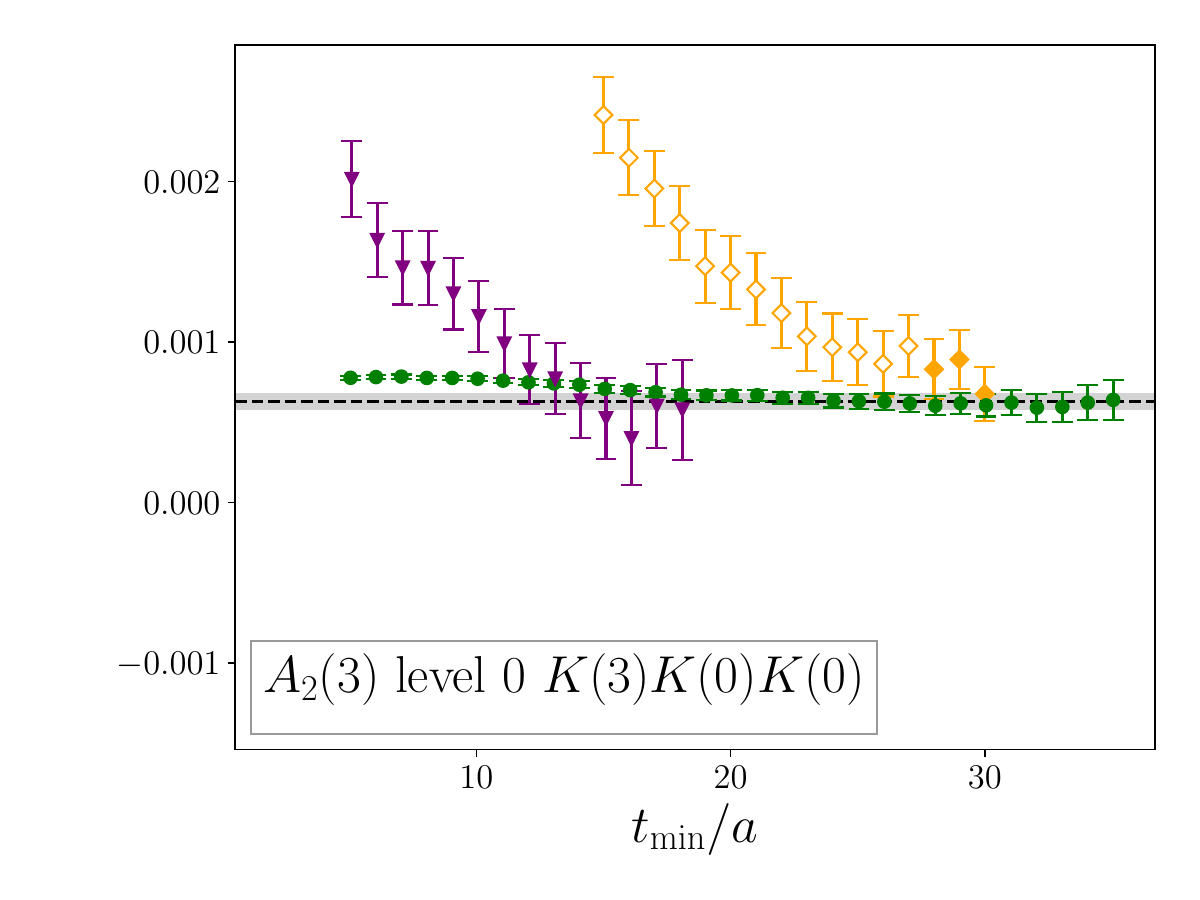}
    \includegraphics[width=0.32\textwidth]{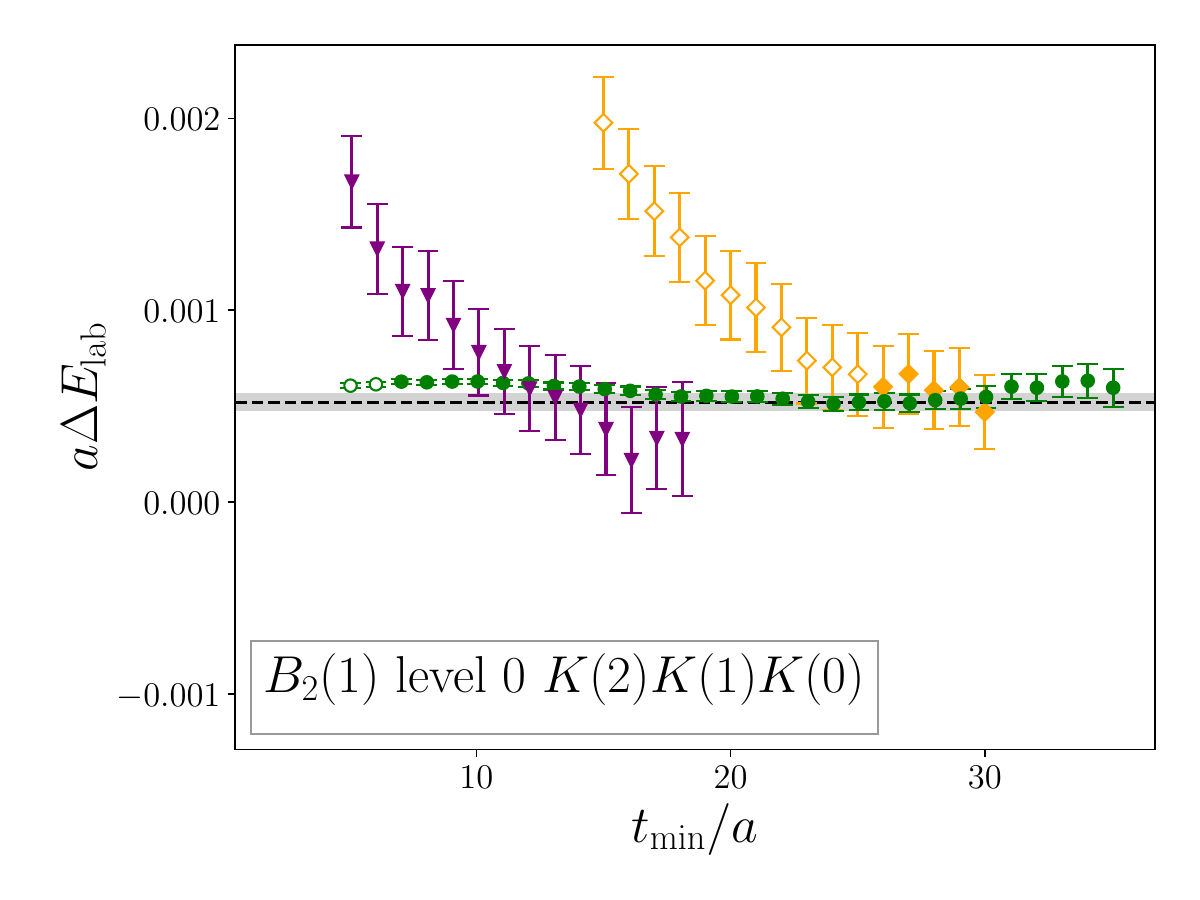}
    \vspace{-0.15cm}
    \caption{As in \Cref{fig:twohadsE}, but for three-meson energies}
    \label{fig:threehadsE}
\end{figure*}

\subsubsection{Multi-meson energies}

Multi-hadron scattering amplitudes can be determined from the multi-hadron finite-volume spectrum. Each energy level in the spectrum leads to a constraint on the multi-hadron interactions, and so larger sets are desirable. However, smaller gaps between states and larger statistical errors make this more challenging than in the single-hadron case. The standard procedure\footnote{
The Lanczos method~\cite{Wagman:2024rid,Hackett:2024xnx} is another promising method, but still needs careful investigation of the systematics involved.} to reliably constrain the low-lying energies is the variational method~\cite{Luscher:1990ck,Blossier:2009kd}, in which the generalized-eigenvalue problem (GEVP) is solved on a matrix of correlators built from sets of operators with the same quantum numbers (flavor, irrep, and total momentum). This method provides a procedure for extracting excited states without using multi-exponential fits. This is the same approach followed in our previous works~\cite{Blanton:2021llb,Draper:2023boj}.

Given a set of $N$ operators, these are used to construct an $N \times N$ correlation matrix $C_{ij}(t_{\rm sep})$.\footnote{We use $t$ in place of $t_{\rm sep}$ for convenience in what follows.}
The GEVP then consists of solving the following equation
\begin{equation}
  C(t) \upsilon_n (t, t_0) = \lambda_n(t, t_0) C(t_0) \upsilon_n (t, t_0) ,
\end{equation}
where $t_0$ is the metric time. For $t_0 \geq t/2$, the generalized eigenvalues behave asymptotically as
\begin{equation}
  \lambda_n(t, t_0) = |A_n|^2 e^{-E_n (t - t_0)} \big[1 + \mathcal{O}(e^{-\Delta_n t})\big] ,
\end{equation}
where $n \in \{0,\ldots,N-1$\},  $E_n$ is the $n$th eigenenergy, and $\Delta_n \equiv E_N - E_n$ is the distance to the first omitted state.

In order to avoid unwanted crossing of eigenvalues/eigenvectors between timeslices or resamplings, we first solve the GEVP on the mean and at a single time separation $t_d$, where $t_0 < t_d \lesssim 2 t_0$. We then use the eigenvectors to rotate the original correlator matrix on all resamplings and at all other times.
The rotated correlator is checked to make sure it remains diagonal at all times.
Then, the $n$th diagonal element of the rotated correlator, denoted $\hat{C}_n(t)$, retains the leading behavior of the $n$th eigenvalue above $\lambda_n(t,t_0)$.
We check for stability in the energies as $t_d$ and $t_0$ are varied.

Rotated correlators, $\hat C_n(t)$, are used to build ratios
\begin{equation}
  R_n(t) = \frac{\hat{C}_n (t)}{\prod_i C_{f_i}(\textbf{p}_i^2, t)} ,
  \label{eq:ratio_corr}
\end{equation}
where $C_{f_i}(\textbf{p}_i^2, t)$ is a single-hadron correlator with flavor $f_i$ averaged over all momenta that are equivalent under allowed rotations of a cube.
The denominator is composed of a product of single-hadron correlators that most closely resembles the $n$th energy eigenstate.
To this end, we calculate the overlaps $|\langle \Omega|\mathcal{O}_i | n \rangle|^2$.
Once we have determined which operator has the most overlap with the $n$th state, the single-meson correlators entering the denominator of the ratio are chosen to correspond to the single-meson operators entering the dominant operator.
As a small aside, since the different operators may not have a consistent normalization, only the ratios
\begin{equation}
    \frac{|\langle \Omega|\mathcal{O}_i | n \rangle|^2}{|\langle \Omega|\mathcal{O}_i | m \rangle|^2}
\end{equation}
are meaningful.
This means that one can only answer the question of which state dominates for a given operator, not which operator dominates for a given state.
This can lead to ambiguities if say two different operators both overlap more onto a particular state than any others; this does not happen often.
In the few cases it does happen, we check to make sure the results of the choice of the denominator in the ratio of \cref{eq:ratio_corr} is consistent with either operator being considered the dominant one.
However, we emphasize that any choice for the operators in the denominator will lead to the same spectrum so long as the asymptotic regime has been reached.

The asymptotic behavior of the ratio correlator is
\begin{equation}
\lim _{t \rightarrow \infty} R_n(t) \propto e^{-\Delta E_{\mathrm{lab}}^n t},
\label{eq:labshiftfit}
\end{equation}
and therefore they provide direct access to the lab-frame energy shift, $\Delta E_{\mathrm{lab}}^n$, which are the quantities that are fit using the quantization conditions.
The main advantage of using the ratio correlator, compared to fitting directly to the rotated correlator $\hat{C}_n(t)$, arises from a correlated cancellation of uncertainties and a partial cancellation of inelastic excited states in weakly interacting systems.
The main disadvantage is the non-monotonic behavior of the effective energy of the ratio correlator: terms contributing to $R_n(t)$ can enter with different signs, potentially leading to a slow approach to the ground state.
This can lead to a misidentification of the onset of saturation by a single state, and therefore the systematics are checked against non-ratio extractions of the energies. 

Once the lab-frame energy shifts have been obtained from the ratio correlators, the full energy can be reconstructed by adding the non-interacting energy back:
\begin{equation}
    E_{\rm lab} = \Delta E_{\rm lab} + \sum_{i} \sqrt{m_{f_i}^2 + \boldsymbol{p}_{f_i}^2} .
    \label{eq:energy_shift_conv}
\end{equation}
This provides a first-order correction of the discretization effects due to the dispersion relation~\cite{Hansen:2024cai}.
Finally, the center-of-mass (c.m.) energies $E^\ast$ can be obtained from
\begin{equation}
    E^\ast = \sqrt{E_{\rm lab}^2 - \textbf{P}^2} .
\end{equation}

In \Cref{fig:twohadsE,fig:threehadsE}, we show several examples of the $t_{\rm min}$ dependence of two- and three-meson lab-frame energy shifts, respectively. We include comparisons to the results from single- and two-exponential fits directly to the rotated correlator $\hat{C}_n (t)$. Note that fits directly to the rotated correlator give us $E_{\rm lab}$, which are then converted to $\Delta E_{\rm lab}$ using \cref{eq:energy_shift_conv} in order to make direct comparisons. The choice of $t_{\rm min}$ is made such that it is at least a few time separations larger than the onset of the plateau in the effective energy, while not being so large such that correlated fluctuations (indicating the start of the loss of the signal) have become significant. We further demand consistency with the other fit models. The dependence of the results on $t_{\rm max}$ is very small, and in the vast majority of cases we choose the largest time separation computed, i.e. $t_{\rm max}/a = 40$. In a few cases, the last few time slices have large errors and we make $t_{\rm max}$ a little smaller.

\subsection{Overview of the spectrum}
\label{sec:spectrum_overview}

\begin{figure*}[th!]
    \includegraphics[width=\linewidth]{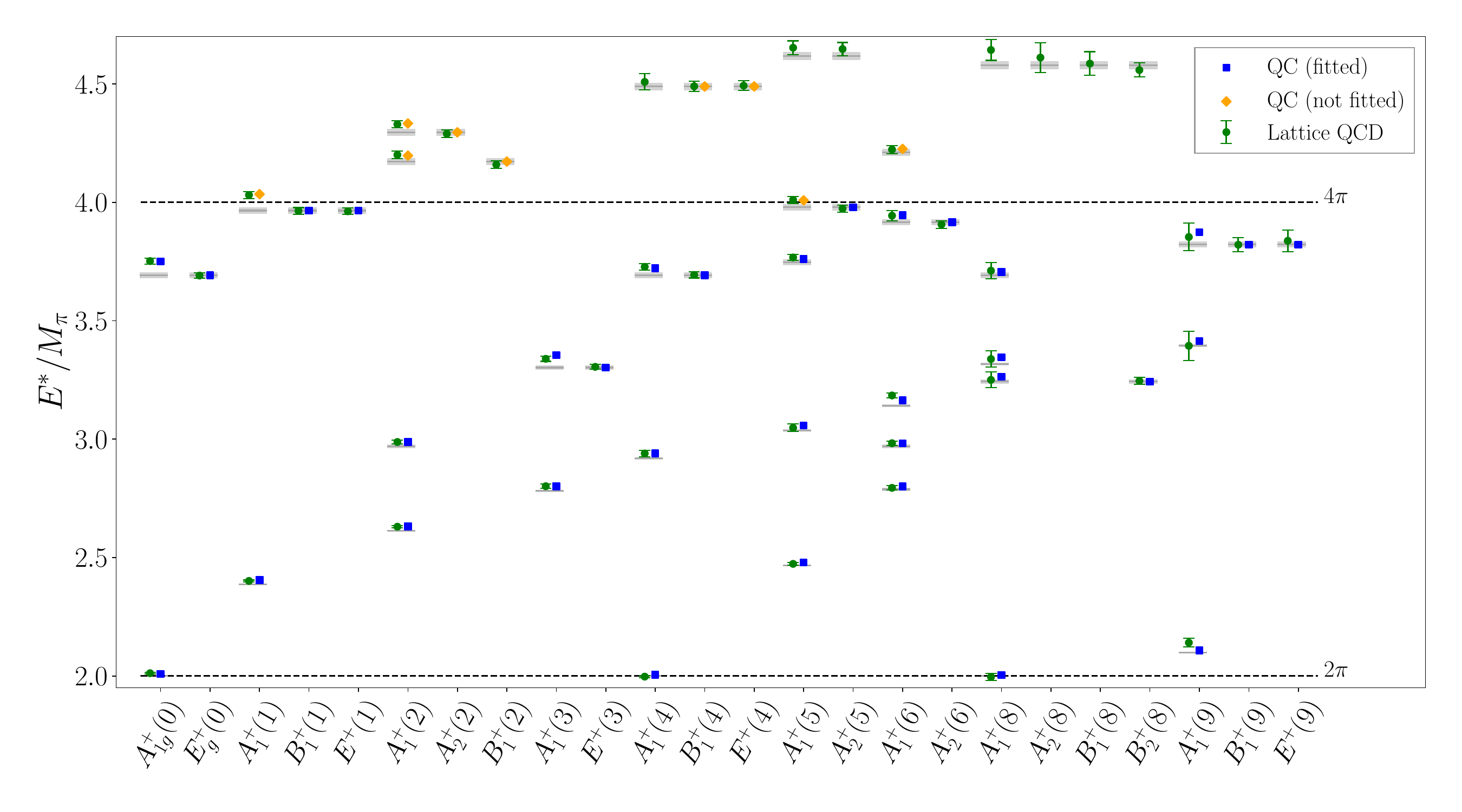}
    \vspace{-0.3cm}
    \caption{Energy spectrum of the $\pi\pi$ system for the E250 ensemble. LQCD c.m.~frame energies are shown as green circles with errors, with the grey bars corresponding to the noninteracting energies (see text). In some cases, these lie right under the lattice energies.
    Labels for the $x$-axis indicate the finite-volume irreps and (in parentheses) the momentum frame squared. Relevant thresholds are 
    shown with dashed, horizontal black lines. 
    Predicted energies (blue squares or orange diamonds) are obtained from the simultaneous fit to $2\pi$ and $3\pi$ levels (using the QC2 and QC3, respectively) given in the last column of \Cref{tab:ppp-params-E250}. Only levels below the $4\pi$ threshold are included in the fits. Orange diamonds are predictions for levels not included in the fit.}
    \vspace{-0.2cm}
    \label{fig:pipioverview}
\end{figure*}

\begin{figure*}[h!]
    \includegraphics[width=\linewidth]{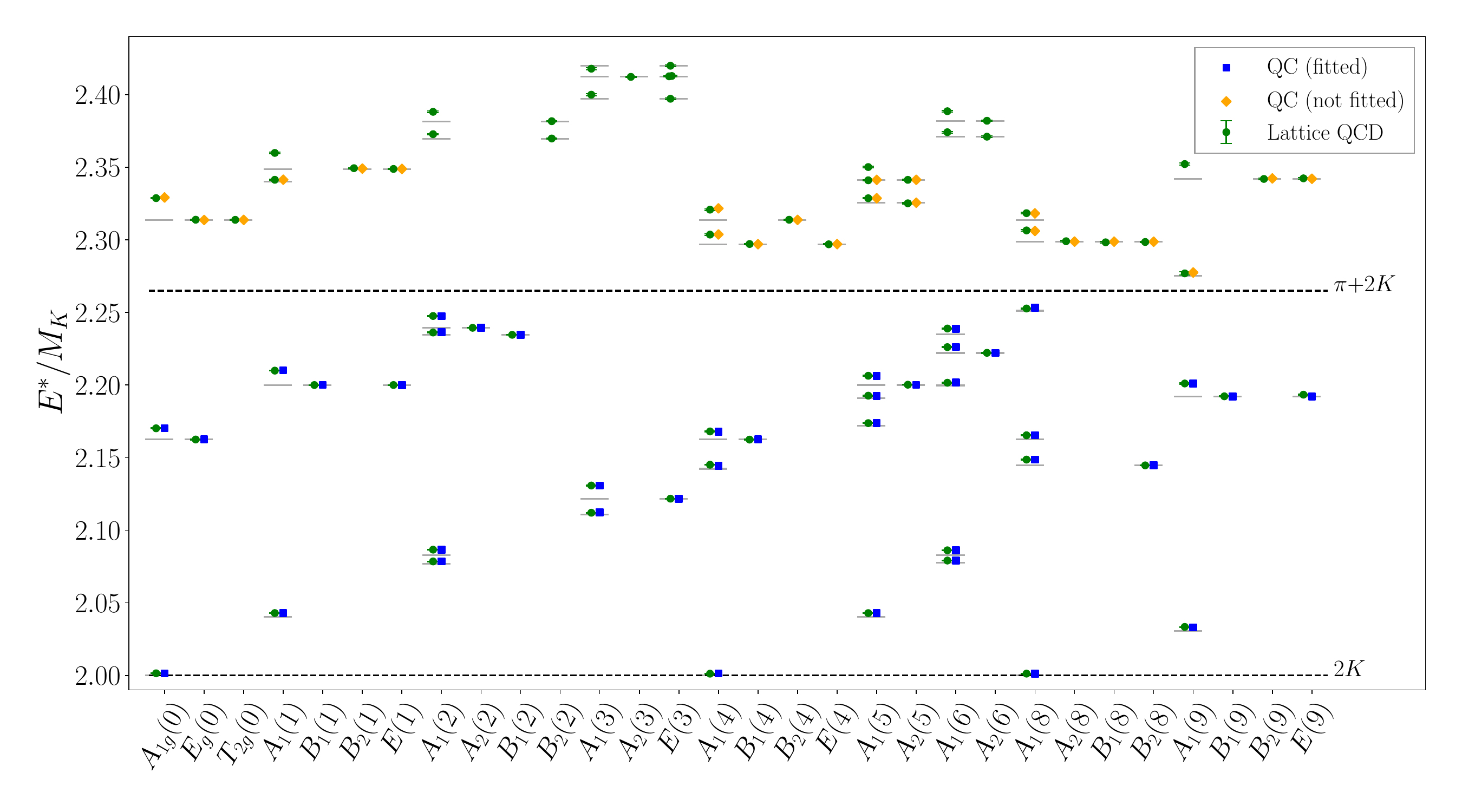}
    \caption{Energy spectrum of the $KK$ system for the E250 ensemble. All other notation as \Cref{fig:pipioverview}.
    Predicted energies are obtained from the fit in the last column of  \Cref{tab:kkk-params-E250}.}
    \label{fig:kkoverview}
\end{figure*}

\begin{figure*}[h!]
    \includegraphics[width=\linewidth]{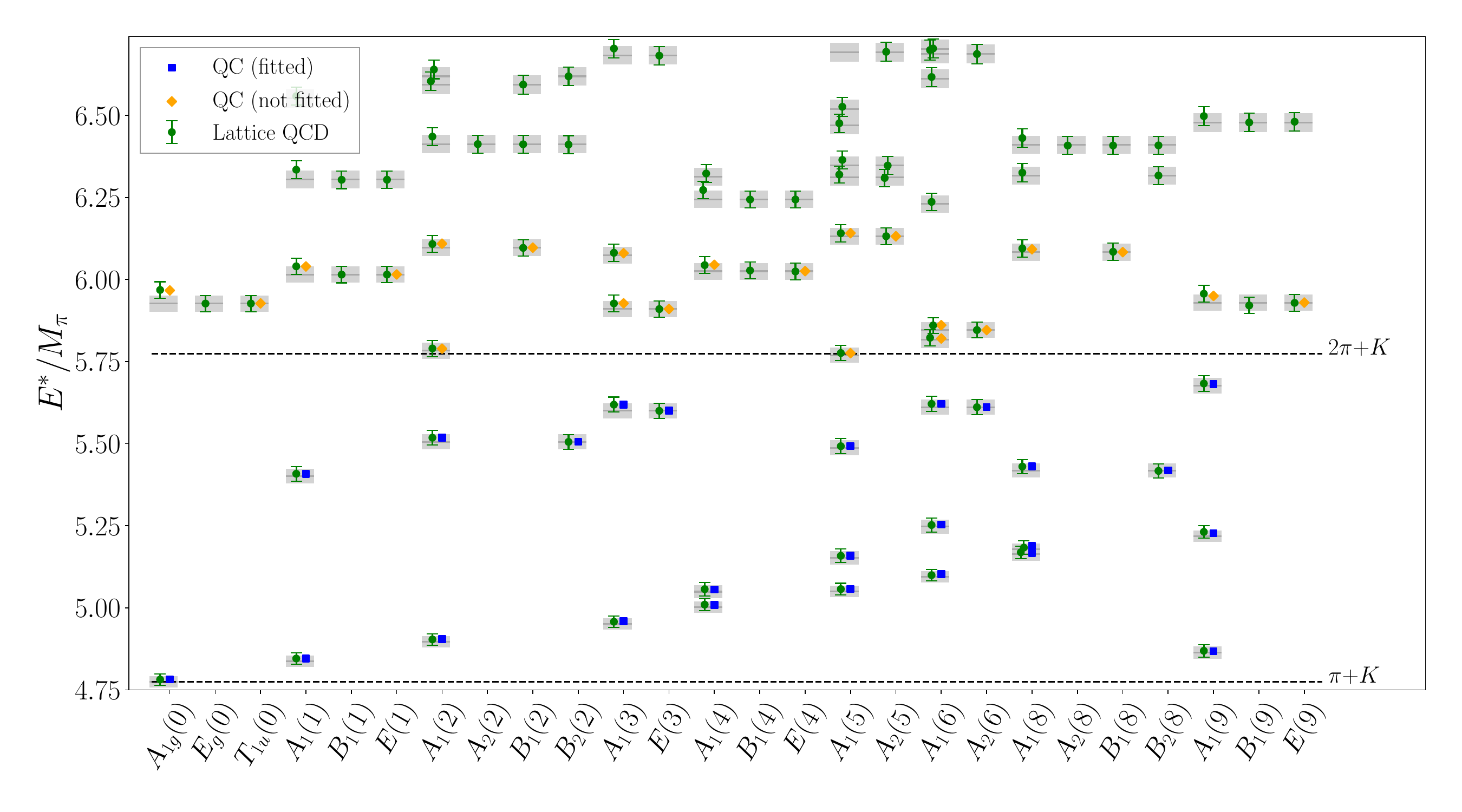}
    \caption{Energy spectrum of the $\pi K$ system for the E250 ensemble. All other notation as \Cref{fig:pipioverview}.
    Predicted energies are obtained from the $s$ and $p$ wave fit from \Cref{tab:ppK-params-E250}.}
    \label{fig:pkoverview}
\end{figure*}

\begin{figure*}[h!]
    \includegraphics[width=\linewidth]{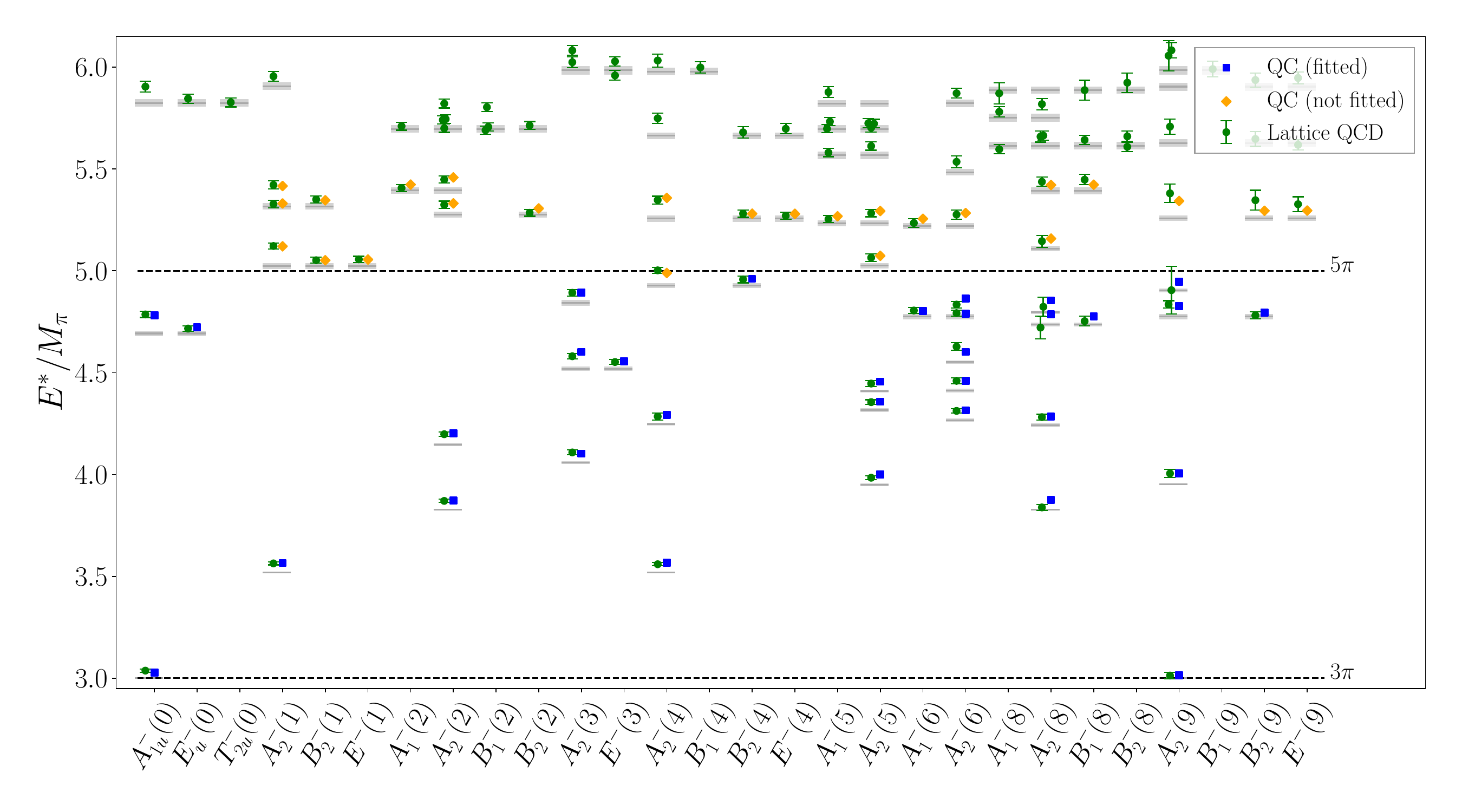}
    \caption{Energy spectrum of the $3\pi$ system for the E250 ensemble. All other notation as \Cref{fig:pipioverview}.
   Predicted energies are obtained from the fit in the last column of  \Cref{tab:ppp-params-E250}.
   We stress that this fit is to not only these levels, but also those in \Cref{fig:pipioverview}.}
    \label{fig:pipipioverview}
\end{figure*}

\begin{figure*}[h!]
    \includegraphics[width=\linewidth]{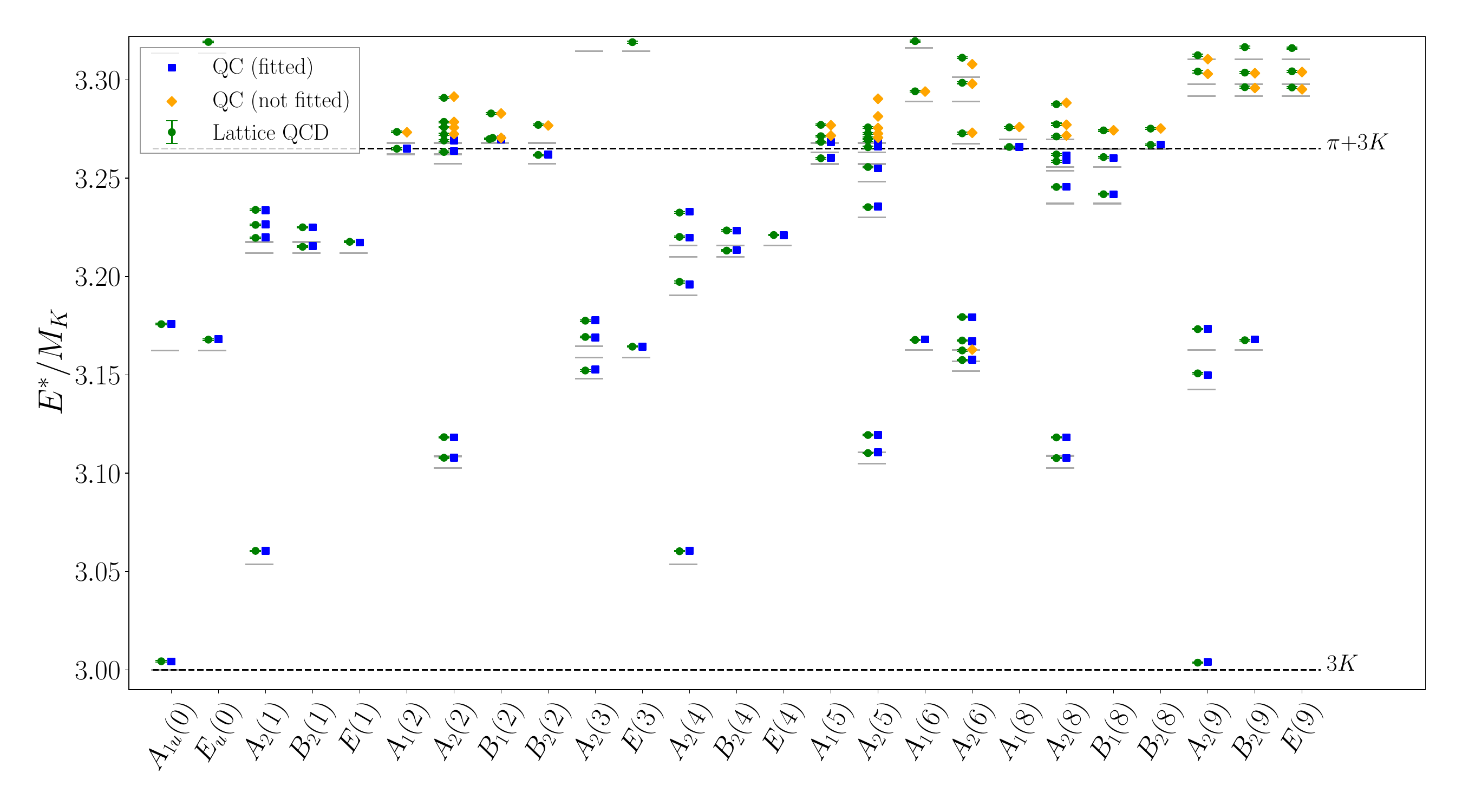}
    \caption{Energy spectrum of the $3K$ system for the E250 ensemble. All other notation as \Cref{fig:pipioverview}.
    Predicted energies are obtained from the fit in the last column of  \Cref{tab:kkk-params-E250}.
    We stress that this fit is to not only these levels, but also those in \Cref{fig:kkoverview}.}
    \label{fig:kkkoverview}
\end{figure*}

\begin{figure*}[h!]
    \includegraphics[width=\linewidth]{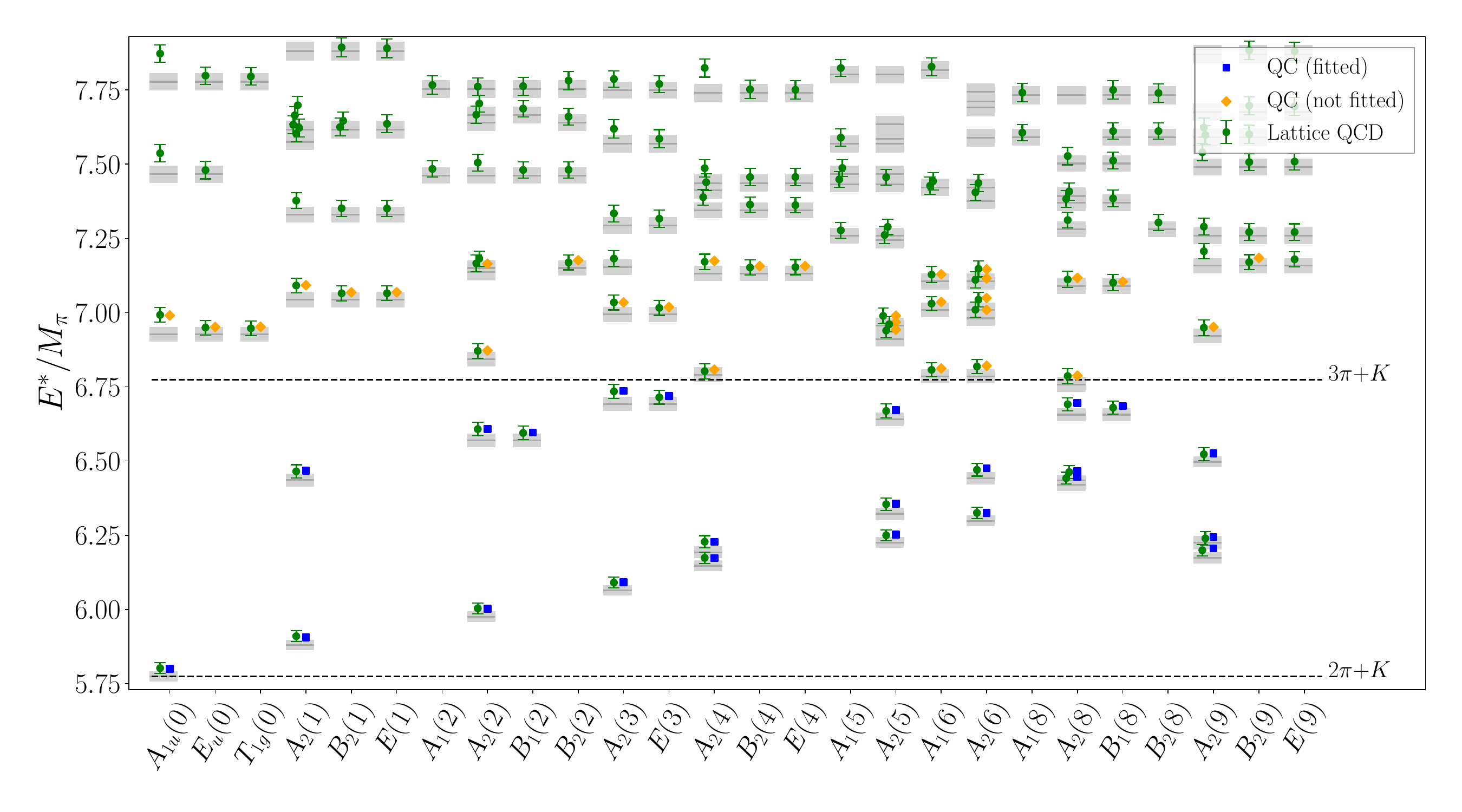}
    \caption{Energy spectrum of the $\pi \pi K$ system for the E250 ensemble. All other notation as \Cref{fig:pipioverview}.
    Predicted energies are obtained from the $s$ and $p$ wave fit from \Cref{tab:ppK-params-E250}.
     We stress that this fit is to not only these levels, but also those in \Cref{fig:pipioverview,fig:pkoverview}.}
    \label{fig:ppkoverview}
\end{figure*}

\begin{figure*}[h!]
    \includegraphics[width=\linewidth]{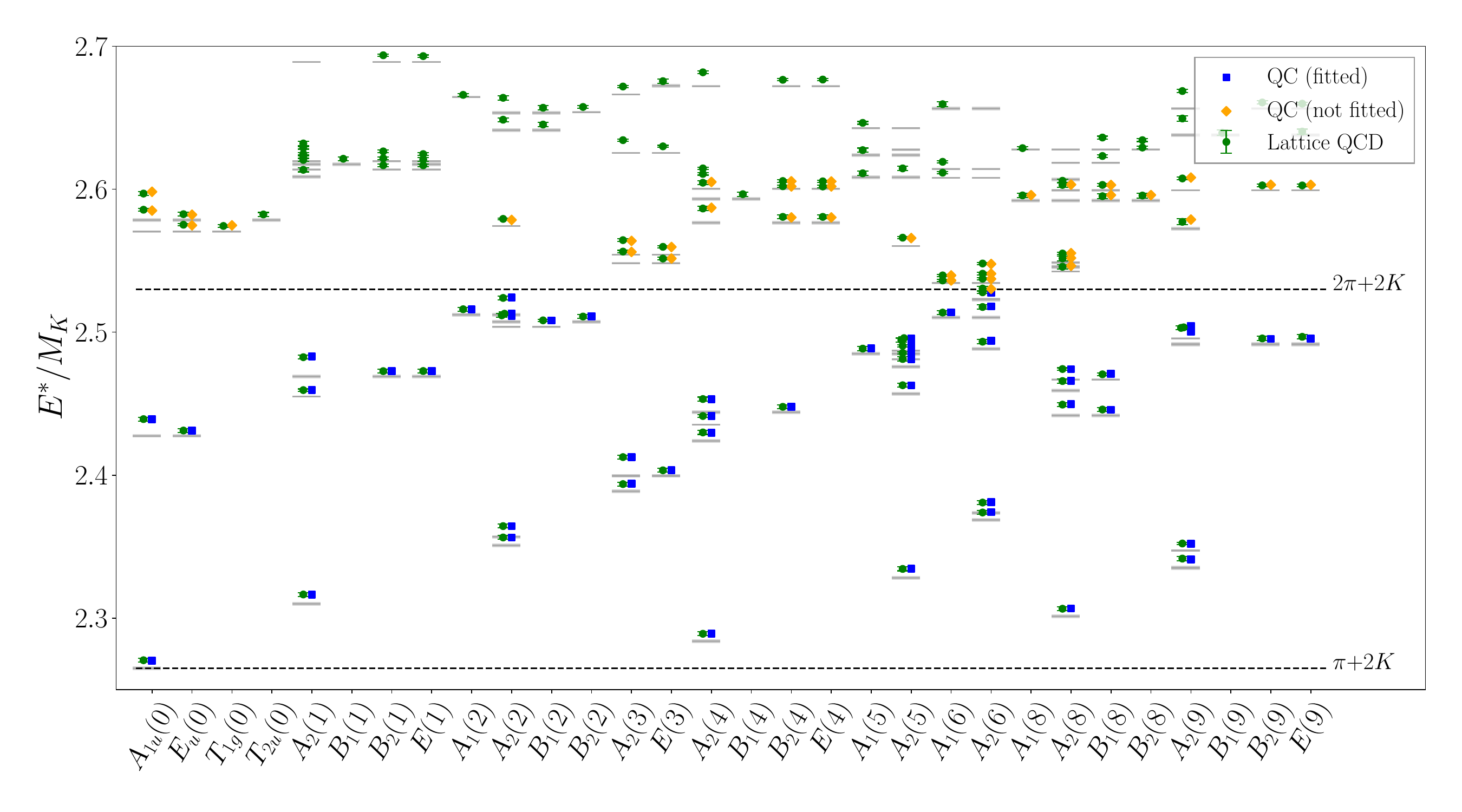}
    \caption{Energy spectrum of the $KK\pi$ system for the E250 ensemble. All other notation as \Cref{fig:pipioverview}.
    Predicted energies are obtained from the $s$ and $p$ wave fit from \Cref{tab:KKp-params-E250}.
     We stress that this fit is to not only these levels, but also those in \Cref{fig:kkoverview,fig:pkoverview}.}
    \label{fig:kkpoverview}
\end{figure*}

Here we provide an overview of the energies used in this work for all two- and three-meson systems. The spectra are shown in \Cref{fig:pipioverview,fig:pipipioverview,fig:pkoverview,fig:kkoverview,fig:kkkoverview,fig:ppkoverview,fig:kkpoverview}. All figures follow the same structure: LQCD energies in the c.m. frame are shown as green circles with their uncertainties, and the associated non-interacting energies are shown as gray bands. All errors are estimated using jackknife resampling. All relevant thresholds are shown as dashed horizontal black lines. We also show predictions from the finite-volume formalism that will be explained below for selected fits: blue squares represent those included in the fit, while orange diamonds are those not included in the fit. As can be seen, overall good consistency between measured and predicted energies is observed, even beyond the inelastic thresholds. We stress, as discussed further below, that the fits include the full correlated error matrix, and not just the diagonal errors shown in the figures.


\section{Formalism}
\label{sec:formalism}

This section describes the pipeline used to obtain scattering amplitudes from the finite-volume spectrum.

\subsection{Finite-volume quantization conditions}
\label{sec:QCs}

The finite-volume multi-hadron spectrum contains information about scattering amplitudes. In particular, the finite-volume energies deviate from those of a non-interacting theory due to power-law volume-dependent energy shifts,\footnote{%
The finite-volume formalism neglects $\exp(-M_\pi L)$ effects.}
and these shifts can constrain scattering amplitudes through the finite-volume formalism in the form of quantization conditions. This formalism has been developed for two-hadron~\cite{Luscher:1986pf,Kim:2005gf,Briceno:2014oea}, and three-hadron systems~\cite{Hansen:2014eka,Hansen:2015zga,Hammer:2017uqm,Hammer:2017kms,Mai:2017bge}. While they share some parallels, the three-hadron case is significantly more complex. Here, we summarize how the quantization conditions can be used to access the two- and three-hadron K matrices that parametrize the two- and three-hadron interactions. 

First, we describe the two-particle quantization condition (QC2). The inputs to the quantization condition are the momentum of the two-hadron system, $\bm{P}$, the size of the box, $L$, a kinematic function labelled $F$ (see Ref.~\cite{Kim:2005gf}), which contains finite-volume effects, and the two-hadron K matrix, $\cK_2$. The QC2 takes the form
\begin{equation}
\det_{\ell m} \left[ F\left( E_2, \bm{P}, L \right)^{-1} + \cK_2(E_2^*) \right] = 0\,,
\label{eq:QC2}
\end{equation}
whose solutions determine the two-particle energies, $E_2$. 
The corresponding c.m.~frame energies are given as $E_2^* = \sqrt{E_2^2 - \bm{P}^2}$. We assume a cubic spatial box of volume $L^3$, and so the allowed total momenta are restricted to be $\bm{P} = (2 \pi / L) \bm{d}$, where $\bm{d} \in \mathbb{Z}^3$. The matrices in \Cref{eq:QC2} have angular momentum indices. To reach a finite-dimensional matrix, all waves with $\ell >\ell_{\rm max}$ are neglected, which is justified by the suppression  of higher partial waves near threshold.
In this way, each individual energy level provides a constraint on the two-body K matrices for all partial waves up to and including $\ell=\ell_{\rm max}$.

The three-particle quantization condition (QC3) determines the finite-volume three-hadron spectrum:
\begin{equation}
\det_{i \boldsymbol{k} \ell m } \left[ {F}_3\left( E, \bm{P}, L \right)^{-1} + {\cK}_{\text{df},3}(E^*) \right] = 0\,,
\label{eq:QC3}
\end{equation}
where $\kdf$ is the three-particle K matrix, $E$ is the three-particle lab-frame energy, and
${E^* = \sqrt{E^2 - \bm{P}^2}}$ is the corresponding c.m. frame energy. Despite its similarity to the QC2, the QC3 is significantly more complex.
In particular, we note that whereas $F$ in the two-particle quantization condition is a purely kinematic function, ${F}_3$ depends on $F$, on an additional kinematic function $G$, 
and on the two-particle K matrix, $\cK_2$. 
In this way, finite-volume energies of three-hadron systems provide constraints on both the two- and three-body K matrices. Another key difference is the set of matrix indices: the QC3 has two additional indices compared to the QC2.
To explain this, we first note that, in the three-particle formalism, the system is grouped as a ``pair'' and a ``spectator''. The first new index, $i$, labels the choice of the spectator particle, for example $i=\pi$ or $K$ in $\pi\pi K$ systems. 
(This index takes only one value in the $3\pi$ and $3K$ systems.)
The second new index, $\boldsymbol{k}$ labels the finite-volume momentum of the spectator. 
Finally, the $\ell$ and $m$ indices of the QC2 remain, but now denote the two-body partial waves of the pair.
The sum over $\bm k$ is truncated to a finite set by a cutoff function $H(\boldsymbol{k})$ that is an intrinsic part of the formalism.
In addition to this truncation, the pair partial waves 
are set to zero for $\ell >\ell_{\rm max}$, as for the QC2. Explicit expressions for the QC3 are given in Ref.~\cite{Blanton:2019igq} for three identical particles ($3\pi^+, 3K^+$) and Ref.~\cite{Blanton:2021eyf} for systems with two identical scalars and distinct third particle ($K^+K^+\pi^+, \pi^+\pi^+K^+$). 

We now return to the cutoff function $H(\bm k)$. 
For fixed $E$ and $\bm P$, as the spectator momentum $\bm k$ varies, so does the invariant mass of the pair:
\begin{equation}
    \sigma_k = \left(E-\sqrt{m_i^2 + \boldsymbol{k}^2}\right)^2 - (\boldsymbol{P} - \boldsymbol{k})^2.
\end{equation}
The cutoff function is actually a function of $\sigma_k$, and
interpolates smoothly between $1$ and $0$. 
It equals unity when $\sigma_k$ lies above the pair's threshold,
remains unity for a short distance below the threshold,
and reaches zero when $\sigma_k = \sigma_{\rm min}$, remaining zero for lower values.
The explicit form of the function we use is (see, e.g., Ref.~\cite{Blanton:2021mih})
    %%%%%
    \begin{align}
    H(\bm k) &= J(x[\sigma_k])\,, 
    \qquad x[\sigma_k] = 
    \frac{\sigma_k - \sigma_{\rm min}}{\sigma_{\rm th} - \sigma_{\rm min}}
    \label{eq:Hdef}
    \\
    %%%%%%
    J(x) & =
    \begin{cases}
    0 \,, & x \le 0 \, , \\
    %%%
    \exp \left( - \frac{1}{x} \exp \left [-\frac{1}{1-x} \right] \right ) \,, 
    & 0<x < 1 \, , \\
    %%%
    1 \,, & 1\le x \,.
    \end{cases}
    \label{eq:cutoff-J}
    %%%%%
    \end{align}
Here $\sigma_{\rm th}=4 M_\pi^2$ for a $\pi\pi$ channel,
$(M_\pi+M_K)^2$ for a $\pi K$ channel, and $4M_K^2$ for a $KK$ channel.

In previous work, various choices for $\sigma_{\rm min}$ have been made,
depending on the composition of the pair being considered.
For a $\pi\pi$ or $KK$ pair, the choice $\sigma_{\rm min} = 0$ was made in Refs.~\cite{Blanton:2021llb,Draper:2023boj}.
For a $\pi K$ pair, the choice $\sigma_{\rm min}=M_K^2-M_\pi^2$ was made
in Ref.~\cite{Draper:2023boj}, in order to avoid the nonanalyticity that appears in the $\pi K$ amplitude at the left-hand cut due to two-pion exchange in the $t$ channel.
Here we maintain these choices for the $\pi\pi$ and $\pi K$ channels, but change that for the $KK$ system to $\sigma_{\rm min} = 4(M_K^2-M_\pi^2)$, so as to avoid the two-pion exchange left-hand cut in $KK$ scattering.
In this way, we avoid left-hand cut singularities in all three two-particle channels.
A side benefit of this choice is that the number of values of $\bm k$ that contribute for a $KK$ pair is much reduced, leading to faster numerical solutions to the QC3.

\subsection{Scattering amplitudes from K matrices}
\label{sec:inteqs}

Given the two- and three-body K matrices, one can construct the corresponding $2 \to 2$ and $3 \to 3$ scattering amplitudes. In the two-body case, the partial-wave amplitude, $\cM_{2,\ell}$, is recovered from the partial-wave unitarity relation,
    %%%%%%
    \begin{equation}
    \mathcal M_{2,\ell}^{-1}(E_2^*) =  \mathcal K_{2,\ell}^{-1}(E_2^*) - i\rho(E_2^*),
    \label{eq:M2iell}
    \end{equation}
    %%%%%%
where $\rho(E_2^*)$ is the two-body phase-space defined as $\rho(E_2^*) = \eta q/(8\pi E_2^*)$, with the symmetry factor given by $\eta=1/2$ ($\eta=1$) for identical (distinguishable) particles. Here, $q$ is the magnitude of the relative momentum of the particles in the two-body c.m.~frame,
    %%%%%%
    \begin{align}
    q = \frac{\lambda^{1/2}((E_2^*)^2, M_1^2, M_2^2)}{2 E_2^\star} \, ,
    \end{align}
    %%%%%%
where $M_1$ and $M_2$ are the masses of scattered particles, while $\lambda(x,y,z) = x^2 + y^2 + z^2 - 2(xy + yz + zx)$ is the K\"all\'en triangle function. The partial-wave projected amplitude is related to the full amplitude through the standard relation,
    \begin{align}
    \cM_2(s_2,t_2) = \sum_{\ell=0}^\infty (2\ell+1) \, \cM_{2,\ell}(E^*_2) \, P_\ell(\cos \theta^*) \, ,
    \end{align}
where $s_2, t_2$ are the two-body Mandelstam variables, $P_\ell$ is the Legendre polynomial, and $\theta^*$ is the scattering angle in the two-body c.m.~frame.

The three-to-three elastic amplitude, $\cM_3$, depends on eight, rather than two, kinematical variables. It is obtained by solving a set of relativistic on-shell integral equations that take $\cM_2$ and $\cK_{{\rm df},3}$ as an input. These equations were first derived in Ref.~\cite{Hansen:2015zga} for three identical bosons,
and extended to $\pi\pi K$ and $K K \pi$ systems in Ref.~\cite{Blanton:2021mih}. They satisfy three-body S-matrix unitarity~\cite{Briceno:2019muc, Jackura:2022gib} and share many similarities with the relativistic three-body equations proposed in the past~\cite{Blankenbecler:1965gx, Taylor:1966zza, Aaron:1968aoz, Brayshaw:1978tx, Lindesay:1980ib}, as well as with dynamical equations derived within other modern three-body finite-volume formalisms~\cite{Mai:2017bge, Mai:2017vot, Jackura:2019bmu, Blanton:2020jnm, Sadasivan:2020syi, Sadasivan:2021emk, Feng:2024wyg}.

We give here a conceptual overview of these equations, relegating technical details and certain definitions to \Cref{app:pwproj} and external references.
Furthermore, for clarity of presentation, we introduce the integral equations here only for the non-degenerate systems, i.e. $\pi\pi K$ and $KK\pi$. 
Up to kinematic differences, these have exactly the same form as those for the $DD\pi$ system, which have been described in full detail in Ref.~\cite{Dawid:2024dgy}.
Those for degenerate particles are simpler, and are obtained from those given below by dropping the indices corresponding to spectator flavor.
They are given in full detail in Refs.~\cite{Hansen:2015zga,Jackura:2020bsk}. Without loss of generality, we consider $\cM_3$ only in the overall c.m.~frame.

The $3\to3$ scattering process is an elastic reaction in which three particles of initial momenta ${\{\bm k\} = \{\bm k_1, \bm k_{1'}, \bm k_2 \}}$ collide and emerge with final momenta $\{\bm p\} = \{\bm p_1, \bm p_{1'}, \bm p_2 \}$. Here, indices $1$ and $1'$ label the two identical particles, e.g. the two pions in the $\pi\pi K$ system. The probability of the reaction is given by the fully connected three-body amplitude, $\cM_3(\{\bm p\},\{\bm k\})$, whose precise definition is given in Ref.~\cite{Blanton:2021mih}. Following the convention of the three-body quantization condition, we group particles in the external states into pairs and spectators and decompose $\cM_3$ into the so-called unsymmetrized pair-spectator amplitudes, denoted $\cM^{(u,u)}_{3,ij}$. 
These are characterized by having 
final and initial spectators with flavors given by the indices $i,j \in \{1,2\}$, respectively, and by the rule that, when expressed in terms of Feynman diagrams,
the last (first) pairwise interaction involves the final (initial) pair (rather than the
corresponding spectator).
The full $\cM_3$ amplitude is obtained by summing $\cM^{(u,u)}_{3,ij}$ over all possible pair-spectator choices, as described explicitly below.

The connected pair-spectator amplitude is a sum of two terms describing a set of different three-body processes
    %%%%%
    \begin{align}
    \label{eq:decomp}
    \cM_{3,ij}^{(u,u)} = \cD_{ij}^{(u,u)} + \cM_{\text{df}, 3; ij}^{(u,u)} \, .
    \end{align}
    %%%%%
The first term is the so-called ladder amplitude, given by the solution of the integral equation,
\begin{align}
    \begin{split}
    \cD_{ij}^{(u,u)}(\bm p_i,\bm k_j) =   
    %%%
    - \cM_{2}^{(i)}(\bm p_i) \, 
    G_{ij}(\bm p_i,\bm k_j) \, 
    \cM_{2}^{(j)}(\bm k_j) \\- \sum_{n=1}^2 
    \int_{\bm q_n} 
    \cM_{2}^{(i)}(\bm p_i) \, 
    G_{in}(\bm p_i,\bm q_n) \, 
    \cD^{(u,u)}_{nj}(\bm q_n,\bm k_j) \, .
    \end{split}
    \label{eq:ladder}
\end{align}
It describes all contributions to the scattering process that do not depend on the dynamics encoded in the three-body K matrix. 
The objects in \Cref{eq:ladder} have implicit matrix indices describing the angular momentum of the pair; for details see Ref.~\cite{Dawid:2024dgy}.

The amplitude depends explicitly on the spectator momenta, $\bm p_i$ and $\bm k_j$, as well as implicitly on the angular momenta of the final and initial pair, and the total c.m. energy $E$. A full description is given in \Cref{app:pwproj}. 
The kernel of the integral equation contains two objects. The first is the one-particle-exchange (OPE) amplitude, $G_{ij}$, describing the exchange of a particle between the external pairs corresponding to spectators $i$ and $j$. The second one is the two-body on-shell amplitude, $\cM_2^{(i)}$, describing interactions within the pair corresponding to spectator $i$. Thus, $\cD^{(u,u)}_{ij}$ depends on the two-body K matrices via $\cM_2$. Finally, the integral is defined as,
    \begin{align}
    \int_{\bm q_n} \equiv \int \frac{d \bm q_n}{(2\pi)^3 2\omega_{q_n}} \,,
    \ \ \omega_{q_n} = \sqrt{q_n^2 + M_n^2}\,,
    \end{align}
where the Lorentz-invariant measure contains the on-shell energy of the intermediate spectator of flavor $n \in \{1,2\}$.

The second term in \Cref{eq:decomp}---the unsymmetrized, divergence-free amplitude---is given by,
\begin{align}
    \begin{split}
    \label{eq:Mdf3}
    &\cM_{{\rm df},3;ij}^{(u,u)}(\bm p_i,\bm k_j) = \\ &\sum_{n,l=1}^2
    \int\limits_{\bm q_n} 
    \int\limits_{\bm q_l}   
    \cL_{in}^{(u)}(\bm p_i,\bm q_n) \, 
    \cT_{nl}(\bm q_n, \bm q_l) \, 
    \cR_{lj}^{(u)}(\bm q_l,\bm k_j) \, .
    \end{split}
\end{align}
and describes a complicated set of three-body processes involving short-range interactions encoded in the three-body K matrix. The so-called left and right ``endcap'' functions are,
\begin{align}
    \begin{split}
    \cL^{(u)}_{in}(\bm p_i,\bm q_n) =   \Big( \tfrac{1}{3} - \cM_2^{(i)}(\bm p_i) \, 
    \tilde{\rho}_i(\bm p_i) \Big) 
    \overline\delta(\bm p_i \!-\!\bm q_n) \, \delta_{in} \\ 
    - \cD_{in}^{(u,u)}(\bm p_i,\bm q_n) \, 
    \tilde{\rho}_n(\bm q_n) \, ,
    \end{split}
    \label{eq:CLu}
    \\
    \begin{split}
    \cR^{(u)}_{lj}(\bm q_l,\bm k_j) =   
    \Big( \tfrac{1}{3} - \tilde{\rho}_j(\bm k_j)\cM_2^{(j)}(\bm k_j) \, 
     \Big) 
    \overline\delta(\bm q_l \!-\!\bm k_j) \, \delta_{lj} \\ 
    - \tilde{\rho}_l(\bm q_l) \cD_{lj}^{(u,u)}(\bm q_l,\bm k_j) \, 
     \, ,
    \end{split}   
    \label{eq:CRu}
\end{align}
where,
\begin{equation}
\overline\delta(\bm p_i \!-\!\bm q_n) = 2\omega_{q_n} (2\pi)^3
\delta^{3}(\bm p_i \!-\! \bm q_n)\,.
\end{equation}
The quantity $\tilde{\rho}_n(\bm q_n)$ is the standard two-body phase space modified by a regulating function~\cite{Dawid:2024dgy}. The factors $\cL$ and $\cR$ describe initial- and final-state rescatterings due to two-body interactions and pairwise one-particle exchanges. The $\cT_{nl}$ amplitude is related to the three-body K matrix by the integral equation,
\begin{align}
    \begin{split}
    \cT_{nl}(\bm q_n,\bm q_l) & = \cK_{{\rm df}, 3}^{(nl)}(\bm q_n,\bm q_l) \\ 
    & - \sum_{n',l'=1}^2 
    \int\limits_{\bm q_{n'}} 
    \int\limits_{\bm q_{l'}} 
    \cK_{{\rm df, 3}}^{(nn')}(\bm q_n, \bm q_{n'}) \, 
    \tilde{\rho}_{n'}(\bm q_{n'}) \\
    & \qquad \qquad \quad \times 
    \cL^{(u)}_{n'l'}(\bm q_{n'},\bm q_{l'}) \, 
    \cT_{l'l}(\bm q_{l'},\bm q_l) \, .
    \end{split}
    \label{eq:Teq}
\end{align}
This incorporates the effect of short-range three-body interactions,
interwoven with those involving two particles.

In practice, as in the two-body case, one simplifies the above formulas by performing partial-wave projection~\cite{Jackura:2023qtp, Dawid:2024dgy, Briceno:2024ehy}. 
This reduces the three-dimensional integrals over the intermediate spectator momenta to one-dimensional integrals over their magnitudes. Partial-wave projection 
replaces the dependence on the directions of the spectators with that on the
relative angular momentum, $\ell$, between the spectator and the pair. The angular momentum of the pair, $s$, remains as a variable. The resulting amplitude is a matrix in the orbital basis (also known as the $J\ell s$ basis), 
    \begin{align}
    \cM_{3,ij}^{(u,u)}(\bm p_i,\bm k_j) \xrightarrow[{\rm p.w.~proj.}]{} \cM_{3,i\ell_i's'_i;j\ell_j s_j}^{(u,u) J}(p_i,k_j) \,,
    \end{align}
where $s'_i,s_j$ are ``spins'' of external pairs (orbital angular momentum of two particles forming a pair in their c.m.~frame), $\ell'_i,\ell_j$ are orbital angular momenta of the external pair-spectator states, and $J$ is the total angular momentum of the three-body system. The exact form of the transformation and further details are discussed in~\Cref{sec:PW}.
Results for the projection onto the $J\ell s$ basis of the OPE amplitude G, and of $\Kdf$, are given, respectively, in \Cref{app:PWG,app:PWK3}.

Finally, to reconstruct the full three-body amplitude one has to sum over different choices of the spectators, as well as over $J,\ell,s$ indices. The full expression is given by,
    %%%%%
    \begin{align}
    \cM_3(\{\bm p\}; \{\bm k\}) = \sum_J \cM_3^J(\{\bm p\}; \{\bm k\}) \, ,  
    \label{eq:symmetrization}
    \end{align}
    %%%%%
where, for the purpose of this work, we defined the individual $J$ contributions\footnote{%
Note that these are different from the partial-wave projections of the three-body amplitude introduced in Eq.~(12) of Ref.~\cite{Jackura:2018xnx}.}
%Note that these are different from the definite $J$ partial-wave projections of the three-body amplitude~\cite{Jackura:2018xnx}.}
%
as,
    %%%%%
    \begin{multline} 
    \cM_3^J(\{\bm p\}; \{\bm k\}) = \sum_{i,j} \sum_{\sigma_i, \sigma_j} \sum_{\ell'_i, s'_i} \sum_{\ell_j, s_j} \mathcal X_{ij} 
    \\\times \cM_{3,i \ell'_i s'_i; j \ell_j s_j}^{(u,u) J}(p_{\sigma_i},k_{\sigma_j}) 
    %%%
    \cZ^{J}_{\ell'_i s'_i; \ell_j s_j}(\theta_{ij}, \hat{\bm{q}}_{p_i}^\star, \hat{\bm{q}}_{k_j}^\star ) \, .
    \label{eq:symmetrization2}
    \end{multline}
    %%%%%
The sum over $\sigma_i$ accounts for the fact that, if the final flavor is $i=1$, then there are two choices of spectator momenta, $p_1$ and $p_{1'}$,
and similarly for the initial state. 
This is encoded by $\sigma_1$ having two possible values, $1$ and $1'$.
On the other hand, if $i=2$, then there is only a single choice $\sigma_2=2$.
Symmetry factors are $\mathcal X_{11} = 1$, $\mathcal X_{12} = \mathcal X_{21} = \sqrt{2}$, and $\mathcal X_{22} = 2$, $\theta_{ij}$ is the angle between the final $i$th and initial $j$th spectator in the total~c.m.~frame, and $\hat{\bm{q}}_{p_i}^\star$, $\hat{\bm{q}}_{k_j}^\star$ are orientations of relative momenta of external pairs in their rest frames. The projector $\cZ$ contains the angular dependence of the amplitude and its explicit form is provided in \Cref{app:symm}.

\subsection{Parametrization of K matrices}
\label{sec:models}

The mapping from energies to two- and three-particle $K$-matrix elements is, in general, not one-to-one, due to the presence of multiple partial waves and 
(in the three-particle case) of several two-particle channels.
Thus one must use parametrizations of the K matrices as functions of the kinematic quantities and determine their parameters by global fits.
We use parametrizations that both satisfy all the relevant symmetries---Lorentz invariance, $C$, $P$ and $T$ symmetries, and particle-exchange symmetry if appropriate---and are smooth functions in the kinematic range that we access. In particular, since we consider systems at maximal isospin, we do not expect nearby resonances or bound states. Thus we expand the K matrices about the threshold, and investigate the sensitivity of fits to the number of terms that are kept.

The two-particle systems that we consider---$\pi\pi$, $KK$ and $\pi K$, all at maximal isospin---do not exhibit flavor mixing, and so the corresponding K matrices can be written simply in terms of phase shifts
\begin{equation}
   \mathcal K_{2,\ell}^{-1} = \frac{\eta}{ 8\pi E_2^* q^{2\ell} }  q^{2\ell+1} \cot \delta_\ell(q)\,.
   \label{eq:K2inv}
\end{equation}
Here $\eta$ is as in \Cref{eq:M2iell}, and $q$ is the momentum of both particles in their c.m.~frame, and satisfies,
\begin{equation}
    E_2^* = \sqrt{M_1^2+q^2} + \sqrt{M_2^2+q^2}\,,
    \label{eq:q2def}
\end{equation}
where $M_1$ and $M_2$ are the masses of the particles in the pair. 
In \Cref{eq:K2inv}, we have pulled out the quantity $q^{2\ell+1} \cot \delta_\ell$, since this is nonsingular at the threshold, and is a smooth function of $q^2$.

We use two families of parametrizations for the $s$-wave phase shift. First, the effective-range expansion (ERE):
\begin{equation}
\frac{q}{M} \cot \delta_{0}(q) = \sum_{n=0}^{ n < n_{\rm max}} b_n \left( \frac{q^2}{M^{2}}\right)^n\,,
\label{eq:ERE}
\end{equation}
where $b_n$ are fit parameters and $M$ is the mass of one of the particles in the pair. If $M=M_\pi$ then we refer to the resulting parameters as being in ``pion units'', while $M=M_K$ leads to ``kaon units''. 
This model is abbreviated as ERE2, ERE3, or ERE4, where the number is $n_{\rm max}$, or equivalently the number of terms that are kept in the truncated expansion.
In this work, we use these parametrizations for $KK$ systems.

The second model incorporates the Adler zero expected from chiral perturbation theory (ChPT):
\begin{equation}
\frac{q}{M_{1}} \cot \delta_{0}(q) = \frac{M_{1} E^{*}_{2}}{E^{*2}_{2} \!-\! z^2 (M_{1}^{2} \!+\! M_{2}^{2})} 
\sum_{n=0}^{n < n_{\rm max}} B_n \left( \frac{q^2}{M_{1}^{2}}\right)^n\,,
\label{eq:Adler}
\end{equation}
where $B_n$ and $z^2$ are fit parameters, and we explicitly take into account that there could be nondegenerate mesons. At LO in ChPT, $z^2=1$.
This fit form is used for both $\pi\pi$ and $\pi K$ systems, for which we expect ChPT to work well.
While we have tried fits with more parameters, the final fits below all have $n_{\rm max}=2$.
We refer to these fit forms as ADLER2 if $z^2=1$ is chosen,
and ADLER2z if $z^2$ is taken as a third fit parameter.

In order to make contact with the literature, it is convenient to relate the fit parameters in \Cref{eq:ERE,eq:Adler} to the $s$-wave scattering length $a_0$ and $s$-wave effective range $r_0$, which are defined using the convention
\begin{equation}
    q^{2\ell+1} \cot \delta_\ell = -\frac{1}{a_\ell} + \frac{1}{2} r_{\ell} q^2 + \cdots
\end{equation}
For the ERE parametrizations, the relations are 
\begin{equation}
    M a_0 = -\frac{1}{b_0}, \quad M^2 a_0 r_0 =  -\frac{2 b_1}{b_0},
    \label{eq:a0ERE}
\end{equation}
while for the ADLER parametrizations, they are
\begin{align}
a_0 M_1 &=  - \frac{(M_1+M_2)^2 - z^2 (M_1^2+M_2^2)}{M_1(M_1+M_2)B_0}\,,
\label{eq:a0Adler}
\\
a_0 r_0 M_1^2 & = \frac{M_1}{M_2} \frac{ (M_1+M_2)^2+ z^2 (M_1^2+M_2^2)}{(M_1+M_2)^2 - z^2 (M_1^2+M_2^2)}
- 2 \frac{B_1}{B_0} \,.
\label{eq:r0Adler}
\end{align}

Higher partial waves must also be considered. Due to Bose symmetry, for identical mesons ($\pi\pi$ and $KK$), the next contributing wave is $d$-wave. We parametrize this with a single parameter $D_0$,
\begin{equation}
    \frac{q^5}{M^5} \cot \delta_2=\frac{E_2^*}{2 M} \frac1{D_0}-1\,,
    \label{eq:cotdelta_d}
\end{equation}
where the term $-1$ is included to avoid unphysical subthreshold poles as in Ref.~\cite{Blanton:2021llb}. The $d$-wave scattering length is given by
$M^5 a_2=-D_0/(D_0-1)$. We ignore  higher partial waves.
For $\pi K$ systems, the next contributing wave is the $p$ wave. We use the following form:
\begin{equation}
     \frac{q^3}{M_1^3} \cot \delta_1(q)=\frac{E_2^*}{M_1+M_2} \frac{1}{P_0^{\pi K}},
     \label{eq:cotdelta_p}
\end{equation}
with a single fit parameter, $P_0^{\pi K}$, related to the scattering length as $M_\pi^3 a^{\pi K}_1 = - P^{\pi K}_0$.

We also need to parametrize the three-particle K matrix. For systems of three identical particles, the threshold expansion, incorporating appropriate symmetries, is given in ref.~\cite{Blanton:2019igq}. Up to quadratic order, it is
\begin{equation}
    M^2 \mathcal{K}_{\mathrm{df}, 3}=\mathcal{K}_0+\mathcal{K}_1 \Delta+\mathcal{K}_2 \Delta^2+\mathcal{K}_{\mathrm{A}} \Delta_{\mathrm{A}}+\mathcal{K}_{\mathrm{B}} \Delta_{\mathrm{B}} \,,
    \label{eq:Kdfidentical}
\end{equation}
where $\cK_0, \cK_1, \cK_2, \cK_A, $ and $\cK_B$ are free parameters, and $\Delta,\Delta_A, \Delta_B$ are kinematic functions defined as
\begin{align}
\begin{split}
    \Delta &= \frac{s-E_{\rm th}^2}{E_{\rm th}^2},\quad  s= (p_1+p_2+p_3)^2,
    \\
    \Delta_{\mathrm{A}} &\equiv \sum_i\left(\Delta_i^2+\Delta_i^{\prime 2}\right)-\Delta^2, \quad \Delta_{\mathrm{B}} \equiv \sum_{i, j} \tilde{t}_{i j}^2-\Delta^2 \,.
    \end{split}
    \label{eq:Kdfidenticala}
\end{align}
Here
\begin{align}
\begin{split}
E_{\rm th} = 3M, \quad & \tilde{t}_{i j} \equiv \frac{\left(p_i-k_j\right)^2}{E_{\rm th}^2} \\
\Delta_j \equiv\frac{\left(P-k_j\right)^2-4 M^2}{E_{\rm th}^2}, &\quad
\Delta_i^{\prime} \equiv \frac{\left(P-p_i\right)^2-4 M^2}{E_{\rm th}^2}\,,
   \end{split}
   \label{eq:Kdfidenticalb}
\end{align}
where $p_i$ ($k_i$) are the outgoing (incoming) momenta,
with $i=1-3$. 

For systems of two identical mesons and a third distinct particle, the corresponding expansion was worked out in Ref.~\cite{Blanton:2021mih}. This expansion is less constrained, as the only symmetry that can be imposed is the exchange of the two identical particles. Thus, the number of terms grows faster with the order, and we only consider it up to linear order in the expansion,
\begin{equation}
M_1^2 \cK_{\text{df},3} = \cK_{0} + \cK_{1} \Delta + \cK_{B} \Delta^S_2 + \cK_{E} \tilde{t}_{22}\,,
\label{eq:KdfExpansion}
\end{equation}
where the definitions given above are used, but now with $E_{\rm th} = 2M_1+M_2$, and,
\begin{equation}
    \Delta_2^S \equiv \frac{(p_1+p_{1'})^2- 4 M_1^2}{E_{\rm th}^2} + \frac{(k_1+k_{1'})^2 - 4 M_1^2}{E_{\rm th}^2}\,.
    \label{eq:KdfExpansiona}
\end{equation}
Here, 1 and $1'$ label the particle that appears twice, and 2 the distinct one. 

Note that \Cref{eq:Kdfidentical,eq:KdfExpansion} use the same letters for different terms; these are differentiated by specifying the system, 
e.g., $\cK_B^{3\pi}$ versus $\cK_B^{\pi\pi K}$.

\subsection{Fitting procedure}
\label{sec:fitstrategy}

In order to constrain the parameters in the K matrices, labeled generically as $\bm p$, we minimize the $\chi^2$ function
\begin{align}
\begin{split}
\chi^2\left(\bm p\right)=\sum_{i j}  \Delta X_i (C^{-1})_{ij} \Delta X_j , \\ \Delta X_i =  \Delta E^{\rm lab}_i - \Delta E^{\rm lab, QC}_i (\bm p)\,,
\end{split}
\label{eq:chi2} 
\end{align}
where $i,j$ label the levels included in the fit,
$\Delta E^{\rm lab}_i$ are lab-frame energy shifts 
obtained from the ratio fits described around \Cref{eq:labshiftfit}, 
while $\Delta E^{\rm lab, QC}_i (\bm p)$ are the corresponding shifts predicted by the quantization conditions given parameters $\bm p$. $C$ is the covariance matrix between the LQCD lab-frame shifts, estimated from jackknife resamplings 
in which the energy shifts are determined on each jackknife sample.
Previous work has explored using other quantities to define the $\chi^2$ function, e.g. the shifts in the overall c.m.~frame, and concluded that the choice in \Cref{eq:chi2} is preferred, since it leads to smaller uncertainties in the resulting parameters, and to covariance matrices with lower condition numbers~\cite{Draper:2023boj}.

The $\chi^2$ function is minimized using standard routines in SciPy~\cite{2020SciPy-NMeth}. We find the Nelder-Mead algorithm to be robust in finding the minima. Since each fit can take several hours on a cluster, we estimate the uncertainties of the fit parameters by means of the derivative method, rather than using jackknife.
The covariance of the parameters is computed as
\begin{equation}
V_{n m}=\left(\frac{\partial \Delta E^{\rm lab, QC}_i}{\partial p_n} 
(C^{-1})_{ij} \frac{\partial \Delta E^{\rm lab, QC}_j}{\partial p_m}\right)^{-1}\,,
\end{equation}
with derivatives (approximated by finite differences) taken at the minimum of the $\chi^2$ function.
We have verified that the alternative approach of determining the $\chi^2 + 1$ contour yields errors that are consistent.

We perform fits to both two-particle levels alone
($\pi\pi$, $KK$, and $\pi K$)
and to the combination of two- and three-particle levels.
The combinations that we use are
$\pi\pi + \pi\pi\pi$,  $KK  + KKK$, $\pi\pi + \pi K + \pi \pi K$ , and $KK + \pi K+  + KK\pi$,
i.e. we fit the possible two-particle subchannels along with the three-particle channel.
Since all levels are calculated on the same ensemble, the correlation matrix between these levels can be determined.
The advantage of these combined fits is that all information on two-particle interactions is considered together.
In this regard, we recall that the dominant contributions to the energy shifts in three-particle systems are due to two-particle interactions, and these shifts are larger than in two-particle systems because there are three pairs.
Other types of fit, in which the two-particle channels are fit first, with information then fed into fits to the three-particle levels, were considered in ref.~\cite{Draper:2023boj}, and found to lead to weaker constraints on the parameters.

The fits depend on knowledge of single-hadron masses, which enter, for example, in the calculation of the energy shifts from the quantization conditions. We neglect the uncertainties in these quantities since they are much smaller than those in the lab-frame energy shifts.

\section{Results}
\label{sec:results}

Here, we summarize the fit results and derived threshold parameters for the different systems and fit models. For the sake of legibility, tables involving two-meson results are in \Cref{app:E2502meson}.

\subsection{Fit results for the K matrices}
\label{sec:fits}

We first summarize the results of fits to systems of two and three mesons using the parametrizations described in \Cref{sec:models} and the fit strategy of \Cref{sec:fitstrategy}. We consider only the E250 ensemble, as results on the other ensembles have been presented in previous studies~\cite{Blanton:2021llb,Draper:2023boj}. 
The only exceptions are the fits to $KK\pi$ and $KKK$ levels, where, as explained above, we are using a different cutoff function in the $KK$ subchannels---see \Cref{eq:Hdef} and surrounding discussion.
Since $\Kdf$ depends on the choice of cutoff function, in order to have consistent chiral extrapolations, we have repeated the $KK\pi$ and $KKK$ fits with the new function on the D200, N200, and N203 ensembles. The results are collected in \Cref{app:reananlysis}.
We also note that we have done many more fits than are displayed below.
Fits not shown have investigated the importance of additional parameters.
For each class of fits, we display only those with the highest significance.

We begin with systems of pions. We first consider the two-pion system in isolation,  with results given in \Cref{tab:pp-params-E250} of \Cref{app:E2502meson}. 
Fits include either $s$-wave interactions alone, or both $s$ and $d$ waves.
In all cases, we use Adler-zero parametrizations of the $s$-wave amplitude, since with our near-physical pion mass we are in the regime where ChPT is reliable.
We find good fits ($p\ge 0.24$) with the two-parameter ADLER2 form, in which the Adler zero is fixed to its leading-order value. Fits with additional parameters do not improve the $p$ values and are not shown.
Including the nontrivial irreps to which the $d$-wave amplitude contributes leads to higher $p$ values, but the resulting $d$-wave amplitude (parametrized by $D_0^{\pi\pi}$) is consistent with zero.
We also find that we can increase the maximum c.m. frame energy of levels included in the fit (labeled ``Cutoff'' in the figure) beyond the inelastic threshold at $E^*=4M_\pi$ and still obtain good fits, in which the errors are slightly reduced.
The insensitivity to inelastic channels is expected since the coupling to the four-pion states is expected to be small~\cite{Garcia-Martin:2011iqs}.

Next, we turn to combined two- and three-pion fits, summarized in \Cref{tab:ppp-params-E250}. 
The additional parameters that enter are those
in $\Kdf$, from which we either keep only the 
purely $s$-wave isotropic contributions ($\cK_0$ and $\cK_1$) or also add in the $\cK_B$ term, which contains also $d$ waves.
Unlike for two pions, 
nontrivial irreps in the three-pion sector are sensitive to the $s$-wave two-particle amplitude.
Thus it makes sense to do a fit with $s$-wave parameters to all the three-pion irreps along with the trivial two-pion irreps; the results are shown in the first column.
Fits to all two-pion irreps including $d$-wave terms are, however, slightly better, although they find a $d$-wave two-pion interaction consistent with zero.
Similarly, the components of $\kdf$ are consistent with zero. This remains true of the entire $\Kdf$ when correlations are included, as will be discussed below.
Finally, we observe that the errors on the two-particle parameters are, in general, slightly reduced when fitting to two- and three-pion levels compared to the corresponding fits to two-pion levels alone.
The central values are, however, consistent.

Turning now to systems of kaons, results for the two-kaon case are shown in \Cref{tab:kk-params-E250} of \Cref{app:E2502meson}. For the $KK$ interaction, we use ERE parametrizations, 
while, as for pions, we do fits with or without $d$ waves. We also consider fits including energies above the strict inelastic threshold ($KK\pi$), since the coupling to this channel is expected to be weak~\cite{Blanton:2021llb}.
Our results indicate that, in contrast to two pions, the $d$-wave amplitude is statistically significant at the $2-3\sigma$ level, and that increasing the cutoff somewhat above the inelastic threshold does not reduce the quality of the fits. 
The $p$ values of all the fits are significantly worse than those for two pions, likely because the errors in the energy levels are smaller while the complexity of the fit form is unchanged. However, we have not found a way of extending the parametrization that leads to improved $p$ values. For example, comparing the last two columns of the table, we see that moving from the ERE2 to the ERE3 fit does not improve the quality. Thus, in fits involving two and three kaons, we consider only ERE2 fits.

The results from $2K/3K$ fits are shown in \Cref{tab:kkk-params-E250}. Here we only include levels up to (slightly above) the inelastic threshold, corresponding to the first and third columns of results for two kaons (\Cref{tab:kk-params-E250}). We draw several conclusions.
First, for the two-kaon parameters, the results are consistent with those from two-kaon fits, but with somewhat reduced uncertainties.
Second, we again find evidence for a nonzero $d$-wave interaction, although with reduced significance compared to the two-kaon case.
Third, we find that a nonzero $\Kdf$ is needed to describe the spectrum, with significant results for $\cK_1^{3K}$ ($4-6\sigma$) and $\cK_B^{3K}$ ($3.5\sigma$).

We now turn to mixed systems of pions and kaons, beginning with the $\pi K$ system, results of fits to which are shown in \Cref{tab:pk-params-E250} of \Cref{app:E2502meson}.
The first two columns show fits up to the inelastic threshold at $E^*=2M_\pi+M_K$, including, respectively, only $A_1$ irrep levels with an $s$-wave interaction, and levels in all irreps with both $s$ and $p$ wave amplitudes.
We find good fits with the ADLER2 form of $s$-wave interaction, but that the $p$-wave amplitude is consistent with zero.
We again expect the coupling to the inelastic channel to be weak, and thus try fits with a raised cutoff.
We find (see the final column) that fits of the same quality are possible only if we allow for the position of the Adler zero to float. The constraints on the $p$-wave amplitude strengthen, but it remains consistent with zero.

Next we consider $\pi\pi K$ systems, which are fit in combination with $\pi\pi$ and $\pi K$ levels,
with results shown in \Cref{tab:ppK-params-E250}.
Given that we found no evidence for $d$-wave interactions in the two-pion channel within errors, we do not include such interactions in these fits. 
We do include $p$-wave interactions, however, both in the $\pi K$ and $\pi\pi K$ K matrices.
All channels are fit only up to their strict inelastic thresholds. We find no significant evidence for either two- or three-particle $p$-wave terms, and furthermore that the isotropic part of $\Kdf$ is consistent with zero.
Indeed, our fit with the highest $p$ value is that in the first column in which only $s$-wave two-particle interactions are used.

Finally, we turn to the $KK\pi$ systems, which are fit in combination with the $KK$ and $\pi K$ levels, with results displayed in \Cref{tab:KKp-params-E250}.
These are our fits with the largest total number of levels, up to 115 if all irreps are included.
They are also by far our worst fits in terms of $p$ values, which we attribute to the smaller statistical errors for quantities involving kaons combined with the fact that we cannot, in practice, add more parameters to our fit forms and achieve stable fits.
One new feature of our fits compared to ref.~\cite{Draper:2023boj}, is that we have allowed the option of including $d$ waves in the $KK$ channel, in addition to $p$ waves in the $\pi K$ channel, and $s$ waves in both channels.\footnote{%
%%%%%
Strictly speaking, this is not consistent with the power-counting scheme one uses in the threshold expansion---one should also include $d$ waves in the $\pi K$ channel and in $\kdf$. Here we are taking a more pragmatic approach based on the observations that $d$ waves are needed in the $KK$ subchannel, while $p$ waves are not significant in the $\pi K$ channel. Furthermore, since the $\Kdf$ contribution is subleading in a $1/L$ expansion, higher order terms in $\Kdf$ are likely to be numerically small.
%%%%%
} 
In practice, the best fit we find is that in the first column of the table, where only $s$-wave two- and three-particle interactions are included, and only two-particle levels in the $A_1$ irrep (though all three-particle irreps) are fit. In this fit, the $\Kdf$ parameters $\cK_0$ and $\cK_1$ are not individually significant, but, due to the strong (anti-)correlation between these parameters, the full isotropic part of $\Kdf$ has a significance of 99.5\% (equivalent to $2.8\sigma$ for a single variable). This can be seen in \Cref{fig:Kdfconstrain}, and will be discussed further below.

\begin{table*}[h!]
\centering
\begin{tabular}{|c|c|c|c|c|c|}
\hline
Ensemble &  \multicolumn{3}{c|}{E250}  \\ \hline \hline
Cutoffs & $4M_\pi$/$5M_\pi$ & $4M_\pi$/$5M_\pi$  & $4M_\pi$/$5M_\pi$
\\ \hline
Description& $s$ wave (all irreps) &   $s,d$ waves  & $s,d$ waves  \\ \hline
$\chi^2$ & 65.56 &  68.94 & 68.60 \\ \hline
DOF     & 24+32-4=52 &  34+32-5=61  & 34+32-6=60 \\ \hline 
$p$  & 0.098 & 0.23 & 0.21 \\ \hline
\hline
$B_0^{\pi\pi}$& -23.6(2.9) &   -23.2(2.6) & -22.3(2.7)
\\ \hline
$B_1^{\pi\pi}$  & -4.0(1.3) &  -4.3(1.3)& -4.7(1.3)
\\ \hline
$z^2_{\pi\pi}$&   $1$ (fixed)&   $1$ (fixed)&  $1$ (fixed)
\\ \hline
$D_0^{\pi\pi}$ &   $0$ (fixed) & $-0.1(3.0)\cdot 10^{-4}$ & $-0.3(3.0)\cdot 10^{-4}$
\\ \hline
$\mathcal K_0^{3\pi}$ & -240(270) &   -230(270)& -160(280)
\\ \hline
$\mathcal K_1^{3\pi}$   & 230(210) &   220(200)  & 290(220)  \\ \hline
$\mathcal K_B^{3\pi}$   & 0 (fixed) &   0 (fixed) & 300(340)  \\ \hline
\end{tabular}
\caption{Results of combined fits to two- and three-pion levels on ensemble E250, using pion units.
Cutoffs are for the two- and three-pion fits, respectively. 
The DOF entries list in order the number of two-pion and three-pion levels fit minus the number of parameters.
The ``$s$ wave (all irreps)'' fit is to three-pion levels in all irreps but only to $A_1$ two-pion irreps.
The $s,d$ wave fits include all irreps for both two and
three pions.
Fit parameters are as in the models described in \Cref{sec:models}, using the ADLER2 form for the two-pion  interaction.
}
\label{tab:ppp-params-E250}
\end{table*}

\begin{table*}[t]
\centering
\begin{tabular}{|c|c|c|c|}
\hline
Ensemble &  \multicolumn{3}{c|}{E250}  \\ \hline \hline
Cutoffs & $2.27 M_K/3.27 M_K$  & $2.27M_K/3.27M_K$  &  $2.27M_K/3.27M_K$  
\\ \hline
Description & $s$ wave (symmetric irreps)  & $s$ wave (all irreps) & $s,d$ waves   \\ \hline
$\chi^2$ & 84.79  & 120.31 & 129.57  \\ \hline
DOF     &  28+34-4=58 & 28+53-4=77 & 40+53-6=87   \\ \hline 
$p$ & 0.012 &0.0012 & 0.002 \\ \hline \hline
$b_0^{KK}$ & -2.673(41)  & -2.665(39) & -2.637(41) 
\\ \hline
$b_1^{KK}$ & 0.87(18) & 0.85(17)  & 0.70(19) 
\\ \hline
$D_0^{KK}$   & 0 (fixed) & 0 (fixed) & -0.034(26) 
\\ \hline
$\mathcal K_0^{3K}$ & $-4.2(7.0)\cdot 10^3$ & $4.2(6.2)\cdot 10^3$   & $4.8(5.7)\cdot 10^3$ 
\\ \hline
$\mathcal K_1^{3K}$ & $-1.25(0.48)\cdot 10^5$ & $-1.75(0.44)\cdot 10^5$ & $-2.96(0.50)\cdot 10^5$    \\ \hline
$\mathcal K_B^{3K}$ & 0 (fixed) &   0 (fixed)  &   $-1.96(0.54)\cdot 10^6$
 \\ \hline
\end{tabular}
\caption{Results of fits to two- and three-kaon levels on ensemble E250, using kaon units.
Cutoffs are for the two- and three-kaon fits, respectively, and lie slightly above the inelastic thresholds at $E^*=2.265$ and $3.265$.
The DOF entries list the number of two-kaon and three-kaon levels fit minus the number of parameters.
The ``$s$ wave (symmetric irreps)" fit is to two-kaon (three-kaon) levels in the $A_1$ ($A_2$) irreps only,
The ``$s$ wave (all irreps)" fit is to three-kaon levels in all irreps, but only to $A_1$ two-kaon irreps. 
The $s,d$ wave fits include all irreps for both two and
three kaons, except that level 1 from frame $P^2=6(2\pi/L)^2$ in irrep $A_2$ has been removed due to instabilities in the root-finding routine.
Fit parameters are as in the models described in \Cref{sec:models}, using the ERE2 form for the two-kaon interaction.
}
\vspace{-0.12cm}
\label{tab:kkk-params-E250}
\end{table*}

\begin{table*}[t!]
\centering
\begin{tabular}{|c|c|c|c|}
\hline
Ensemble &  \multicolumn{3}{c|}{E250}   \\ \hline \hline
Cutoffs & $2M_\pi$+$M_K$/$4M_\pi$/$3M_\pi+M_K$  &  $2M_\pi$+$M_K$/ $4M_\pi$/$3M_\pi+M_K$ & $2M_\pi+M_K$/$4M_\pi$/$3M_\pi+M_K$ 
\\ \hline
description &  $s$ wave (all irreps)  &$s$ wave (all irreps)  &  $s,p$ waves  \\ \hline
$\chi^2$ & 77.82 & 76.69 & 77.53 \\ \hline
DOF     & $21+24+23-4=64$ & $21 + 24 + 23 - 6=62$  &  $25 + 24 + 23 - 9=63$     \\ \hline 
p   & 0.115   & 0.099  &   0.103    \\ \hline \hline
$B_0^{\pi K}$ & -26.3(1.1) & -26.4(1.2)  & -25.7(1.2)     
\\ \hline
$B_1^{\pi K}$ & -4.81(65) & -4.72(67) & -4.95(70) 
\\ \hline
$z^2_{\pi K}$ & $1$ (fixed)& $1$ (fixed) & $1$ (fixed) 
\\ \hline 
$P_0^{\pi K}$ & $0$ (fixed) & $0$ (fixed)& $-4.2(6.6)\cdot 10^{-4}$ 
\\ \hline \hline
$B_0^{\pi\pi}$ & -23.2(2.4) & -23.1(2.6) & -21.9(2.5)    
\\ \hline
$B_1^{\pi\pi}$ & -3.78(1.14) & -3.75(1.2) & -4.2(1.2)  
\\ \hline
$z^2_{\pi\pi}$ & $1$ (fixed)& $1$ (fixed) & $1$ (fixed)  
\\ \hline \hline
$\mathcal K_0^{\pi\pi K}$ & 0 (fixed) & -160(180) & -200(240) 
\\ \hline
$\mathcal K_1^{\pi\pi K}$  &  0 (fixed) & 550(520) & 100(920) \\ \hline
$\mathcal K_B^{\pi\pi K}$ &  0 (fixed) & 0 (fixed) &  1800(2500)      \\ \hline
$\mathcal K_E^{\pi\pi K}$ &  0 (fixed) & 0 (fixed) &  0(1500)     \\ \hline
\end{tabular}
\caption{Results of fits to $\pi K+\pi\pi+\pi\pi K$ levels on ensemble E250, using pion units. 
Notation as in \Cref{tab:ppp-params-E250},
except that cutoffs and DOF were ordered as 
$\pi K$, $\pi\pi$, and $\pi\pi K$. 
Fit parameters as in the models described in \Cref{sec:models}, using the ADLER2 forms for both
$\pi\pi$ and $\pi K$ interactions.
}
\label{tab:ppK-params-E250}
\end{table*}

\begin{table*}[t!]
\centering
\begin{tabular}{|c|c|c|c|c|c|}
\hline
Ensemble & \multicolumn{3}{c|}{E250}  \\ \hline \hline
Cutoffs & $M_K$+2$M_\pi$/2$M_K$+$M_\pi$/2$M_K$+2$M_\pi$
& $M_K$+2$M_\pi$/2$M_K$+$M_\pi$/2$M_K$+2$M_\pi$
& $M_K$+2$M_\pi$/2$M_K$+$M_\pi$/2$M_K$+2$M_\pi$\\ \hline
description&   $s$ wave (all irreps)  & $s,p$ waves  & $s,p,d$ waves \\ \hline
$\chi^2$ &  179.15 &   191.79 & 204.42 \\ \hline
DOF & $21 + 28 + 50 - 6=93$     
& $25+28+50-9=94$  &  $25+40+50-10=105$  \\ \hline
p     &   $2\times 10^{-7}$ & $1.1\times 10^{-8}$  & $2.2\times 10^{-8}$ \\ \hline  \hline
$B_0^{\pi K}$ & -2.010(92)     & -1.95(10) &  -1.94(10)
\\ \hline
$B_1^{\pi K}$ & -3.59(75)    &-3.80(80)  & -3.77(79) 
\\ \hline
$z^2_{\pi K}$ & $1$ (fixed)&  $1$ (fixed)&  $1$ (fixed)
\\ \hline 
$P_0^{\pi K}$ & $0$ (fixed)&   0.031(38)&  0.011(38) 
\\ \hline \hline
$b_0^{KK}$  & -2.626(40)  &  -2.626(40)   & -2.599(40) 
\\ \hline
$b_1^{KK}$  & 0.66(18)&  0.67(18) & 0.53(19) 
\\ \hline
$D_0^{KK}$   & 0 (fixed) &   0 (fixed) & -0.042(25)
\\ \hline \hline
$\mathcal K_0^{KK\pi}$ &  -800(1200) &  -700(1200) & -600(1200)
\\ \hline
$\mathcal K_1^{KK\pi}$ &   -5100(6600) &  $-1(2) \cdot 10^4$& $-2(2) \cdot 10^4$  \\ \hline
$\mathcal K_B^{KK\pi}$  & 0 (fixed) &    $0.3(1.3) \cdot 10^4$ &  $0.8(1.3) \cdot 10^4$  \\ \hline
$\mathcal K_E^{KK\pi}$ & 0 (fixed) &    $-9(7) \cdot 10^4$&  $ -11(7) \cdot 10^4$  \\ \hline
\end{tabular}
\caption{Results of fits to $K \pi +KK+ K K\pi$ levels on ensemble E250, using kaon units.
Notation as in \Cref{tab:ppp-params-E250},
except that cutoffs and DOF were ordered as 
$K \pi$, $KK$, and $KK\pi$. 
Fit parameters as in the models described in \Cref{sec:models}, using the ADLER2 and ERE2 forms for
$\pi K$ and $KK$ K matrices, respectively. We stress that the numerical values of the $\pi K$ fit parameters cannot be directly compared to those in \Cref{tab:pk-params-E250,tab:ppK-params-E250}, since they are in pion units.
}
\label{tab:KKp-params-E250}
\end{table*}

\subsection{Two-body phase shifts}
\label{sec:phaseshifts}

\begin{figure*}[th!]
     \centering
     \subfloat[\label{fig:spp}]{%
     \includegraphics[width=0.49\textwidth]{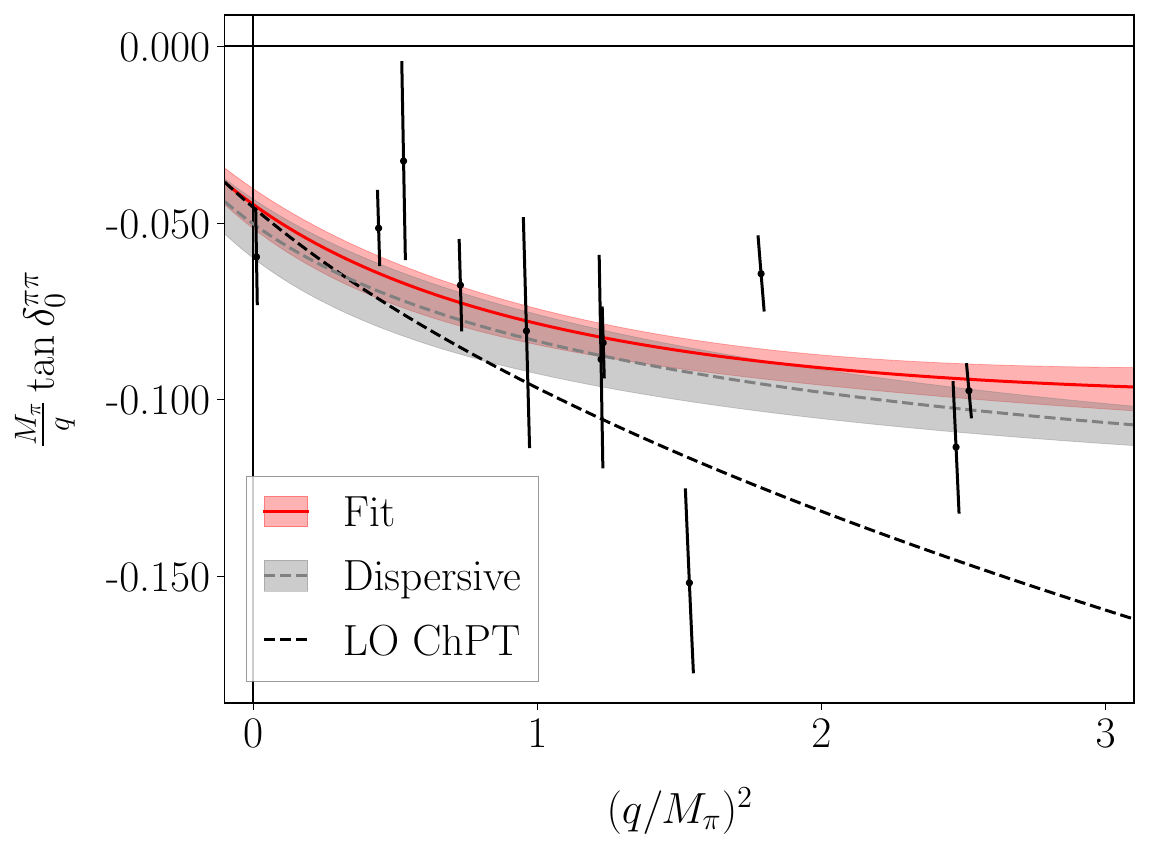}
    }
    \hfill
    \subfloat[\label{fig:dpp}]{%
     \includegraphics[width=0.49\textwidth]{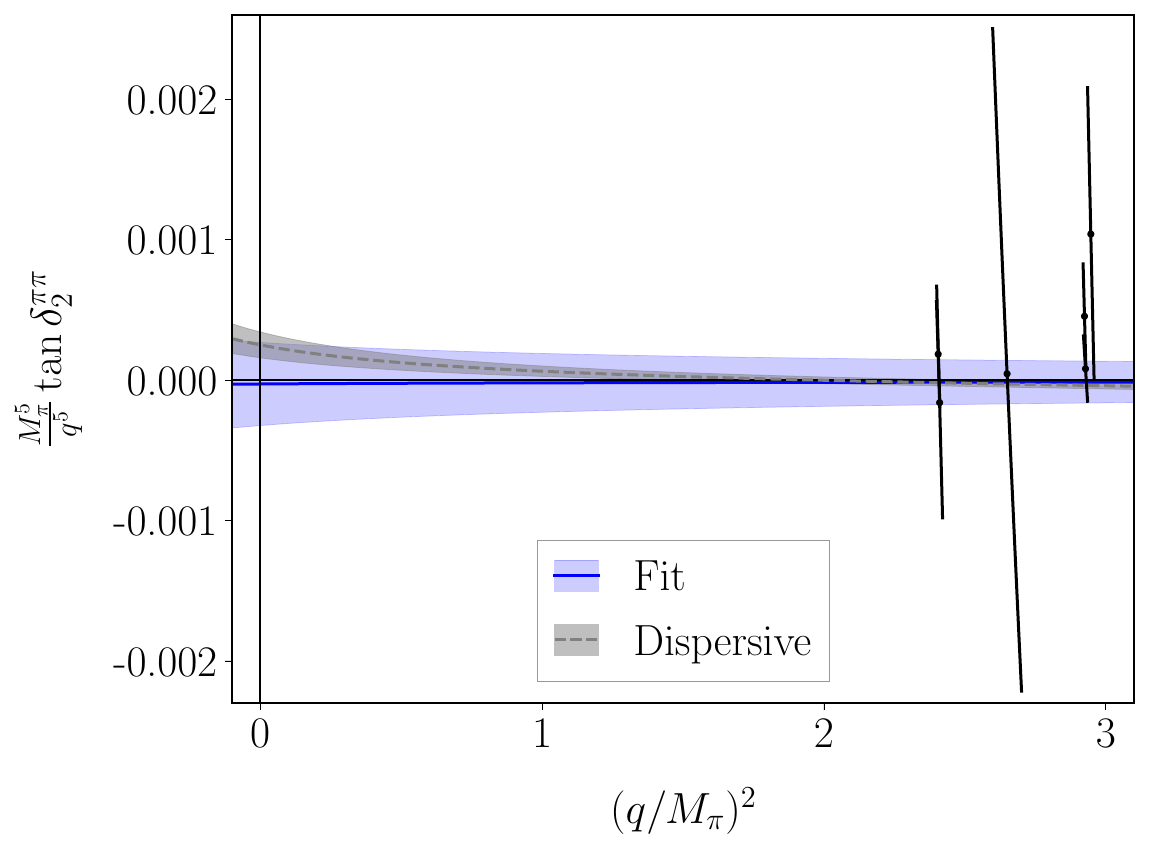}
    }
    \vspace{-0.12cm}
    \caption{Phase shift (plotted as $\tan \delta/q^{2\ell+1}$ as a function of the c.m.~frame momentum squared for the $\pi^+\pi^+$ system in an $s$ (left) or $d$ (right) wave. The results are for the E250 ensemble.
    The maximum value of $q^2$ corresponds to the inelastic (four pion) threshold.
    %All quantities are expressed in pion units. 
    The ``Fit'' results use two-meson parameters from the combined two- and three-hadron fit with $s$ and $d$ waves and including $\cK_B$ from \Cref{tab:ppp-params-E250}. 
    Dispersive results are from Ref.~\cite{Garcia-Martin:2011iqs}. The dashed line in the left panel gives the LO ChPT prediction. The data points are discussed in the text. 
    }
    \label{fig:pp_phase}
    \vspace{-0.2cm}
\end{figure*}

\begin{figure*}[th!]
     \centering
     \subfloat[\label{fig:skk}]{%
     \includegraphics[width=0.49\textwidth]{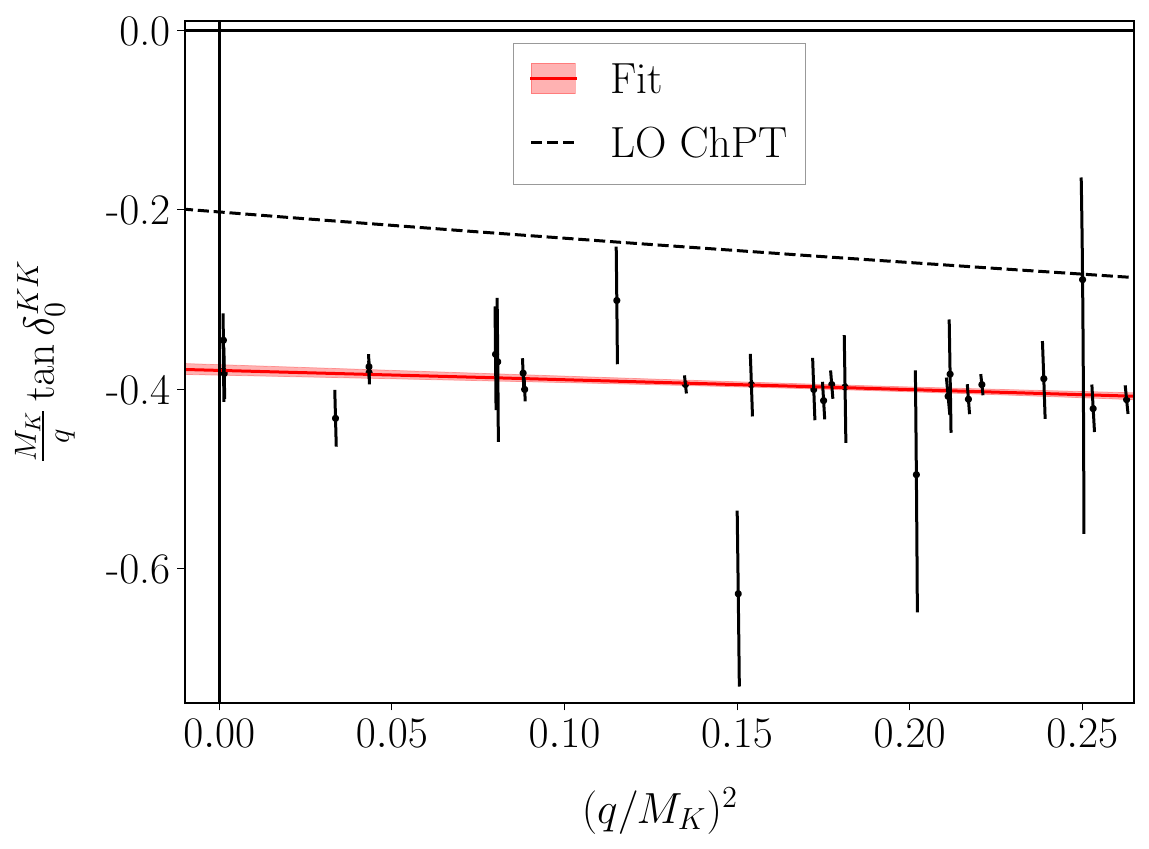}
    }
    \hfill
    \subfloat[\label{fig:dkk}]{%
     \includegraphics[width=0.49\textwidth]{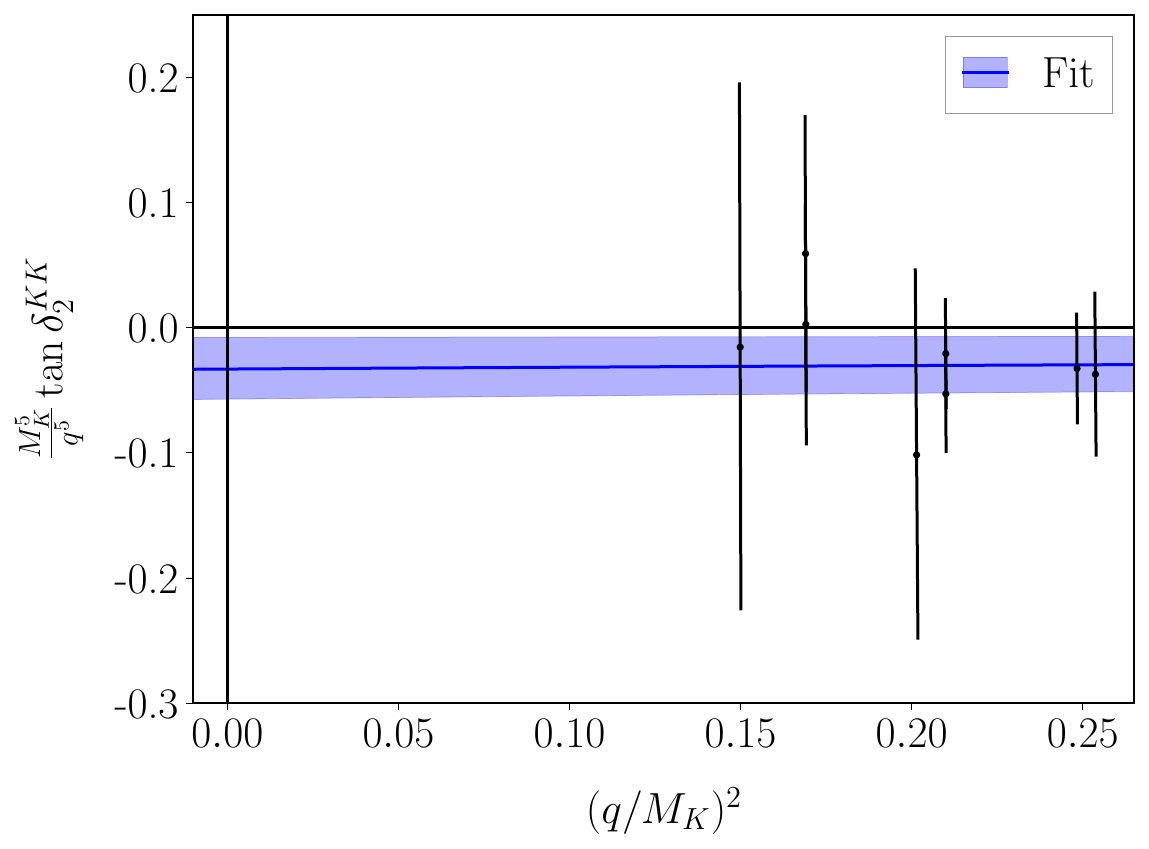}
    }
    \vspace{-0.12cm}
    \caption{Phase shifts for the $K^+K^+$ system in an $s$ (left) or $d$ (right) wave. Notation as in \Cref{fig:pp_phase}.
        The maximum value of $q^2$ corresponds to the inelastic ($KK\pi$) threshold.
    Two-meson parameters have been taken from the combined two- and three-hadron fit with $s$ and $d$ waves from \Cref{tab:kkk-params-E250}.}
    \label{fig:kk_phase}
    \vspace{-0.12cm}
\end{figure*}

\begin{figure*}[th!]
     \centering
     \subfloat[\label{fig:spk}]{%
     \includegraphics[width=0.49\textwidth]{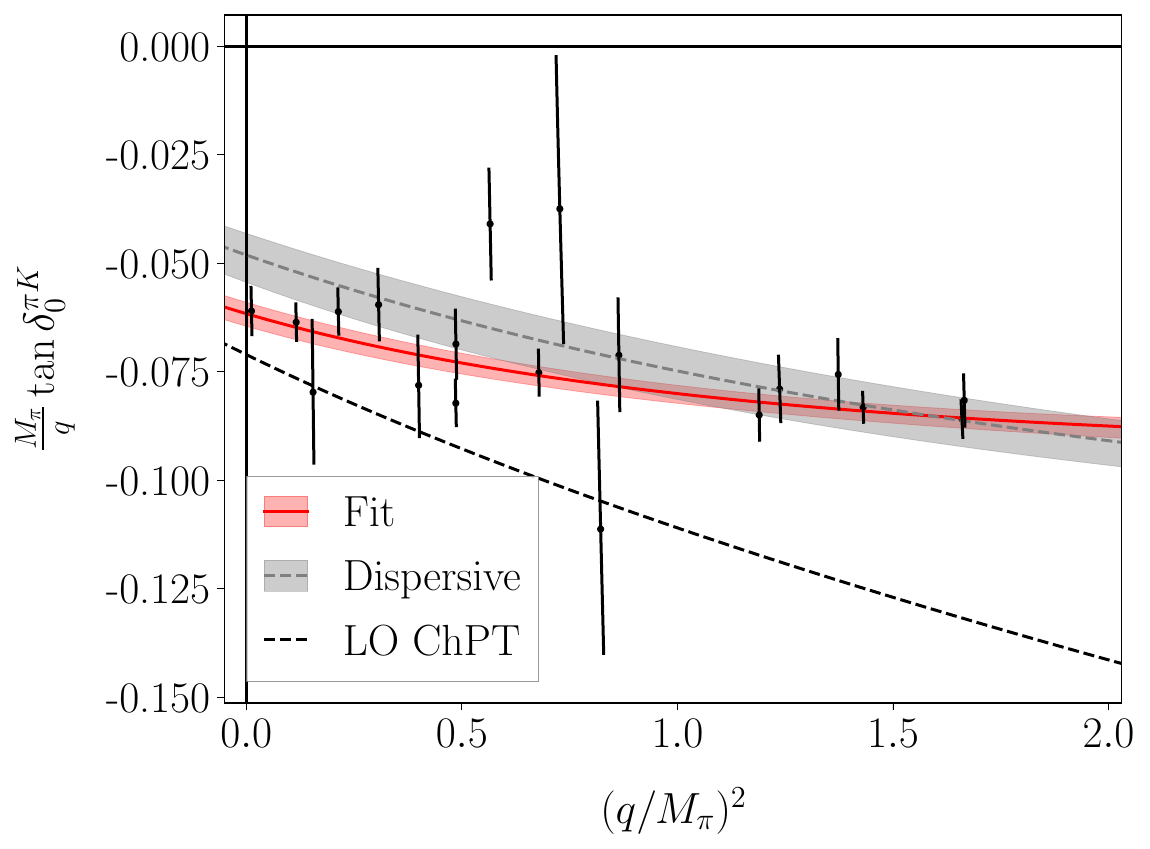}
    }
    \hfill
    \subfloat[\label{fig:ppk}]{%
     \includegraphics[width=0.49\textwidth]{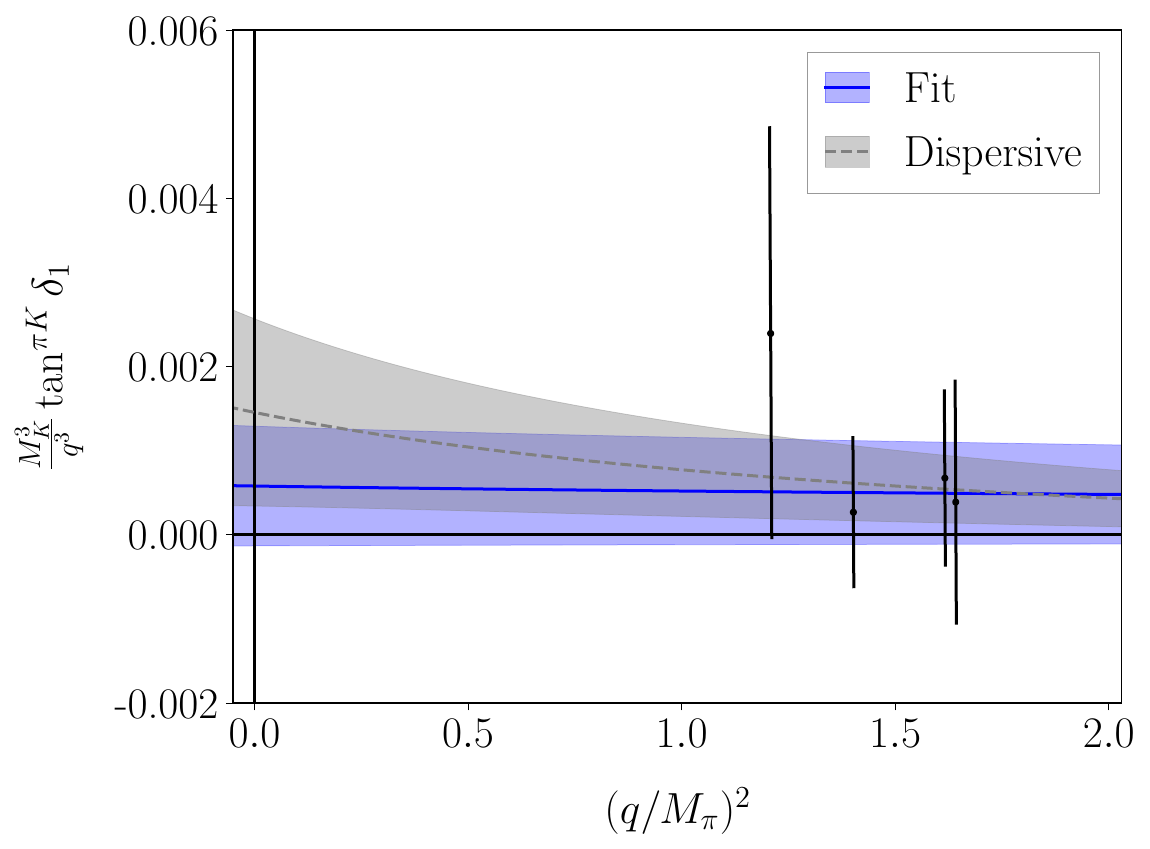}
    }
    \vspace{-0.2cm}
    \caption{Phase shifts for the $\pi^+ K^+$ system in an $s$ (left) or $p$ (right) wave. 
    The maximum value of $q^2$ corresponds to the inelastic ($\pi\pi K$) threshold.
    Two-meson parameters have been taken from the combined two- and three-hadron fit with $s$ and $p$ waves from \Cref{tab:ppK-params-E250} for $s$ wave, and from the combined two- and three-hadron fit with $s$ and $p$ waves from \Cref{tab:KKp-params-E250} for $p$ wave. Dispersive results are from Ref.~\cite{Pelaez:2020gnd}. 
    Other notation as in \Cref{fig:pp_phase}.}
    \vspace{-0.12cm}
    \label{fig:pk_phase}
\end{figure*}

In this section, we show plots of the two-meson phase shifts that follow from the fits presented in the previous section. These are plotted 
as a function of $q^2$, where $q$ is the pair c.m.~frame momentum, defined in \Cref{eq:q2def}. We plot $(1/q^{2\ell +1})  \tan \delta_\ell$, instead of the more conventional $q^{2\ell +1} \cot \delta_{\ell}$, since, for the weakly interacting systems that we study here, the phase shift can vanish, leading to poles in $\cot \delta_\ell$. 
Results for all channels are shown in \Cref{fig:pp_phase,fig:pk_phase,fig:kk_phase};
we discuss them in turn.

For the $\pi^+\pi^+$ system, we show results for both
$s$- and $d$-wave phase shifts in \Cref{fig:pp_phase}.
The colored bands denoted ``Fit'' are plots of (the inverse of) \Cref{eq:Adler} (left panel) and
\Cref{eq:cotdelta_d} (right panel), using the results for
$B_0^{\pi\pi}$, $B_1^{\pi\pi}$ and $D_0^{\pi\pi}$ 
from the $s,d$ fit including $\cK_B^{3\pi}$ in \Cref{tab:ppp-params-E250},
with the error bands incorporating the effects of correlations between these fit parameters.
The most important comparison in the plots is to the results from the dispersive analysis of Ref.~\cite{Garcia-Martin:2011iqs}, shown by the gray curves.
The latter can be considered as effectively the experimental result, since it incorporates experimental input constrained by analyticity and crossing symmetry.
As can be seen, our $s$-wave results agree with the dispersive analysis within approximately one standard deviation. Given that we have not accounted for the systematic errors due to finite lattice spacing or the fact that our pion mass differs slightly from the physical value, we consider this a good agreement.
We stress that the statistical uncertainty in our result is comparable to, or even smaller than, those in the dispersive result. We also show the LO ChPT result (see e.g. Ref.~\cite{Bijnens:2011fm}), which shows the expected behavior of agreement at the threshold and increasing discrepancies with increasing $q^2$.
By contrast, for the $d$-wave phase shift, our results are consistent with zero, with larger uncertainties than dispersive studies, but are nevertheless consistent with the dispersive results. Clearly, a substantial reduction in statistical errors will be needed to obtain a nonzero result for this quantity.

In both panels, we include also data points that require further explanation. If one uses the QC2 keeping only $s$-wave interactions, then there is a one-to-one correspondence between energy shifts of individual two-pion levels (in the $A_1$ irrep) and values of $\delta_0$. Using this, 
one obtains the data points in the left panel. We stress that the ``Fit'' curve  is based on substantially more information than that in the ``data'' points, and is not a fit to these points. Nevertheless, we display these points as they give an indication of the spread in values of $q^2$ accessed by our energy levels, and also of the reduction in errors obtained by fitting to multiple levels (both $\pi\pi$ and $3\pi$, in all irreps).
Similarly, for the $d$-wave plot, a one-to-one mapping can be obtained from two-pion levels in nontrivial irreps to the phase shift, and leads to the data points shown.  We note that these levels give information only at the largest values of $q^2$.

For the $KK$ system, shown in \Cref{fig:kk_phase}, no dispersive prediction is available.
Thus, to our knowledge, this is the first theoretical determination of maximal-isospin $KK$ scattering at the physical point. As can be seen, a significant deviation from LO ChPT is present. This breakdown is not unexpected for kaon system, and we return to it below when discussing other quantities.
For the $d$-wave phase shift, the LO ChPT prediction vanishes. We find a result that is nonzero at the $1-2\sigma$ level.
The data points in this figure result from one-to-one mapping using the QC2 applied to the two-kaon levels in $A_1$ irreps in the $s$-wave-only approximation for the left panel, and to those in nontrivial irreps in the $d$-wave-only approximation for the right panel.

Finally, \Cref{fig:pk_phase} shows the results for $s$- and $p$-wave $\pi^+K^+$ phase shifts, with dispersive results from Ref.~\cite{Pelaez:2020gnd}. 
We note that our statistical errors are substantially smaller than those of the dispersive analysis.
We observe some tension between our results and those from the dispersive analysis in the threshold region.
A possible explanation is that the dispersive analysis includes data that only starts at larger energies
(see Fig.~6 in Ref.~\cite{Pelaez:2020gnd}). 
Our $p$-wave results are consistent with zero, though slightly favor a positive scattering phase, a sign that agrees with the dispersive analysis.
In this plot, the data points are obtained from the $A_1$ irrep $\pi K$ levels (left panel) and those in nontrivial irreps (right panel).

\subsection{Two-meson threshold parameters}
\label{sec:threshold}

The fits of \Cref{sec:fits} provide multiple determinations of the scattering lengths and effective ranges in the two-meson systems, using \Cref{eq:a0ERE,eq:a0Adler,eq:r0Adler,eq:cotdelta_d,eq:cotdelta_p}.
We collect the results (with errors accounting for correlations between the underlying fit parameters) in \Cref{tab:ppparams-E250-table,tab:KKparams-E250-table,tab:pKparams-E250-table}, listed in \Cref{app:E2502meson}.

For the sake of comparison, we also include results for the scattering lengths obtained from the energy shift of a single energy level---the threshold state---applying a truncated form of the threshold expansion of ref.~\cite{Luscher:1986pf} (explicit expressions can be found in Ref.~\cite{Sharpe:2017jej}). 
Historically, the threshold expansion method was the first used to obtain estimates of the scattering lengths.
Our results illustrate the significant reduction in statistical error that one obtains by fitting to the extended spectrum.

We combine the scattering lengths and effective ranges using model averaging procedures~\cite{Borsanyi:2020mff,Jay:2020jkz,Neil:2023pgt,Pefkou:2021fni}.
Each determination of a single quantity can be assigned a weight $w_i$, normalized such that $\sum_i w_i=1$. The model-averaged value is then obtained using
\begin{equation}
    \langle \mathcal O \rangle = {\sum_i w_i \mathcal O_i },
\end{equation}
where $ \mathcal O_i $ is the estimate of the quantity $\mathcal O$ for the $i$th model. In the above equation, brackets indicate the model-averaged quantity. The uncertainty in a model-averaged quantity has two terms:
\begin{equation}
    \delta   \langle \mathcal O \rangle = \sqrt{ \delta   \langle \mathcal O \rangle^2_{\rm stat} + \delta \langle \mathcal O \rangle^2_{\rm syst}},
\end{equation}
where
\begin{align}
     \delta   \langle \mathcal O \rangle^2_{\rm stat} &= \sum_i w_i  \delta  \mathcal O_i^2, \\
     \delta   \langle \mathcal O \rangle^2_{\rm syst} &= \sum_i w_i \mathcal O_i^2 - \left( \sum_i w_i \mathcal O_i \right)^2,
\end{align}
where $\delta  \mathcal O_i$ is the statistical uncertainty in the $i$th model. 

One choice is to assign the weights based on the Akaike Information Criterion (AIC). 
Specifically, we use the result of Ref.~\cite{Jay:2020jkz};
\begin{equation}
    w_i = N \exp{(-{\rm AIC}/2)}\,, \quad {\rm AIC} = \chi^2 - 2 \, {\rm DOF}\,.
\end{equation}
  where the constant $N$ is chosen so that the weights are normalized.
    An alternative averaging procedure is given in Ref.~\cite{Pefkou:2021fni} in terms of the $p$ values of the various fits, $p_i$. Explicitly,
\begin{equation}
w_i = N' {\rm p_i}/{\delta  \mathcal O_i^2}\,,
\end{equation} 
where $N'$ is again a normalization constant.

For both of these choices, it is usually envisaged that one is averaging over a range of fits to similar quantities, with variations of choices of parameters and fit ranges. Our situation is somewhat different, given that we are combining fits to different data sets (e.g. $\pi\pi$ vs. $\pi\pi+3\pi$).
It is possible to bias the average by including more fits of one type than of another. To minimize this bias, we use roughly equal numbers of the best fits of each type.

The results from applying both the AIC- and $p$-based methods to our datasets are collected in
\Cref{tab:a0summary,tab:r0summary},
where we show statistical and systematic uncertainties separately. We stress that the latter is due only to the variation across fits. Discretization errors will be discussed below.
We have included also results from ensembles other than E250 using the model-averaging methodology. These results replace those given in refs.~\cite{Blanton:2021llb, Draper:2023boj} using a more primitive method.\footnote{%
%%%%%
The results for two-pion quantities are obtained from Ref.~\cite{Blanton:2021llb}, which used a different $\chi^2$ function based on the c.m. energies. 
See \Cref{app:reananlysis} for discussion of why the results remain valid.
}
Comparing the results using the two averaging methods, we see that the central values are consistent, usually well within the errors. The only exception is $a_0^{\pi K} r_0^{\pi K}$ on the E250 ensemble, where the difference approaches $3\sigma$. This is an example where a single measurement dominates the $p$-value average and is an outlier. There are also some small differences in the error estimates obtained by the two methods. In the plots below we use the results from the $p$-value averages. We compare these results to those in the literature in \Cref{sec:a0_ChPT} below.

\begin{table*}[h!]
\centering
\begin{tabular}{|c|c|c|c|c|c|c|}
\hline
Ensemble & $M_\pi a_0^{\pi\pi}$ &  $M_K a_0^{KK}$ & $M_\pi a_0^{\pi K}$ & $M_\pi^5 a_{2}^{\pi\pi}$& $M_K^5 a_{2}^{KK}$  & $M_\pi^3 a_{1}^{\pi K}$
\\ \hline
E250 (AIC)& 0.0436(49)(9) &  0.379(6)(2) & 0.0610(30)(05) & $0.2(3.0)(0.2) \cdot 10^{-4}$ & 0.039(17)(4) &  $-1.6(3.1)(0.1)\cdot 10^{-4}$ \\ \hline
E250 ($p$)& 0.0443(55)(12) & 0.379(6)(3) & 0.0603(29)(24) & $-1.0(3.2)(1.0) \cdot 10^{-4}$& 0.043(16)(5) &  $-1.1(4.3)(1.6)\cdot 10^{-4}$ \\ \hline
D200 (AIC) & 0.0886(52)(24) & 0.3677(52)(25) & 0.109(3)(0) & {0.0016(7)(0)} & 0.039(3)(0) & -0.0005(6)(0) \\ \hline
D200 ($p$) & 0.0881(51)(23) & 0.3676(54)(31) & {0.107(4)(0)} & {0.0016(7)(0)}& 0.039(3)(0) & -0.0012(7)(1) \\ \hline
N200 (AIC) &{0.1562(46)(13)} & {0.3375(48)(17)} & --- & {0.0042(9)(0)} & 0.0263(14)(0) & --- \\ \hline
N200 ($p$) & {0.1524(54)(40)}& 0.3370(49)(24) & --- &  {0.0038(10)(3)} & 0.0263(16)(0) & --- \\ \hline
N203 (AIC) &0.2083(41)(17) & {0.2970(43)(0)} & {0.212(5)(0)} & 0.0073(8)(1) & {0.019(1)(0)} & {-0.004(4)(0)}  \\ \hline
N203 ($p$) & 0.2082(49)(8) & 0.3012(46)(32) & 0.207(6)(2) & 0.0066(9)(3) & 0.019(1)(0) & -0.003(3)(1)  \\ \hline
\end{tabular}
\caption{
Final results for $s$-, $p$ and $d$-wave scattering lengths, denoted $a_0$, $a_{1}$ and $a_{2}$ respectively. Errors are respectively statistical and systematic, based on model averaging using the AIC and
$p$-value based methods.
The data used to obtain these averages is that
in \Cref{tab:ppparams-E250-table,tab:pKparams-E250-table,tab:KKparams-E250-table,tab:KKparams-reanalysis},
together (for the $\pi\pi$ results) 
with Tables 4-6 in Ref.~\cite{Blanton:2021llb}.
}
\label{tab:a0summary}
\end{table*}

\begin{table*}[h!]
\centering
\begin{tabular}{|c|c|c|c|}
\hline
Ensemble & $M_\pi^2 a_0^{\pi\pi} r_0^{\pi\pi}$ & $M_K^2 a_0^{KK} r_0^{KK}$ & $M_\pi^2 a_0^{\pi K} r_0^{\pi K}$
\\ \hline
E250 (AIC) & 2.62(15)(3)   & 0.54(19)(10) & 0.66(12)(04)
\\ \hline
E250 ($p$) &2.62(15)(3)  & 0.50(12)(8)  & 1.00(07)(12)
\\ \hline
D200 (AIC) & 2.42(26)(22) & 0.48(13)(12)  & 1.22(5)(0)
\\ \hline
D200 ($p$) & 2.58(13)(10) & 0.36(08)(10)  & 1.22(6)(0)
\\ \hline
N200 (AIC) & 2.35(16)(6) & 1.14(16)(14) & ---  
\\ \hline
N200 ($p$) & 2.38(13)(4) & 1.05(13)(17) & ---  
\\ \hline
N203 (AIC) &  {2.03(13)(9)}& {1.47(14)(1)} & {1.61(8)(1)}
\\ \hline
N203 ($p$) &  1.93(17)(21)& 1.21(12)(18) & 1.64(8)(2)
\\ \hline
\end{tabular}
\caption{
Final results for products $a_0 r_0$. Notation as
in \Cref{tab:a0summary}.
}
\vspace{-0.12cm}
\label{tab:r0summary}
\end{table*}

\subsection{Constraints on the three-meson K matrix}
\label{sec:kdfconstraints}

We now turn to $\kdf$, the K matrix that describes three-particle interactions,
focusing on the statistical significance by which it differs from zero.
Previous work has found evidence for nonzero values of $\kdf$ for some systems and pion masses; here we extend the study to near-physical masses, and also present the global picture from all four ensembles.
In this regard, it is important to keep in mind that $\kdf$ is not a physically measurable quantity, since it depends on the choice of cutoff function.
For that reason, if we wish to investigate the quark mass dependence of $\Kdf$, we must ensure that the cutoff function itself varies smoothly with quark mass.
This is one reason why we have repeated the determination of $\Kdf$ from $KK\pi$ and $3K$ systems on the ensembles other than E250 using the same class of cutoff function as that we use on E250 (see \Cref{app:reananlysis}).

Given the cutoff dependence of $\Kdf$, it might seem that there is nothing special about the point $\Kdf=0$. This is not the case for weakly interacting systems, however. In particular, in ChPT, the LO contribution is scheme-independent, as discussed in refs~\cite{Blanton:2021llb,Draper:2023boj}. 
Thus, at this order, a nonzero value of $\Kdf$ is physically meaningful.
Scheme dependence enters only at NLO, at which order, for example,
the contributions for the $3\pi^+$ system to the coefficients
$\cK_0$, $\cK_1$, $\cK_A$, and $\cK_B$ in $\Kdf$ (see in \Cref{eq:Kdfidentical})
are of $\cO(M_\pi^6/F_\pi^6)$.
In the case of $3\pi^+$, this scheme dependence has been explicitly calculated and is numerically small~\cite{Baeza-Ballesteros:2023ljl}.
We stress that, when we show comparisons with ChPT below in \Cref{sec:kdf_chiral}, we use the same cutoff function in ChPT as we do in the QC3.

Another reason for considering the statistical significance of results from $\kdf$
is that being able to extract nonzero results is numerically challenging, and we wish to gauge how accurately one needs to determine energy levels in order to tease out values for $\kdf$. The challenge arises from the fact that a three-particle interaction leads to energy shifts that generically scale like $1/L^6$, much suppressed compared to the effects of two-particle interactions that scale as $1/L^3$. 

Thus motivated, we turn to results.
The tables of fits given above, \Cref{tab:ppp-params-E250,tab:kkk-params-E250,tab:ppK-params-E250,tab:KKp-params-E250}, provide errors for individual components of $\kdf$.
In ChPT, the first two constants appearing in the threshold expansion for identical-particle systems, \Cref{eq:Kdfidentical}, behave at LO as~\cite{Blanton:2021llb}
\begin{equation}
\cK_0^{3\pi}, \cK_1^{3\pi} \propto x_\pi^4\,, \ \
\cK_0^{3K}, \cK_1^{3K}\propto x_K^4\,,
\label{eq:K0K1IDChPT}
\end{equation}
where $x_\pi=M_\pi/F$, $x_K=M_K/F$, and $F$ can be either $F_\pi$ or $F_K$ at the
considered accuracy.
For mixed systems, the corresponding constants in \Cref{eq:KdfExpansion} behave as~\cite{Blanton:2021eyf}
\begin{equation}
\begin{split}
\cK_0^{\pi\pi K} &\propto x_\pi^3 x_K\,,\ \ 
\cK_1^{\pi\pi K} \propto x_\pi^2 x_K^2\,,
\\
\cK_0^{KK\pi} &\sim \cK_1^{KK\pi} \propto x_K^4 \,.
\end{split}
\label{eq:K0K1mixedChPT}
\end{equation}
Here we have dropped contributions suppressed by powers of $M_\pi/M_K$.
Higher order terms in the threshold expansion appear at NLO in ChPT (explicit results only available for $3\pi$~\cite{Baeza-Ballesteros:2023ljl}).
For the other systems, we note that, in the chiral limit, contributions from kaon loops (or insertions of $m_s$) dominate over those from pion loops (or insertions of $m_u$ or $m_d$).
This leads to the expectations
\begin{align}
\cK_2^{3\pi} &\sim \cK_A^{3\pi} \sim  \cK_B^{3\pi} \propto x_\pi^6\,, 
\label{eq:KB3piChPT}
\\ 
\cK_2^{3K} &\sim \cK_A^{3K}\sim \cK_B^{3K} \propto x_K^6 \,, \\
\cK_B^{\pi\pi K} &\sim \cK_E^{\pi\pi K} \propto x_\pi^2 x_K^4\,,\\
\cK_B^{KK\pi} &\sim \cK_E^{KK\pi} \propto x_K^6\,.
\end{align}
Thus, on the E250 ensemble, for which $x_\pi^2 \ll x_K^2$, the systems in which we are most likely to find significant signals are $3K$ and $KK\pi$.
In fact, we find that the only components with values differing from zero by more than $2\sigma$ are $\cK_1^{3K}$ and $\cK_B^{3K}$.

This is not the full story, however, as the tables
do not display the (typically rather large) correlations between these components. 
Thus it is instructive to determine the significance of
the full $\kdf$, and of its isotropic component $\mathcal K^{\rm iso}$, including these correlations.
These results are collected in \Cref{tab:Kdfsignificance,tab:Kisosignificance}. Since we are assuming a multivariate Gaussian distribution, the confidence interval of a given number of $\sigma$s can differ substantially from that for a single variable. For instance, in a two-variable distribution, $1\sigma$ means 39.3\% rather than the usual 68.3\% of a single-variable Gaussian distribution. Thus, in addition to the number of $\sigma$s, we quote the significance, and also the effective single-variable number of sigmas, which we denote $\sigma_{\rm eff}$.
The overall conclusion is that, on the E250 ensemble,
the only significant results for $\kdf$ are for the
$3K$ and $KK\pi$ systems, which is in qualitative accord with the ChPT expectations given above. For $\cK^{\rm iso}$, only the $3K$ system has a significant result on the E250 ensemble.
We also observe, in almost all cases, that, as the pion mass increases the significance of $\kdf$ and $\cK^{\rm iso}$ on all ensembles for all systems increases, again consistent with the general expectations of ChPT.

\begin{table*}[h!]
\centering
\begin{tabular}{|c|c|c|c|c|}
\hline
Ensemble & Fit & $\sigma$ &
Significance & $\sigma_{\rm eff}$
\\ \hline \hline
E250 & $3\pi$, $s+d$ & 1.2 &  0.30 & 0.4
\\ \hline 
D200 & $3\pi$, $s+d$ & 1.0 &  0.19 & 0.3
\\ \hline 
N200 & $3\pi$, $s+d$ & 3.8 &  1.00 & 3.0
\\ \hline 
N203 & $3\pi$, $s+d$ & 4.5 &  1.00 & 3.8
\\ \hline \hline
E250 & $3K$, $s+d$ & 11.8 & 1.00 & 11.4
\\ \hline
D200 & $3K$, $s+d$ & 8.8 & 1.00 & 8.3
\\ \hline 
N200 & $3K$, ERE3, $s+d$ & 10.7 & 1.00 & 10.2
\\ \hline
N203 & $3K$, ERE4, $s+d$ & 11.6 & 1.00 & 11.2
\\ \hline \hline
E250 & $\pi\pi K$, $s+p$ & 1.2 & 0.16 & 0.2
\\ \hline
D200 & $\pi\pi K$, $s+p$ & 3.3 & 0.97 & 2.2
\\ \hline
N203 & $\pi\pi K$, $s+p$ & 3.6 & 0.99 & 2.5
\\ \hline\hline
E250 & $KK\pi$, $s+p+d$ & 3.5 & 0.99 & 2.5
\\ \hline
D200 & $KK\pi$, $s+p+d$ & 5.9 & 1.00 & 5.0
\\ \hline
N203 & $KK\pi$, $s+p+d$ & 1.6 & 0.37 & 0.5
\\ \hline
\end{tabular}
\caption{
Significance of $\Kdf$ in various fits. The chosen $3\pi$ and $3K$ fits are those using a three-parameter form for $\Kdf$, using $\cK^{\rm iso,0}$, $\cK^{\rm iso,1}$ and $\cK^B$. 
The chosen $\pi\pi K$ and $KK \pi$ fits use a four-parameter form for $\Kdf$, using using $\cK^{\rm iso,0}$, $\cK^{\rm iso,1}$, $\cK^B$, and $\cK^E$.
Results for $3\pi$, and $\pi\pi K$ fits on ensembles
other than E250 are from Refs.~\cite{Blanton:2021llb} and \cite{Draper:2023boj}, respectively.
The significance is given by the number of $\sigma$ in the
three- or four-parameter Gaussian form, by the numerical significance, and by the equivalent single-parameter Gaussian $\sigma_{\rm eff}$.
See text for further discussion.
}
\vspace{-0.12cm}
\label{tab:Kdfsignificance}
\end{table*}

\begin{table*}[th!]
\centering
\begin{tabular}{|c|c|c|c|c|}
\hline
Ensemble & Fit & $\sigma$ &
Significance & $\sigma_{\rm eff}$
\\ \hline \hline
E250 & $3\pi$, $s+d$ & 1.2 &  0.51 & 0.7
\\ \hline 
D200 & $3\pi$, $s+d$ & 0.8 &  0.26 & 0.3
\\ \hline 
N200 & $3\pi$, $s+d$ & 3.8 &  1.00 & 3.4
\\ \hline 
N203 & $3\pi$, $s+d$ & 4.5 &  1.00 & 4.2
\\ \hline \hline
E250 & $3K$, $s+d$ & 8.0 & 1.00 & 7.7
\\ \hline
D200 & $3K$, $s+d$ & 7.8 & 1.00 & 7.5
\\ \hline 
N200 & $3K$, ERE3, $s+d$ & 7.5 & 1.00 & 7.2
\\ \hline
N203 & $3K$, ERE4, $s+d$ & 10.1 & 1.00 & 9.8
\\ \hline \hline
E250 & $\pi\pi K$, $s+p$ & 0.9 & 0.33 & 0.4
\\ \hline
D200 & $\pi\pi K$, $s+p$ & 2.8 & 0.98 & 2.3
\\ \hline
N203 & $\pi\pi K$, $s+p$ & 3.1 & 0.99 & 2.6
\\ \hline\hline
E250 & $KK\pi$, $s+p+d$ & 1.3 & 0.57 & 0.8
\\ \hline
D200 & $KK\pi$, $s+p+d$ & 5.0 & 1.00 & 4.6
\\ \hline
N203 & $KK\pi$, $s+p+d$ & 0.5 & 0.12 & 0.15
\\ \hline
\end{tabular}
\caption{
Significance of $\cK^{\rm iso}$ in various fits. The chosen fits are as in \Cref{tab:Kdfsignificance},
in all of which a two-parameter form is used for $\cK^{\rm iso}$.
The significance is given by the number of $\sigma$ in the
two-parameter Gaussian form, by the numerical significance, and by the equivalent single-parameter Gaussian $\sigma_{\rm eff}$.
}
\vspace{-0.12cm}
\label{tab:Kisosignificance}
\end{table*}

\begin{figure}[h!]
    \includegraphics[width=8cm]{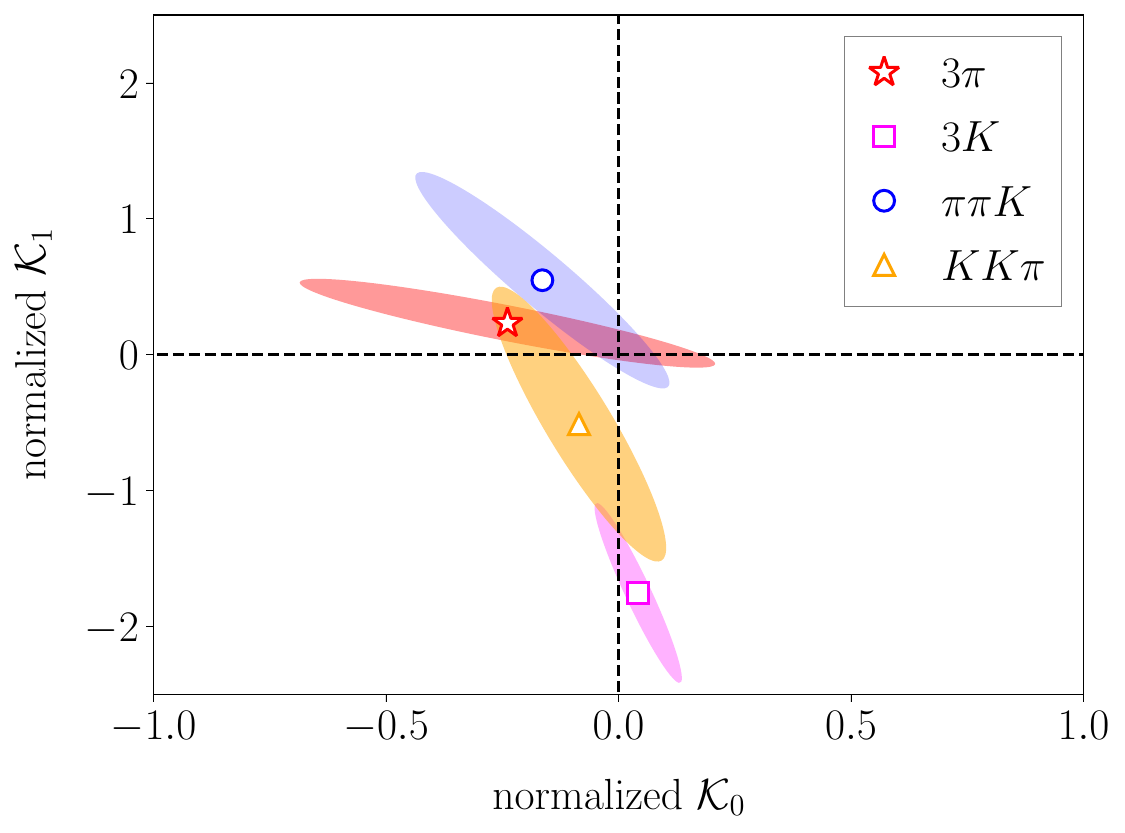}
    \vspace{-0.12cm}
    \caption{Representation of the significance of the isotropic component $\kdf$ for the three-meson systems. 
    The ellipses correspond to the 68.3\% confidence interval.
    The normalizations of $\cK_0$ and $\cK_1$ have been rescaled (with different factors for the different three-meson systems) so that they all fit in a single figure. Thus, only the distance to the origin is relevant. 
    }
    \vspace{-0.12cm}
    \label{fig:Kdfconstrain}
\end{figure}

We close this section by presenting \Cref{fig:Kdfconstrain},
which displays the significance of $\cK^{\rm iso}$ for all four systems on the E250 ensemble, in a way that illustrates graphically the impact of correlations.
This plot is based on results from the ``$s$ wave (all irreps)'' fits in \Cref{tab:ppp-params-E250,tab:kkk-params-E250,tab:ppK-params-E250,tab:KKp-params-E250}.
We find that, while for the $3\pi$ and $\pi\pi K$ systems
$\cK^{\rm iso}$ has a significance of only $\approx 65\%$, that for the $3K$ and $KK\pi$ systems is much higher.
Again, this is broadly consistent with the expectations of ChPT.

\section{Comparisons to ChPT}
\label{sec:comp_ChPT}

The main goal of this section is to compare our results for two- and three-particle scattering parameters with the pion-mass dependence predicted by ChPT.
For this, we use the results from all four ensembles that we have collected above.
The fits presented here generalize those of Refs.~\cite{Blanton:2021llb, Draper:2023boj}
by the inclusion of our new results near the physical point, and by the use of the updated fits presented in \Cref{app:reananlysis}.

While our results are at a single lattice spacing, we estimate the discretization effects using ChPT augmented by discretization terms. Moreover, while the E250 ensemble is very close to physical quark masses, some small mistuning effects can be present. We estimate their size using continuum ChPT.

For some two-meson quantities, we can compare our results at physical quark masses to those in the literature, 
many of which are collected in the FLAG review~\cite{FlavourLatticeAveragingGroupFLAG:2024oxs}.

In the fits below, we use the ratios $M_\pi/F_\pi$ from 
\Cref{tab:masses_decay_constants} for the independent variables.
We neglect their uncertainties, as they are significantly smaller than those in the quantities being fit.
We also need the ratios $M_K/F_K$ from this table in some fits.
For $SU(3)$ ChPT fits to results involving $\pi\pi$, $\pi K$ and $KK$ quantities, we neglect the correlations between these quantities on individual ensembles.

\subsection{$s$-wave scattering lengths}
\label{sec:a0_ChPT}

\begin{figure}[h!]
    \includegraphics[width=8cm]{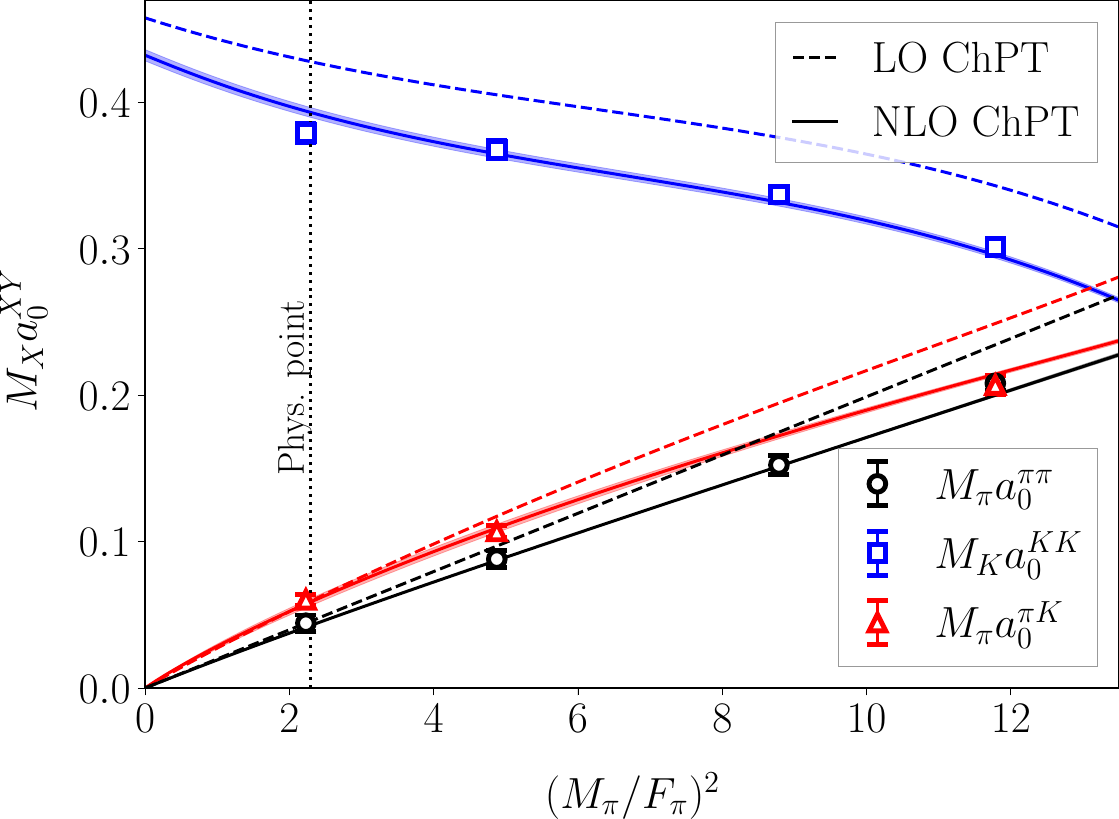}
    \vspace{-0.15cm}
    \caption{Two-meson $s$-wave scattering lengths for all four ensembles used in this work. The LO ChPT prediction, and a NLO ChPT fit are included. In order to plot the NLO curve, an interpolating function for $M_K/M_\pi$ and $M_K/F_K$ as a function of $M_\pi/F_\pi$ has been constructed using \Cref{tab:masses_decay_constants}.
    The physical point value in $M_\pi/F_\pi$ is included as a vertical line. }
    \label{fig:a0all}
    \vspace{-0.12cm}
\end{figure}

We begin with the $s$-wave scattering lengths of the two-meson systems,  results for which are displayed in \Cref{fig:a0all}. We use the averages in \Cref{tab:a0summary} obtained with the $p$-value method. 
We fit all eleven results to the NLO prediction of $SU(3)$ ChPT 
[given in Eqs.~(3.14)-(3.21) of Ref.~\cite{Draper:2023boj}], 
which depends on two low-energy coefficients (LECs). 
The fit results are
\begin{align}
\begin{split}
     L_5(4\pi F_\pi) &= 0.1(7) \cdot 10^{-3}\,, 
     \\  
     L_{\pi\pi}(4\pi F_\pi) &= -0.98(2) \cdot 10^{-3}\,, \\
      \chi^2/{\rm DOF} &= 12.4/(11\!-\!2)\,,\ \ p = 0.19\,,
\end{split}
\label{eq:L5ChPT}
\end{align}
where, as indicated, LECs are quoted at the scale $\mu = 4\pi F_\pi$.\footnote{%
This choice of scale is advantageous in chiral fits, as it allows one to express all quantities as functions of $M_\pi/F_\pi$ or $M_K/F_K$. It induces some mass dependence in the chiral logarithms, but this is of higher order in the chiral expansion. 
}
The resulting fits are shown in the figure, along with the LO ChPT prediction.
Note that only for the $\pi\pi$ and $\pi K$ scattering lengths do we expect the LO and NLO results to agree in the chiral limit.
The results for the LECs improve on those found in Ref.~\cite{Draper:2023boj},
which included only the D200 and N203 ensembles, and did not include $d$
 waves in the $\pi\pi$ and $KK$ subchannels.
The results were $L_5=0.0(1.5)\cdot 10^{-3}$ and $L_{\pi\pi}=-0.88(4) \cdot 10^{-3}$.
Thus the inclusion of the physical-point data leads to a halving of the errors, and, in the case of $L_{\pi\pi}$, to a shift in the central value by about $2\sigma$.
We can also compare $L_{\pi\pi}$ to that obtained in
Ref.~\cite{Blanton:2021llb}, where the fit was to $\pi\pi$ and $KK$ data only on the D200, N200, and N203 ensembles, and gave $L_{\pi\pi}=-1.13(3) \cdot 10^{-3}$.
The  nearly $4\sigma$ shift in the central value 
suggests that the fitting systematics were underestimated, and again shows the importance of the inclusion of the physical-point data.

Discretization effects in these results can be estimated by introducing additional LECs into ChPT~\cite{Sharpe:1998xm}---for Wilson-like fermions this method is Wilson ChPT (WChPT)~\cite{Bar:2003mh}.
The necessary results are presented in Ref.~\cite{Draper:2023boj}, 
and it turns out that, at LO in discretization errors, only a single combination of LECs enters for all three scattering lengths. We denote this combination $W=2w_6' + w'_8$.
The inclusion of $W$ leads to nonvanishing values for the $\pi\pi$ and $\pi K$ scattering lengths in the chiral limit.
Fitting to the expressions given in Ref.~\cite{Draper:2023boj} 
[see Eqs.~(3.29)-(3.31) of that work]
we find
\begin{align}
\begin{split}
     L_5(4\pi F_\pi) &= -0.3(9) \cdot 10^{-3}, \\  L_{\pi\pi}(4\pi F_\pi) &= -0.94(4) \cdot 10^{-3}, \\
       W&=-0.22(17) ,\\
      \chi^2/{\rm DOF} &= 10.7/(11\!-\!3),\ \ p=0.22\,.
\end{split}
\label{eq:L5WChPT}
\end{align}
The fit is marginally improved by the inclusion of $W$ 
[compare to \Cref{eq:L5ChPT}],
although the resulting value lies only slightly more than $1\sigma$ away from zero.
The errors on the results for $L_5$ and $L_{\pi\pi}$ increase, while the central values are consistent.
Although the inclusion of discretization errors is only approximate (as higher-order terms in $a^2$ are dropped), the comparison of the ChPT and WChPT fits suggests that discretization errors are small relative to the statistical errors for $L_5$,
and comparable to the statistical errors for $L_{\pi\pi}$.
We take the results from \Cref{eq:L5WChPT} as our best estimate of these quantities.

Our result for $L_5$ can be compared to those in the latest FLAG review~\cite{FlavourLatticeAveragingGroupFLAG:2024oxs}.
In this regard, it is important to note that the standard method of determining $L_5$ uses the NLO prediction for $f_K/f_\pi$, a quantity that involves only single particles and thus, in lattice calculations, has much smaller errors than our determination.
Thus the comparison with FLAG serves as a cross-check on our methods. 
FLAG quotes LECs at the conventional value for the renormalization scale, $\mu=770\;$MeV; our result from \Cref{eq:L5WChPT} run to this scale is $10^3 L_5(\mu = 770~{\rm MeV}) = 0.7(9)$. 
The lattice determinations quoted in FLAG are from Ref.~\cite{MILC:2010hzw} ($2+1$ flavors), which finds $10^3 L_5=0.95(41)$, and from Ref.~\cite{Dowdall:2013rya} ($2+1+1$ flavors), which finds $10^3 L_5=1.19(25)$.
We see that our result is consistent with these, but has larger errors, as expected.

As is clear from \Cref{eq:L5ChPT,eq:L5WChPT},
we are able to determine $L_{\pi\pi}$ with good precision:
the combined statistical and (approximate) discretization error is less than 5\%.
This LEC is related to the standard set of $SU(3)$ LECs by
\begin{equation}
L_{\pi\pi} = 2 L_1 + 2 L_2 + L_3 - 2 L_4 - L_5 + 2 L_6 + L_8\,,
\end{equation}
and to the quantity $L_{\rm scatt}$ used in the FLAG report by
\begin{equation}
L_{\pi\pi} = L_{\rm scatt} - \tfrac12 L_5\,.
\end{equation}
To our knowledge, the only determination of $L_{\rm scatt}$ is that given in Ref.~\cite{Helmes:2018nug}, where the result $10^3 L_{\rm scatt}(\sqrt2 F_\pi)=3.8(3)$
is quoted (using the notation $L_{\pi K}$ instead of $L_{\rm scatt}$). Running our result to other scales we find
\begin{equation}
    \begin{split}
    L_{\pi\pi}(\mu = 770\, {\rm MeV}) = -0.44(4)\cdot 10^{-3}\,,
    \\
    L_{\pi\pi}(\mu = \sqrt2 F_\pi) = 1.75(4) \cdot 10^{-3}\,.
    \end{split}
\end{equation}
If we determine $L_{\rm scatt}$ using the FLAG estimate $10^3 L_5=1.19(25)$, as was done in Ref.~\cite{Helmes:2018nug}, then we obtain
\begin{equation}
    \begin{split}
    L_{\rm scatt}(\mu = 770\, {\rm MeV}) = 0.16(13)\cdot 10^{-3}\,,
    \\
    L_{\rm scatt}(\mu = \sqrt2 F_\pi) = 4.45(13) \cdot 10^{-3}\,.
    \end{split}
\end{equation}
Thus our result is consistent with that of Ref.~\cite{Helmes:2018nug} at the $2\sigma$ level. We stress that the dominant error in $L_{\rm scatt}$ comes from that in $L_5$ (even when using the FLAG result), and so advocate quoting $L_{\pi\pi}$ instead.

We can also fit the pion data to the NLO $SU(2)$ ChPT expression.
This involves a single combination of LECs, $\ell_{\pi\pi}$---see Eq.~(3.15) of Ref.~\cite{Draper:2023boj}---where here the notation matches that in the FLAG report.
In this case, we use the same renormalization scale as in FLAG, namely 
$\mu=\sqrt2 F_\pi$. 
The fit to the continuum NLO ChPT form yields
\begin{equation}
    \ell_{\pi\pi} = 7.4(5)\,,\ \chi^2/{\rm DOF} = 0.43/(4\!-\!1)\,,\ p=0.93\,,
    \label{eq:lppChPT}
\end{equation}
while that to the corresponding WChPT expression, which includes the same new LEC $W$ as in the $SU(3)$ fits above, gives
\begin{equation}
    \begin{split}
    \ell_{\pi\pi} &= 7.5(8)\,,\  W=-0.04(22)\,,
    \\
    \chi^2/{\rm DOF} &= 0.40/(4\!-\!2)\,,\ p=0.82\,.
    \end{split}
    \label{eq:lppWChPT}
\end{equation}
Here we see no evidence for discretization errors---the chiral extrapolation is consistent with $a_0^{\pi\pi}$ vanishing in the chiral limit.
Nevertheless, to be conservative, we take the WChPT fit result as our preferred value.
To determine the impact of the new E250 result at the physical point, we have repeated the fits dropping this point, finding $\ell_{\pi\pi}=7.4(5)$ for the ChPT fit,\footnote{%
%%%%
This differs slightly from the result $\ell_{\pi\pi}=7.6(4)$ quoted in Ref.~\cite{Blanton:2021llb} for the same fit, the difference being due to our new method of averaging fits.
}
and $\ell_{\pi\pi}=7.2(1.0)$ and $W=0.07(32)$ for the WChPT fit.
Thus including the physical point makes no difference if one constrains the fit to have the correct chiral behavior, but does lead to a reduction in the error if one leaves the intercept free.
 FLAG finds that only the calculation of Ref.~\cite{Helmes:2015gla} meets their criteria for inclusion in their average for $\ell_{\pi \pi}$. The result from this work is $\ell_{\pi\pi}=3.79(61)({}^{1.34}_{-0.11})$, which differs from our results by $2-3\sigma$. There are several differences between our work and Ref.~\cite{Helmes:2015gla} regarding discretization effects, finite-volume effects, and quark masses. Moreover, Ref.~\cite{Helmes:2015gla} used the ground-state energy, while we use the full spectrum. Thus, it is difficult to isolate the source of the discrepancy.

The fits described above allow us to determine the scattering lengths at the physical point. We do so by inserting into our chiral fits the physical masses and decay constants, which we take to be
\begin{equation}
\begin{split}
M_\pi&= 139.57018\,{\rm MeV}\,,\
F_\pi= \frac{130.41\, {\rm MeV}}{\sqrt2}\,,\\ 
M_K &= 493.677\,{\rm MeV}\,,\ 
F_K = \frac{156.1\, {\rm MeV}}{\sqrt2}\,.
\end{split}
\label{eq:physicalMpi}
\end{equation}
The results are collected in \Cref{tab:physa0}. We also include in the table the FLAG estimates, along with the original references. 
These are ``estimates'' rather than ``averages'' since only a single result 
satisfies the FLAG criteria for each of the quantities.

In \Cref{tab:physa0}, we also include our ``direct'' results from the E250 ensemble. These have larger error bars than those obtained from chiral fits. They also have slightly mistuned quark masses. In order to estimate the mistuning error, we use LO ChPT,
\begin{equation}
\begin{split}
    \Delta_{m_q}(M_X a_0^{XX}) &= \frac1{16\pi}\Delta(x_X^2)\,,
    \\
    \Delta_{m_q}(M_\pi a_0^{\pi K}) &= \frac1{8\pi} 
    \Delta\left( \frac{x_\pi x_K}{1+M_K/M_\pi} \right)\,,
    \end{split}
\end{equation}
with $X \in \{ \pi, K\}$, $x_X = M_X/F_X$, 
and 
\begin{equation}
    \Delta[Q] = {\rm abs}(Q_{\rm E250} - Q_{\rm phys}),
\end{equation}
where E250 labels the values in \Cref{tab:masses_decay_constants} and ``phys'' the values in \Cref{eq:physicalMpi}.
Although this leads to a shift of definite sign, we treat it as a two-sided uncertainty.
The results are included in \Cref{tab:physa0}. As can be seen, the size of this effect turns out to be most significant for the $KK$ system, 
indicating that, for $a_0^{KK}$, results from the chiral fits provide better estimates. 

We conclude this section with a summary of our findings from the chiral fits of $s$-wave scattering lengths. We have computed results for the three scattering lengths ($a_0^{\pi\pi}$, $a_0^{\pi K}$ and $a_0^{KK}$ ). While our results are at a single lattice spacing, we have investigated discretization effects using WChPT, and our findings suggest that they are small. We have also estimated the effects of the quark-mass mistuning on our direct E250 results, finding them to be small except for the $KK$ system.
Incorporating this additional error, we find good consistency between direct results
and those from the chiral fits.
A final comment on the small size of the uncertainties in the $s$-wave $\pi\pi$ scattering lengths is appropriate. We stress that these include only statistical effects, 
and lack a complete error budget. Uncontrolled sources of error are those due to higher-order ChPT corrections, discretization errors, and exponentially suppressed finite-volume effects.

\begin{table}[h!]
\centering
\begin{tabular}{|c|c|c|c|}
\hline
Method \vphantom{\Big[}& $M_\pi a_0^{\pi\pi}$ & $M_K a_0^{KK}$ & $M_\pi a_0^{\pi K}$
\\ \hline \hline
Direct & 0.0443(56)(13) &  0.379(7)(31) & 0.060(4)(2) \\ \hline
SU(3) ChPT  & 0.04267(4) & 0.360(3)  & 0.058(2) \\ \hline
SU(3) WChPT & 0.04260(7) & 0.355(5) & 0.057(3) \\ \hline
SU(2) ChPT  & 0.04295(17) & ---  & --- \\ \hline
SU(2) WChPT & 0.0429(3) & --- & --- \\ \hline\hline
FLAG~\cite{FlavourLatticeAveragingGroupFLAG:2024oxs} & 0.0441(4)~\cite{Helmes:2015gla} & 0.388(20)~\cite{Helmes:2017smr} 
&0.059(2)~\cite{Helmes:2018nug} \\ \hline
\end{tabular}
\caption{ Summary of determinations of the physical-point scattering lengths using three different methods: direct calculation on the E250 ensemble (first uncertainty is statistical, second is from quark-mass mistuning), 
chiral fits without discretization effects (ChPT), and chiral fits including the leading discretization effects (WChPT). FLAG estimates are shown for comparison.
}
\vspace{-0.2cm}
\label{tab:physa0}
\end{table}

\subsection{$s$-wave effective ranges}
\label{sec:effrange}

We now discuss the results for the effective ranges. It is convenient to package these into the quantities $M_X^2 a_{0}^{XY} r_{0}^{XY} $, with $X,Y \in {\pi, K}$. This is because the LO ChPT predictions for these quantities are simple,
\begin{equation}
    M_X^2 a_{0}^{XY} r_{0}^{XY} = 1 + \frac{M_X}{M_Y} + \frac{M^2_X}{M^2_Y}.
\end{equation}
The NLO $SU(2)$ ChPT prediction exists for pions~\cite{NPLQCD:2011htk}, while no closed form in $SU(3)$ ChPT is available.

\begin{figure}[h!]
    \includegraphics[width=8cm]{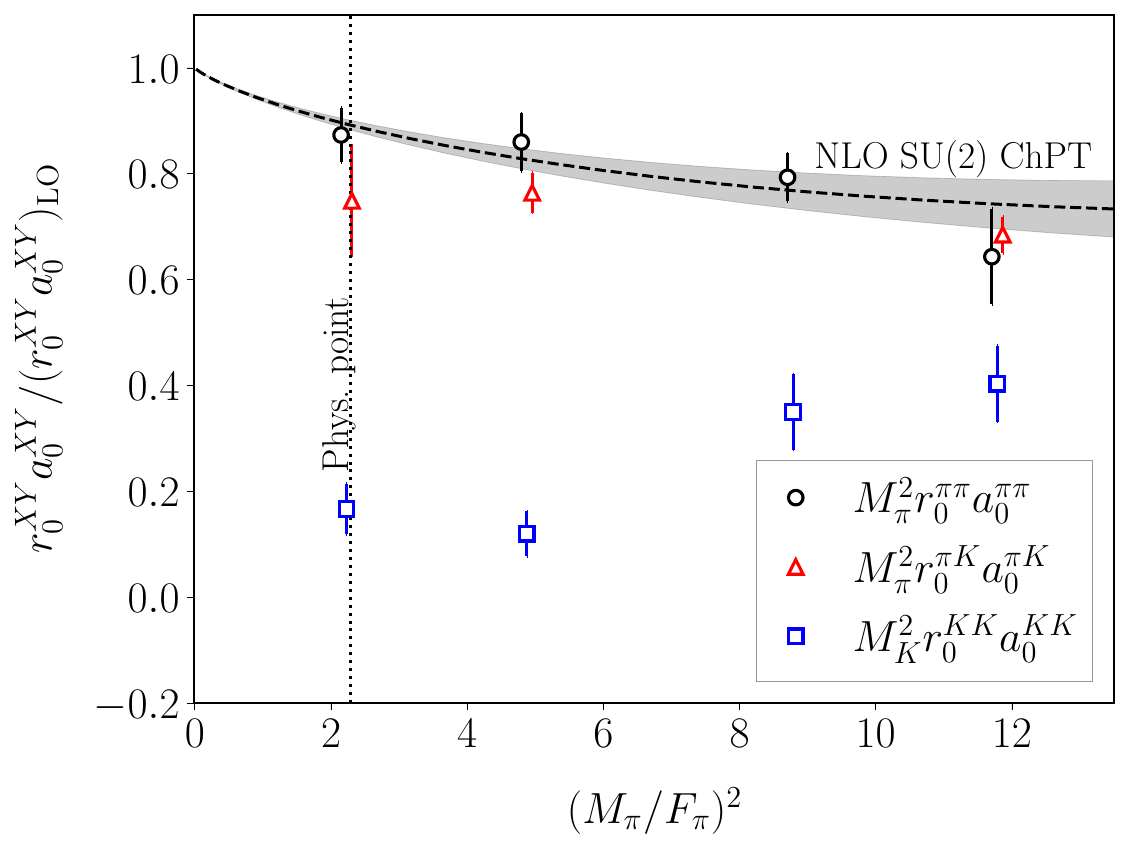}
    \vspace{-0.12cm}
    \caption{Results for the product of the $s$-wave effective range and scattering length, normalized by the LO ChPT prediction.
    The band shows the results of a fit of the two-pion results to the prediction of  $SU(2)$ NLO ChPT.
    The physical point is indicated by the dashed vertical line.
    }
    \vspace{-0.12cm}
    \label{fig:r0all}
\end{figure}

The results are shown in \Cref{fig:r0all},
where we have normalized them by the LO ChPT prediction.
Only the $\pi\pi$ result is required to equal unity in the chiral limit;
for the $\pi K$ and $KK$ systems the plot shows the extent of $SU(3)$ breaking,
which is very large for the $KK$ case.
We have fit the $\pi\pi$ results to the NLO ChPT form 
[using the form in Eq.~(3.16) of Ref.~\cite{Blanton:2021llb}]
finding
\begin{equation}
C_3 = -0.296(24), \ \ \chi^2/{\rm DOF} = 2.05/3\,, \ \ p=0.56\,.
\end{equation}
Here $C_3$ is a combination of $SU(2)$ LECs, and is evaluated at the renormalization scale $\mu=\sqrt{2}F_\pi$. 
Evaluating the chiral fit at the physical point, we find $M_\pi^2 a_0^{\pi\pi} r^{\pi\pi}_0 = 2.68(3)$, which is consistent with the direct determination on E250, $M_\pi^2 a_0^{\pi\pi} r^{\pi\pi}_0 =2.62(16)$, but with a much-reduced error.
To our knowledge, no other chiral fits to this quantity appear in the literature.

\subsection{Scattering lengths in higher partial waves}

\begin{figure}[h!]
    \includegraphics[width=8cm]{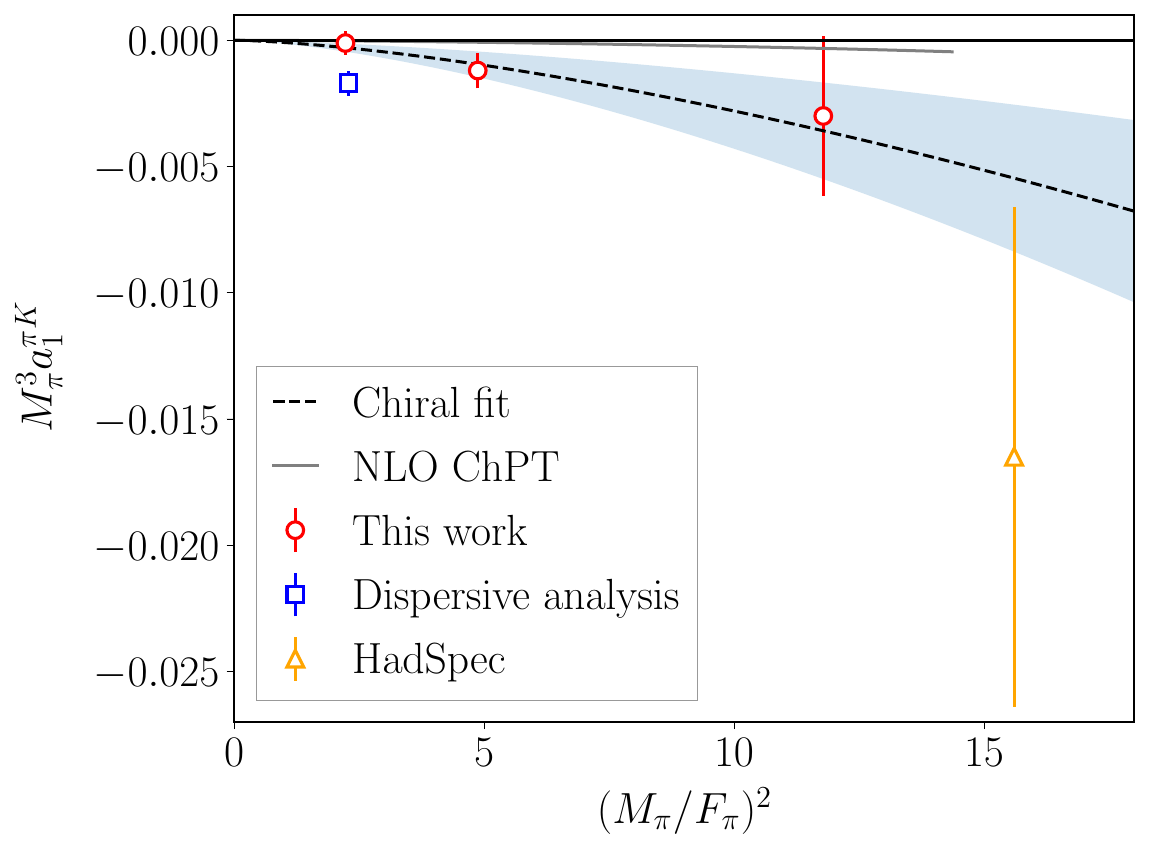}
    \vspace{-0.12cm}
    \caption{ Results for the $p$-wave scattering lengths as a function of the pion mass.
    The three ensembles used in this work are shown as red circles. LQCD results by the HadSpec collaboration are shown as orange triangles~\cite{Wilson:2014cna}. The dispersive result of Ref.~\cite{Pelaez:2020gnd} is given as a blue square.  }
    \vspace{-0.12cm}
    \label{fig:P0piK}
\end{figure}

We next investigate the chiral dependence of scattering lengths in higher partial waves: $p$ wave for $\pi K$, and $d$ wave for $\pi\pi$ and $KK$.
The results for the $p$-wave $\pi K$ scattering length are shown in \Cref{fig:P0piK}. As can be seen, only one ensemble obtains a value that is more than $1\sigma$ away from zero. However, the central values of all results are negative, suggesting that the interaction is attractive at threshold. This is in qualitative agreement with the dispersive results of Ref.~\cite{Pelaez:2020gnd}, although their central value differs by about $3\sigma$ with ours on the E250 ensemble. Similar tension has been observed in the low-energy $s$-wave phase shift displayed in \Cref{fig:pk_phase}, and both might be explained by the lack of experimental data in the low-energy region. 
In this sense, our results complement those available from experiments.

The leading chiral dependence of this quantity is $M^3_\pi a_1^{\pi K} \propto (M_\pi/F_\pi)^3$, and we observe that such behavior qualitatively describes the points in \Cref{fig:P0piK}. However, as can be seen, this fit disagrees at about the $2\sigma$ level from the NLO $SU(3)$ ChPT prediction (given in Appendix C of Ref.~\cite{Draper:2023boj}).

\begin{figure*}[th!]
     \centering
     \subfloat[\label{fig:a2pp}]{%
     \includegraphics[width=0.49\textwidth]{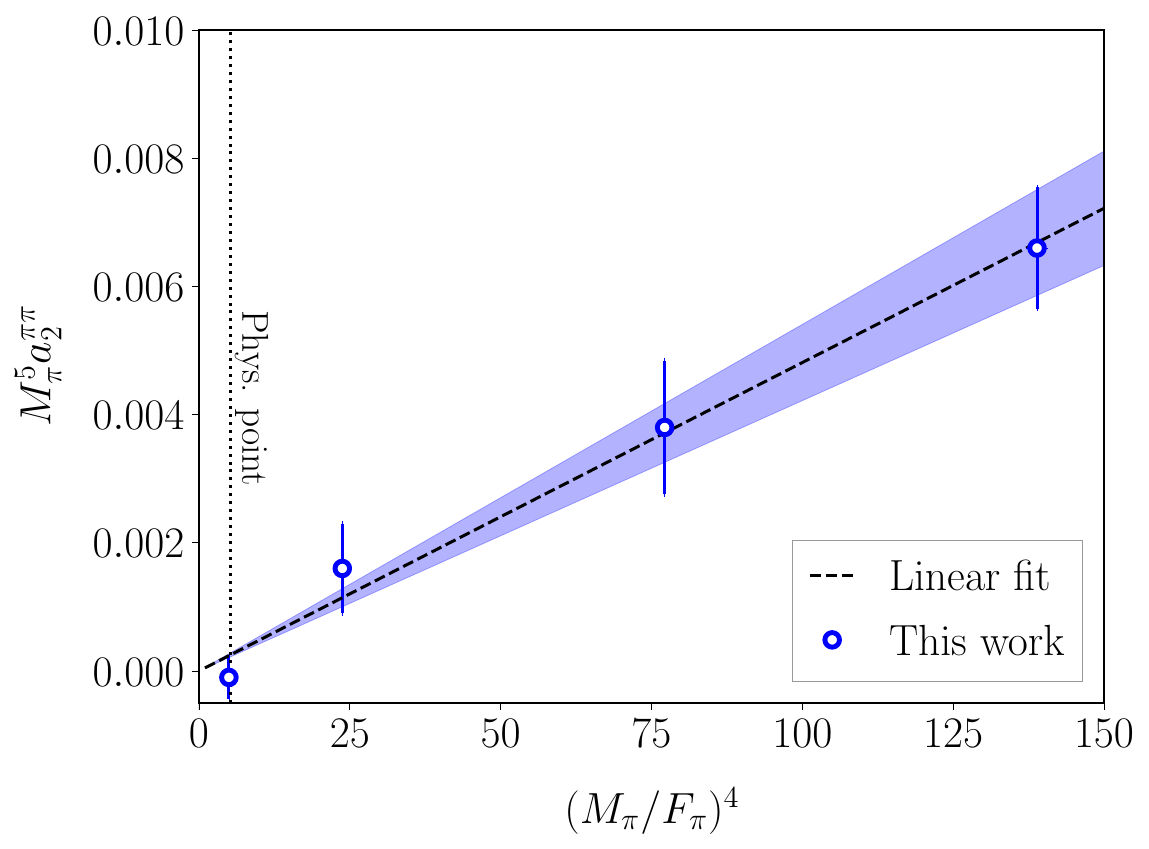}
    }
    \hfill
    \subfloat[\label{fig:a2kk}]{%
     \includegraphics[width=0.49\textwidth]{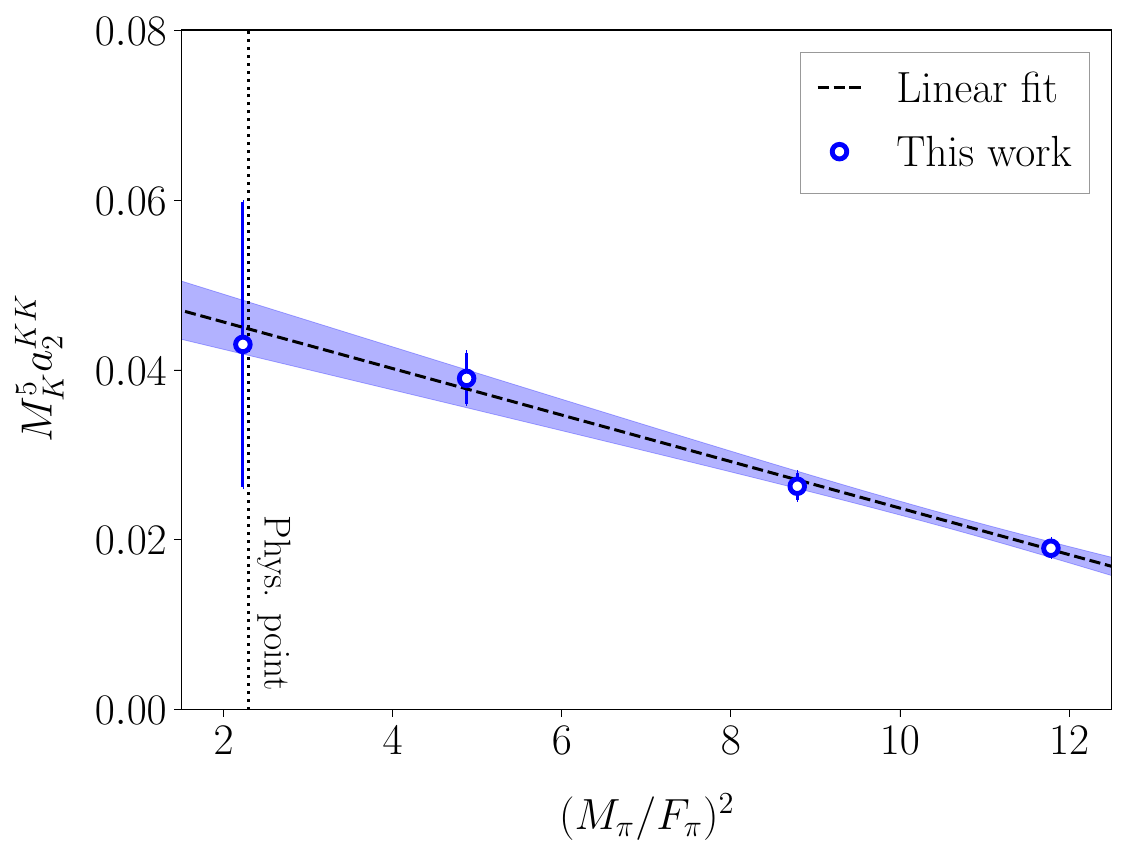}
    }
    %\vspace{-0.12cm}
    \caption{Results for the $d$-wave scattering lengths for the two-pion (left) and two-kaon (right) systems. The results are fit to a linear dependence on the quantity plotted on the $x$-axis, with the intercept fixed to zero in the case of pions, as discussed in the text. The physical point is indicated as a vertical line. }
    \label{fig:dwavea2}
    %\vspace{-0.12cm}
\end{figure*}

The final two-meson quantities that we consider  are
the $d$-wave scattering lengths for identical mesons. 
Our results are shown in \Cref{fig:dwavea2}. 
With the exception of that for $a_2^{\pi\pi}$ on the E250 ensemble,
all results have good statistical significance.
This exception is expected, however, because chiral symmetry implies that the scattering length is small for physical pion masses, and thus hard to determine.
In particular, 
$a_2^{\pi\pi}$ vanishes at LO in $SU(2)$ ChPT, and thus, up to logarithms,
behaves as $  M^5_\pi a_2^{\pi\pi} \propto \left({M_\pi}/{F_\pi}\right)^4$.
In the left panel of the figure, we show a fit to this form, which works well.
There is, however, a possible discrepancy with the dispersive result for this quantity at the physical point, $M_\pi^5 a_2^{\pi\pi} = -1.85(18)\cdot 10^{-4}$~\cite{Garcia-Martin:2011iqs}. 
While this is compatible within $1\sigma$ with our E250 result, given the large errors in the latter,
the sign is opposite to that on our heavier-pion ensembles, 
and thus to the prediction of our naive chiral fit.  
As discussed in Ref.~\cite{Blanton:2021llb},
this appears to be related to the fact that $d$-wave phase shift predicted by the dispersive analysis changes sign slightly above the threshold.
Thus our naive fit is unreliable in the vicinity of the physical point,
and, we refrain from quoting an extrapolated value.

For the $KK$ system,
$a_2^{KK}$ is not protected by chiral symmetry, so we expect a linear dependence on $(M_\pi/F_\pi)^2$, up to logarithms. The right panel in the figure shows that such a linear fit describes the data well,
and leads to the physical-point value ${M_K^5 a_2^{KK}=0.045(3)}$.

\subsection{The three-pion K matrix}
\label{sec:kdf_chiral}

Here we discuss the chiral dependence of the three-particle K matrix for the $3\pi^+$ system. While this is a scheme-dependent, and thus unphysical, quantity, it can be calculated in ChPT using the same scheme (i.e. the same cutoff function $H$). This allows for a direct comparison with our lattice results without needing to solve the integral equations that lead to $\cM_3$.

The LO ChPT result was originally computed in Ref.~\cite{Blanton:2019vdk}, 
where it was observed that the LO result was in fact scheme-independent.
However, it was found that there were large differences between this prediction and the lattice results. The NLO result was subsequently calculated in Ref.~\cite{Baeza-Ballesteros:2023ljl} 
(and generalized to all isospins in Ref.~\cite{Baeza-Ballesteros:2024mii}) using the NLO ChPT three-pion scattering amplitudes from Refs.~\cite{Bijnens:2022zsq,Bijnens:2021hpq}. 
Very large NLO corrections were observed in the quantities that are nonzero at LO, namely $\cK_0$ and $\cK_1$ in \Cref{eq:Kdfidentical}.
Thus the disagreement of lattice results with LO ChPT was understood, while, at the same time, the range of pion masses for which ChPT is convergent was seen to be small, barely extending to the physical quark mass.

\begin{figure*}[th!]
     \centering
     \subfloat[\label{fig:K0}]{%
     \includegraphics[width=0.49\textwidth]{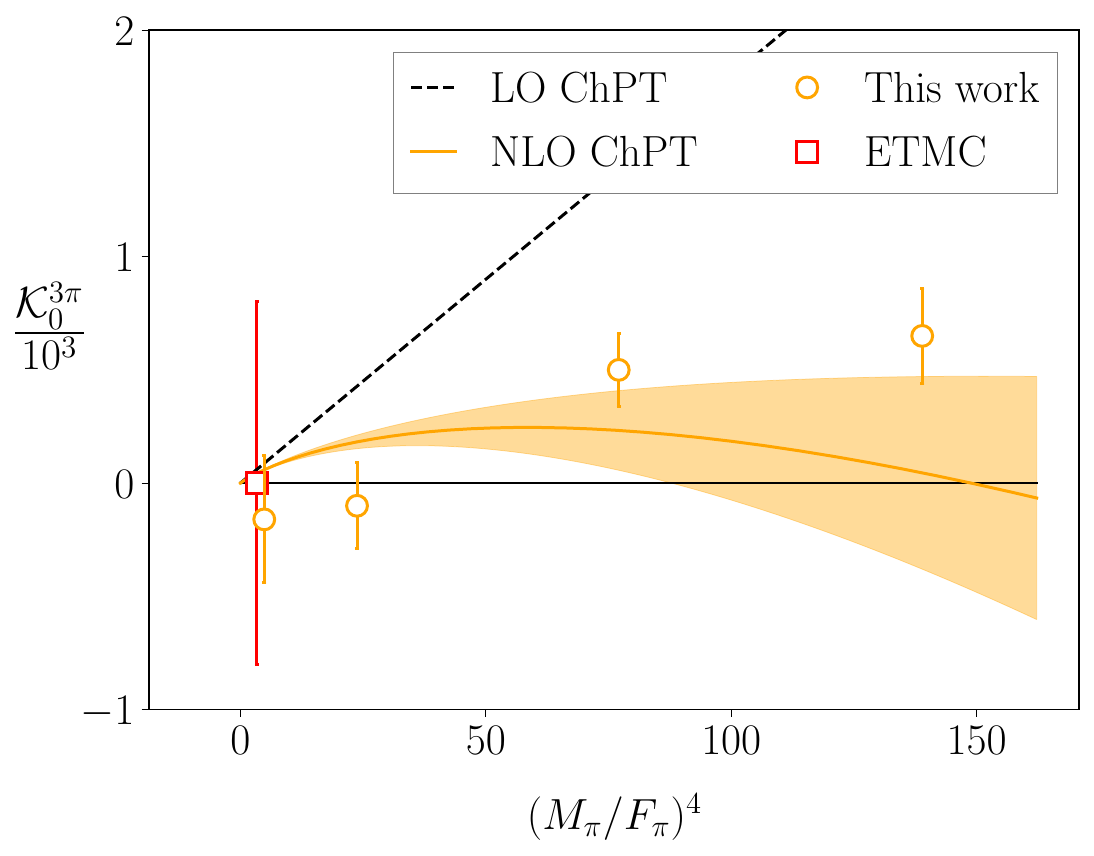}
    }
    \hfill
    \subfloat[\label{fig:K1}]{%
     \includegraphics[width=0.49\textwidth]{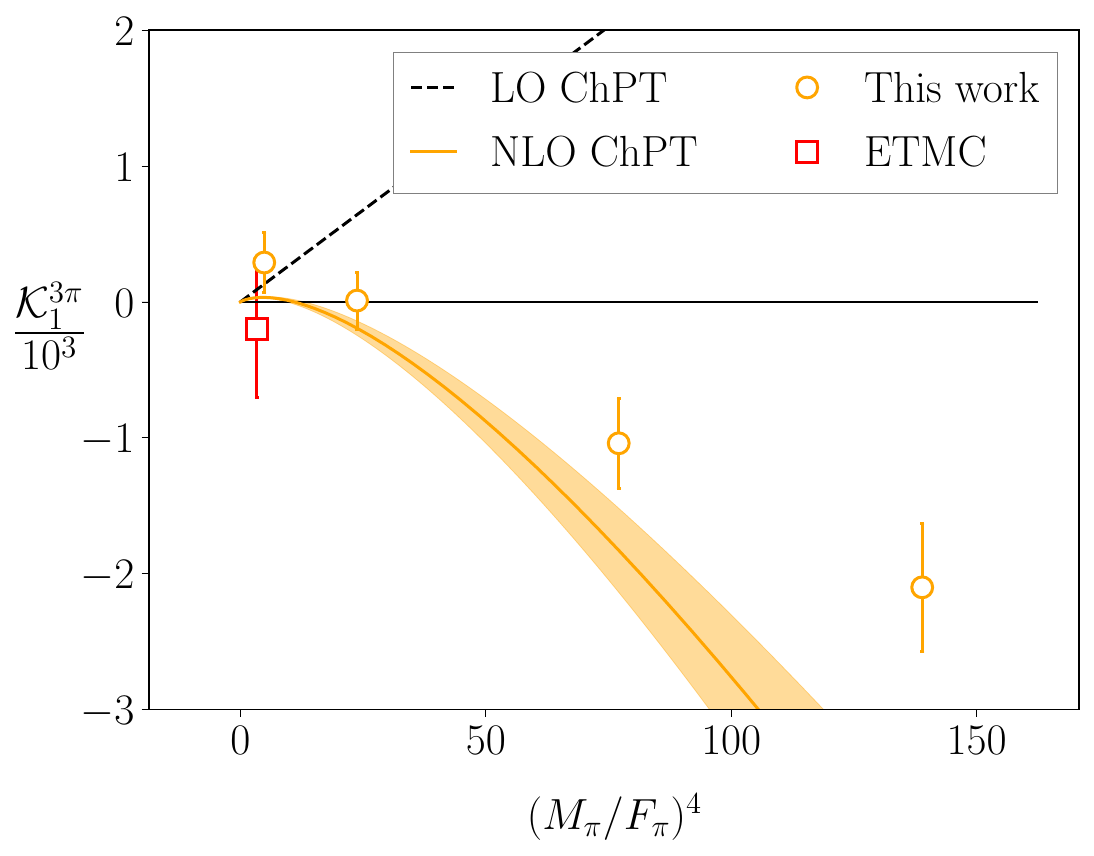}
    }
    \vspace{-0.12cm}
    \caption{Results for the coefficients $\cK_0$ (left) and $\cK_1$ (right) term in the threshold expansion of the three-pion $\kdf$.
    Orange circles are the results of this work, using fits that include $s$ and $d$ waves, and $\cK_B$ (see \Cref{tab:ppp-params-E250}, and Ref.~\cite{Blanton:2021llb}). 
    The results of Ref.~\cite{Fischer:2020jzp} are shown as red squares (and come from a fit that include only $s$ waves). LO~\cite{Blanton:2019vdk} and NLO~\cite{Baeza-Ballesteros:2023ljl} predictions from ChPT are also shown.
    The error band arises from uncertainties in LECs. Note that the horizontal axis is proportional to $M_\pi^4$.}
    \vspace{-0.2cm}
    \label{fig:K01}
\end{figure*}

It is thus of considerable interest to confirm this picture by determining $\cK_0$ and $\cK_1$ at the physical point, as we have done in this work. Given the expected chiral behavior, \Cref{eq:K0K1IDChPT}, however, we expect that $\kdf$ will be small and difficult to extract with statistical significance. This is indeed the case, as seen above. Nevertheless, we can test whether the results we obtain are consistent, within errors, with the expected chiral behavior.
The results of this test are shown in \Cref{fig:K01}, 
where we see that indeed consistency is observed.

\begin{figure}[h!]
    \includegraphics[width=8cm]{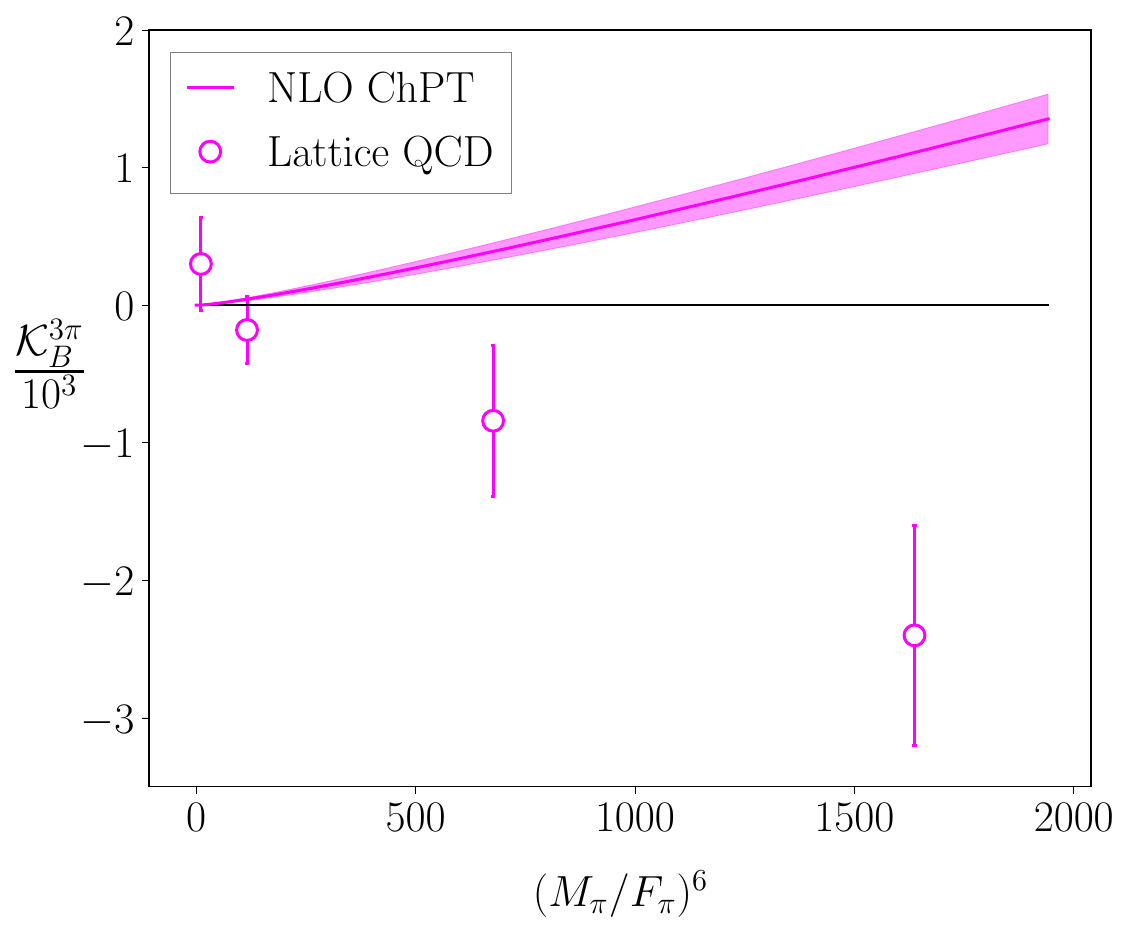}
    \vspace{-0.12cm}
    \caption{Results for the $\cK_B$ term in the threshold expansion of the three-pion $\kdf$ from fits that include $s$ and $d$ waves. The band indicates the NLO ChPT prediction from Ref.~\cite{Baeza-Ballesteros:2023ljl}, with the uncertainty arising from that in the LECs. Note that the horizontal axis is proportional to $M_\pi^6$.}
    \vspace{-0.2cm}
    \label{fig:KB}
\end{figure}

The other coefficient that we determine, $\cK_B$, begins at NLO in ChPT, with a result determined in Ref.~\cite{Baeza-Ballesteros:2023ljl}.
A comparison with this prediction is shown in \Cref{fig:KB}.
Here the situation is analogous to that for $\cK_0$, in which the leading nontrivial chiral prediction has the opposite sign from the lattice data.
It will require a NNLO calculation to determine whether higher orders lead to a sign change around the physical pion mass.
What we do see, however, is that our new result at the physical point is consistent with the chiral prediction.

\section{Three-meson scattering amplitudes}
\label{sec:amplitudes}

In this section, we present results for three-meson scattering amplitudes. Unlike the intermediate K matrices, these are physical quantities that one could compare to the experiment, were three-particle scattering achievable. They are obtained by solving the three-body integral equations, as described in \Cref{sec:inteqs}, using the fit results for the two- and three-body K matrices in \Cref{sec:fits} as input. We show results for all the systems studied in this work, i.e.~$3\pi^+$, $\pi^+\pi^+K^+$, $K^+K^+\pi^+$, $3K^+$, including the quark-mass and partial-wave dependence. 

We begin with a comment on the kinematics. Although the presented formalism is relativistic and valid in general reference frames, for simplicity (and without loss of generality) we consider a three-to-three scattering process in the overall rest frame. Second, as discussed in \Cref{app:threebodykinematics}, the most general, elastic three-hadron amplitudes depend on eight kinematic variables. This multi-variable nature allows for many choices of two-dimensional representations; here, we impose relations between kinematical variables such that the studied amplitudes are always presented as functions of a single variable. This corresponds to selecting a kinematic configuration in which certain orientations of the three-momenta are fixed. One such configuration is the {\it equilateral configuration}, where all three momenta have the same magnitude and lie along bisectors of an equilateral triangle. A more general choice is the {\it isosceles configuration}, where the magnitudes of two of the momenta are equal and lie at a relative angle of $\alpha \neq 120^\circ$, thus lying along bisectors of an isosceles triangle. A more detailed description is given in \Cref{app:threebodykinematics}. In particular, \Cref{fig:configs} depicts the most general kinematic configuration used in this work. We also present the individual, fixed-$J$ contributions to the overall scattering amplitude, as defined in \Cref{eq:symmetrization}.

In this section, we fix the incoming momenta to always lie in an equilateral configuration and explore different equilateral or isosceles choices for the outgoing momenta. For simplicity, we take the initial and final reaction planes to be the same. The $J=0$ amplitudes do not depend on the relative orientation of two planes, but this choice simplifies the description of the $J=1,2$ amplitudes. For instance, the chosen kinematic configuration of incoming momenta makes the $J=1$ amplitudes vanish for systems composed of identical particles ($3\pi$ and $3K$). In general, the resulting amplitudes then depend on the overall c.m.~energy, denoted $E$ in this section, on the angle $\alpha$ describing the final-state configuration, and on the angle between the initial and final ``triangles'', $\theta_{22}$ (defined precisely in \Cref{app:threebodykinematics}). If we fix two of these variables, we can display the dependence on the remaining one.

\begin{figure*}[th!]
     \centering
     \subfloat[\label{fig:amplitudeJ0_equi1} ``equilateral'' configuration]{%
     \includegraphics[width=0.49\textwidth]{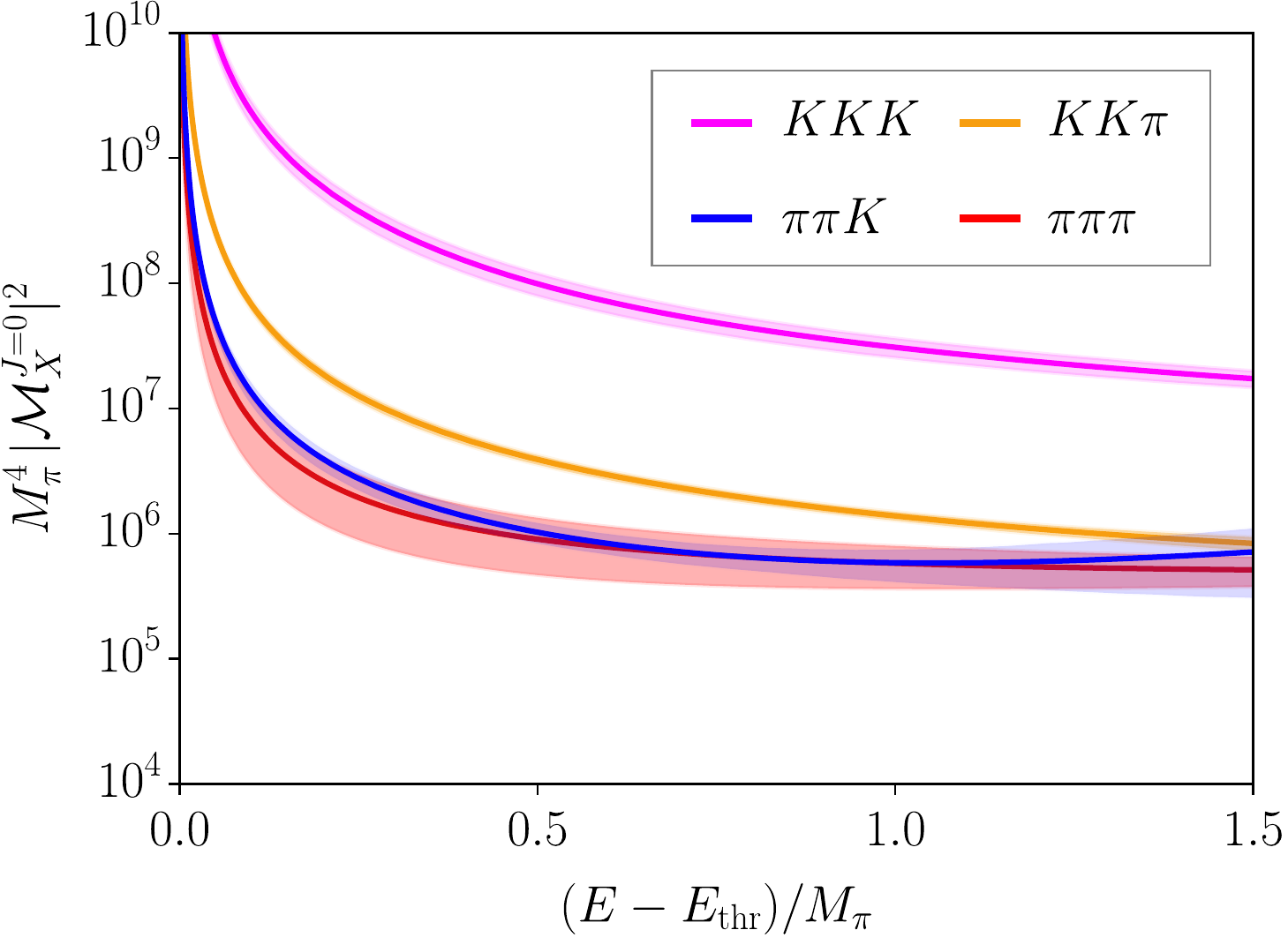}
    }
    \hfill
    \subfloat[\label{fig:amplitudeJ0_iso} ``isosceles'' configuration, $\alpha = 175^\circ$]{%
     \includegraphics[width=0.49\textwidth]{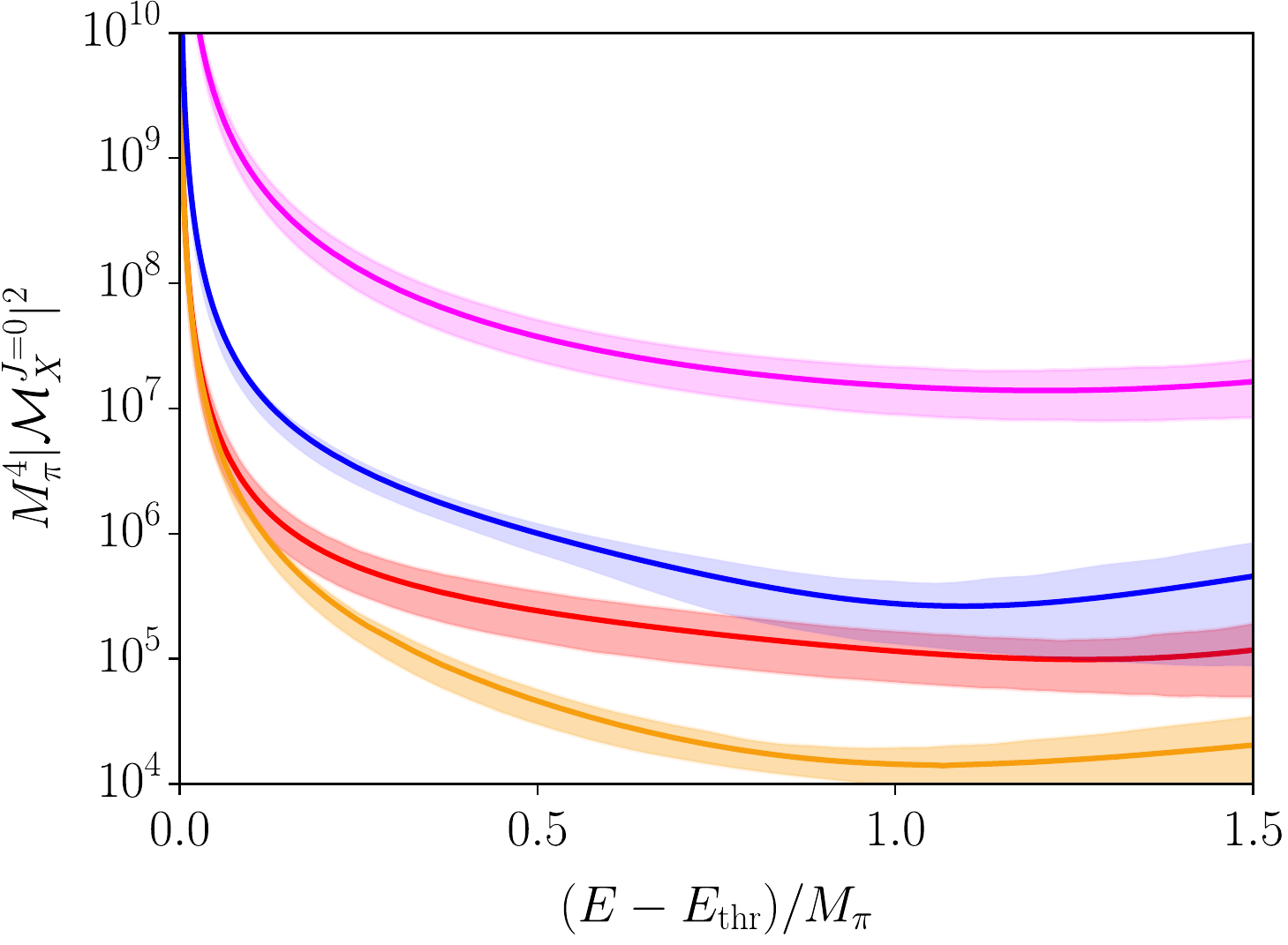}
    }
    \vspace{-0.12cm}
    \caption{Absolute values squared of the $J^P=0^-$ three-body amplitudes as a function of the energy relative to the threshold in the equilateral (left) and the $\alpha = 175^\circ$ isosceles (right) kinematic configurations, based on fits to the spectra obtained on the E250 ensemble. $E_{\rm thr}$ is the three-meson threshold energy, e.g. $E_{\rm thr} = 2M_\pi +M_K$ for the $\pi\pi K$ system. All quantities have been normalized by the pion mass. Bands represent statistical errors propagated from the two- and three-body fit parameters.
    }
       \vspace{-0.12cm}
    \label{fig:amplitudeJ0}
\end{figure*}

\begin{figure}[th!]
     \centering
    \includegraphics[width=0.49\textwidth]{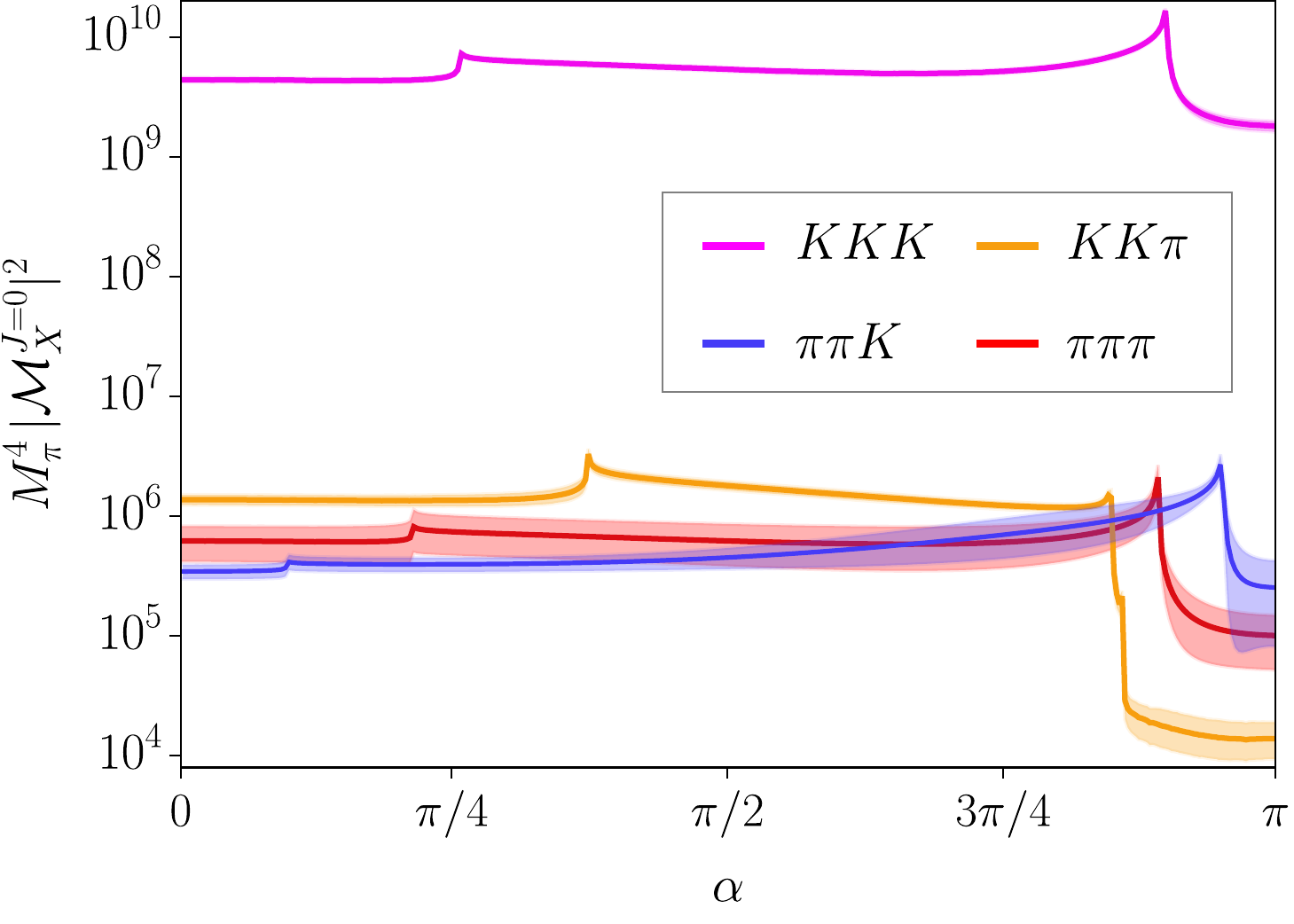}
    \vspace{-0.2cm}
    \caption{Absolute values squared of the $J^P=0^-$ three-body amplitudes as a function of the $\alpha$ angle in the isosceles configuration at $E = E_{\rm thr} + M_\pi$, based on fits to the spectra obtained on the E250 ensemble. All amplitudes have been normalized by the pion mass. 
    }
    \vspace{-0.32cm}
    \label{fig:amplitudeJ0alpha}
\end{figure}

\begin{table*}
   \centering
   \setlength{\tabcolsep}{12pt}
   \renewcommand{\arraystretch}{1.5}
   \begin{tabular}{c c c c c c c}
     \toprule
     System & Pair & $J^P = 0^-$ & $J^P = 1^+$ & $J^P = 2^-$ \\
     \midrule
     $\pi^+ \pi^+ \pi^+$ & $\pi^+ \pi^+$  & $(0,0)$, $(2,2)$ & ---  & $(2,0)$, $(0,2)$, $(2,2)$   \\ \hline 
     \multirow{2}{*}{$\pi^+\pi^+ K^+$} & $\pi^+ K^+$ & $(0,0)$, $(1,1)$  & $(0,1)$, $(1,0)$, $(2,1)$  & $(2,0)$, $(1,1)$   \\
      & $\pi^+\pi^+ $ & $(0,0)$ & $(1,0)$ & $(2,0)$ \\
      \hline 
     \multirow{2}{*}{$\pi^+ K^+ K^+$} & $\pi^+ K^+$ & $(0,0)$, $(1,1)$  & $(0,1)$, $(1,0)$, $(2,1)$  & $(2,0)$, $(1,1)$   \\
      & $K^+ K^+$ & $(0,0)$, $(2,2)$ & $(1,0)$, $(1,2)$  & $(2,0)$, $(0,2)$, $(2,2)$ \\
      \hline
      $K^+K^+K^+$ & $K^+K^+$ & $(0,0)$, $(2,2)$ & ---  & $(2,0)$, $(0,2)$, $(2,2)$   \\
     \bottomrule
   \end{tabular}
   \caption{Pair-spectator $LS$ partial waves included in the solutions of the three-body integral equations for each three-particle system. We denote them by $(\ell,s)$, where $s$ is a ``spin'' of a pair and $\ell$ is the pair-spectator orbital angular momentum.
   }
   \vspace{-0.3cm}
   \label{tab:partial-waves}
\end{table*}

\subsection{Three-hadron amplitudes at the physical point}

\begin{figure*}[th!]
     \centering
     \subfloat[\label{fig:amplitudeJ1_equi}]{%
     \includegraphics[width=0.49\textwidth]{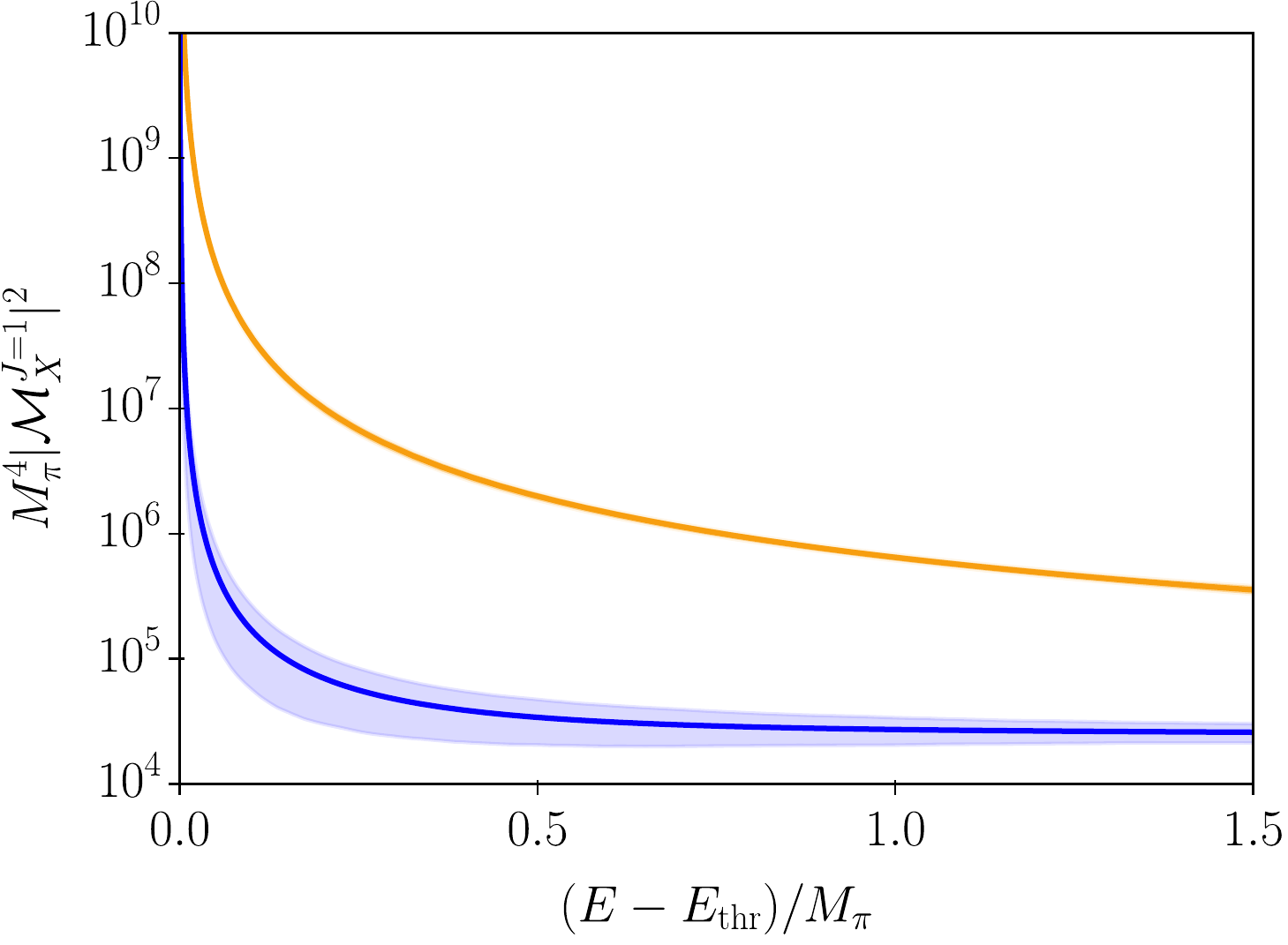}
    }
    \hfill
    \subfloat[\label{fig:amplitudeJ2_equi}]{%
     \includegraphics[width=0.49\textwidth]{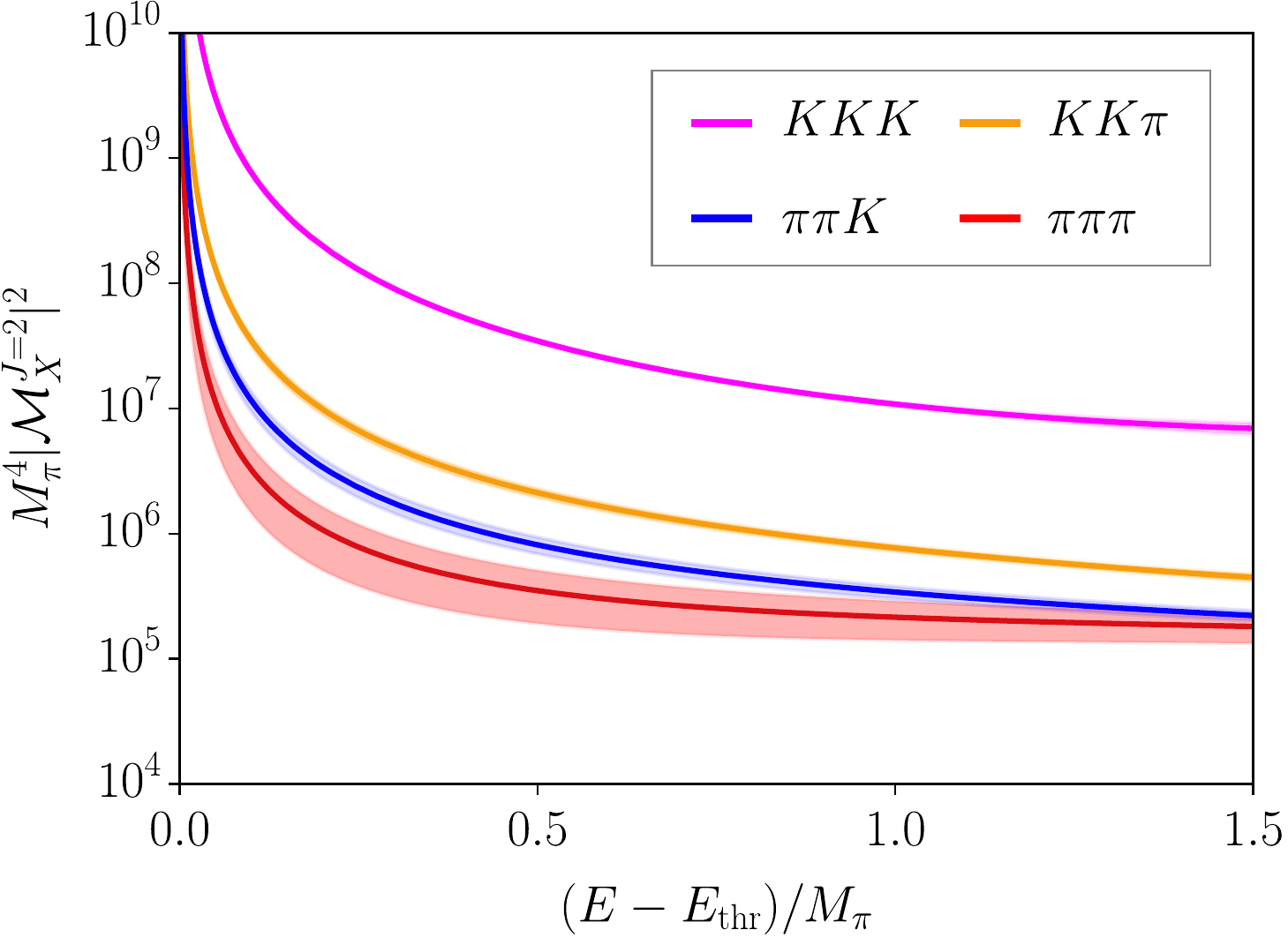}
    }
    %\vspace{-0.2cm}
    \caption{  Absolute values squared of the $J^P=1^+$ (left) and $J^P=2^-$ (right) three-particle amplitudes in the equilateral configuration, as a function of the energy relative to threshold.
    Other notation as in \Cref{fig:amplitudeJ0}.
   }
   %\vspace{-0.2cm}
    \label{fig:amplitudeJ12}
\end{figure*}

We begin by showing results for the three-meson amplitudes with approximately physical quark masses obtained from the E250 ensemble spectra. For each three-body system, we choose a representative model of the two- and three-body $K$ matrices as input to the integral equations. Specifically, we obtain the results discussed in this section using scattering parameters from the last columns of~\Cref{tab:ppp-params-E250} in the case of $3\pi$,~\Cref{tab:kkk-params-E250} in the case of $3K$,~\Cref{tab:ppK-params-E250} in the case of $2\pi K$, and~\Cref{tab:KKp-params-E250} in the case of $2 K \pi$. When solving integral equations, for each $J^P$, we include all non-trivial contributions from pair-spectator partial waves characterized by combinations of angular momenta $(\ell, s)$ such that $\ell,s \leq 2$. We summarize the included waves in~\Cref{tab:partial-waves}. 

The displayed uncertainties have been propagated from the errors in the two- and three-body K matrices and are purely statistical. We do not have a complete error budget, as we work at only one lattice spacing and volume, with slightly mistuned quark masses, and truncated forms of the K matrices. Nevertheless, we have found evidence (discussed in \Cref{sec:a0_ChPT}) that the discretization and mistuning errors are smaller than those shown. Moreover, model dependence of the results for the amplitudes is not studied systematically, but we verified that the differences between different considered models (associated with the inclusion of higher partial waves in two-body subchannels) affect the result very little.

We first display results for the $J^P=0^-$ component of these amplitudes. For this choice, the amplitudes do not depend on $\theta_{22}$ and, for fixed $\alpha$, are functions of $E$ only. Results for $|\cM_3|^2$ are shown in \Cref{fig:amplitudeJ0} for both the final-state equilateral and isosceles configurations (the latter with $\alpha=175^\circ$). We show all amplitudes up to the maximum energy $E_{\rm max} = E_{\rm thr} + 1.5 M_\pi$. While this lies beyond the inelastic threshold for systems with kaons, ${E_{\rm thr} + M_\pi}$, we note that the fits to the $\pi\pi K$, $KK\pi$, and $3K$ spectra continue to work well above the nominal inelastic threshold, as seen in \Cref{sec:spectrum_overview}.

Two general features are noteworthy. First, the $3K$ amplitude is larger than that for systems that include a pion. This is to be expected, based on the suppression of pion interactions as one approaches the chiral limit. Second, all amplitudes diverge at the threshold---a behavior expected for the chosen kinematic configuration of incoming/outgoing momenta due to the one-particle exchange processes. Indeed, within the integral equations, the amplitudes are well-approximated by the LO terms in the iterative expansion of Eqs.~\eqref{eq:ladder} and~\eqref{eq:Teq}. In this approximation, the OPE pair-spectator amplitude gives the most pronounced contribution, as seen in our solutions. In the mixed systems, two types of physical exchanges are allowed since two different particles can be exchanged between external pairs---a pion or a kaon.

Comparing the two kinematic configurations in \Cref{fig:amplitudeJ0}, it can be seen that the hierarchy between $3\pi$, $\pi\pi K$ and $KK\pi$ amplitudes changes. In particular, the $KK\pi$ amplitude is larger than those for $3\pi$ and $\pi\pi K$ in the equilateral case, but smaller for the isosceles configuration. This indicates a nontrivial interplay between kinematic configurations and K-matrix parameters, an effect magnified by our extreme choice of $\alpha=175^\circ$. In this example, the change in the $2K\pi$ amplitude is associated with the energy closing for the physical one-kaon exchange between the external $\pi K$ and $KK$ pairs. 

The uncertainties display a complicated dependence on the flavor of the system and on $E$, which results from the interplay between the errors in the underlying lattice spectra and the choice of fit. Two general features are that errors increase with $E$, and that they are larger for the isosceles configuration. A possible explanation for both features is that the $p$- and $d$-wave parameters, which have larger errors than those for $s$ waves, contribute more significantly at higher energies and to the asymmetric isosceles configuration. Moreover, at higher energies where the $\mathcal M_{\rm df,3}$ part of the three-body amplitude becomes comparable with the ladder amplitude, errors from the three-body K matrices begin contributing significantly to the total uncertainty of the solutions.

Up to this point, we have fixed the orientations of the particles and varied the overall energy. We now discuss the $\alpha$ dependence of the amplitudes at a fixed energy, $E = E_{\rm thr}+M_\pi$, see \Cref{fig:amplitudeJ0alpha}. We assume $180^\circ \geq \alpha \geq 0^\circ$. This corresponds to incoming particles having momenta of the order of the pion mass (for the $2K\pi$ state, ${k_{\rm eq} \approx 1.25 \, M_\pi}$, while for the $3\pi$ state, $k_{\rm eq} \approx 0.88 \, M_\pi$.) 

The momenta of outgoing particles depend on the angle $\alpha$ and differ depending on which bisector of the isosceles triangle they form. For instance, for the $2K\pi$ system, the momenta of kaons vary from $p_{\rm is} \approx 0.77 \, M_\pi$ at $\alpha = 0^\circ$ to $p_{\rm is} \approx 1.55 \, M_\pi$ at $\alpha = 180^\circ$, while the momentum of pion decreases from $p_{2} \approx 1.55 \, M_\pi$ at $\alpha = 0^\circ$ to zero at $\alpha = 180^\circ$. We note that $\alpha = 0^\circ$ corresponds to a ``back-to-back'' configuration between the momentum of a pair of identical particles and a remaining particle. When the angle $\alpha$ approaches $180^\circ$, the two identical particles fly ``back-to-back'' along a certain line, with the momentum of the third orthogonal to that direction and its magnitude approaching zero.

Despite the simplicity of the studied configuration, the non-trivial dependence of the particle's momenta on the angle $\alpha$ brings out a new feature: a cusp-like behavior associated with one-particle exchanges. These arise because the intermediate particle in the definite-$J$ OPE contribution can become on-shell only in specific ranges of $\alpha$. The amplitudes diverge logarithmically at the endpoints, but, due to the finite precision of the plot and small imaginary parts added in the definition of the OPE amplitude (as discussed in~\Cref{app:numerical_solution}), these logarithmic singularities appear as finite cusps in the figure.
These discontinuities (although in different variables) are also visible in the Dalitz-like plots in Ref.~\cite{Hansen:2020otl}. Let us also stress that the infinities associated with logarithmic singularities in the definite-$J$ amplitudes are a mathematical consequence of projecting the three-body amplitude into partial waves. They would be converted into a pole of the propagator were we to sum all the partial waves, so that the full amplitude would diverge.
The divergence of physical three-hadron amplitudes was recognized long ago~\cite{Rubin:1966zz}, and it was also understood how it is regularized by the use of wave packets~\cite{PhysRevA.16.2264,PhysRevA.16.2276}.

For the non-identical cases, for which the exchanged particle can be either a pion or a kaon, there can be separate ranges for which on-shell exchange is possible. 
Let us discuss the $KK\pi$ system as an example. At the considered energy, physical pion exchange between two $\pi K $ pairs is possible for all values of $\alpha$. By contrast, kaon exchange between the initial $KK$ and final $\pi K$ pair is kinematically allowed for $67^\circ < \alpha < 152^\circ $, leading to the central ``plateau region'' between the two cusps in the $KK\pi$ results in \Cref{fig:amplitudeJ0alpha}. On the other hand, the asymmetry between initial (equilateral) and final (isosceles) configurations means that kaon exchange between an initial $\pi K$ and final $KK$ pair is possible for a different range of angles, $\alpha < 155^\circ$. The upper limit leads to the ``double step'' feature in the $KK\pi$ curve above $\alpha=3\pi/4$.

We now consider higher total angular momenta, specifically $J^P=1^+$ and $J^P=2^-$. As an example, results for the squared amplitudes in the equilateral configuration are shown in \Cref{fig:amplitudeJ12}. The amplitude for $J^P=1^+$ is non-zero only for systems with non-identical particles due to the symmetric nature of the equilateral configuration of incoming momenta. Other features are similar to those of the $J^P=0^-$ case: there is divergence at the threshold, and the amplitudes decrease with energy. Away from the divergence, both amplitudes in higher-partial waves are smaller in magnitude than in the $J^P=0^-$ case, with the $J^P=1^+$ amplitudes being the smallest.

We recall from \Cref{eq:decomp} that the unsymmetrized amplitude consists of ladder and divergence-free parts. The former depends only on two-meson interactions, while the latter is nonzero for nonvanishing $\Kdf$ only. One can symmetrize these two components separately, as done in~\Cref{eq:symmetrization}, leading respectively to $\cD^J$ and $\cM_{\rm df,3}^J$. Although this separation is cutoff-dependent, it is interesting to investigate how these components contribute to the amplitude. Within a chosen regularization scheme, it provides an intuition about the relative strength of the long- and short-range effects in the scattering process.

\begin{figure*}[th!]
     \centering
     \subfloat[~$3\pi$ amplitude]{%
     \includegraphics[width=0.49\textwidth]{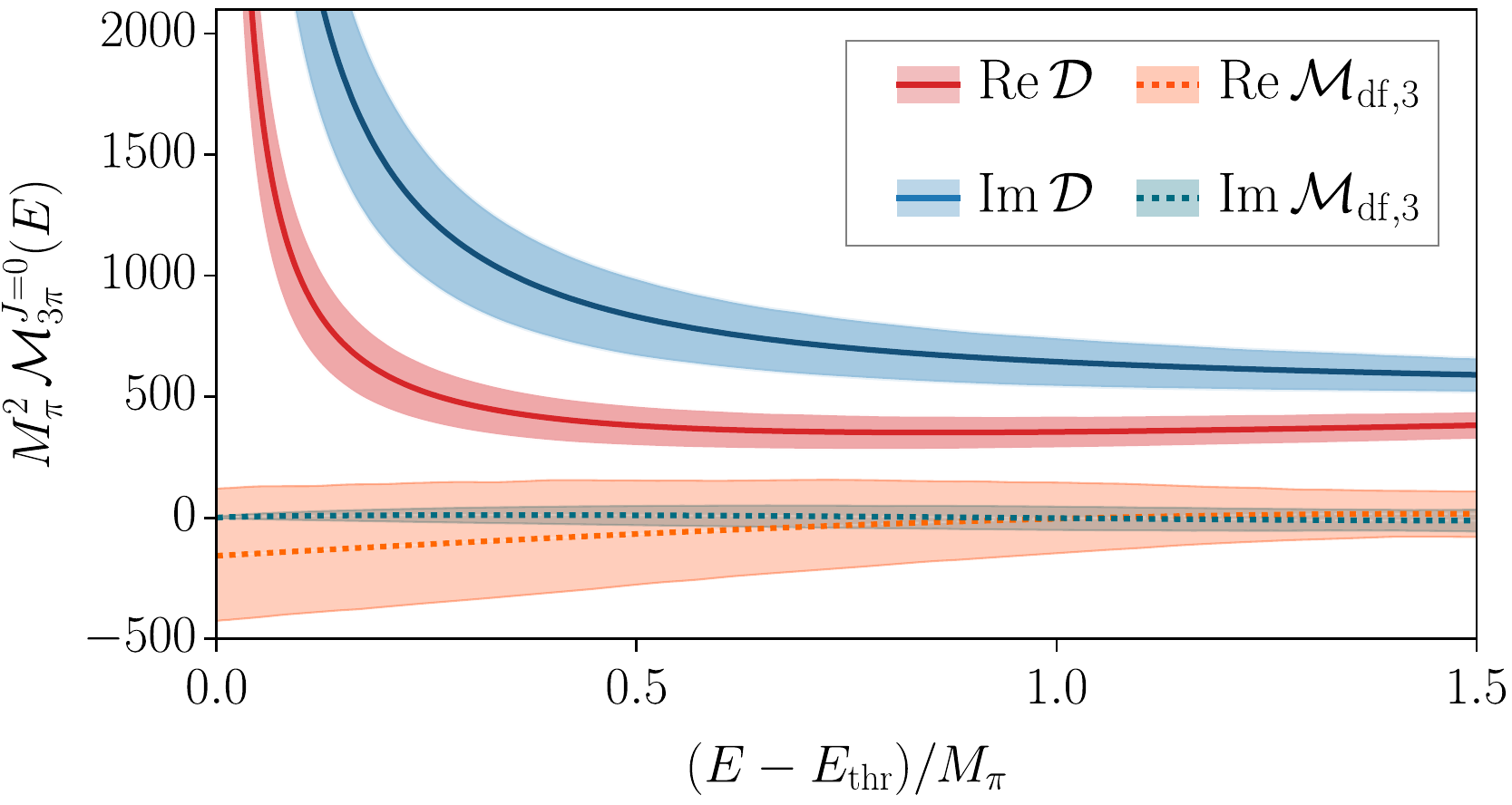}
    }
    \vspace{-0.12cm}
    \hfill
    \subfloat[~$\pi\pi K$ amplitude]{%
     \includegraphics[width=0.49\textwidth]{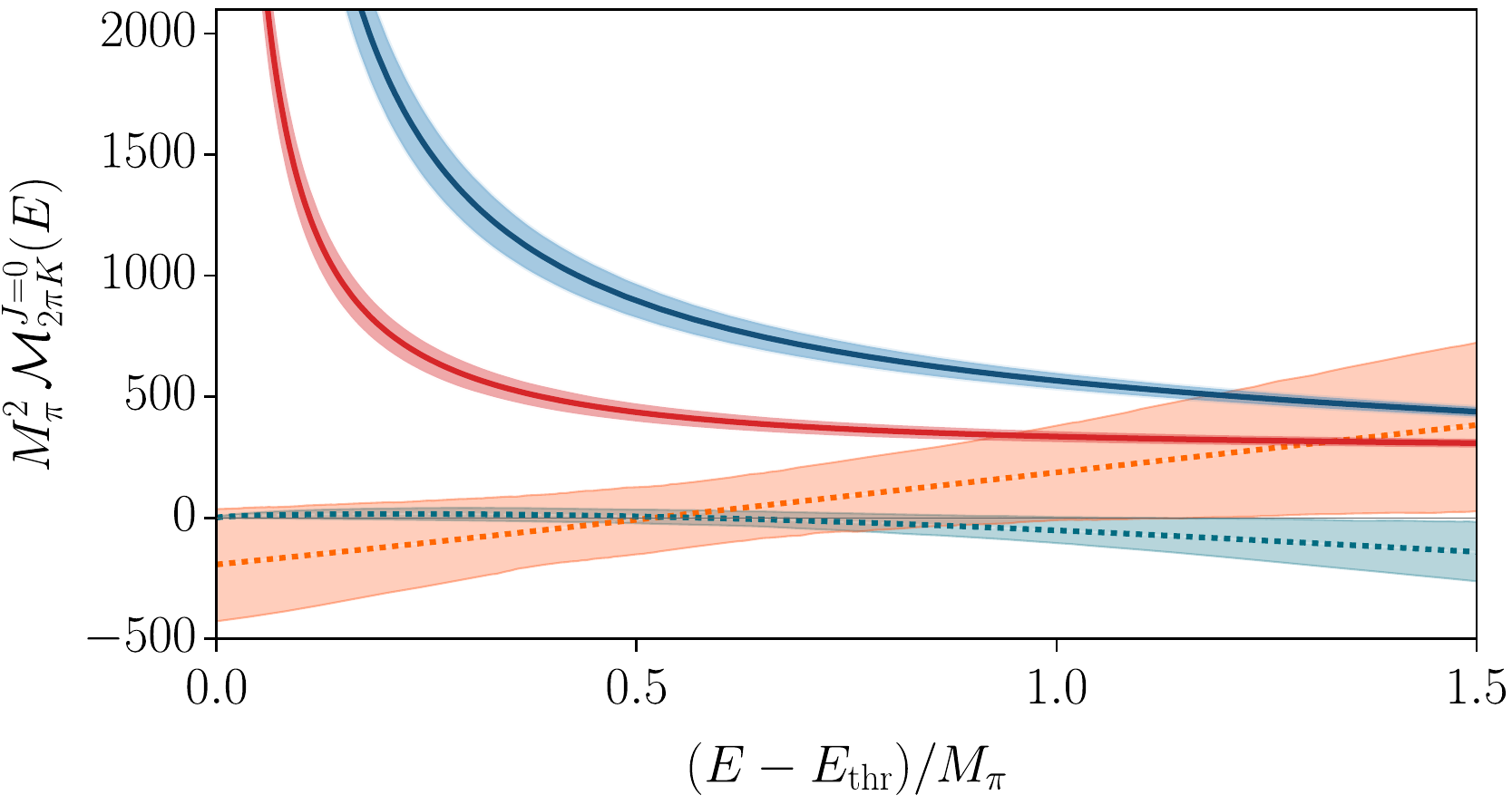}
    }
    \vspace{-0.12cm}
    \hfill
    \subfloat[~$K K \pi$ amplitude]{%
     \includegraphics[width=0.49\textwidth]{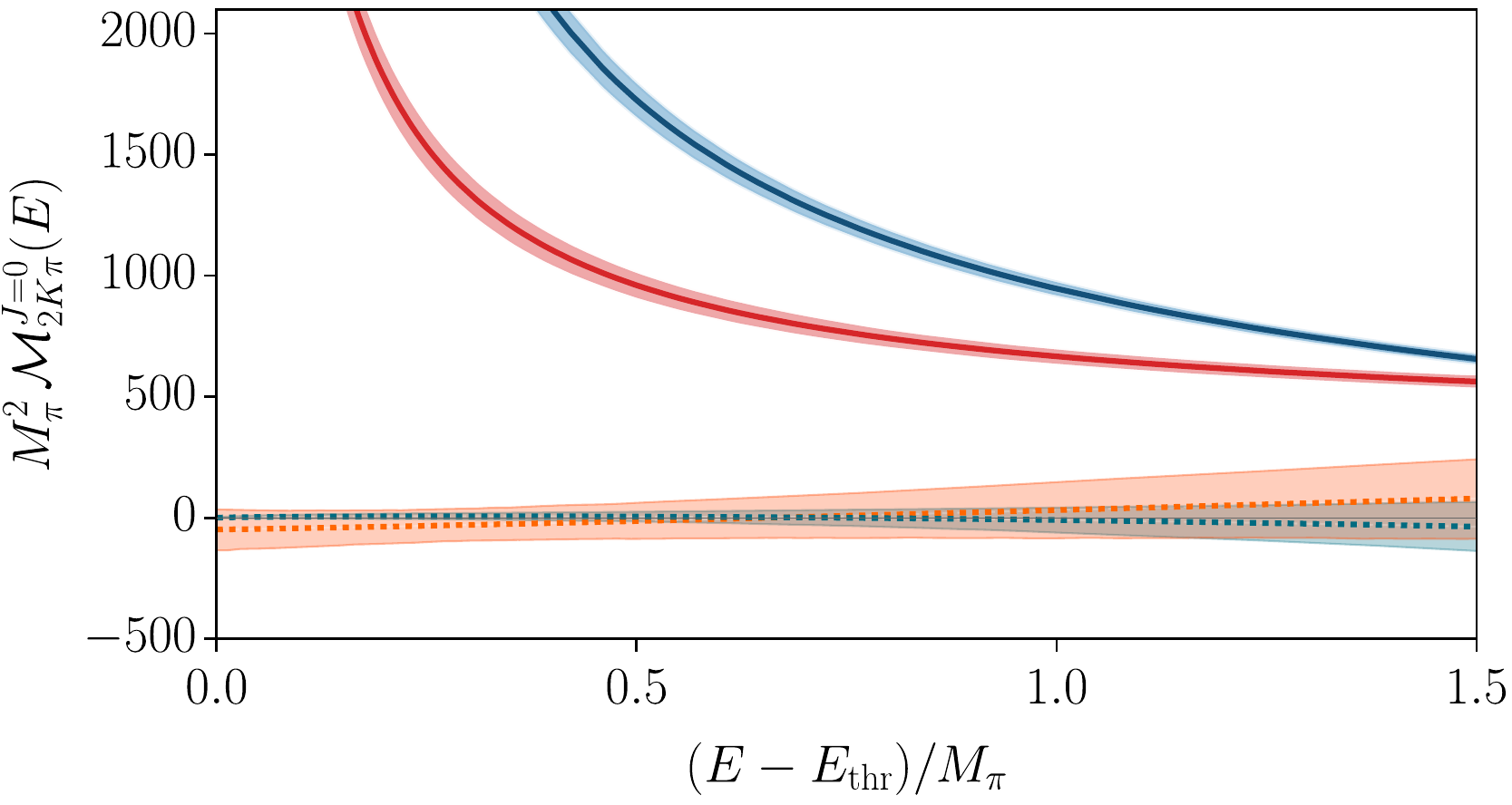}
    }
    \vspace{-0.12cm}
    \hfill
    \subfloat[~$3K$ amplitude]{%
     \includegraphics[width=0.49\textwidth]{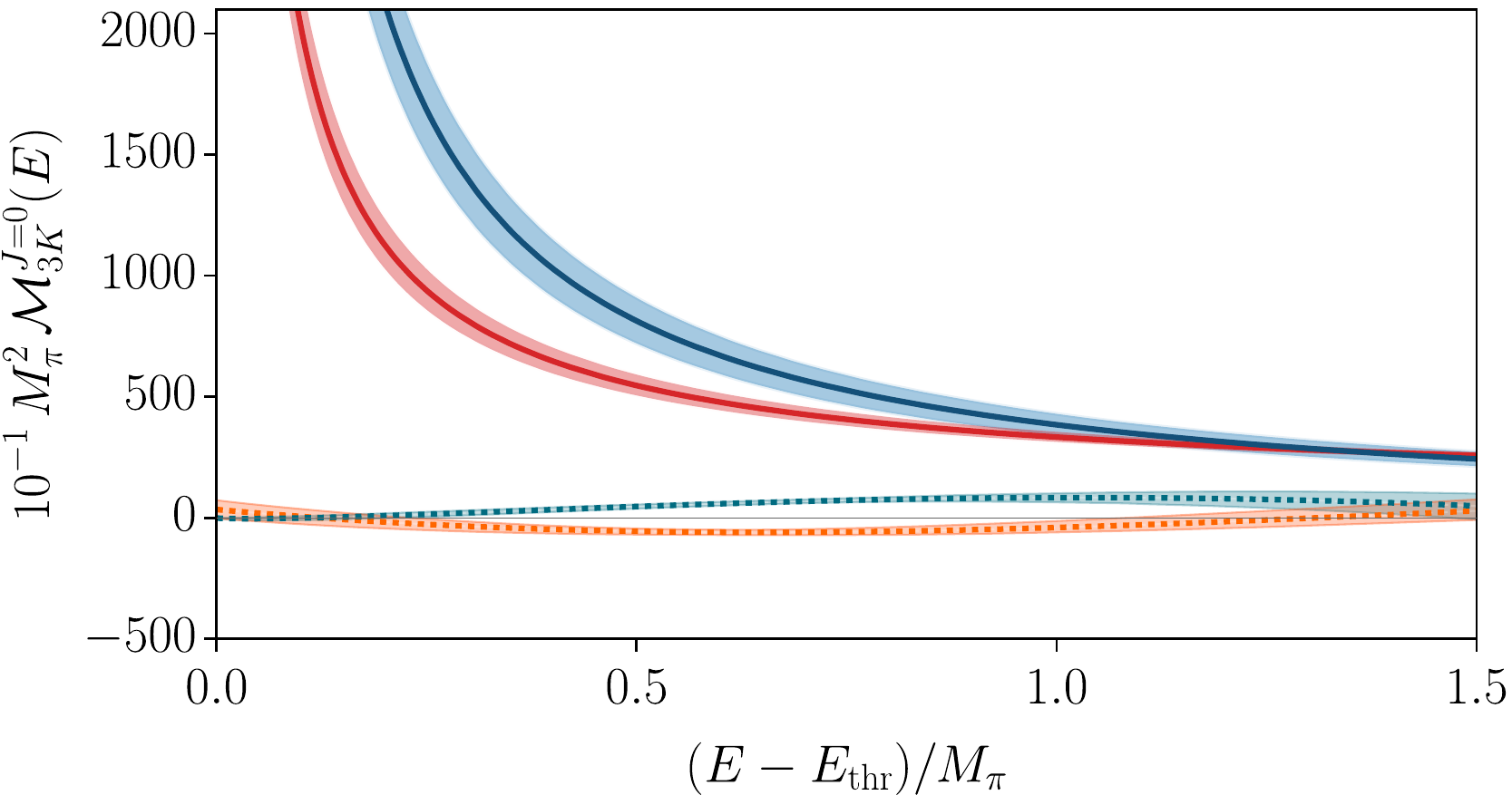}
    }
    \vspace{-0.12cm}
    \caption{Real and imaginary parts of the components of the $J^P=0^-$ three-meson amplitudes in the equilateral configuration as a function of the energy relative to threshold. Contributions from the ladder $\cD$ and divergence-free, $\cM_{\rm df,3}$, parts are shown. 
    The results for $3K$ have been rescaled, as indicated in the $y$-axis label, so as to use the same vertical scale in all panels.
    }
    %\vspace{-0.2cm}
    \label{fig:Kmixed1}
\end{figure*}

In \Cref{fig:Kmixed1}, we show the real and imaginary parts of these components for all $J^P=0^-$ three-meson amplitudes in the equilateral configuration as a function of $E$. Only the ladder component diverges at the threshold, and we find that the imaginary part (originating from the $i\epsilon$ prescription in the OPE propagators) dominates near the threshold. 

The ladder component is significantly larger than 
the divergence-free contribution, which is consistent with zero except for the $3K$ case. This agrees with the pattern of results for $\Kdf$ illustrated in \Cref{fig:Kdfconstrain}.
The conclusion is that the contribution from three-particle interactions is subdominant except possibly at the highest energies. In agreement with expectations, we do not observe any resonance-like enhancements in these amplitudes.

The analogous plots for higher values of $J$ do not exhibit any new features and are not shown.
However, the $J>0$ amplitudes show a non-trivial dependence on the angular orientation of external momenta. To show an example of this dependence, we vary the relative orientation between the initial and final triangle configurations by changing $\theta_{22}$ at a fixed value of energy, $E = E_{\rm thr} + M_\pi$, and $\alpha = 120^\circ$. The results is shown in~\Cref{fig:theta22} for the $\pi\pi K$ and $KK\pi$ amplitudes. This illustrates the relative magnitude of the different waves in the specific kinematic configurations considered, as well as the expected periodicity of the amplitude. Since the squared magnitude is shown, the fact that various components of the $J^P=1^+$ amplitude change sign between $\theta_{22}=0$ and $\pi$ is not apparent. Note that even when real and imaginary parts of $\cD^J$ and $\cM_{\rm df, 3}^J$ all have periodic behavior, their $\theta_{22}$ dependence might be shifted by different non-zero phases, leading to non-trivial interference between these components. In particular, there are angles $\theta_{22}$ at which $\cM_{\rm df, 3}^J$ is non-zero, while $\cD^J$ vanishes, and vice versa. The full $J=1$ amplitude does not vanish at $\theta_{22}=\pi/2$ and $3\pi/2$, but reaches its minimal values.

\begin{figure*}[th!]
     \centering
     \subfloat[~$3\pi$ amplitude]{%
     \includegraphics[width=0.49\textwidth]{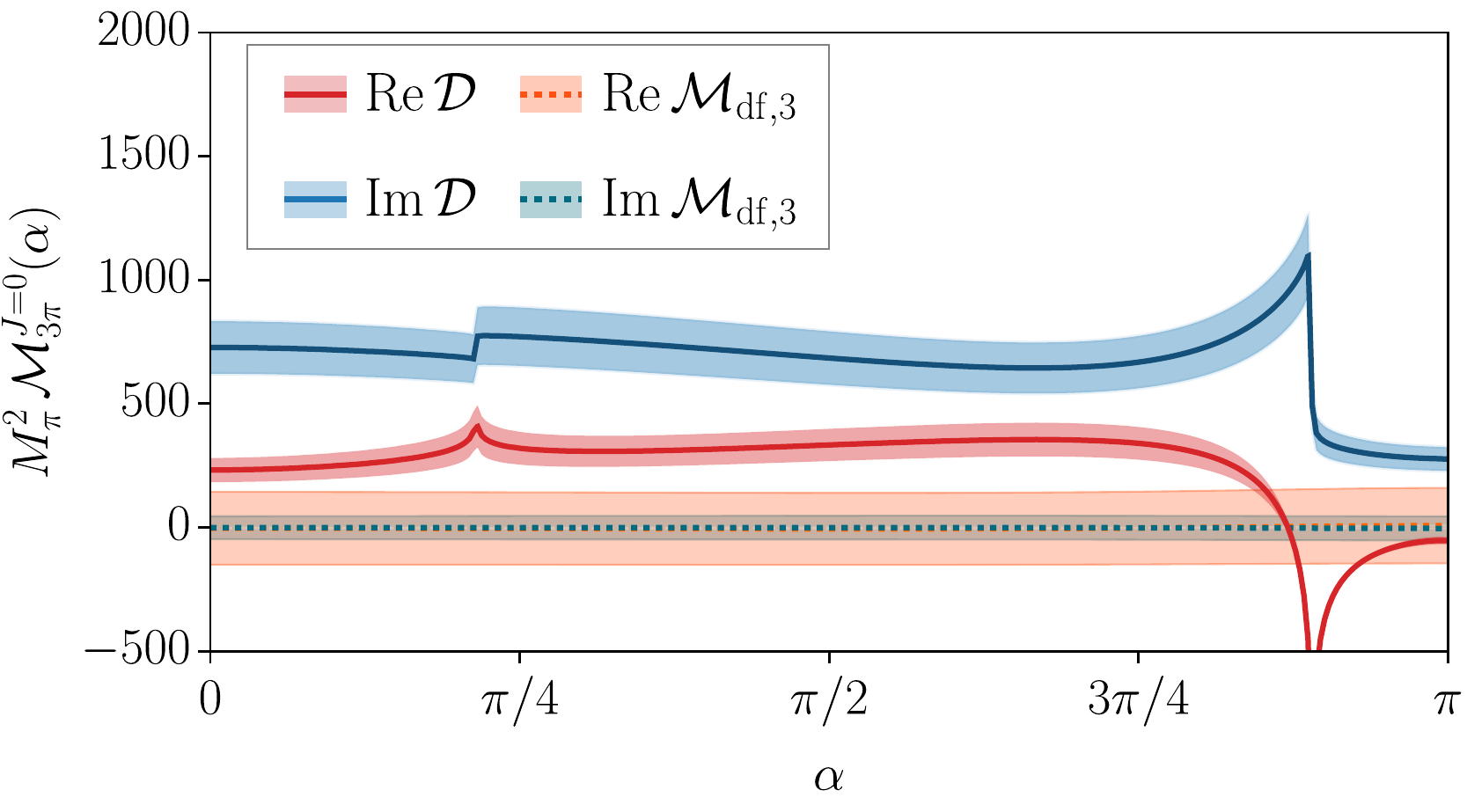}
    }
    \vspace{-0.12cm}
    \hfill
    \subfloat[~$\pi\pi K$ amplitude]{%
     \includegraphics[width=0.49\textwidth]{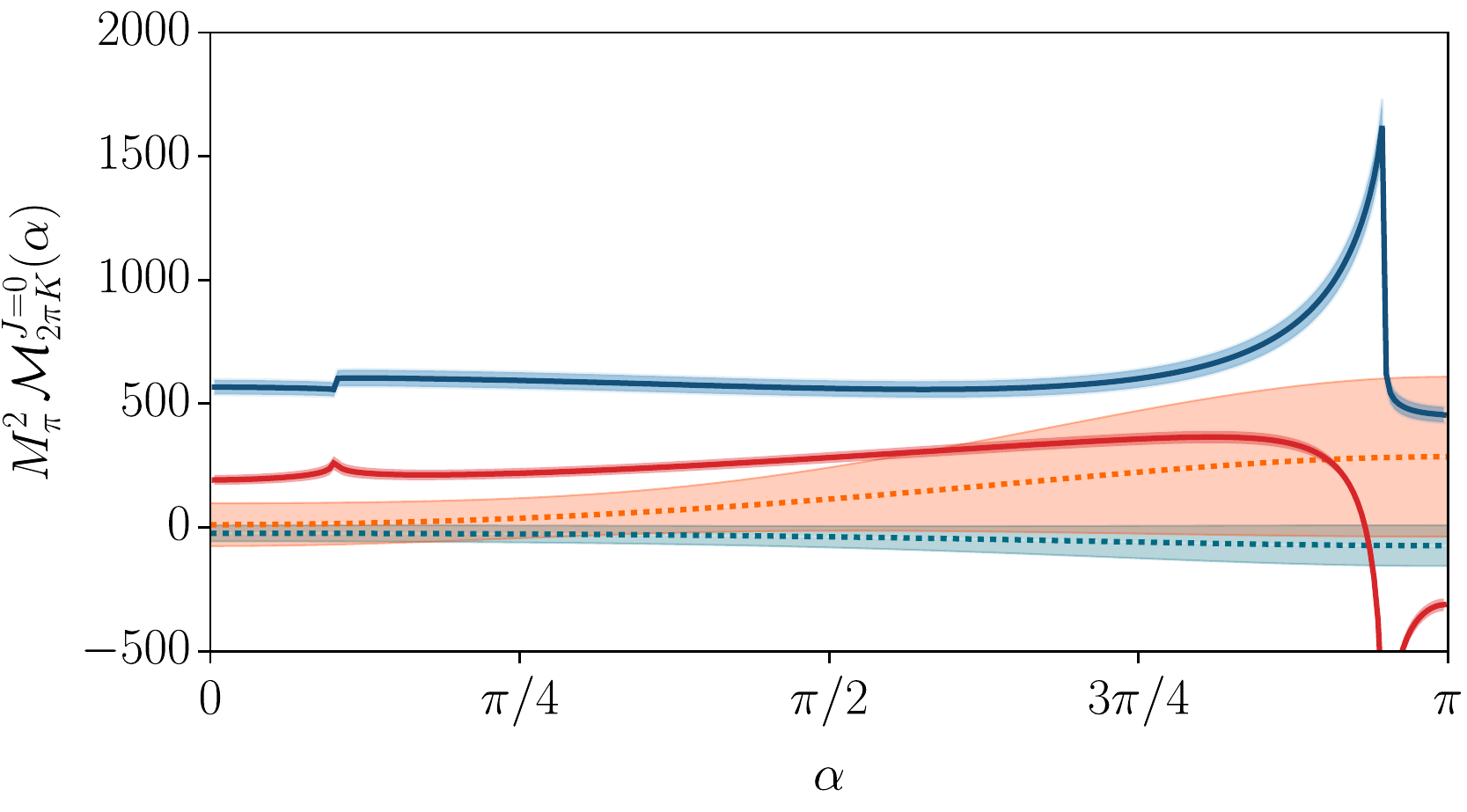}
    }
    \vspace{-0.12cm}
    \hfill
    \subfloat[~$K K \pi$ amplitude]{%
     \includegraphics[width=0.49\textwidth]{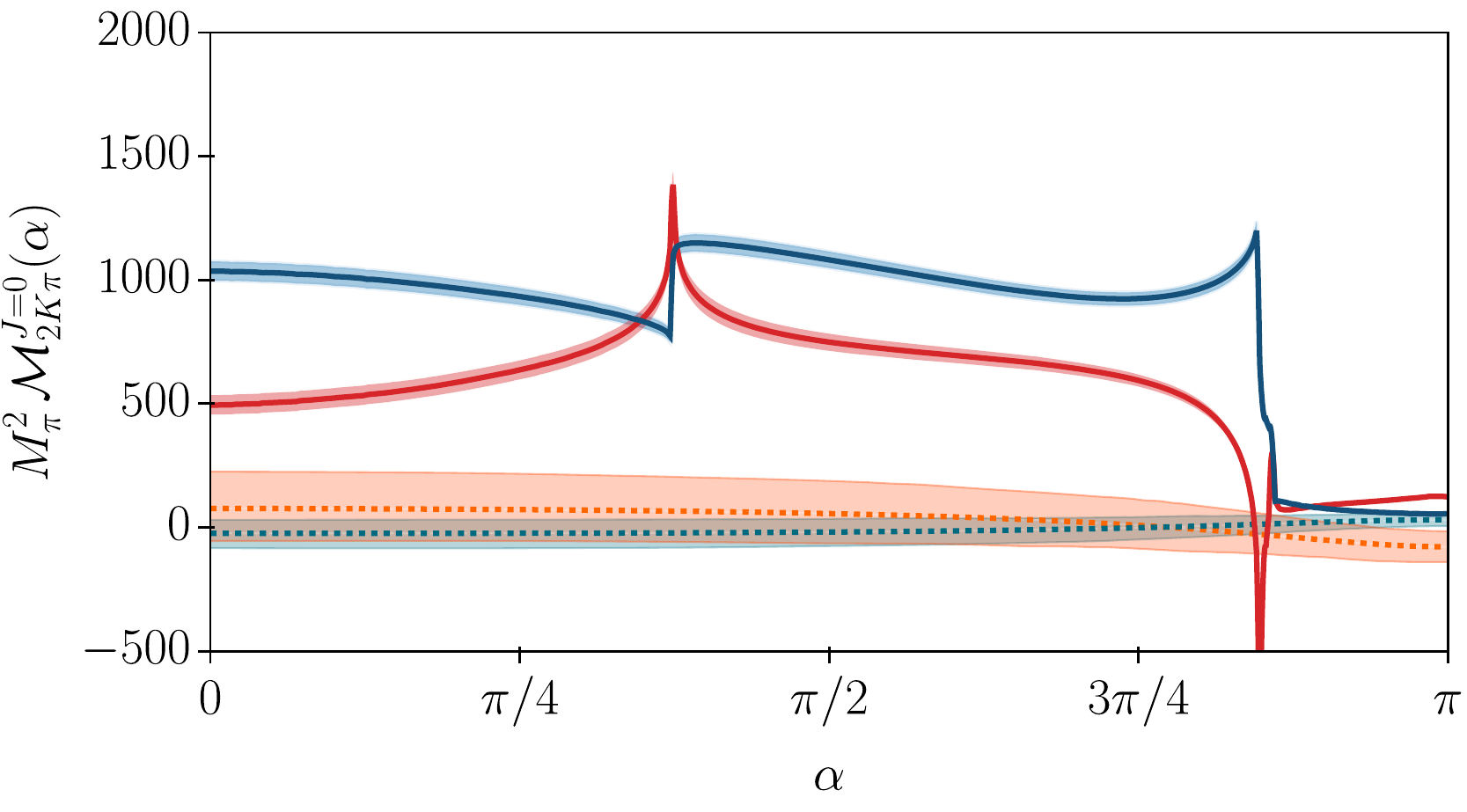}
    }
    \vspace{-0.12cm}
    \hfill
    \subfloat[~$3K$ amplitude]{%
     \includegraphics[width=0.49\textwidth]{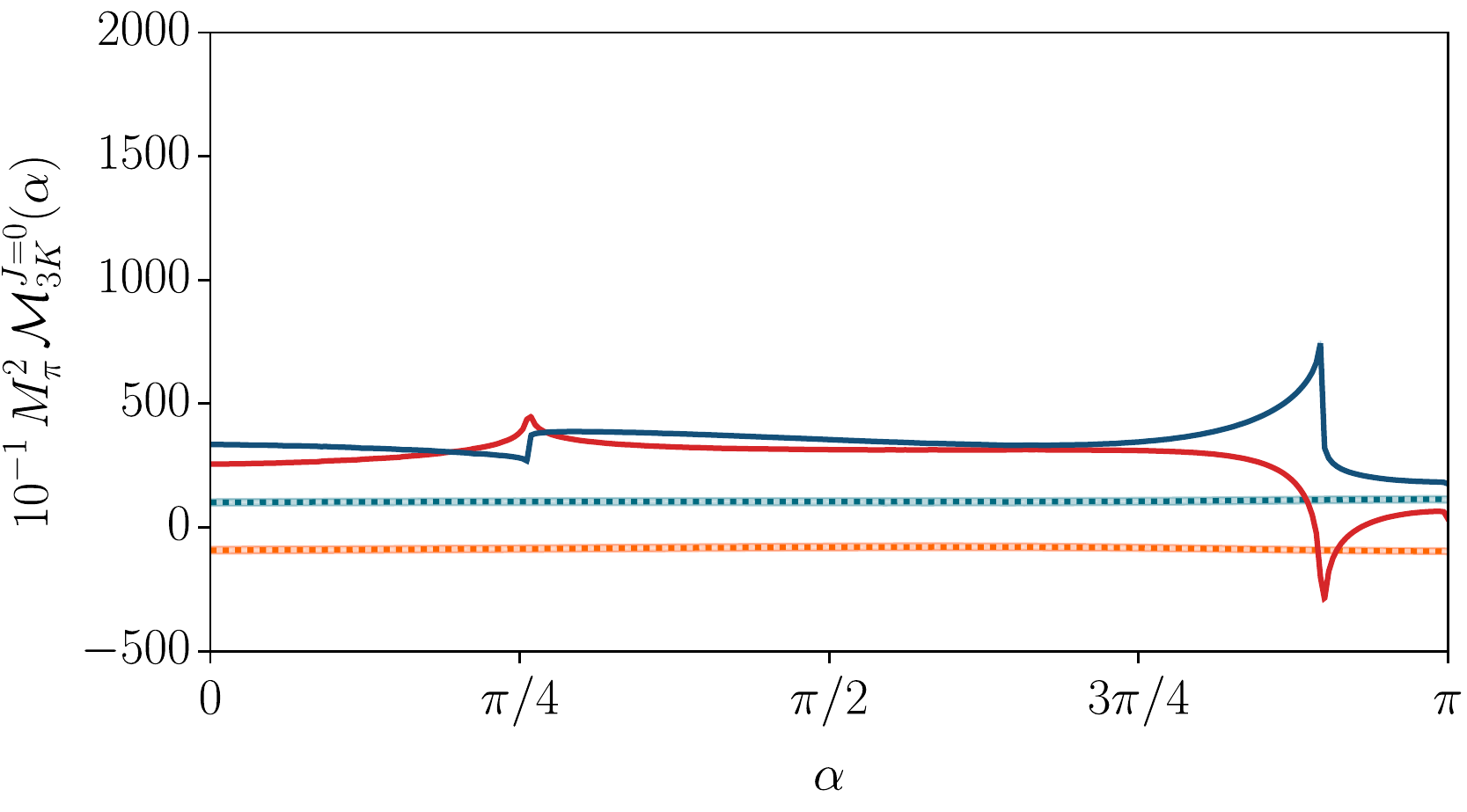}
    }
    \vspace{-0.12cm}
    \caption{Real and imaginary parts of the components of the $J^P=0^-$ three-meson amplitudes in the isosceles configuration, considered as a function of the angle $\alpha$. The total energy is fixed to $E = E_{\rm thr} + M_\pi$. The result for $3K$ has been rescaled, as indicated in the $y$-axis labels, so as to use the same vertical scale in all panels.
    }
    \vspace{-0.2cm}
    \label{fig:alpha_J0}
\end{figure*}

\begin{figure*}[th!]
    \centering
    \hfill
    \subfloat[~$\pi\pi K$ amplitude]{%
     \includegraphics[width=0.48\linewidth]{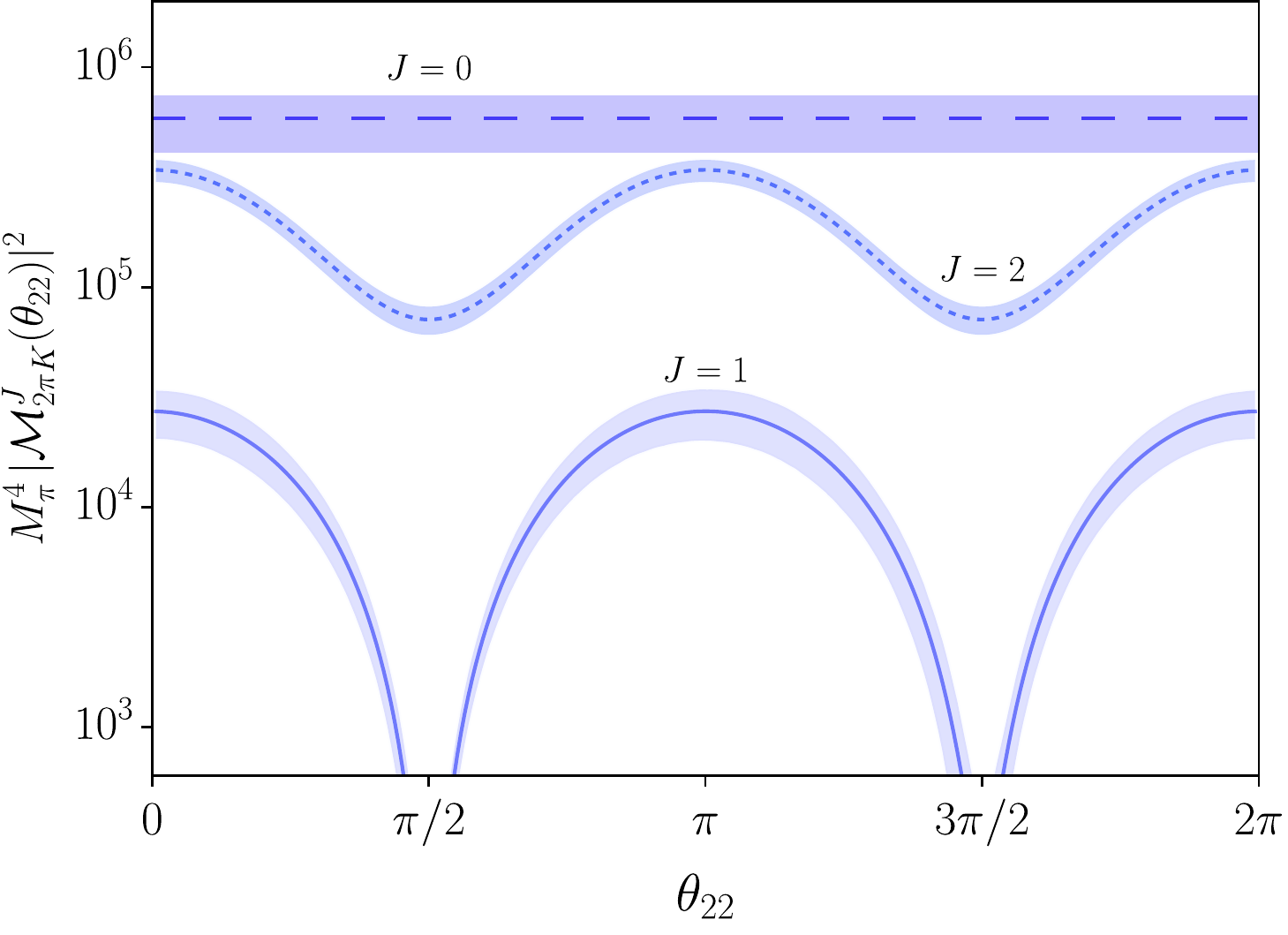}
    }    
     \subfloat[~$KK\pi$ amplitude]{%
     \includegraphics[width=0.48\linewidth]{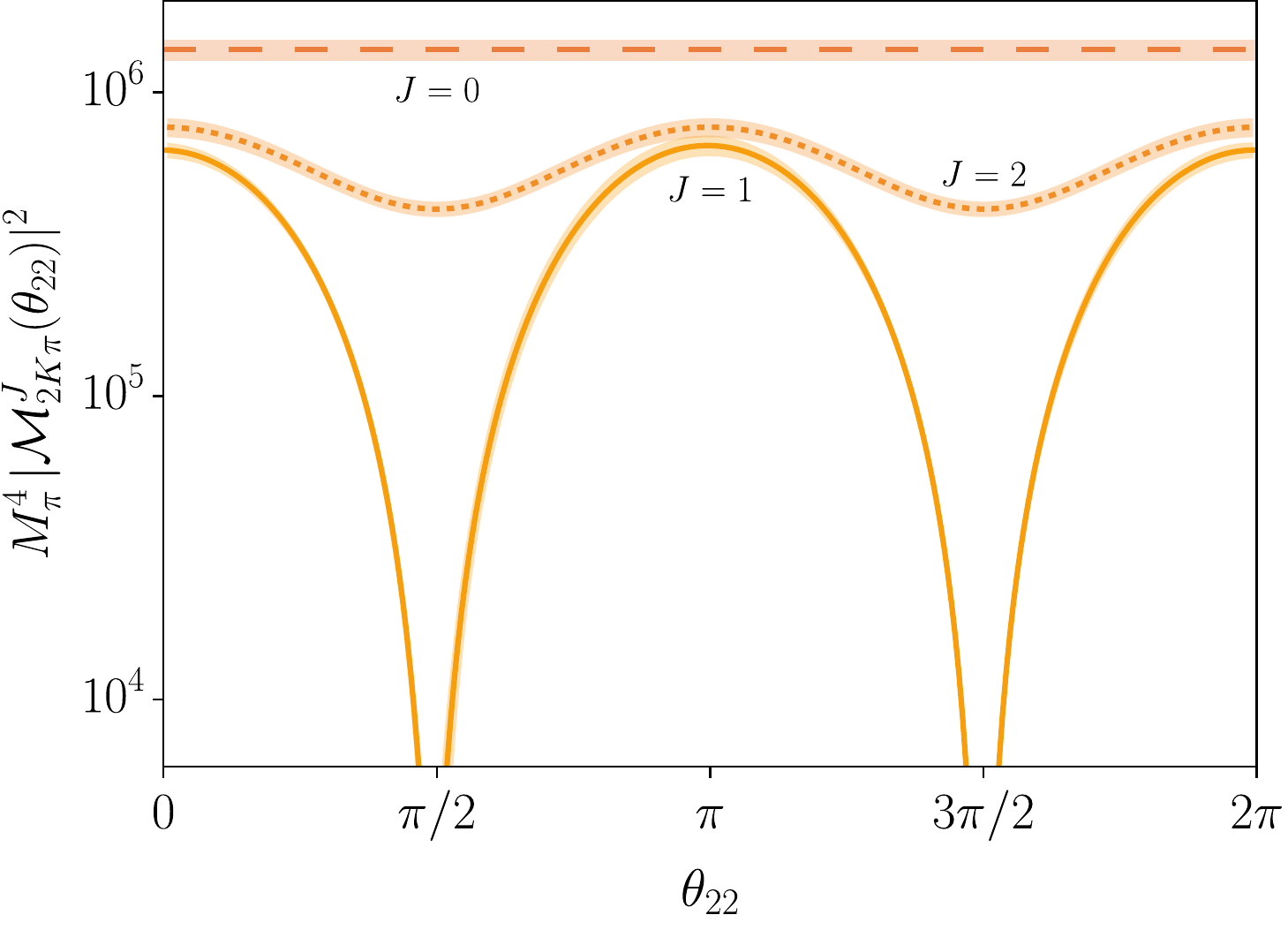}
    }
    \vspace{-0.2cm}
    \caption{Example of $\theta_{22}$ dependence for the $\pi^+\pi^+ K^+$ (left) 
    and $K^+ K^+ \pi^+$ (right) squared amplitudes in the equilateral configuration at $E = E_{\rm thr} + M_\pi$. Results with $J^P=0^-$, $1^+$ and $2^-$ are compared.
    }
    \vspace{-0.2cm}
    \label{fig:theta22}
\end{figure*}

\subsection{Chiral dependence of three-meson amplitudes}

Since we have access to two- and three-body K matrices computed on different ensembles along the $2m_{ud} + m_s = {\rm const }$ quark-mass trajectory, we also investigate the chiral dependence of three-meson amplitudes. Recall that along this trajectory, the kaon becomes heavier as the pion gets lighter. 

\begin{figure*}[th!]
     \centering
     \subfloat[~$3\pi$ amplitude]{%
     \includegraphics[width=0.49\textwidth]{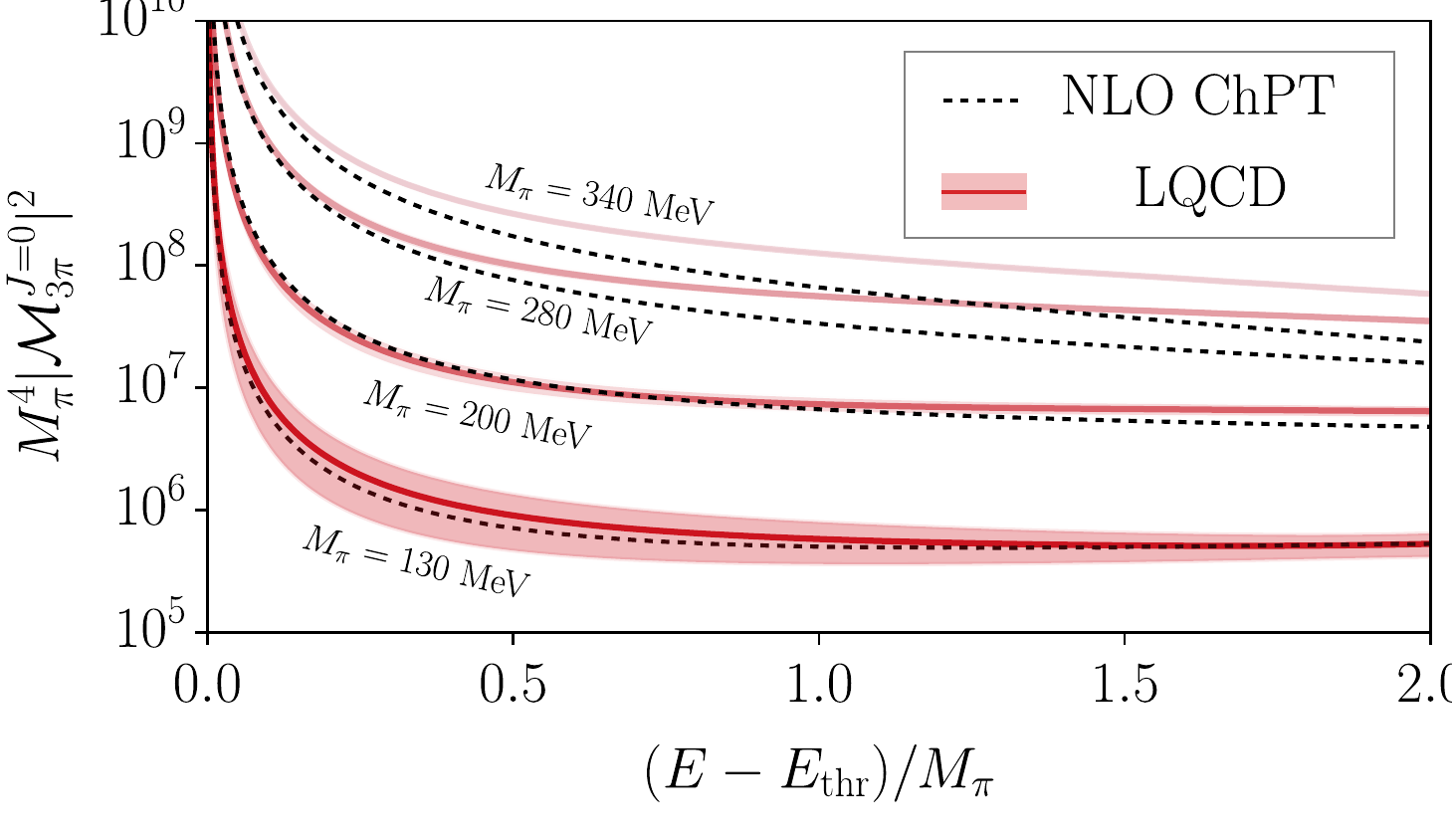}
    }
    \vspace{-0.12cm}
    \hfill
    \subfloat[~$\pi \pi K$ amplitude]{%
     \includegraphics[width=0.49\textwidth]{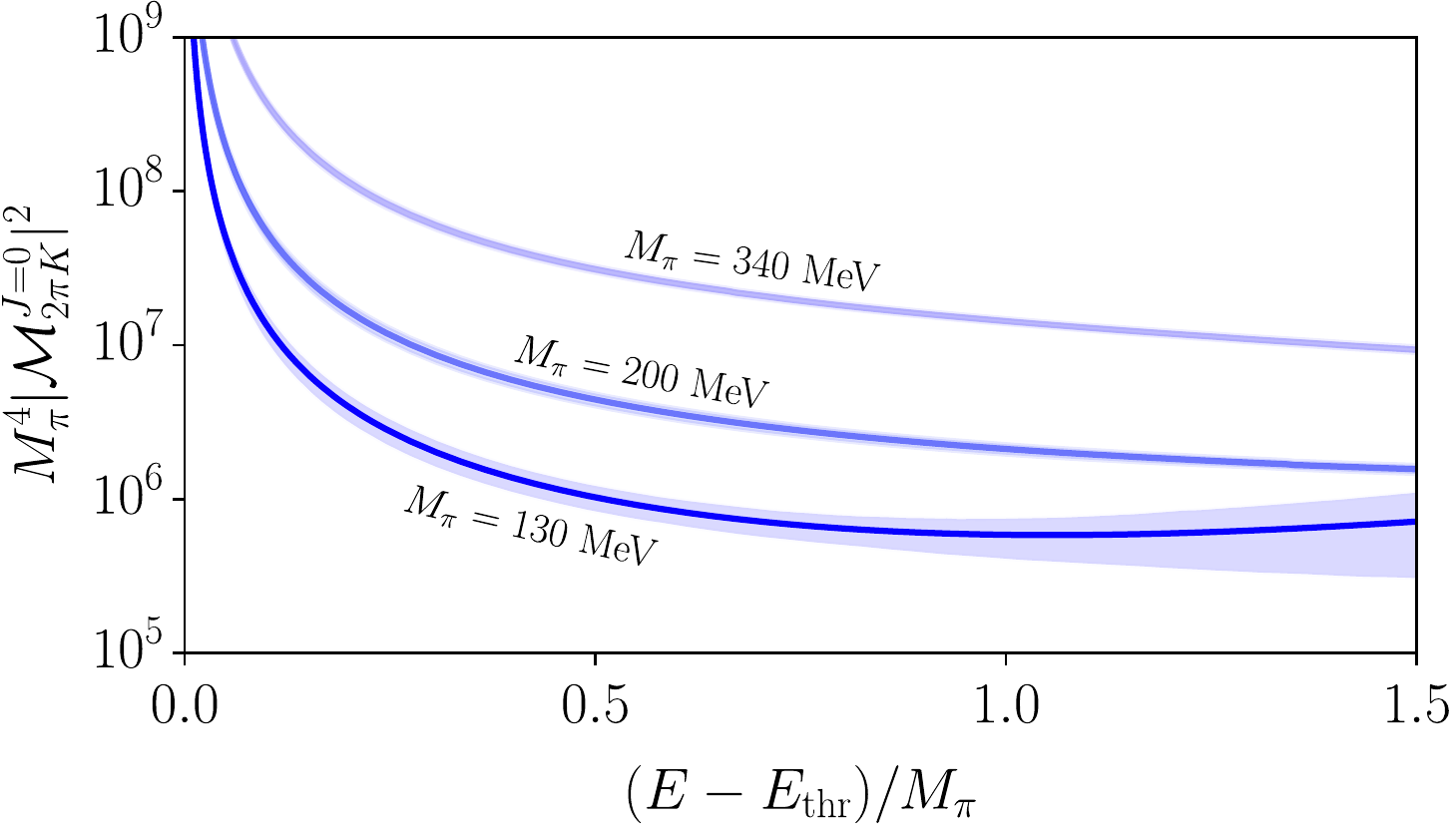}
    }
    \vspace{-0.12cm}
    \hfill
    \subfloat[~$K K \pi$ amplitude]{%
     \includegraphics[width=0.49\textwidth]{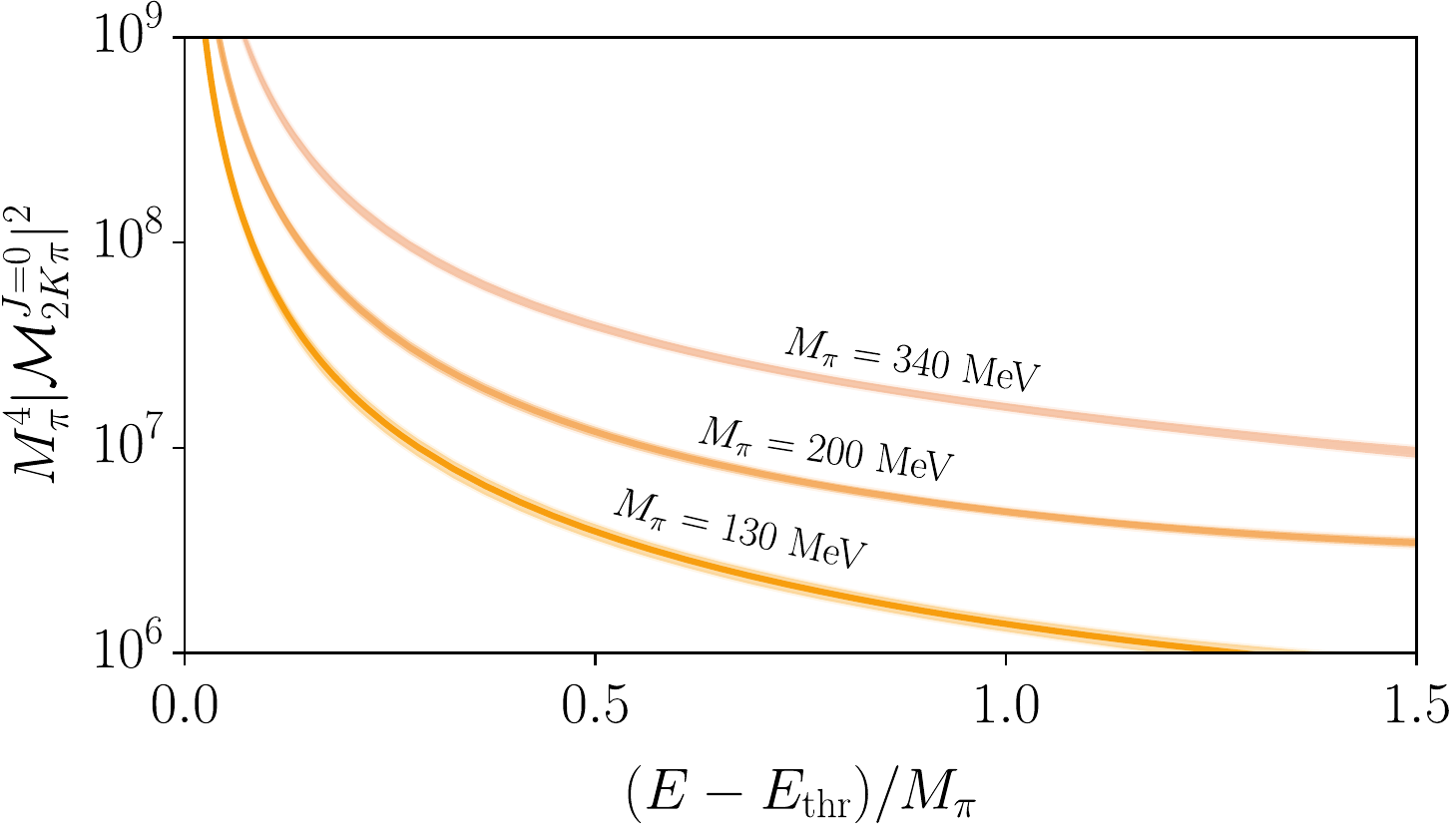}
    }
    \vspace{-0.12cm}
    \hfill
    \subfloat[~$3K$ amplitude]{%
     \includegraphics[width=0.49\textwidth]{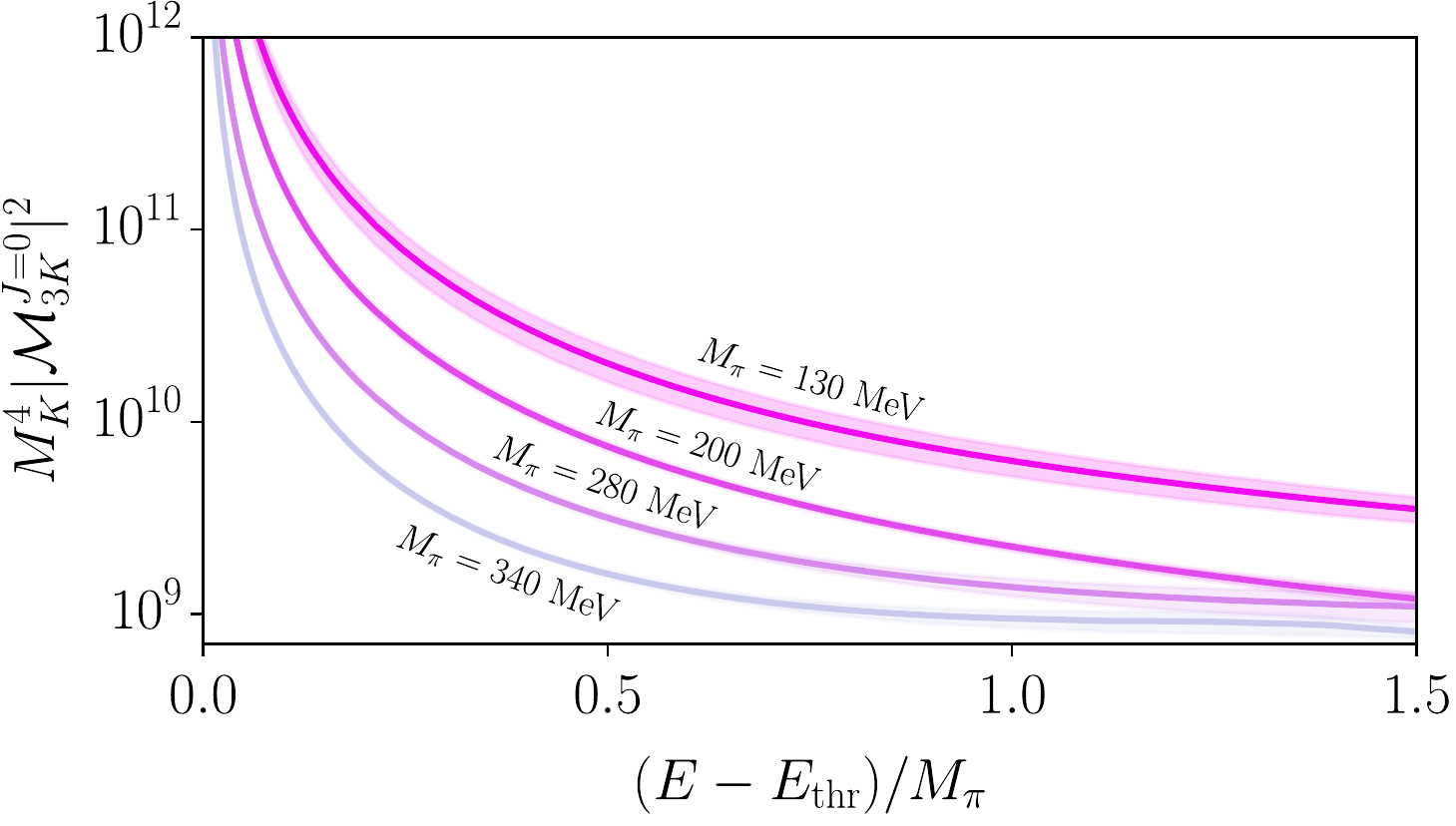}
    }
    \vspace{-0.12cm}
    \caption{Absolute value squared of the $J^P=0^-$ three-meson amplitudes in the equilateral configuration for four pion masses, as a function of the energy relative to threshold. In the $3\pi$ case, the result is compared to the NLO ChPT prediction. 
    } 
    \vspace{-0.2cm}
    \label{fig:amps_J0_masses}
\end{figure*}

For this comparison, we focus on the $J^P=0^-$ amplitude using the equilateral configuration. While similar plots can be made for the isosceles case, and larger partial waves, their qualitative behavior is similar. The results are displayed in \Cref{fig:amps_J0_masses} for all three meson systems.

We first comment on the $3\pi$ amplitude, shown in the upper-left panel of \Cref{fig:amps_J0_masses} for four different quark masses. ChPT predicts that the magnitude of the amplitude decreases as the pion mass is lowered. To illustrate this, we also show the NLO ChPT predictions. These are obtained by solving the integral equations using, as inputs, the two- and three-particle K matrices at NLO in ChPT from Refs.~\cite{Bijnens:2021hpq,Bijnens:2022zsq,Baeza-Ballesteros:2023ljl,Baeza-Ballesteros:2024mii}. The two-body K matrices were translated into unitary two-body on-shell amplitudes through the Inverse Amplitude Method~\cite{GomezNicola:2001as}. As can be seen, the NLO ChPT three-pion amplitude achieves a good agreement with the lattice result on the physical point ensemble, while the deviation increases with an increasing pion mass. Furthermore, the agreement is better close to the threshold, which is explained by the fact that ChPT is an expansion in both the pion mass and the pion squared momentum.

Results for the $\pi\pi K$ and $KK\pi$ amplitudes are also shown in \Cref{fig:amps_J0_masses}, in this case only for the three available quark-mass values. They exhibit a similar behavior to the three-pion amplitudes. Specifically, as the pion mass is reduced, these amplitudes decrease, which is again consistent with the expectation from ChPT. There are no NLO ChPT predictions available for the $\pi\pi^+ K^+$, $K^+K^+\pi^+$ systems. 

Lastly, the $3K$ amplitude is shown in the lower left corner of \Cref{fig:amps_J0_masses}, and it is normalized by $M_K$. In this case, the behavior along the chiral trajectory is different: as kaons become heavier towards the physical point, their interactions also strengthen; this is indeed what is observed in our results. As for the mixed states, there are no NLO ChPT predictions available for the $3K$ system.

\section{Conclusion}
\label{sec:concl}

This work has determined three-hadron scattering amplitudes from LQCD at physical quark masses for systems of two and three mesons at maximal isospin. Our calculations use four different ensembles, including one with $M_\pi=130\;$MeV. For several systems,
in particular $2K^+$, $3K^+$, $\pi^+\pi^+K^+$, and $K^+K^+\pi^+$, this marks the first direct calculation with quark masses that bracket the physical values (albeit in the approximation of exact isospin symmetry). 
In addition, we have computed the three-meson scattering amplitudes at the physical point for the first time.

The lattice setup follows that used in previous works at heavier quark masses~\cite{Blanton:2021llb,Draper:2023boj}. Specifically, we use Laplacian-Heaviside smearing~\cite{Peardon:2009gh} to define interpolating operators. Correlation functions are constructed using noise dilution projectors following the stochastic LapH method~\cite{Morningstar:2011ka}. Euclidean-time correlation functions are then analyzed by solving the GEVP and performing fits to the rotated correlators. This procedure is summarized in \Cref{sec:lattice}.
The key feature is that a large number of energy levels can be determined with good statistical accuracy.

The extracted energies are then analyzed using the finite-volume formalism, explained in \Cref{sec:QCs}. Specifically, parameters in the two- and three-particle K matrices are constrained by directly fitting the predictions of two- and three-particle quantization conditions to the LQCD energies, as described in \Cref{sec:fitstrategy}. For the physical-point ensemble, the fit results are shown in \Cref{sec:fits}. In \Cref{app:reananlysis}, we also reanalyze our previous spectrum using a consistent methodology and choice of cutoff. Several aspects of the chiral dependence of the K matrices are discussed in \Cref{sec:kdf_chiral}, including some plots in \Cref{app:reananlysis}. Overall we find good qualitative agreement when comparing to ChPT. 

A byproduct of this work is the accurate computation of two-meson amplitudes.
We determine threshold parameters, as well as the energy-dependent phase shift, for the $2\pi^+$, $\pi^+ K^+$, and $2K^+$ systems.
The inclusion of three-particle energies leads to a more accurate determination of these quantities since the dominant contribution to the energy shifts comes from two-particle interactions, and furthermore, there are three pairs in a three-particle system.
When available, we compare to experimental results, in particular for $\pi^+\pi^+$ and $\pi^+ K^+$ systems. In these cases, we find agreement with phenomenological fits to experimental data at the $1-2\sigma$ level. For the threshold parameters, our results provide competitive statistical uncertainties when compared to those summarized in the FLAG report~\cite{FlavourLatticeAveragingGroupFLAG:2024oxs}, although (as discussed below) our results do not contain a complete error budget.
We note that the inclusion of the ensemble with $M_\pi=130\;$MeV reduces the
errors in threshold parameters by about a factor of two compared to our previous results based on chiral extrapolations.

A major focus of this work is the computation of three-particle scattering amplitudes from first principles. This requires solving integral equations in which the K matrices, obtained from fits to the spectrum, are inputs. The necessary steps are summarized in \Cref{sec:inteqs}. Further technical details are provided in \Cref{app:pwproj}, including detailed results for the partial-wave projections of the building blocks.  In \Cref{sec:amplitudes}, we show results for the resulting amplitudes with several choices of kinematic configurations and overall angular momentum, and comment on their features. We also display the chiral dependence of these amplitudes. For the case of $3\pi^+$, we compare the full amplitude to ChPT and find an increasingly good agreement with decreasing pion mass.

Some systematic uncertainties remain unquantified. First, although exponentially suppressed finite-volume effects are expected to be small due to $M_\pi L > 4$, they are certainly present at the percent level. 
They could be further quantified by working on additional, larger volumes.
Second, chiral fits are performed using NLO ChPT expressions, which neglects higher-order contributions, which is a concern for the ensembles with larger pion masses. Third, our calculations are conducted at a single lattice spacing. Using Wilson Chiral Perturbation Theory, we estimate discretization errors to be small, but a detailed study of these effects will require additional lattice spacings.
Finally, our physical-point simulation has slightly detuned light-quark masses, and does not include isospin breaking. Despite these limitations, the small statistical uncertainties highlight the strong constraining power of our results.

This work has successfully demonstrated the feasibility of studying physical three-meson systems at maximal isospin. Building on this milestone, future efforts will focus on exploring the properties of three-meson resonances at the physical point.

\begin{acknowledgments}
We acknowledge helpful discussions with Jorge Baeza-Ballesteros, Andrew Jackura, Maxim Mai, and Arkaitz Rodas.
We thank our colleagues within the CLS consortium for sharing ensembles,
and Marco C\`e for providing data for the meson decay constants on E250.

The work of SMD, ZTD, and SRS is partially supported by U.S. Department of Energy grant No. DE-SC0011637. FRL acknowledges partial support by the Mauricio and Carlota Botton Fellowship and by the USDOE Contract No. DE-SC0011090 and DE-SC0021006. CM and SS acknowledge support from the U.S. NSF under award PHY-2209167.

Quark propagators were computed on Frontera at the Texas Advanced Computing Center (U.S. NSF, award OAC-1818253) and Hawk at the High Performance Computing Center in Stuttgart. Meson functions and contractions were carried out on ``Mogon II" at JGU Mainz.

\end{acknowledgments}

\clearpage
\appendix

\section{Interpolating operators on E250}
\label{app:operators}

In this appendix we specify the interpolating operators utilized in this work for the E250 ensemble. Each table corresponds to a different overall flavor combination. Additionally, due to space constraints on each page, some of the tables are further divided by the total momentum. For each total momentum $\bm{P} = \frac{2\pi}{L} \bm{d}$ --- where $\bm{d}$ is a vector of integers --- total irrep $\Lambda$, and total flavor considered, the operators entering the correlator matrix of \Cref{eq:spec_decomp} can be enumerated by the number of appearances of the irrep $\Lambda$ in the rows with the $\bm{d}_{\rm ref}$ that can be obtained from $\bm{d}$ through allowed rotations of a cube.
Our notation for the irreps listed can be found in Ref.~\cite{Morningstar:2013bda}.
For every $\bm{d}_{\rm ref}$ listed, we also constructed corresponding operators for all $\bm{d}$ that can be obtained through these allowed rotations of a cube.
Each operator corresponds to a particular multi-hadron free energy level. For each free energy, the tables list the corresponding momentum squared of each hadron, the total energy, and the irreps of the corresponding operators.

The $\pi\pi\pi$ operators are listed in \Cref{tab:ppp_psq0-4_ops,tab:ppp_psq5-9_ops}, the $K\pi\pi$ operators are listed in \Cref{tab:kpp_psq0-2_ops,tab:kpp_psq3-9_ops}, the $KK\pi$ operators are listed in \Cref{tab:kkp_psq0-6_ops,tab:kkp_psq8-9_ops}, the $KKK$ operators are listed in \Cref{tab:kkk_psq0-4_ops,tab:kkk_psq5-9_ops}, the $\pi\pi$ operators are listed in \Cref{tab:pp_psq0-5_ops,tab:pp_psq6-9_ops}, the $K\pi$ operators are listed in \Cref{tab:kp_psq0-3_ops,tab:kp_psq4-9_ops}, and the $KK$ operators are listed in \Cref{tab:kk_psq0-6_ops,tab:kk_psq8-9_ops}.

\begin{table}[!bp]
\centering
\begin{tabular}{c|c|c|c}
$\bm{d}_{\rm ref}$&$[d^2_{\pi_1}, d^2_{\pi_2}, d^2_{\pi_3}]$&$E^{\rm free}/M_{\pi}$&operators\\%
\hline%
(0, 0, 0)&{[}0, 0, 0{]}&3.0&$A_{1u}^-$\\%
&{[}1, 1, 0{]}&4.69237&$A_{1u}^- \oplus E_u^-$\\%
&{[}2, 2, 0{]}&5.8236&$A_{1u}^- \oplus E_u^- \oplus T_{2u}^-$\\%
\hline%
(0, 0, 1)&{[}1, 0, 0{]}&3.51919&$A_2^-$\\%
&{[}2, 1, 0{]}&5.02374&$A_2^- \oplus B_2^- \oplus E^-$\\%
&{[}1, 1, 1{]}&5.31668&$2 A_2^- \oplus B_2^-$\\%
&{[}4, 1, 0{]}&5.90663&$A_2^-$\\%
\hline%
(0, 1, 1)&{[}2, 0, 0{]}&3.82717&$A_2^-$\\%
&{[}1, 1, 0{]}&4.14747&$A_2^-$\\%
&{[}3, 1, 0{]}&5.27585&$A_2^- \oplus B_2^-$\\%
&{[}2, 2, 0{]}&5.39422&$A_1^- \oplus A_2^-$\\%
&{[}2, 1, 1{]}&5.69597&$A_1^- \oplus 4 A_2^- \oplus 3 B_1^- \oplus B_2^-$\\%
\hline%
(1, 1, 1)&{[}3, 0, 0{]}&4.05855&$A_2^-$\\%
&{[}2, 1, 0{]}&4.51898&$A_2^- \oplus E^-$\\%
&{[}1, 1, 1{]}&4.84256&$A_2^-$\\%
&{[}3, 1, 1{]}&5.98437&$2 A_2^- \oplus 2 E^-$\\%
\hline%
(0, 0, 2)&{[}1, 1, 0{]}&3.51919&$A_2^-$\\%
&{[}4, 0, 0{]}&4.24778&$A_2^-$\\%
&{[}2, 2, 0{]}&4.92755&$A_2^- \oplus B_2^-$\\%
&{[}2, 1, 1{]}&5.25617&$A_2^- \oplus B_2^- \oplus E^-$\\%
&{[}5, 1, 0{]}&5.66272&$A_2^- \oplus B_2^- \oplus E^-$\\%
&{[}3, 3, 0{]}&5.97821&$A_2^- \oplus B_1^-$\\%
\end{tabular}%
\caption{\label{tab:ppp_psq0-4_ops} $\pi\pi\pi$ operators with $\bm{d}_{\rm ref}^2 \leq 4$ used in this work.}
\end{table}%
\begin{table}[!bp]
\centering
\begin{tabular}{c|c|c|c}
$\bm{d}_{\rm ref}$&$[d^2_{\pi_1}, d^2_{\pi_2}, d^2_{\pi_3}]$&$E^{\rm free}/M_{\pi}$&operators\\%
\hline%
(0, 1, 2)&{[}2, 1, 0{]}&3.95025&$A_2^-$\\%
&{[}1, 1, 1{]}&4.31666&$A_2^-$\\%
&{[}5, 0, 0{]}&4.4097&$A_2^-$\\%
&{[}4, 1, 0{]}&5.0254&$A_2^-$\\%
&{[}3, 2, 0{]}&5.23387&$A_1^- \oplus A_2^-$\\%
&{[}3, 1, 1{]}&5.56739&$A_1^- \oplus A_2^-$\\%
&{[}2, 2, 1{]}&5.69597&$2 A_1^- \oplus 5 A_2^-$\\%
&{[}6, 1, 0{]}&5.821&$A_1^- \oplus A_2^-$\\%
\hline%
(1, 1, 2)&{[}3, 1, 0{]}&4.26627&$A_2^-$\\%
&{[}2, 2, 0{]}&4.4118&$A_2^-$\\%
&{[}6, 0, 0{]}&4.55223&$A_2^-$\\%
&{[}2, 1, 1{]}&4.77603&$A_1^- \oplus 2 A_2^-$\\%
&{[}5, 1, 0{]}&5.22011&$A_1^- \oplus A_2^-$\\%
&{[}4, 2, 0{]}&5.48405&$A_2^-$\\%
&{[}4, 1, 1{]}&5.82218&$A_2^-$\\%
\hline%
(0, 2, 2)&{[}2, 2, 0{]}&3.82717&$A_2^-$\\%
&{[}2, 1, 1{]}&4.2419&$A_2^-$\\%
&{[}5, 1, 0{]}&4.73633&$A_2^- \oplus B_1^-$\\%
&{[}8, 0, 0{]}&4.79663&$A_2^-$\\%
&{[}3, 3, 0{]}&5.10934&$A_2^-$\\%
&{[}4, 1, 1{]}&5.39268&$A_2^- \oplus B_1^-$\\%
&{[}3, 2, 1{]}&5.61355&$A_1^- \oplus 2 A_2^- \oplus B_1^- \oplus 2 B_2^-$\\%
&{[}2, 2, 2{]}&5.75186&$A_1^- \oplus A_2^-$\\%
&{[}6, 2, 0{]}&5.88601&$A_1^- \oplus A_2^- \oplus B_1^- \oplus B_2^-$\\%
\hline%
(0, 0, 3)&{[}1, 1, 1{]}&3.0&$A_2^-$\\%
&{[}4, 1, 0{]}&3.95236&$A_2^-$\\%
&{[}2, 2, 1{]}&4.77603&$A_2^- \oplus B_2^-$\\%
&{[}9, 0, 0{]}&4.90383&$A_2^-$\\%
&{[}5, 2, 0{]}&5.25826&$A_2^- \oplus B_2^- \oplus E^-$\\%
&{[}5, 1, 1{]}&5.62752&$A_2^- \oplus B_2^- \oplus E^-$\\%
&{[}4, 2, 1{]}&5.90409&$A_2^- \oplus B_2^- \oplus E^-$\\%
&{[}3, 3, 1{]}&5.98437&$A_2^- \oplus B_1^-$\\%
\end{tabular}%
\caption{\label{tab:ppp_psq5-9_ops} $\pi\pi\pi$ operators with $5 \leq \bm{d}_{\rm ref}^2 \leq 9$ used in this work.}
\end{table}%

\begin{table}[!bp]
\centering
\begin{tabular}{c|c|c|c}
$\bm{d}_{\rm ref}$&$[d^2_{K}, d^2_{\pi_1}, d^2_{\pi_2}]$&$E^{\rm free}/M_{\pi}$&operators\\%
\hline%
(0, 0, 0)&{[}0, 0, 0{]}&5.77412&$A_{1u}$\\%
&{[}1, 1, 0{]}&6.92691&$A_{1u} \oplus E_u \oplus T_{1g}$\\%
&{[}0, 1, 1{]}&7.46648&$A_{1u} \oplus E_u$\\%
&{[}2, 2, 0{]}&7.77766&$A_{1u} \oplus E_u \oplus T_{1g}$\\%
\hline%
(0, 0, 1)&{[}1, 0, 0{]}&5.87936&$A_2$\\%
&{[}0, 1, 0{]}&6.43583&$A_2$\\%
&{[}2, 1, 0{]}&7.04309&$A_2 \oplus B_2 \oplus E$\\%
&{[}1, 2, 0{]}&7.33005&$A_2 \oplus B_2 \oplus E$\\%
&{[}4, 1, 0{]}&7.57532&$A_2$\\%
&{[}1, 1, 1{]}&7.6166&$4 A_2 \oplus 2 B_2 \oplus E$\\%
&{[}0, 2, 1{]}&7.88075&$A_2 \oplus B_2 \oplus E$\\%
\hline%
(0, 1, 1)&{[}2, 0, 0{]}&5.97557&$A_2$\\%
&{[}1, 1, 0{]}&6.57003&$A_2 \oplus B_1$\\%
&{[}0, 2, 0{]}&6.84256&$A_2$\\%
&{[}0, 1, 1{]}&7.13664&$A_2$\\%
&{[}3, 1, 0{]}&7.15042&$A_2 \oplus B_2$\\%
&{[}2, 2, 0{]}&7.46158&$A_1 \oplus A_2 \oplus B_1 \oplus B_2$\\%
&{[}1, 3, 0{]}&7.63969&$A_2 \oplus B_2$\\%
&{[}5, 1, 0{]}&7.66515&$A_2 \oplus B_1$\\%
&{[}2, 1, 1{]}&7.75359&$A_1 \oplus A_2 \oplus B_1 \oplus B_2$\\%
\end{tabular}%
\caption{\label{tab:kpp_psq0-2_ops} $K\pi\pi$ operators with $\bm{d}_{\rm ref}^2 \leq 2$ used in this work.}
\end{table}%
\begin{table}[!bp]
\centering
\begin{tabular}{c|c|c|c}
$\bm{d}_{\rm ref}$&$[d^2_{K}, d^2_{\pi_1}, d^2_{\pi_2}]$&$E^{\rm free}/M_{\pi}$&operators\\%
\hline%
(1, 1, 1)&{[}3, 0, 0{]}&6.06448&$A_2$\\%
&{[}2, 1, 0{]}&6.69241&$A_2 \oplus E$\\%
&{[}1, 2, 0{]}&6.99377&$A_2 \oplus E$\\%
&{[}0, 3, 0{]}&7.15375&$A_2$\\%
&{[}1, 1, 1{]}&7.29355&$A_2 \oplus E$\\%
&{[}0, 2, 1{]}&7.56898&$A_2 \oplus E$\\%
&{[}6, 1, 0{]}&7.74998&$A_2 \oplus E$\\%
\hline%
(0, 0, 2)&{[}4, 0, 0{]}&6.14734&$A_2$\\%
&{[}1, 1, 0{]}&6.19262&$A_2$\\%
&{[}0, 1, 1{]}&6.79079&$A_2$\\%
&{[}2, 2, 0{]}&7.13151&$A_2 \oplus B_2 \oplus E$\\%
&{[}5, 1, 0{]}&7.34423&$A_2 \oplus B_2 \oplus E$\\%
&{[}0, 4, 0{]}&7.41135&$A_2$\\%
&{[}2, 1, 1{]}&7.43649&$A_2 \oplus B_2 \oplus E$\\%
&{[}1, 2, 1{]}&7.73954&$A_2 \oplus B_2 \oplus E$\\%
\hline%
(0, 1, 2)&{[}5, 0, 0{]}&6.2251&$A_2$\\%
&{[}2, 1, 0{]}&6.32231&$A_2$\\%
&{[}1, 2, 0{]}&6.64048&$A_2$\\%
&{[}4, 1, 0{]}&6.91027&$A_2$\\%
&{[}1, 1, 1{]}&6.9555&$2 A_2$\\%
&{[}0, 2, 1{]}&7.2438&$A_2$\\%
&{[}3, 2, 0{]}&7.25841&$A_1 \oplus A_2$\\%
&{[}6, 1, 0{]}&7.43273&$A_1 \oplus A_2$\\%
&{[}2, 3, 0{]}&7.46685&$A_1 \oplus A_2$\\%
&{[}3, 1, 1{]}&7.56821&$A_1 \oplus A_2$\\%
&{[}1, 4, 0{]}&7.58558&$A_2$\\%
&{[}0, 5, 0{]}&7.63376&$A_2$\\%
&{[}5, 2, 0{]}&7.80185&$A_1 \oplus A_2$\\%
\hline%
(1, 1, 2)&{[}6, 0, 0{]}&6.29847&$A_2$\\%
&{[}3, 1, 0{]}&6.44166&$A_2$\\%
&{[}2, 2, 0{]}&6.7854&$A_1 \oplus A_2$\\%
&{[}1, 3, 0{]}&6.98078&$A_2$\\%
&{[}5, 1, 0{]}&7.00863&$A_1 \oplus A_2$\\%
&{[}2, 1, 1{]}&7.10526&$A_1 \oplus 2 A_2$\\%
&{[}4, 2, 0{]}&7.37637&$A_2$\\%
&{[}1, 2, 1{]}&7.42184&$2 A_1 \oplus 3 A_2$\\%
&{[}0, 3, 1{]}&7.58948&$A_2$\\%
&{[}4, 1, 1{]}&7.69064&$A_2$\\%
&{[}0, 2, 2{]}&7.71171&$A_2$\\%
&{[}2, 4, 0{]}&7.74409&$A_2$\\%
&{[}1, 5, 0{]}&7.81679&$A_1$\\%
\hline%
(0, 2, 2)&{[}2, 2, 0{]}&6.42066&$A_2$\\%
&{[}8, 0, 0{]}&6.43424&$A_2$\\%
&{[}5, 1, 0{]}&6.65614&$A_2 \oplus B_1$\\%
&{[}2, 1, 1{]}&6.7578&$A_2$\\%
&{[}1, 2, 1{]}&7.08992&$A_2 \oplus B_1$\\%
&{[}3, 3, 0{]}&7.28061&$A_2 \oplus B_2$\\%
&{[}4, 1, 1{]}&7.37083&$A_2 \oplus B_1$\\%
&{[}0, 2, 2{]}&7.39281&$A_2$\\%
&{[}1, 5, 0{]}&7.50236&$A_2 \oplus B_1$\\%
&{[}6, 2, 0{]}&7.59076&$A_1 \oplus A_2 \oplus B_1 \oplus B_2$\\%
&{[}3, 2, 1{]}&7.73246&$A_1 \oplus A_2 \oplus B_1 \oplus B_2$\\%
\hline%
(0, 0, 3)&{[}4, 1, 0{]}&6.174&$A_2$\\%
&{[}1, 1, 1{]}&6.22459&$A_2$\\%
&{[}9, 0, 0{]}&6.49747&$A_2$\\%
&{[}1, 4, 0{]}&6.92152&$A_2$\\%
&{[}5, 2, 0{]}&7.15788&$A_2 \oplus B_2 \oplus E$\\%
&{[}2, 2, 1{]}&7.25914&$A_2 \oplus B_2 \oplus E$\\%
&{[}5, 1, 1{]}&7.49035&$A_2 \oplus B_2 \oplus E$\\%
&{[}0, 4, 1{]}&7.5631&$A_2$\\%
&{[}1, 2, 2{]}&7.59023&$A_2 \oplus B_2$\\%
&{[}2, 5, 0{]}&7.67561&$A_2 \oplus B_2 \oplus E$\\%
&{[}4, 2, 1{]}&7.87054&$A_2 \oplus B_2 \oplus E$\\%
\end{tabular}%
\caption{\label{tab:kpp_psq3-9_ops} $K\pi\pi$ operators with $3 \leq \bm{d}_{\rm ref}^2 \leq 9$ used in this work.}
\end{table}%

\begin{table}[!bp]
\centering
\begin{tabular}{c|c|c|c}
$\bm{d}_{\rm ref}$&$[d^2_{K_1}, d^2_{K_2}, d^2_{\pi}]$&$E^{\rm free}/M_{\pi}$&operators\\%
\hline%
(0, 0, 0)&{[}0, 0, 0{]}&2.26496&$A_{1u}$\\%
&{[}1, 1, 0{]}&2.42744&$A_{1u} \oplus E_u$\\%
&{[}1, 0, 1{]}&2.57041&$A_{1u} \oplus E_u \oplus T_{1g}$\\%
&{[}2, 2, 0{]}&2.57854&$A_{1u} \oplus E_u \oplus T_{2u}$\\%
&{[}3, 3, 0{]}&2.72036&$A_{1u} \oplus T_{2u}$\\%
&{[}2, 1, 1{]}&2.7272&$E_u \oplus T_{1g}$\\%
\hline%
(0, 0, 1)&{[}1, 0, 0{]}&2.30989&$A_2$\\%
&{[}0, 0, 1{]}&2.45497&$A_2$\\%
&{[}2, 1, 0{]}&2.46899&$A_2 \oplus B_2 \oplus E$\\%
&{[}4, 1, 0{]}&2.60873&$A_2$\\%
&{[}2, 0, 1{]}&2.61381&$A_2 \oplus B_2 \oplus E$\\%
&{[}3, 2, 0{]}&2.61735&$A_2 \oplus B_1 \oplus E$\\%
&{[}1, 1, 1{]}&2.61958&$4 A_2 \oplus 2 B_2 \oplus E$\\%
&{[}1, 0, 2{]}&2.68902&$A_2 \oplus B_2 \oplus E$\\%
&{[}3, 1, 1{]}&2.76773&$B_1$\\%
\hline%
(0, 1, 1)&{[}2, 0, 0{]}&2.3509&$A_2$\\%
&{[}1, 1, 0{]}&2.35676&$A_2$\\%
&{[}1, 0, 1{]}&2.50377&$A_2 \oplus B_1$\\%
&{[}3, 1, 0{]}&2.50735&$A_2 \oplus B_2$\\%
&{[}2, 2, 0{]}&2.51211&$A_1 \oplus A_2$\\%
&{[}0, 0, 2{]}&2.57417&$A_2$\\%
&{[}5, 1, 0{]}&2.64141&$A_2 \oplus B_1$\\%
&{[}4, 2, 0{]}&2.65351&$A_2 \oplus B_1$\\%
&{[}3, 0, 1{]}&2.65391&$B_2$\\%
&{[}2, 1, 1{]}&2.66448&$A_1$\\%
\hline%
(1, 1, 1)&{[}3, 0, 0{]}&2.38875&$A_2$\\%
&{[}2, 1, 0{]}&2.39953&$A_2 \oplus E$\\%
&{[}2, 0, 1{]}&2.5483&$A_2 \oplus E$\\%
&{[}1, 1, 1{]}&2.55421&$A_2 \oplus E$\\%
&{[}1, 0, 2{]}&2.62539&$A_2 \oplus E$\\%
&{[}0, 0, 3{]}&2.66642&$A_2$\\%
&{[}6, 1, 0{]}&2.67225&$E$\\%
\hline%
(0, 0, 2)&{[}1, 1, 0{]}&2.28389&$A_2$\\%
&{[}4, 0, 0{]}&2.42398&$A_2$\\%
&{[}1, 0, 1{]}&2.4353&$A_2$\\%
&{[}2, 2, 0{]}&2.44388&$A_2 \oplus B_2$\\%
&{[}5, 1, 0{]}&2.5766&$A_2 \oplus B_2 \oplus E$\\%
&{[}3, 3, 0{]}&2.59307&$A_2 \oplus B_1$\\%
&{[}2, 1, 1{]}&2.60025&$2 A_2 \oplus 2 B_2 \oplus 2 E$\\%
&{[}2, 0, 2{]}&2.67214&$A_2 \oplus B_2 \oplus E$\\%
&{[}6, 2, 0{]}&2.71945&$B_1$\\%
\hline%
(0, 1, 2)&{[}2, 1, 0{]}&2.328&$A_2$\\%
&{[}5, 0, 0{]}&2.45702&$A_2$\\%
&{[}4, 1, 0{]}&2.47571&$A_2$\\%
&{[}2, 0, 1{]}&2.48107&$A_2$\\%
&{[}3, 2, 0{]}&2.48479&$A_1 \oplus A_2$\\%
&{[}1, 1, 1{]}&2.48714&$2 A_2$\\%
&{[}1, 0, 2{]}&2.56018&$A_2$\\%
&{[}6, 1, 0{]}&2.60821&$A_1 \oplus A_2$\\%
&{[}5, 2, 0{]}&2.62377&$A_1 \oplus 2 A_2$\\%
&{[}4, 0, 1{]}&2.62768&$A_2$\\%
&{[}3, 1, 1{]}&2.64273&$A_1 \oplus A_2$\\%
\hline%
(1, 1, 2)&{[}3, 1, 0{]}&2.36865&$A_2$\\%
&{[}2, 2, 0{]}&2.37369&$A_2$\\%
&{[}6, 0, 0{]}&2.48816&$A_2$\\%
&{[}5, 1, 0{]}&2.51012&$A_1 \oplus A_2$\\%
&{[}4, 2, 0{]}&2.52285&$A_2$\\%
&{[}3, 0, 1{]}&2.52327&$A_2$\\%
&{[}2, 1, 1{]}&2.53439&$2 A_1 \oplus 3 A_2$\\%
&{[}2, 0, 2{]}&2.6081&$A_1 \oplus A_2$\\%
&{[}1, 1, 2{]}&2.6142&$A_1 \oplus 2 A_2$\\%
&{[}6, 2, 0{]}&2.65655&$A_1 \oplus A_2$\\%
\end{tabular}%
\caption{\label{tab:kkp_psq0-6_ops} $KK\pi$ operators with $\bm{d}_{\rm ref}^2 \leq 6$ used in this work.}
\end{table}%
\begin{table}[!bp]
\centering
\begin{tabular}{c|c|c|c}
$\bm{d}_{\rm ref}$&$[d^2_{K_1}, d^2_{K_2}, d^2_{\pi}]$&$E^{\rm free}/M_{\pi}$&operators\\%
\hline%
(0, 2, 2)&{[}2, 2, 0{]}&2.30135&$A_2$\\%
&{[}5, 1, 0{]}&2.44183&$A_2 \oplus B_1$\\%
&{[}3, 3, 0{]}&2.45921&$A_2$\\%
&{[}2, 1, 1{]}&2.46677&$A_2 \oplus B_1$\\%
&{[}2, 0, 2{]}&2.54244&$A_2$\\%
&{[}8, 0, 0{]}&2.54571&$A_2$\\%
&{[}1, 1, 2{]}&2.5487&$A_2$\\%
&{[}6, 2, 0{]}&2.59212&$A_1 \oplus A_2 \oplus B_1 \oplus B_2$\\%
&{[}5, 0, 1{]}&2.59929&$A_2 \oplus B_1$\\%
&{[}4, 4, 0{]}&2.60674&$A_2$\\%
&{[}4, 1, 1{]}&2.61847&$A_2 \oplus B_1$\\%
&{[}3, 2, 1{]}&2.62779&$A_1 \oplus 2 A_2 \oplus B_1 \oplus 2 B_2$\\%
&{[}3, 1, 2{]}&2.70886&$A_1 \oplus A_2 \oplus B_1 \oplus B_2$\\%
\hline%
(0, 0, 3)&{[}4, 1, 0{]}&2.33512&$A_2$\\%
&{[}1, 1, 1{]}&2.34724&$A_2$\\%
&{[}5, 2, 0{]}&2.49155&$A_2 \oplus B_2 \oplus E$\\%
&{[}4, 0, 1{]}&2.49567&$A_2$\\%
&{[}9, 0, 0{]}&2.57248&$A_2$\\%
&{[}2, 1, 2{]}&2.59919&$A_2 \oplus B_2 \oplus E$\\%
&{[}6, 3, 0{]}&2.63779&$A_2 \oplus B_1 \oplus E$\\%
&{[}5, 1, 1{]}&2.65641&$A_2 \oplus B_2 \oplus E$\\%
&{[}3, 2, 2{]}&2.76043&$B_1$\\%
\end{tabular}%
\caption{\label{tab:kkp_psq8-9_ops} $KK\pi$ operators with $8 \leq \bm{d}_{\rm ref}^2 \leq 9$ used in this work.}
\end{table}%

\begin{table}[!bp]
\centering
\begin{tabular}{c|c|c|c}
$\bm{d}_{\rm ref}$&$[d^2_{K_1}, d^2_{K_2}, d^2_{K_3}]$&$E^{\rm free}/M_{K}$&operators\\%
\hline%
(0, 0, 0)&{[}0, 0, 0{]}&3.0&$A_{1u}$\\%
&{[}1, 1, 0{]}&3.16248&$A_{1u} \oplus E_u$\\%
&{[}2, 2, 0{]}&3.31358&$A_{1u} \oplus E_u$\\%
\hline%
(0, 0, 1)&{[}1, 0, 0{]}&3.05368&$A_2$\\%
&{[}2, 1, 0{]}&3.21182&$A_2 \oplus B_2 \oplus E$\\%
&{[}1, 1, 1{]}&3.21755&$2 A_2 \oplus B_2$\\%
&{[}4, 1, 0{]}&3.35084&$A_2$\\%
&{[}3, 2, 0{]}&3.35942&$E$\\%
&{[}2, 2, 1{]}&3.36983&$B_2$\\%
\hline%
(0, 1, 1)&{[}2, 0, 0{]}&3.10277&$A_2$\\%
&{[}1, 1, 0{]}&3.10856&$A_2$\\%
&{[}3, 1, 0{]}&3.25744&$A_2 \oplus B_2$\\%
&{[}2, 2, 0{]}&3.26215&$A_1 \oplus A_2$\\%
&{[}2, 1, 1{]}&3.26794&$A_1 \oplus 4 A_2 \oplus 3 B_1 \oplus B_2$\\%
&{[}5, 1, 0{]}&3.39019&$A_2 \oplus B_1$\\%
&{[}3, 2, 1{]}&3.4166&$A_1 \oplus B_2$\\%
\hline%
(1, 1, 1)&{[}3, 0, 0{]}&3.14814&$A_2$\\%
&{[}2, 1, 0{]}&3.15873&$A_2 \oplus E$\\%
&{[}1, 1, 1{]}&3.16457&$A_2$\\%
&{[}3, 1, 1{]}&3.31453&$A_2 \oplus E$\\%
\hline%
(0, 0, 2)&{[}1, 1, 0{]}&3.05368&$A_2$\\%
&{[}4, 0, 0{]}&3.19044&$A_2$\\%
&{[}2, 2, 0{]}&3.2099&$A_2 \oplus B_2$\\%
&{[}2, 1, 1{]}&3.21578&$A_2 \oplus B_2 \oplus E$\\%
&{[}5, 1, 0{]}&3.33994&$A_2 \oplus B_2 \oplus E$\\%
\end{tabular}%
\caption{\label{tab:kkk_psq0-4_ops} $3K$ operators with $\bm{d}_{\rm ref}^2 \leq 4$ used in this work.}
\end{table}%
\begin{table}[!bp]
\centering
\begin{tabular}{c|c|c|c}
$\bm{d}_{\rm ref}$&$[d^2_{K_1}, d^2_{K_2}, d^2_{K_3}]$&$E^{\rm free}/M_{K}$&operators\\%
\hline%
(0, 1, 2)&{[}2, 1, 0{]}&3.10474&$A_2$\\%
&{[}1, 1, 1{]}&3.11068&$A_2$\\%
&{[}5, 0, 0{]}&3.23014&$A_2$\\%
&{[}4, 1, 0{]}&3.24835&$A_2$\\%
&{[}3, 2, 0{]}&3.2572&$A_1 \oplus A_2$\\%
&{[}3, 1, 1{]}&3.26311&$A_1 \oplus A_2$\\%
&{[}2, 2, 1{]}&3.26794&$2 A_1 \oplus 5 A_2$\\%
&{[}6, 1, 0{]}&3.37766&$A_1 \oplus A_2$\\%
\hline%
(1, 1, 2)&{[}3, 1, 0{]}&3.15192&$A_2$\\%
&{[}2, 2, 0{]}&3.15679&$A_2$\\%
&{[}2, 1, 1{]}&3.16276&$A_1 \oplus 2 A_2$\\%
&{[}6, 0, 0{]}&3.26762&$A_2$\\%
&{[}5, 1, 0{]}&3.28893&$A_1 \oplus A_2$\\%
&{[}4, 2, 0{]}&3.30129&$A_2$\\%
&{[}3, 2, 1{]}&3.31614&$A_1$\\%
\hline%
(0, 2, 2)&{[}2, 2, 0{]}&3.10277&$A_2$\\%
&{[}2, 1, 1{]}&3.10884&$A_2$\\%
&{[}5, 1, 0{]}&3.23711&$A_2 \oplus B_1$\\%
&{[}3, 3, 0{]}&3.25379&$A_2$\\%
&{[}4, 1, 1{]}&3.25571&$A_2 \oplus B_1$\\%
&{[}3, 2, 1{]}&3.26476&$A_1 \oplus 2 A_2 \oplus B_1 \oplus 2 B_2$\\%
&{[}2, 2, 2{]}&3.26968&$A_1 \oplus A_2$\\%
&{[}8, 0, 0{]}&3.33697&$A_2$\\%
&{[}6, 2, 0{]}&3.38171&$A_1 \oplus B_1 \oplus B_2$\\%
\hline%
(0, 0, 3)&{[}1, 1, 1{]}&3.0&$A_2$\\%
&{[}4, 1, 0{]}&3.14252&$A_2$\\%
&{[}2, 2, 1{]}&3.16276&$A_2 \oplus B_2$\\%
&{[}5, 2, 0{]}&3.29169&$A_2 \oplus B_2 \oplus E$\\%
&{[}5, 1, 1{]}&3.29777&$A_2 \oplus B_2 \oplus E$\\%
&{[}4, 2, 1{]}&3.31039&$A_2 \oplus B_2 \oplus E$\\%
\end{tabular}%
\caption{\label{tab:kkk_psq5-9_ops} The list of $3K$ operators with $5 \leq \bm{d}_{\rm ref}^2 \leq 9$ used in this work.}
\end{table}%

\begin{table}[!bp]
\centering
\begin{tabular}{c|c|c|c}
$\bm{d}_{\rm ref}$&$[d^2_{\pi_1}, d^2_{\pi_2}]$&$E^{\rm free}/M_{\pi}$&operators\\%
\hline%
(0, 0, 0)&{[}0, 0{]}&2.0&$A_{1g}^+$\\%
&{[}1, 1{]}&3.69237&$A_{1g}^+ \oplus E_g^+$\\%
&{[}2, 2{]}&4.8236&$A_{1g}^+ \oplus E_g^+ \oplus T_{2g}^+$\\%
\hline%
(0, 0, 1)&{[}1, 0{]}&2.38587&$A_1^+$\\%
&{[}2, 1{]}&3.9651&$A_1^+ \oplus B_1^+ \oplus E^+$\\%
&{[}4, 1{]}&4.8656&$A_1^+$\\%
\hline%
(0, 1, 1)&{[}2, 0{]}&2.6122&$A_1^+$\\%
&{[}1, 1{]}&2.96931&$A_1^+$\\%
&{[}3, 1{]}&4.17209&$A_1^+ \oplus B_1^+$\\%
&{[}2, 2{]}&4.29539&$A_1^+ \oplus A_2^+$\\%
\hline%
(1, 1, 1)&{[}3, 0{]}&2.78135&$A_1^+$\\%
&{[}2, 1{]}&3.30231&$A_1^+ \oplus E^+$\\%
\hline%
(0, 0, 2)&{[}1, 1{]}&2.0&$A_1^+$\\%
&{[}4, 0{]}&2.91922&$A_1^+$\\%
&{[}2, 2{]}&3.69237&$A_1^+ \oplus B_1^+$\\%
&{[}5, 1{]}&4.48902&$A_1^+ \oplus B_1^+ \oplus E^+$\\%
&{[}3, 3{]}&4.8236&$A_1^+ \oplus B_2^+$\\%
\hline%
(0, 1, 2)&{[}2, 1{]}&2.46748&$A_1^+$\\%
&{[}5, 0{]}&3.03689&$A_1^+$\\%
&{[}4, 1{]}&3.74706&$A_1^+$\\%
&{[}3, 2{]}&3.97918&$A_1^+ \oplus A_2^+$\\%
&{[}6, 1{]}&4.61848&$A_1^+ \oplus A_2^+$\\%
&{[}5, 2{]}&4.92307&$2 A_1^+ \oplus A_2^+$\\%
\end{tabular}%
\caption{\label{tab:pp_psq0-5_ops} $\pi\pi$ operators with $\bm{d}_{\rm ref}^2 \leq 5$ used in this work.}
\end{table}%
\begin{table}[!bp]
\centering
\begin{tabular}{c|c|c|c}
$\bm{d}_{\rm ref}$&$[d^2_{\pi_1}, d^2_{\pi_2}]$&$E^{\rm free}/M_{\pi}$&operators\\%
\hline%
(1, 1, 2)&{[}3, 1{]}&2.78797&$A_1^+$\\%
&{[}2, 2{]}&2.96931&$A_1^+$\\%
&{[}6, 0{]}&3.14028&$A_1^+$\\%
&{[}5, 1{]}&3.91593&$A_1^+ \oplus A_2^+$\\%
&{[}4, 2{]}&4.21062&$A_1^+$\\%
\hline%
(0, 2, 2)&{[}2, 2{]}&2.0&$A_1^+$\\%
&{[}5, 1{]}&3.2431&$A_1^+ \oplus B_2^+$\\%
&{[}8, 0{]}&3.3172&$A_1^+$\\%
&{[}3, 3{]}&3.69237&$A_1^+$\\%
&{[}6, 2{]}&4.57822&$A_1^+ \oplus A_2^+ \oplus B_1^+ \oplus B_2^+$\\%
&{[}4, 4{]}&4.8236&$A_1^+$\\%
\hline%
(0, 0, 3)&{[}4, 1{]}&2.09927&$A_1^+$\\%
&{[}9, 0{]}&3.39467&$A_1^+$\\%
&{[}5, 2{]}&3.82139&$A_1^+ \oplus B_1^+ \oplus E^+$\\%
&{[}6, 3{]}&4.9544&$A_1^+ \oplus B_2^+ \oplus E^+$\\%
\end{tabular}%
\caption{\label{tab:pp_psq6-9_ops} $\pi\pi$ operators with $6 \leq \bm{d}_{\rm ref}^2 \leq 9$ used in this work.}
\end{table}%

\begin{table}[!bp]
\centering
\begin{tabular}{c|c|c|c}
$\bm{d}_{\rm ref}$&$[d^2_{K}, d^2_{\pi}]$&$E^{\rm free}/M_{\pi}$&operators\\%
\hline%
(0, 0, 0)&{[}0, 0{]}&4.77412&$A_{1g}$\\%
&{[}1, 1{]}&5.92691&$A_{1g} \oplus E_g \oplus T_{1u}$\\%
&{[}2, 2{]}&6.77766&$A_{1g} \oplus E_g \oplus T_{1u} \oplus T_{2g} \oplus T_{2u}$\\%
&{[}3, 3{]}&7.50144&$A_{1g} \oplus T_{1u} \oplus T_{2g}$\\%
&{[}4, 4{]}&8.14738&$E_g$\\%
&{[}5, 5{]}&8.73834&$T_{2u}$\\%
\hline%
(0, 0, 1)&{[}1, 0{]}&4.83791&$A_1$\\%
&{[}0, 1{]}&5.40179&$A_1$\\%
&{[}2, 1{]}&6.01507&$A_1 \oplus B_1 \oplus E$\\%
&{[}1, 2{]}&6.30433&$A_1 \oplus B_1 \oplus E$\\%
&{[}4, 1{]}&6.55135&$A_1$\\%
&{[}3, 2{]}&6.87223&$A_1 \oplus B_2 \oplus E$\\%
&{[}2, 3{]}&7.06539&$A_1 \oplus B_2 \oplus E$\\%
&{[}1, 4{]}&7.17575&$A_1$\\%
&{[}5, 2{]}&7.37732&$A_1 \oplus B_1 \oplus E$\\%
&{[}2, 5{]}&7.82482&$A_1 \oplus B_1 \oplus E$\\%
&{[}6, 3{]}&8.07692&$B_2$\\%
\hline%
(0, 1, 1)&{[}2, 0{]}&4.8965&$A_1$\\%
&{[}1, 1{]}&5.50559&$A_1 \oplus B_2$\\%
&{[}0, 2{]}&5.78349&$A_1$\\%
&{[}3, 1{]}&6.09666&$A_1 \oplus B_1$\\%
&{[}2, 2{]}&6.41248&$A_1 \oplus A_2 \oplus B_1 \oplus B_2$\\%
&{[}1, 3{]}&6.59298&$A_1 \oplus B_1$\\%
&{[}5, 1{]}&6.61878&$A_1 \oplus B_2$\\%
&{[}4, 2{]}&6.96045&$A_1 \oplus B_2$\\%
&{[}2, 4{]}&7.30417&$A_1 \oplus B_2$\\%
&{[}1, 5{]}&7.37235&$A_1 \oplus B_2$\\%
&{[}6, 2{]}&7.45204&$A_1 \oplus A_2 \oplus B_1 \oplus B_2$\\%
&{[}5, 3{]}&7.6878&$A_1 \oplus A_2 \oplus B_1 \oplus B_2$\\%
\hline%
(1, 1, 1)&{[}3, 0{]}&4.95085&$A_1$\\%
&{[}2, 1{]}&5.60038&$A_1 \oplus E$\\%
&{[}1, 2{]}&5.90997&$A_1 \oplus E$\\%
&{[}0, 3{]}&6.07387&$A_1$\\%
&{[}6, 1{]}&6.68255&$A_1 \oplus E$\\%
&{[}5, 2{]}&7.0433&$A_1 \oplus A_2 \oplus 2 E$\\%
&{[}4, 3{]}&7.27364&$A_1 \oplus E$\\%
&{[}3, 4{]}&7.42269&$A_1 \oplus E$\\%
&{[}2, 5{]}&7.51072&$A_1 \oplus A_2 \oplus E$\\%
\end{tabular}%
\caption{\label{tab:kp_psq0-3_ops} $K\pi$ operators with $\bm{d}_{\rm ref}^2 \leq 3$ used in this work.}
\end{table}%
\begin{table}[!bp]
\centering
\begin{tabular}{c|c|c|c}
$\bm{d}_{\rm ref}$&$[d^2_{K}, d^2_{\pi}]$&$E^{\rm free}/M_{\pi}$&operators\\%
\hline%
(0, 0, 2)&{[}4, 0{]}&5.00169&$A_1$\\%
&{[}1, 1{]}&5.04923&$A_1$\\%
&{[}2, 2{]}&6.02521&$A_1 \oplus B_1 \oplus E$\\%
&{[}5, 1{]}&6.24431&$A_1 \oplus B_1 \oplus E$\\%
&{[}0, 4{]}&6.31333&$A_1$\\%
&{[}3, 3{]}&6.8292&$A_1 \oplus B_2 \oplus E$\\%
&{[}1, 5{]}&7.03809&$A_1 \oplus B_1 \oplus E$\\%
&{[}6, 2{]}&7.12153&$A_1 \oplus B_2 \oplus E$\\%
&{[}2, 6{]}&7.6941&$A_1 \oplus B_2 \oplus E$\\%
&{[}5, 5{]}&8.16854&$B_1$\\%
\hline%
(0, 1, 2)&{[}5, 0{]}&5.04955&$A_1$\\%
&{[}2, 1{]}&5.15243&$A_1$\\%
&{[}1, 2{]}&5.48735&$A_1$\\%
&{[}4, 1{]}&5.76945&$A_1$\\%
&{[}3, 2{]}&6.1314&$A_1 \oplus A_2$\\%
&{[}6, 1{]}&6.31187&$A_1 \oplus A_2$\\%
&{[}2, 3{]}&6.34714&$A_1 \oplus A_2$\\%
&{[}1, 4{]}&6.46976&$A_1$\\%
&{[}0, 5{]}&6.51946&$A_1$\\%
&{[}5, 2{]}&6.69263&$2 A_1 \oplus A_2$\\%
&{[}2, 5{]}&7.18291&$2 A_1 \oplus A_2$\\%
&{[}1, 6{]}&7.22087&$A_1 \oplus A_2$\\%
&{[}6, 3{]}&7.45675&$A_1 \oplus A_2$\\%
&{[}5, 4{]}&7.63641&$A_1$\\%
&{[}3, 6{]}&7.82963&$A_2$\\%
\hline%
(1, 1, 2)&{[}6, 0{]}&5.09484&$A_1$\\%
&{[}3, 1{]}&5.24744&$A_1$\\%
&{[}2, 2{]}&5.61127&$A_1 \oplus A_2$\\%
&{[}1, 3{]}&5.81669&$A_1$\\%
&{[}5, 1{]}&5.84591&$A_1 \oplus A_2$\\%
&{[}4, 2{]}&6.23011&$A_1$\\%
&{[}2, 4{]}&6.61191&$A_1$\\%
&{[}1, 5{]}&6.68715&$A_1 \oplus A_2$\\%
&{[}0, 6{]}&6.70177&$A_1$\\%
&{[}6, 2{]}&6.77491&$A_1 \oplus A_2$\\%
&{[}5, 3{]}&7.03339&$A_1 \oplus A_2$\\%
&{[}3, 5{]}&7.31622&$A_1 \oplus A_2$\\%
&{[}2, 6{]}&7.37445&$A_1 \oplus A_2$\\%
&{[}6, 4{]}&7.73389&$A_1 \oplus A_2$\\%
\hline%
(0, 2, 2)&{[}2, 2{]}&5.16426&$A_1$\\%
&{[}8, 0{]}&5.17897&$A_1$\\%
&{[}5, 1{]}&5.41829&$A_1 \oplus B_2$\\%
&{[}3, 3{]}&6.08312&$A_1 \oplus B_1$\\%
&{[}1, 5{]}&6.31674&$A_1 \oplus B_2$\\%
&{[}6, 2{]}&6.40957&$A_1 \oplus A_2 \oplus B_1 \oplus B_2$\\%
&{[}4, 4{]}&6.86386&$A_1 \oplus B_2$\\%
&{[}0, 8{]}&7.01608&$A_1$\\%
&{[}2, 6{]}&7.04029&$A_1 \oplus A_2 \oplus B_1 \oplus B_2$\\%
&{[}5, 5{]}&7.55589&$A_1 \oplus A_2 \oplus B_1 \oplus B_2$\\%
\hline%
(0, 0, 3)&{[}4, 1{]}&4.86343&$A_1$\\%
&{[}9, 0{]}&5.21829&$A_1$\\%
&{[}1, 4{]}&5.67664&$A_1$\\%
&{[}5, 2{]}&5.92939&$A_1 \oplus B_1 \oplus E$\\%
&{[}2, 5{]}&6.4777&$A_1 \oplus B_1 \oplus E$\\%
&{[}6, 3{]}&6.78009&$A_1 \oplus B_2 \oplus E$\\%
&{[}0, 9{]}&7.15456&$A_1$\\%
&{[}3, 6{]}&7.18815&$A_1 \oplus B_2 \oplus E$\\%
\end{tabular}%
\caption{\label{tab:kp_psq4-9_ops} $K\pi$ operators with $4 \leq \bm{d}_{\rm ref}^2 \leq 9$ used in this work.}
\end{table}%

\begin{table}[!bp]
\centering
\begin{tabular}{c|c|c|c}
$\bm{d}_{\rm ref}$&$[d^2_{K_1}, d^2_{K_2}]$&$E^{\rm free}/M_{K}$&operators\\%
\hline%
(0, 0, 0)&{[}0, 0{]}&2.0&$A_{1g}$\\%
&{[}1, 1{]}&2.16248&$A_{1g} \oplus E_g$\\%
&{[}2, 2{]}&2.31358&$A_{1g} \oplus E_g \oplus T_{2g}$\\%
&{[}3, 3{]}&2.4554&$A_{1g} \oplus T_{2g}$\\%
&{[}4, 4{]}&2.58946&$A_{1g} \oplus E_g$\\%
&{[}5, 5{]}&2.71692&$T_{2g}$\\%
\hline%
(0, 0, 1)&{[}1, 0{]}&2.04022&$A_1$\\%
&{[}2, 1{]}&2.19993&$A_1 \oplus B_1 \oplus E$\\%
&{[}4, 1{]}&2.34012&$A_1$\\%
&{[}3, 2{]}&2.34877&$A_1 \oplus B_2 \oplus E$\\%
&{[}5, 2{]}&2.48141&$A_1 \oplus B_1 \oplus E$\\%
&{[}6, 3{]}&2.61489&$A_1 \oplus B_2 \oplus E$\\%
&{[}5, 4{]}&2.62113&$B_1$\\%
\hline%
(0, 1, 1)&{[}2, 0{]}&2.07692&$A_1$\\%
&{[}1, 1{]}&2.08283&$A_1$\\%
&{[}3, 1{]}&2.23451&$A_1 \oplus B_1$\\%
&{[}2, 2{]}&2.23931&$A_1 \oplus A_2$\\%
&{[}5, 1{]}&2.36938&$A_1 \oplus B_2$\\%
&{[}4, 2{]}&2.38155&$A_1 \oplus B_2$\\%
&{[}6, 2{]}&2.50963&$A_1 \oplus A_2 \oplus B_1 \oplus B_2$\\%
\hline%
(1, 1, 1)&{[}3, 0{]}&2.11078&$A_1$\\%
&{[}2, 1{]}&2.12168&$A_1 \oplus E$\\%
&{[}6, 1{]}&2.397&$A_1 \oplus E$\\%
&{[}5, 2{]}&2.41231&$A_1 \oplus A_2 \oplus 2 E$\\%
&{[}4, 3{]}&2.41979&$A_1 \oplus E$\\%
&{[}8, 3{]}&2.66811&$A_1 \oplus E$\\%
&{[}6, 5{]}&2.68493&$A_2$\\%
\hline%
(0, 0, 2)&{[}1, 1{]}&2.0&$A_1$\\%
&{[}4, 0{]}&2.1423&$A_1$\\%
&{[}2, 2{]}&2.16248&$A_1 \oplus B_1$\\%
&{[}5, 1{]}&2.29691&$A_1 \oplus B_1 \oplus E$\\%
&{[}3, 3{]}&2.31358&$A_1 \oplus B_2$\\%
&{[}6, 2{]}&2.44132&$A_1 \oplus B_2 \oplus E$\\%
&{[}5, 5{]}&2.58946&$A_1 \oplus B_1$\\%
&{[}8, 4{]}&2.70637&$E$\\%
&{[}6, 6{]}&2.71692&$B_2$\\%
\hline%
(0, 1, 2)&{[}2, 1{]}&2.04043&$A_1$\\%
&{[}5, 0{]}&2.17185&$A_1$\\%
&{[}4, 1{]}&2.19085&$A_1$\\%
&{[}3, 2{]}&2.20009&$A_1 \oplus A_2$\\%
&{[}6, 1{]}&2.32539&$A_1 \oplus A_2$\\%
&{[}5, 2{]}&2.34117&$2 A_1 \oplus A_2$\\%
&{[}6, 3{]}&2.4822&$A_1 \oplus A_2$\\%
&{[}5, 4{]}&2.48878&$A_1$\\%
&{[}6, 5{]}&2.6212&$A_1 \oplus A_2$\\%
\hline%
(1, 1, 2)&{[}3, 1{]}&2.07767&$A_1$\\%
&{[}2, 2{]}&2.08283&$A_1$\\%
&{[}6, 0{]}&2.19969&$A_1$\\%
&{[}5, 1{]}&2.22208&$A_1 \oplus A_2$\\%
&{[}4, 2{]}&2.23505&$A_1$\\%
&{[}6, 2{]}&2.37105&$A_1 \oplus A_2$\\%
&{[}5, 3{]}&2.38196&$A_1 \oplus A_2$\\%
&{[}6, 4{]}&2.52024&$A_1 \oplus A_2$\\%
\end{tabular}%
\caption{\label{tab:kk_psq0-6_ops} $KK$ operators with $\bm{d}_{\rm ref}^2 \leq 6$ used in this work.}
\end{table}%
\begin{table}[!bp]
\centering
\begin{tabular}{c|c|c|c}
$\bm{d}_{\rm ref}$&$[d^2_{K_1}, d^2_{K_2}]$&$E^{\rm free}/M_{K}$&operators\\%
\hline%
(0, 2, 2)&{[}2, 2{]}&2.0&$A_1$\\%
&{[}5, 1{]}&2.14464&$A_1 \oplus B_2$\\%
&{[}3, 3{]}&2.16248&$A_1$\\%
&{[}8, 0{]}&2.25115&$A_1$\\%
&{[}6, 2{]}&2.29864&$A_1 \oplus A_2 \oplus B_1 \oplus B_2$\\%
&{[}4, 4{]}&2.31358&$A_1$\\%
&{[}5, 5{]}&2.4554&$A_1 \oplus A_2$\\%
&{[}6, 6{]}&2.58946&$A_1$\\%
&{[}8, 8{]}&2.83865&$A_2$\\%
\hline%
(0, 0, 3)&{[}4, 1{]}&2.03064&$A_1$\\%
&{[}5, 2{]}&2.19197&$A_1 \oplus B_1 \oplus E$\\%
&{[}9, 0{]}&2.27508&$A_1$\\%
&{[}6, 3{]}&2.34201&$A_1 \oplus B_2 \oplus E$\\%
\end{tabular}%
\caption{\label{tab:kk_psq8-9_ops} $KK$ operators with $8 \leq \bm{d}_{\rm ref}^2 \leq 9$ used in this work.}
\end{table}%

\clearpage
\begin{widetext}
\section{Fit results for two mesons on E250}
\label{app:E2502meson}

In this appendix, we collect some of the tables that display two-meson results. These tables are discussed in \Cref{sec:results}. Specifically, \Cref{tab:pp-params-E250,tab:kk-params-E250,tab:pk-params-E250}  are described in \Cref{sec:fits}, and \Cref{tab:ppparams-E250-table,tab:KKparams-E250-table,tab:pKparams-E250-table} in \Cref{sec:threshold}.

\begin{table}[h!]
\centering
\begin{tabular}{|c|c|c|c|c|}
\hline
Ensemble & \multicolumn{4}{c|}{E250}   \\ \hline \hline
Cutoff & $4M_\pi$  &$4.5M_\pi$& $4M_\pi$ & $4.5M_\pi$
\\ \hline
Description & $s$ waves &  $s$ waves &  $s,d$ waves  &   $s$,$d$ waves \\ \hline
$\chi^2$ & 26.23 & 28.21 &30.33 &  36.82 
\\ \hline
DOF     & 24-2=22 & 29-2=27 & 34-3=31   &  43-4=40
\\ \hline 
$p$ & 0.24 & 0.40 & 0.50 & 0.61 \\ \hline
\hline
$B_0^{\pi\pi}$ & -22.0(3.4) & -22.2(2.8) & -21.7(3.1)  & -22.9(2.7)
\\ \hline
$B_1^{\pi\pi}$ & -4.3(1.5) &-4.6(1.1) & -4.4(1.4)& -4.3 (1.1)
\\ \hline
$z^2_{\pi\pi}$ & $1$ (fixed)& $1$ (fixed)& $1$ (fixed) & $1$ (fixed)
\\ \hline
$D_0^{\pi\pi}$   & 0 (fixed) & 0 (fixed) & $0.6(3.4)\cdot 10^{-4}$ & $2.1(3.1) \cdot 10^{-4}$ 
\\ \hline
\end{tabular}
\caption{Results of fits to two-pion levels on ensemble E250, using pion units. The DOF entries list the number of levels fit minus the number of parameters.
$s$-wave fits are to levels in the $A_1$ irreps only,
while those to $s$ and $d$ waves include levels in nontrivial irreps.
Fit parameters as in the models described in \Cref{sec:models}, using the ADLER2 two-pion K matrix.
}
\label{tab:pp-params-E250}
\end{table}

\begin{table}[h!]
\centering
\begin{tabular}{|c|c|c|c|c|c|c|c|}
\hline
Ensemble &  \multicolumn{5}{c|}{E250}  \\ \hline \hline
Cutoff & $2.27M_K$ & $2.40M_K$  &  $2.27M_K$ & $2.40M_K$ & $2.40M_K$
\\ \hline
Description & $s$ wave & $s$ wave & $s,d$ waves & $s,d$ waves & $s,d$ waves  \\ \hline
$\chi^2$ & 45.68 & 72.18  & 56.7 & 100.39 & 99.48  \\ \hline
DOF     & 28-2=26 & 44-2=42  & 40-3=37 & 76-3=73 & 76-4=72   \\ \hline %
$p$ & 0.0099 & 0.0026  & 0.020 & 0.018 & 0.018 \\ \hline  \hline
$b_0^{KK}$ & $-2.641(46)$  &  $-2.656(34)$ &$-2.621(46) $& $-2.621(34)$ & -2.655(53)
\\ \hline
$b_1^{KK}$ & 0.73(20) & 0.73(11) & 0.59(20) & 0.58(11) &  0.92(42)
\\ \hline
$b_2^{KK}$ & 0 (fixed) & 0 (fixed) & 0 (fixed) & 0 (fixed) & -0.69(83)
\\ \hline
$D_0^{KK}$   & 0 (fixed) & 0 (fixed) & $- 0.059(30)$ & $-0.044(13)$  & -0.042(16)
\\ \hline
\end{tabular}
\caption{ Results of fits to two-kaon levels on ensemble E250, using kaon units. The DOF entries list the number of levels fit minus the number of parameters. 
$s$-wave fits are to levels in the $A_1$ irreps only,
while those to $s$ and $d$ waves include levels in nontrivial irreps.
In the $s$-wave fits with cutoff $E=2.40$, levels 5 and 6 (with counting starting at 0) from frame $P^2=5 (2\pi/L)^2$ have been removed, since they form a degenerate pair that are split only by $d$-wave interactions.
The inelastic threshold lies at $2 M_K+M_\pi=2.265 M_K$.
Fit parameters are as in the models described in \Cref{sec:models}, using the ERE2 and ERE3 forms for the two-kaon interaction.
}
\label{tab:kk-params-E250}
\end{table}

\begin{table}[h!]
\centering
\begin{tabular}{|c|c|c|c|c|c|c|}
\hline
Ensemble &  \multicolumn{5}{c|}{E250}  \\ \hline \hline
Cutoff &  $2M_\pi+M_K$  & $2M_\pi+M_K$  &  $3 M_\pi+M_K$ &  $3 M_\pi+M_K$ & $3 M_\pi+M_K$
\\ \hline
description & $s$ wave & $s,p$ waves & $s$ wave  & $s,p$ waves & $s,p$ waves \\ \hline
$\chi^2$ & 22.96 & 25.33 & 74.7 &  101.26 & 89.67 \\ \hline
DOF & 21-2=19 & 25-3=22 & 53-3=50 & 84-3=81 & 84-4=80 \\ \hline
$p$ & 0.239 & 0.281 & 0.017 & 0.063 & 0.215 \\ \hline
\hline
$B_0^{\pi K}$ & -25.5(1.4)  & -25.6(1.4) &  -29.0(1.2) & -28.8(1.1) & -43.3(6.0)
\\ \hline
$B_1^{\pi K}$ & -4.90(75) & -4.88(82)  & -2.82(38) & - 2.80(36) & -0.7 (0.9) 
\\ \hline
$z^2_{\pi K}$ & $1$ (fixed)& $1$ (fixed) & $1$ (fixed) & $1$ (fixed) & 0.67(14)
\\ \hline
$P_0^{\pi K}$   & 0 (fixed) & $0.0(8.4)\cdot 10^{-4}$ & 0 
(fixed) & $2.3(3.2)\cdot 10^{-4}$ & $1.6(3.1) \cdot 10^{-4}$
\\ \hline
\end{tabular}
\caption{Results of fits to $\pi K$ levels on ensemble E250, using pion units.
Fit parameters as in the models described in \Cref{sec:models}, using the ADLER2 and ADLER2z 
forms for the $\pi K$ interaction.
The inelastic threshold is at $2M_\pi+M_K$.
Other notation as \Cref{tab:pp-params-E250}.
}
\label{tab:pk-params-E250}
\end{table}

\begin{table}[h!]
\centering
\begin{tabular}{|c|c|c|c|c|c|c|c|}
\hline
Fit   &$\chi^2/{\rm dof}$ & p-value & $M_\pi a_0^{\pi\pi}$ & $a_0^{\pi\pi}r_0^{\pi\pi}$ & $M_\pi^5 a_2^{\pi\pi}$ % $D_0^{\pi\pi}$
\\ \hline \hline
$\pi \pi$ ($s$ 24) &  26.2/22 & 0.24 & 0.0454(71) & 2.61(20) & ---
\\ \hline
$\pi \pi$ ($s$ 29) &  28.2/27 & 0.4 & 0.0450(57) & 2.59(15) & ---
\\ \hline
$\pi\pi$ ($s\!+\!d$ 34) &  30.3/31 & 0.50 & 0.0462(66) & 2.60(19) & $-0.6(3.4) \cdot 10^{-4}$
\\ \hline
$\pi\pi$ ($s\!+\!d$ 43) &  36.8/40 & 0.61 & 0.0436(52) & 2.62(14) & $-2.1(3.1) \cdot 10^{-4}$
\\ \hline
$\pi\pi\pi$ ($s$, all irreps) &  65.6/52 & 0.098 &  0.0424(52) & 2.66(15) & ---
\\ \hline
$\pi\pi\pi$ ($s$+$d$, $\cK_B=0$) &  68.9/61 & 0.23 & 0.0430(48)  & 2.63(15) & $0.1(3.0) \cdot 10^{-4}$
\\ \hline
$\pi\pi\pi$ ($s$+$d$) &  68.6/60 & 0.21 & 0.0447(53) & 2.58(16) & $0.3(3.0) \cdot 10^{-4}$
\\ \hline
$\pi\pi K$ ($s$, $\cK_{\rm df,3}=0$) & 77.8/64 & 0.115 & 0.0431(44) &  2.67(13)   & ---
\\ \hline
$\pi\pi K$ ($s$) & 76.7/62 & 0.099 & 0.0432(48)  &   2.68(14) & ---
\\ \hline
$\pi\pi K$ ($s$+$p$) & 77.5/63 & 0.10 & 0.0457(51) & 2.62(14)  & ---
\\ \hline \hline
$\pi \pi$ (thr5) &  --- & --- & 0.0594(133) & --- & ---
\\ \hline
\end{tabular}
\caption{
Comparison of $\pi^+ \pi^+$ scattering parameters from different fits on the E250 ensemble. 
The fits are those of \Cref{tab:pp-params-E250,tab:ppp-params-E250,tab:ppK-params-E250}, using the same ordering of fits as in those Tables.
The final row gives the result of fitting
the threshold state to the threshold expansion up to $1/L^5$.
}
\label{tab:ppparams-E250-table}
\end{table}

\begin{table}[h!]
\centering
\begin{tabular}{|c|c|c|c|c|c|c|}
\hline
Fit &$\chi^2/{\rm dof}$ & p-value & $M_K a_0^{KK}$ & $a_0^{KK}r_0^{KK}$ & $M_K^5 a_2^{KK}$ 
\\ \hline \hline
$KK$ ($s$, 28) &  45.68/26 & 0.009 & 0.379(7) & 0.56(14)  & ---
\\ \hline
$KK$ ($s$, 44) &  72.18/42 & 0.0026 & 0.377(5) & 0.55(8)  & ---
\\ \hline
$KK$ ($s+d$,40) & 56.7/37 & 0.02 & 0.382(7) & 0.45(16)  & 0.056(27) 
\\ \hline
$KK$ ($s+d$,76, ERE2) &  100.39/73 & 0.018 & 0.381(5) & 0.45(8)  & 0.042(12) 
\\ \hline
$KK$ ($s+d$,76, ERE3) &  99.48/72 & 0.018  &  0.377(8) & 0.69(32) &  0.040(15) 
\\ \hline
$KKK$ ($s$, symm irreps) & 84.79/58 & 0.012 & 0.374(6) & 0.65(13) & ---
\\ \hline
$KKK$ ($s$, all irreps) & 120.31/77 & 0.0012 & 0.375(6) & 
0.64(12)& ---
\\ \hline
$KKK$ ($s+d$) & 129.6/87 & 0.002 & 0.379(6) & 0.53(14)  & 0.033(24) 
\\ \hline
$K K \pi$ ($s$ wave)&  179.15/93 & $2\cdot 10^{-7}$ & 0.381(6) & 0.50(13)   & --- \\ \hline
$K K \pi$ ($s+p$ waves)&  191.79/94 & $1.1\cdot 10^{-8}$ & 0.381(6) & 0.51(13)   & ---
\\ \hline
$K K \pi$ ($s+p+d$ waves)& 204.42/105 & $2.2\cdot 10^{-8}$ & 0.385(6) & 0.41(14)  & 0.040(23) 
\\ \hline \hline
$KK$ (thr5) & --- & --- & 0.383(27) & ---  & ---
\\ \hline
\end{tabular}
\caption{
As in \Cref{tab:ppparams-E250-table}, but for $K^+K^+$ scattering parameters.
The fits are those of \Cref{tab:kk-params-E250,tab:kkk-params-E250,tab:KKp-params-E250}, using the same ordering of fits as in those Tables.
}
\label{tab:KKparams-E250-table}
\end{table}

\begin{table}[h!]
\centering
\begin{tabular}{|c|c|c|c|c|c|c|c|c|}
\hline
Fit  &$\chi^2/{\rm dof}$ & p-value & $M_K a_0^{\pi K}$ & $a_0^{\pi K}r_0^{\pi K}$ &  $10^4 M_\pi^3 a_1^{\pi K}$ 
\\ \hline \hline
$\pi K$ ($s$, 21) & 22.96/19 & 0.239 & 0.062(3)& 0.95(8)  &  --- 
\\ \hline
$\pi K$ ($s+p$, 25)  & 25.3/22 & 0.28 & 0.062(3) & 0.95(8) &  0(8)
\\ \hline
$\pi K$ ($s$, 53) & 74.7/50 & 0.017 & 0.054(2)& 1.14(3)  &  --- 
\\ \hline
{$\pi K$ ($s+p$,84 )}   &101.3/81 & 0.063 & 0.055(2)  & 1.14(3)  &  -2.3(3.2)
\\ \hline
{$\pi K$ ($s+p$, 84z)}  & 89.7/80 & 0.22 & 0.061(3)  & 0.66(12)  &  -1.6(3.1)
\\ \hline
$\pi \pi K$ ($s$, $\cK_{\rm df,3}=0$)  & 77.82/64 & 0.115 & 0.060(3) &  -0.97(6) & ---
\\ \hline 
$\pi \pi K$ ($s$)  & 76.67/62 & 0.099 & 0.060(3) &  0.98(6) & ---
\\ \hline
$\pi \pi K$ ($s+p$, 72)  & 77.53/63 & 0.103 & 0.061(3) & 0.95(7) & 4.2(6.6)
\\ \hline
$K K \pi$ ($s$)& 179.2/93& $2 \cdot 10^{-7}$ & 0.055(3) & 1.08(6)  &  ---
\\ \hline
$K K \pi$ ($s+p$)& 191.8/94 & $1 \cdot 10^{-8}$ & 0.057(3) &  1.06(7)  &   -5.8(7.1)
\\ \hline
$K K \pi$ ($s+p+d$)& 204.4/105 & $2.2 \cdot 10^{-8}$ & 0.057(3) & 1.06(7)  & -2.0(7.1)
\\ \hline \hline
$\pi K$ (thr5) & --- & --- & 0.061(5) & --- & ---
\\ \hline
\end{tabular}
\caption{
As in \Cref{tab:ppparams-E250-table}, but for $\pi^+ K^+$ scattering parameters, all expressed in pion units.
The fits are those of \Cref{tab:pk-params-E250,tab:ppK-params-E250,tab:KKp-params-E250}, using the same ordering of fits as in those Tables.
}
\label{tab:pKparams-E250-table}
\end{table}
\end{widetext}
\clearpage

\begin{figure*}[th!]
     \centering
     \subfloat[\label{fig:Kisoppk}]{%
     \includegraphics[width=0.49\textwidth]{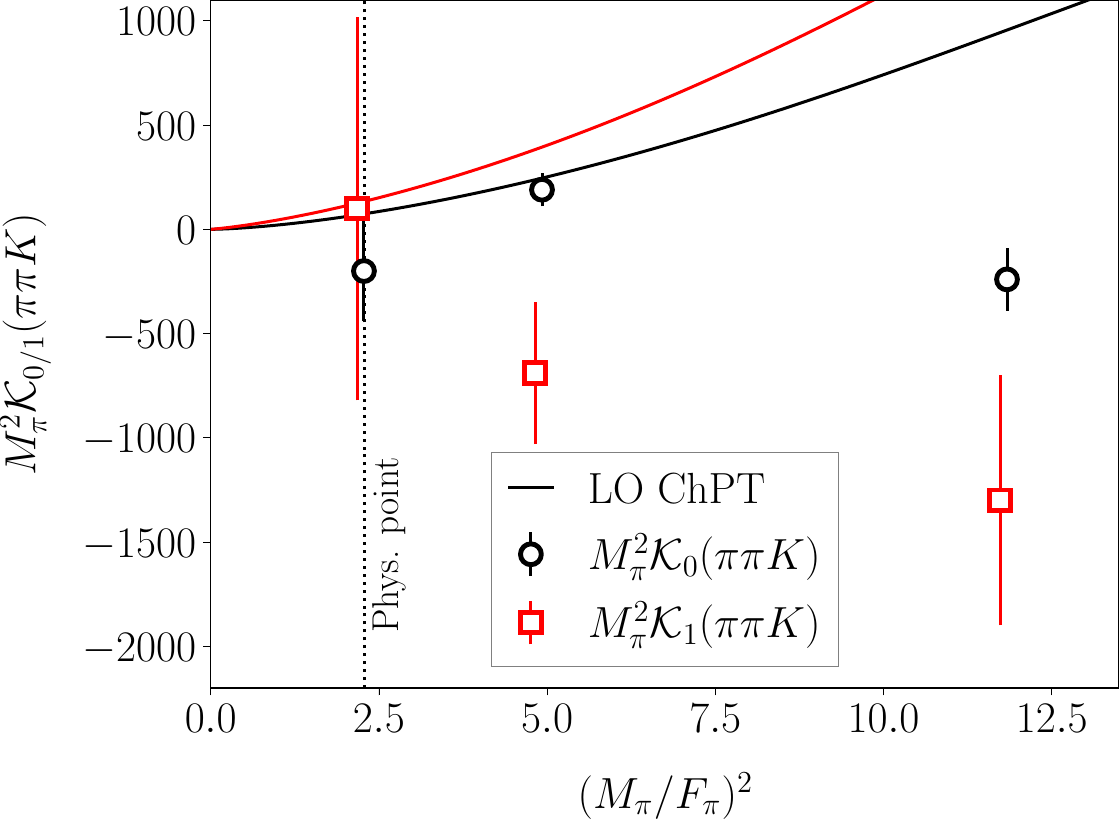}
    }
    \hfill
    \subfloat[\label{fig:Kisokkp}]{%
     \includegraphics[width=0.49\textwidth]{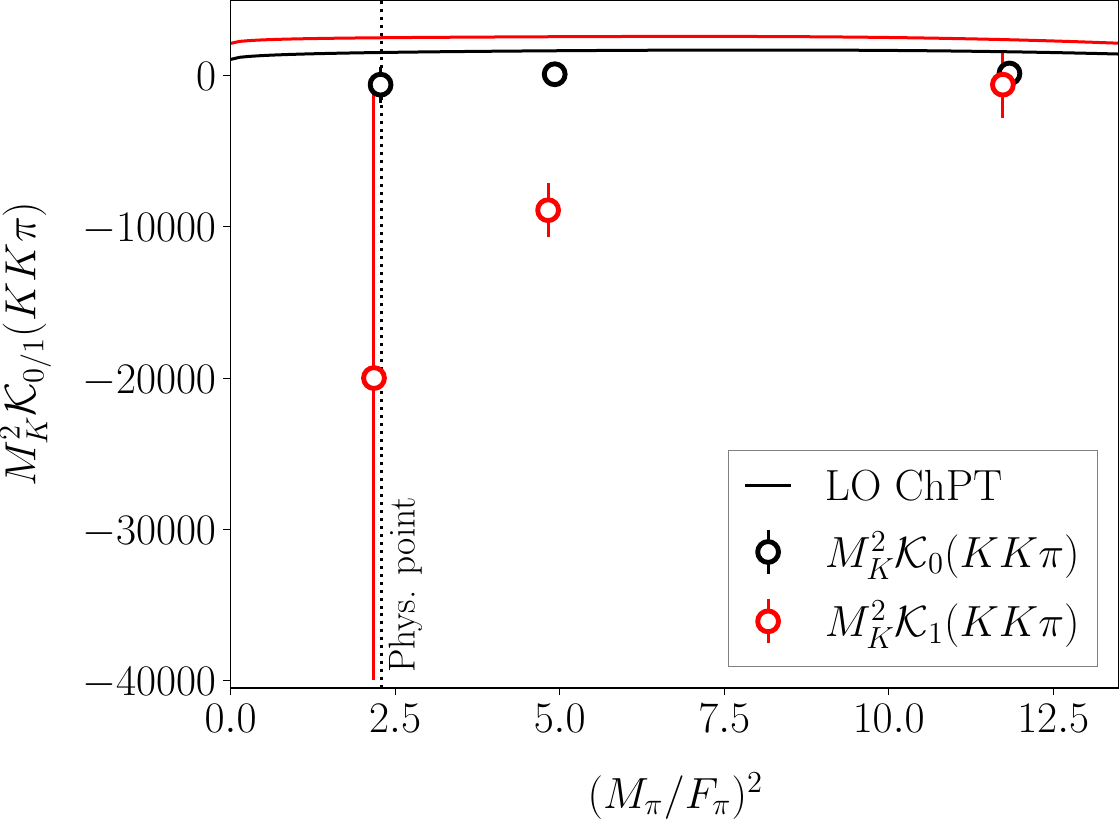}
    }
    \caption{Isotropic terms in the expansion of $\Kdf$ for the $\pi\pi K$ (right) and $KK\pi$ (left) systems. Black circles indicate $\cK_0$, while red squares $\cK_1$. The LO ChPT prediction is shown as solid lines. }
    \label{fig:Kmixed1app}
\end{figure*}

\begin{figure*}[th!]
     \centering
     \subfloat[\label{fig:KEBppk}]{%
     \includegraphics[width=0.49\textwidth]{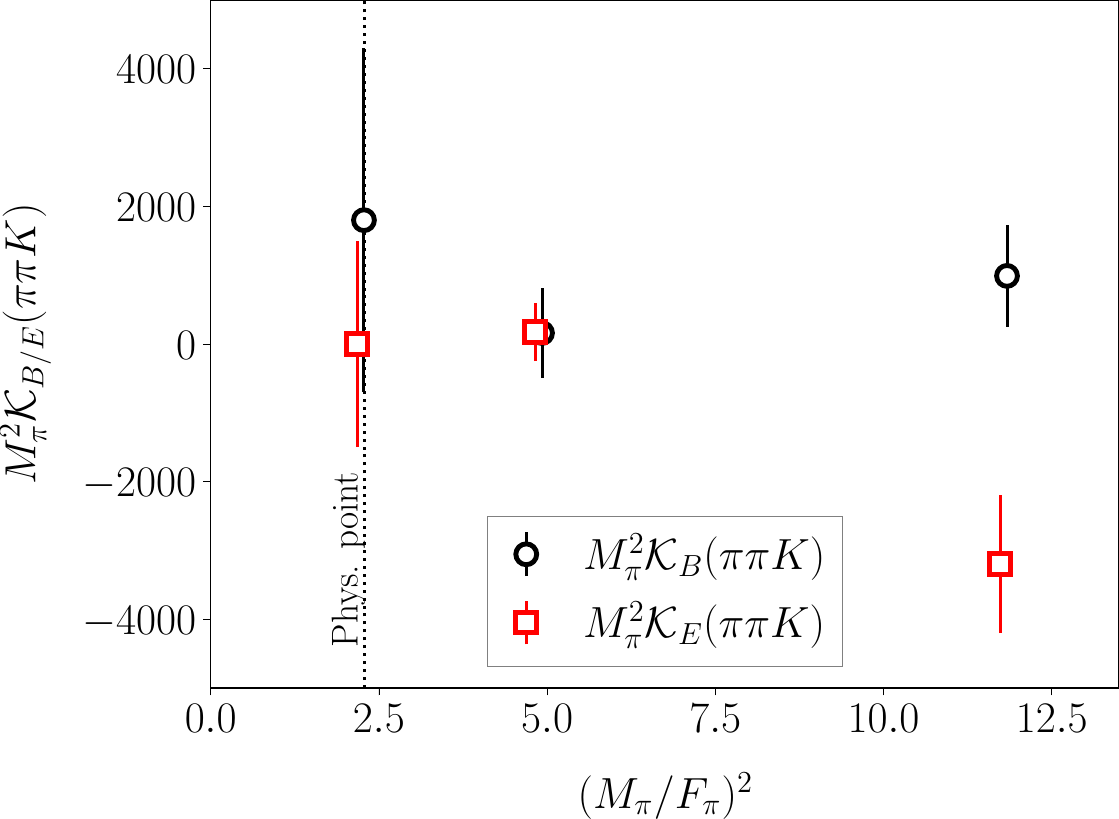}
    }
    \hfill
    \subfloat[\label{fig:KEBkkp}]{%
     \includegraphics[width=0.49\textwidth]{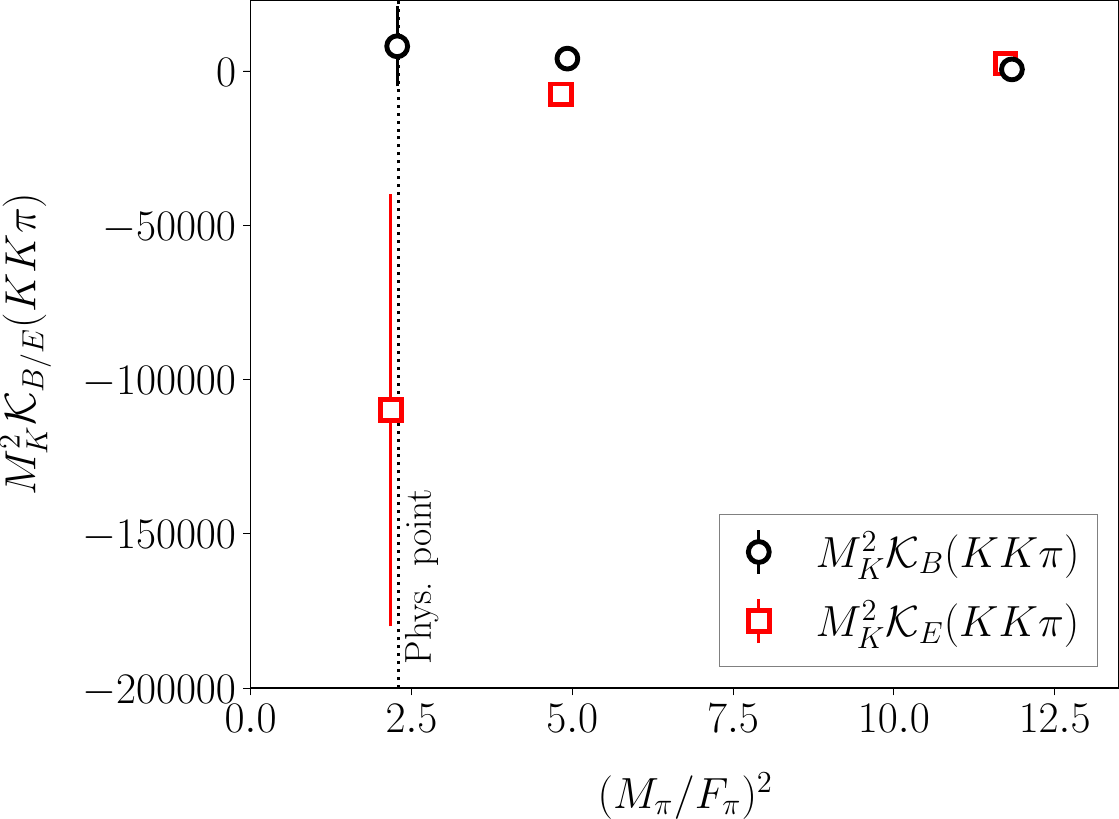}
    }
    \caption{ Additional terms in the expansion of $\Kdf$ for the $\pi\pi K$ (right) and $KK\pi$ (left) systems. Black circles indicate the $\cK_B$ term, while red squares the $\cK_E$ term. }
    \label{fig:Kmixed2}
\end{figure*}

\section{Re-analysis of previous data }
\label{app:reananlysis}

\begin{figure}%[h!]
    \includegraphics[width=8cm]{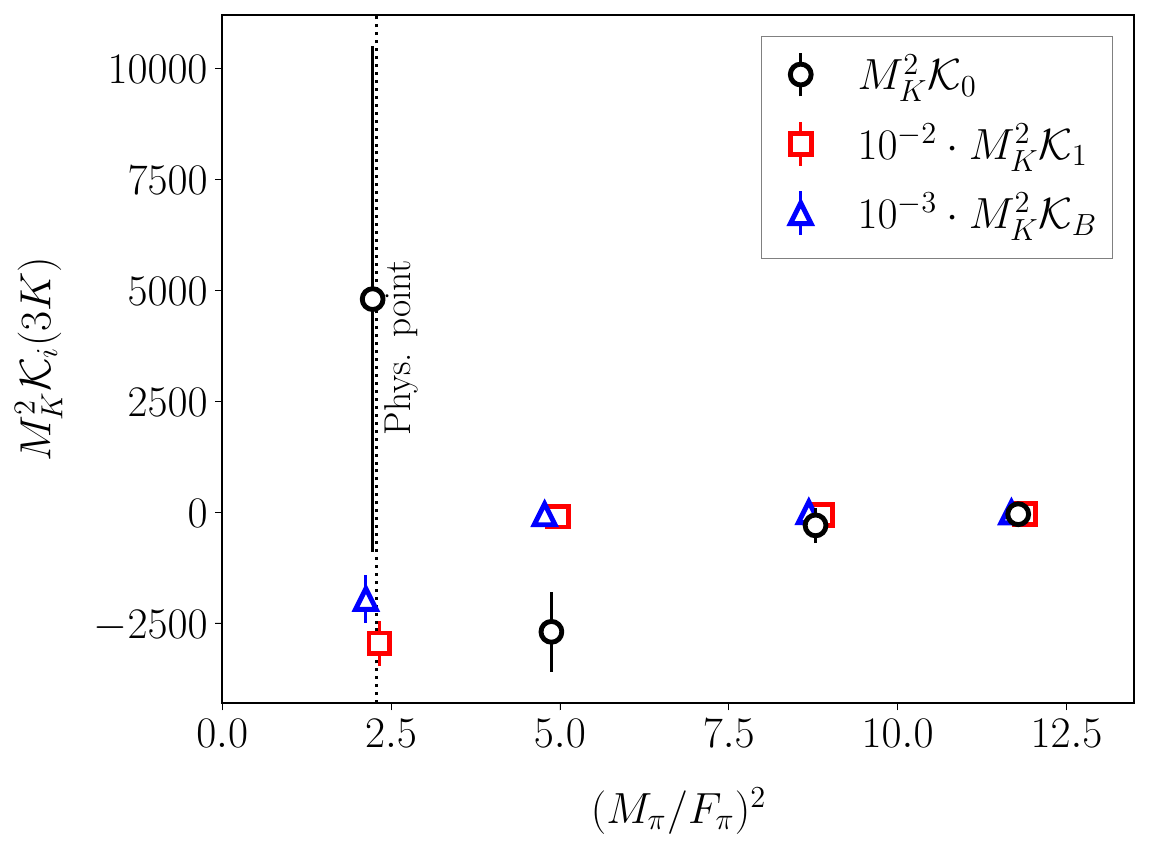}    \caption{ Results for the terms in the expansion of three-kaon $\Kdf$ for the different ensembles. For visibility, some terms have been rescaled. 
    }
    \label{fig:K3kaons}
\end{figure}

The ensembles in \Cref{tab:ensems} with heavier-than-physical quark masses were  analyzed in two previous articles: Ref.~\cite{Blanton:2021llb} for systems of identical particles, and Ref.~\cite{Draper:2023boj} for mixed systems of pions and kaons. 
Our present analysis of the results on the E250 ensemble differs, however, in three ways:
\begin{enumerate}
    \item 
    In Ref.~\cite{Blanton:2021llb} the $\chi^2$ function used in the fits was defined using the the c.m. energy, rather than the lab-frame energy shift. As argued in Ref.~\cite{Draper:2023boj}, the latter may lead to slightly reduced uncertainties.
    \item 
    In Refs.~\cite{Blanton:2021llb,Draper:2023boj}, the cutoff function $H(\bm k)$ in the $KK$ system remained nonzero down to $E_{2,\rm min}=0$. 
    As discussed in the final paragraph of \Cref{sec:QCs}, this does not account for the presence of the left-hand cut due to two-pion exchanges. Here we use $E_{2,\rm min}=2\sqrt{M_K^2-M_\pi^2}$, which avoids this left-hand cut.
    \item 
    In Ref.~\cite{Draper:2023boj}, fits to the $KK+K\pi+KK\pi$ spectra included only $s$ waves in the identical particle subchannels. Here we also include fits with $d$ waves added to these channels (the $s+p+d$ fits in the tables above).
\end{enumerate}
To allow our new results to be combined with those from the other ensembles, we must use a consistent method of analysis across all ensembles.
This is important particularly for chiral extrapolations.
Thus we have reanalyzed the $2K+3K$ and $2K+K\pi +KK\pi$ spectra using the same methods and fits as used here for the E250 ensemble, resolving all three differences described above.
The results are collected in \Cref{tab:KK-params-D200,tab:KKK-params-D200,tab:KKp-params-ERE-D200}.

Using these new fits, we have determined the values for two-particle scattering lengths for the $KK$ and $\pi K$ systems, with results collected in \Cref{tab:KKparams-reanalysis,tab:pKparams-reanalysis}. 
We have used these results to obtain the averages for $\pi K$ and $KK$ quantities given in \Cref{tab:a0summary,tab:r0summary}.

The analysis of $2\pi+ 3\pi$ spectra in Ref.~\cite{Blanton:2021llb}
differs only by the first of the differences listed above.
In this case, however, we have not redone all the fits.
Ref.~\cite{Draper:2023boj} found that, in test cases, the changes in fit parameters were small---central values were little changed, while there was a reduction of uncertainties by around 10\% when using the lab-frame shifts in the $\chi^2$ function.
We have double-checked this by repeating some fits on all the heavier mass ensembles,
reaching the same conclusion.
Thus we are confident that the results of these fits from Ref.~\cite{Blanton:2021llb} remain valid, with conservative errors. We have used them to obtain the averages for $\pi\pi$ quantities given in \Cref{tab:a0summary,tab:r0summary}.

Finally, for completeness, we also show the chiral dependence of the three-body K matrix for $\pi\pi K$ and $K\pi\pi$ in \Cref{fig:Kmixed1app,fig:Kmixed2} and for $3K$ in \Cref{fig:K3kaons}. 

\begin{table*}[t]
\centering
\begin{tabular}{|c|c|c||c|c||c|c|}
\hline
Ensemble& D200 & D200 &N200 & N200 &  N203 & N203 
\\ \hline \hline
Cutoffs & 2.53 & 2.53 & 2.75 & 2.75 & 2.9 & 2.9  
\\ \hline
waves & $s+d$ & $s+d$ & $s+d$ & $s+d$ & $s+d$ & $s+d$ \\ \hline
$\chi^2$ & 54.6 & 51.0 & 49.4 & 45.3  & 41.6 & 36.8 \\ \hline
DOFs   & 40-3=37 & 40-4=36  & 28 -4   & 28 - 5  & 33-4=29 & 33-5=28  \\ \hline 
$p$   & 0.031 & 0.050   & 0.0017 & 0.0037 & 0.061 & 0.124  \\ \hline 
\hline
$B_0^{KK}$ & -2.70(4) & -2.75(5) &  -2.959(43)  &-2.999(48)  & -3.28(5) & -3.31(5)  \\ \hline
$B_1^{KK}$ & 0.44(8) & 0.90(27) &  1.53(17) &  2.14(35) & 1.73(16) & 2.18(29) \\ \hline
$B_2^{KK}$ &  $0$ (fixed) & -0.75(40)  & -1.24(17) &  -3.13(96)  & -1.16(14) & -2.47(68) \\ \hline
$B_3^{KK}$ & $0$ (fixed)& $0$ (fixed) & $0$ (fixed)&  1.50(75)  &  0 (fixed) & 0.95(48) \\ \hline
$D_0^{KK}$ & -0.040(4) & -0.040(4)  & -0.027(2)  & -0.027(2)  & -0.0188(10) & -0.0187(10) \\ \hline
\end{tabular}
\caption{Results of fits to two-kaon levels on ensembles D200, N200, and N203, using ERE3 and ERE4 forms for the $s$-wave $KK$ phase shift, and a cutoff function avoiding the left-hand cut due to two-pion exchange. All results are in units such that $M_K=1$.}
\label{tab:KK-params-D200}
\end{table*}

\begin{table*}[t]
\centering
\begin{tabular}{|c|c|c||c|c||c|c|}
\hline
Ensemble &  D200 & D200 &  N200 & N200  &  N203 & N203 \\ \hline \hline
Cutoffs & 2.53/3.53  & 2.53/3.53   & 2.75/3.75 & 2.75/3.75 & 2.9/3.9 & 2.9/3.9  
\\ \hline
waves &  $s+d$  & $s+d$ &  $s+d$  & $s+d$ &  $s+d$  & $s+d$  \\ \hline
$\chi^2$ & 95.0  & 92.3  & 74.1 & 70.1   &  120.1 & 107.9 \\ \hline
DOFs    &  77-6=71 & 77-7=70 &  28+23-7=44 & 28+23-8=43   &  62-7=55 & 62-8=54 \\ \hline 
p  & 0.030 & 0.038 & 0.003  & 0.0056  & $1\cdot 10^{-6}$ & $2\cdot 10^{-5}$\\ \hline \hline
$B_0^{KK}$  & -2.70(3) & -2.74(4) &-2.938(38)   & -2.972(43)  & -3.21(4) & -3.26(4) \\ \hline
$B_1^{KK}$ & 0.45(8) & 0.79(22) & 1.33(15)  & 1.81(28) & 1.58(12) & -2.16(23) \\ \hline
$B_2^{KK}$  & $0$ (fixed)& -0.54(34) & -1.03(15)  & -2.47(75) & -1.08(11)& -2.66(51)
\\ \hline
$B_3^{KK}$  & $0$ (fixed)& $0$ (fixed) & $0$ (fixed) & 1.14(57) & $0$ (fixed)& 1.13(8)  \\ \hline
$D_0^{KK}$  & -0.041(3)& -0.040(3) & -0.0271(15)  &  -0.0270(15)& -0.018(1)& -0.018(1)
\\ \hline
$\mathcal K^{\rm iso,0}_{KKK}$ & -3400(800)& -2700(900) & -900(350) & -300(400) & -920(200) & -480(240)
\\ \hline
$\mathcal K^{\rm iso,1}_{KKK}$ &  -9400(4300) & -10200(4300) & -5900(1200) &  -7200(1400)   & -3200(670) & -4500(800)  \\ \hline
$\mathcal K^B_{KKK}$ & -57000(22000)  & -50000(22000)  &-4000(6000)  &  -4000(6000) & -5500(2400) 
& -7300(2400) \\ \hline
\end{tabular}
\caption{Results of fits to two- and three-kaon levels on ensembles D200, N200, and N203, using ERE2, ERE3, and ERE4 forms for the $s$-wave $KK$ phase shift, and a cutoff function avoiding the left-hand cut due to two-pion exchange. All results are in units such that $M_K=1$. 
}
\label{tab:KKK-params-D200}
\end{table*}

\begin{table*}[t]
\centering
\begin{tabular}{|c|c|c|c|c|c|}
\hline
Ensemble & D200 &  D200 & D200 & N203 &  N203    \\ \hline \hline
Cutoffs & 2.53/1.95/2.832  & 2.53/1.95/2.832 & 2.53/1.95/2.832   & 2.9/2.66/3.521 & 2.9/2.66/3.521
\\ \hline
waves & $s+p$ & $s+p+d$  & $s+p+d$  & $s+p+d$  & $s+p+d$ \\ \hline
$\chi^2$ & 161.8  & 151.3 & 116.9  & 101.7 &  99.0 \\ \hline
DOFs     & 28+16+29-10=63 & 40+16+29-7=78  & 40+16+29-11=74   & 33+37+35-8=79 & 33+37+35-12=75 \\ \hline 
p & $1.2\cdot 10^{-10}$ & $1.3\cdot 10^{-6}$ & 0.0011 & 0.044 & 0.033 \\ \hline
\hline
$B_0^{KK}$ &-2.87(4) & -2.80(4)& -2.77(4)  & -3.37(5)& -3.37(5)
\\ \hline
$B_1^{KK}$ & 1.41(25)  & 1.26(23)& 0.93(24)  & 2.48(25) & 2.53(26)
\\ \hline
$B_2^{KK}$ &-1.26(38)  & -1.24(35) & -0.72(36) & -2.98(59) & -3.13(60)
\\ \hline
$B_3^{KK}$ & 0 (fixed) & 0 (fixed) & 0 (fixed)    & 1.22(41) & 1.32(42)
\\ \hline
$D_0^{KK}$   & 0 (fixed)& -0.039(3) & -0.040(3)  & -0.019(1)& -0.019(1)
\\ \hline
$B_0^{K\pi}$ & -2.42(7) & -2.40(7) & -2.28(6)  & -3.24(7) & -3.22(7)
\\ \hline
$B_1^{K\pi}$ & -2.27(32)  & -1.88(29) & -2.42(29)  & -2.08(17) & -2.15(17)
\\ \hline
$P_0^{K\pi}$ & 0.027(8) & 0.0057(77)& 0.0067(82)  & 0.008(8) &  0.004(9)
\\ \hline
$\mathcal K^{\rm iso,0}_{KK\pi}$ & 130(280) & 0 (fixed) & 90(270)  & 0 (fixed) & 140(300)
\\ \hline
$\mathcal K^{\rm iso,1}_{KK\pi}$ & -8300(1800) & 0 (fixed)& -8900(1800)  & 0 (fixed)& -600(2200)
\\ \hline
$\mathcal K^B_{KK\pi}$ & 3700(1300) &0 (fixed) &  4000(1200)  & 0 (fixed) &  500(1800)
\\ \hline
$\mathcal K^E_{KK\pi}$ & -8600(4100) & 0 (fixed) &  -7600(4000) & 0 (fixed) & 2400(2500)
\\ \hline
\end{tabular}
\caption{Results of fits to $KK+K\pi+KK\pi$ levels on ensembles D200 and N203, using ERE3 and ERE4 forms for the $s$-wave $KK$ phase shift, the ADLER2 form for the $K\pi$ s-wave phase shift, and a cutoff function avoiding the left-hand cuts due to two-pion exchange. All results are in units such that $M_K=1$. 
}
\label{tab:KKp-params-ERE-D200}
\end{table*}

\begin{table*}[t]
\centering
\begin{tabular}{|c|c|c|c|c|c|c|c|}
\hline
Ensemble & Fit & $\chi^2$/DOF  & p-value & $M_K a_0^{KK}$ & $a_0^{KK}r_0^{KK}$ & $D_0^{KK}$
\\ \hline \hline
D200 & $KK$ ($s+d$, ERE2) & 54.6/37 &0.031 & 0.3703(51) & 0.32(6)& -0.040(4)
\\ \hline
D200 & $KK$ ($s+d$, ERE3) & 51.0/36 &0.050 & 0.3635(63) & 0.66(19)&  -0.040(4)
\\ \hline 
D200 & $KKK$ ($s+d$, ERE2)& 95.0/71 &0.030 & 0.3707(46) & 0.33(5)& -0.041(3)
\\ \hline
D200 & $KKK$ ($s+d$, ERE3)& 92.3/70 &0.038 & 0.3656(56) & 0.58(16)& -0.040(3)
\\ \hline
D200 & $K K \pi$ ($s+p+d$)& 116.9/74 &0.0011 & 0.3611(54) & 0.67(16)& -0.040(3)
\\ \hline
D200 & $KK$ (thr)~\cite{Draper:2023boj} & N/A &N/A & 0.3531(83) & N/A & N/A
\\ \hline 
D200 & $KKK$ (thr)~\cite{Draper:2023boj} &N/A &N/A & 0.3541(83) & N/A & N/A
\\ \hline \hline
N200 & $KK$ ($s+d$,ERE3)  & 49.9/24 &0.0017 &  0.3380(49) & 1.03(10) & -0.027(2)
\\ \hline
N200 & $KK$ ($s+d$,ERE4) & 45.3/23 &0.0037 & 0.3335(53) & 1.42(22) &  -0.027(2)  \\ \hline
N200 & $KKK$ ($s+d$,ERE3)  & 74.1/44 &0.003 & 0.3404(44) & 0.91(9) & -0.0271(15)
\\ \hline
N200 & $KKK$ ($s+d$,ERE4)  & 70.1/43 &0.0056  & 0.3365(49) & 1.22(18) &  -0.0270(15)
\\ \hline
N200 & $KK$ (thr) &N/A  &N/A & 0.3285(62) & N/A &N/A \\ \hline
N200 & $KKK$ (thr) &N/A  &N/A & 0.3301(61)  & N/A &N/A 
\\ \hline \hline
N203 & $KK$ ($s+d$, ERE3) & 41.6/29 &0.06 & 0.3052(45) & 1.05(8)& -0.019(1)\\ \hline
N203 & $KK$ ($s+d$, ERE4) & 36.8/28 & 0.12 & 0.3022(48) & 1.32(16) & -0.019(1)\\ \hline
N203 & $KKK$ ($s+d$, ERE4)  & 107.9/54 &0.00002 & 0.3067(42) &1.32(13) & -0.018(1)\\ \hline
N203 & $K K \pi$ ($s+p+d$, ERE4, $\mathcal K_{\rm df,3}\!\!=\!\!0$) & 101.7/79 &0.044 & 0.2970(43) & 1.47(14)& -0.019(1)\\ \hline
N203 & $K K \pi$ ($s+p+d$, ERE4) &   99.0/75 &0.033 & 0.2970(44) & 1.50(14) & -0.019(1)\\ \hline
N203 & $KK$ (thr)~\cite{Draper:2023boj} &N/A  &N/A & 0.3022(54) & N/A &N/A \\ \hline
N203 & $KKK$ (thr)~\cite{Draper:2023boj} &N/A  &N/A & 0.3060(51) & N/A &N/A \\ \hline
\end{tabular}
\caption{
Comparison of $KK$ scattering parameters from different ERE lab-shift fits on the D200, N200, and N203 ensembles, using kaon units. Results are from the fits in \Cref{tab:KK-params-D200,tab:KKK-params-D200,tab:KKp-params-ERE-D200}, keeping only those with reasonable $p$ values.
For comparison, we show the results (denoted ``thr") from Ref.~\cite{Draper:2023boj} of fitting the
energy of the threshold $KK$ and $KKK$ states to the
$1/L$ expansion keeping up to $1/L^5$ terms. 
}
\label{tab:KKparams-reanalysis}
\end{table*}

\begin{table*}[t]
\centering
\begin{tabular}{|c|c|c|c|c|c|c|c|c|}
\hline
Ensemble & Fit & $\chi^2/$DOF &p-value & $M_\pi a_0^{\pi K}$ & $M_\pi^2 a_0^{\pi K}r_0^{\pi K}$ & $P_0^{\pi K}$
\\ \hline \hline
D200 & $\pi K$ ($s+p$, 16)~\cite{Draper:2023boj} & 15/13  & 0.31 & 0.107(4) & 1.22(6) & 0.0012(7)
\\ \hline
D200 & $\pi K$ ($s+p$, 26)~\cite{Draper:2023boj} & 31/23   &0.12 & 0.107(4) & 1.23(5) & 0.0013(6)
\\ \hline
D200 & $\pi \pi K$ ($s+p$, 59)~\cite{Draper:2023boj} & 112/50 &  0.0000012& 0.110(3) & 1.15(5) & 0.0007(6)
\\ \hline
D200 & $K K \pi$ ($s+p+d$, 85) & 117/74  & 0.0011 & 0.109(3) &  1.22(5) & 0.0005(6)
\\ \hline
D200 & $\pi K$ (thr5)~\cite{Draper:2023boj} & N/A &N/A & 0.106(5) & N/A & N/A
\\ \hline \hline
N203 & $\pi K$ (s+p, 19)~\cite{Draper:2023boj} & 21.1/16 &0.17 & 0.207(6) & 1.63(9)& 0.004(4)
\\ \hline
N203 & $\pi K$ (s+p, 36)~\cite{Draper:2023boj} & 36.1/33 &0.33 & 0.206(6) & 1.65(8)& 0.002(3)
\\ \hline
N203 &  $\pi \pi K$ (s+p, 82)~\cite{Draper:2023boj} &119/73 & 0.0005 & 0.208(4) & 1.69(7) &  0.006(4)
\\ \hline
N203 & $K K \pi$ ($s+p+d$, $\mathcal K_{\rm df,3}\!\!=\!\!0$, 87) & 101.7/79 &0.044 & 0.212(5) &  1.61(8)&0.004(4)
\\ \hline
N203 & $K K \pi$ ($s+p+d$, 87) &  99.0/75  &0.033 & 0.213(5) &  1.58(8) & 0.002(4)
\\ \hline
N203 & $\pi K$ (thr5)~\cite{Draper:2023boj} & N/A &N/A & 0.213(7) & N/A & N/A \\ \hline
\end{tabular}
\caption{
Comparison of $\pi K$ scattering parameters from different Adler2 $s+p$ fits on the D200 and N203 ensembles using pion units. 
The $KK\pi$ results are those from \Cref{tab:KKp-params-ERE-D200}; all others are from Ref.~\cite{Draper:2023boj}.
The ``thr5" rows give the results of
fitting the threshold $\pi K$ state to the threshold expansion up to $1/L^5$. 
}
\label{tab:pKparams-reanalysis}
\end{table*}

\pagebreak

\section{Further details on integral equations }
\label{app:pwproj}

This appendix provides further details on the relativistic three-body integral equations described in outline in \Cref{sec:inteqs}. In particular, we discuss their partial-wave projection, the reconstruction of the full amplitude, and the numerical methods implemented in their solution. It is largely based on Refs.~\cite{Hansen:2015zga, Jackura:2018xnx, Hansen:2020otl, Jackura:2020bsk, Dawid:2023jrj, Jackura:2023qtp, Dawid:2024dgy, Briceno:2024ehy},
with necessary generalizations.

\subsection{Three-body kinematics}
\label{app:threebodykinematics}

In this work, two types of three-body infinite-volume reactions are reconstructed from 
finite-volume spectra obtained using LQCD. The first is a $3 \to 3$ scattering process in which three identical scalar particles of initial momenta $\bm k = \{\bm k_1, \bm k_2, \bm k_3 \}$ collide and emerge with final momenta $\bm p = \{\bm p_1, \bm p_2, \bm p_3 \}$. These are either three pions or three kaons. The second type is that of a ${2 + 1 \to 2+ 1}$ process in which two of the incoming/outgoing particles are indistinguishable while the third is different. As discussed in the main text, in this case we denote the sets of initial momenta as ${\bm k = \{\bm k_1, \bm k_{1'}, \bm k_2 \}}$, and the final momenta as $\bm p = \{\bm p_1, \bm p_{1'}, \bm p_2 \}$. These describe either the $\pi\pi K$ or $KK\pi$  systems.

The objects entering the integral equations are unsymmetrized $3\to3$ amplitudes, in which the particles are separated into a spectator and a pair.
These are denoted $\cM^{(u,u)}_3$ for identical particles,
and (as already noted in the main text) $\cM^{(u,u)}_{3,ij}$ for non-identical particles.
Here the indices $i,j \in \{1,2\}$ in the latter case are needed because two different choices of spectator flavor can be made in the $\pi\pi K$ and $KK\pi$ states. 
The precise definition of the unsymmetrized amplitudes has been discussed for $2+1$ systems in the main text; the definition for identical particles is given in Ref.~\cite{Hansen:2014eka}.
The $2+1$ formalism requires a convention for which is the ``primary'' member of the pair in the case that the members differ. Our convention, following Ref.~\cite{Blanton:2021eyf}, is that the primary member has the same flavor as the spectator. For example, for the $\pi\pi K$ system, if a pion is the spectator, then the primary member of the $\pi K$ pair is also a pion. If the spectator is the kaon, the pair is composed of two identical pions and no choice of primary is necessary.

The unsymmetrized amplitudes, like $\cM_3$, are invariant under Poincar\'e transformations, and involve on shell external particles. They depend on eight kinematical variables, which we choose as follows.
There are three energy-like variables: the total c.m.~energy, denoted $E$, ({we drop the $*$ index} since we work entirely in the c.m.~frame when discussing integral equations), and the magnitudes of momenta of two external spectators, $p$ and $k$. The remaining five variables are angles fixing orientations of (some of) the external momenta. For the first, we choose the angle between the incoming and outgoing spectator momenta in the total c.m.~frame, and call it $\Theta$. The next two variables are the angles defining the orientation of the relative momentum $\hat{q}_p^\star$ of the primary particle in the final pair, $\Omega_p^\star = (\vartheta^\star_p, \varphi^\star_p)$. It is defined in the pair c.m.~frame {(indicated by the $\star$ symbol)} with respect to the direction given by total momentum of the pair in the overall c.m.~frame, $-\hat{p}$. 
The subscripts here indicate the momentum of the corresponding spectator. The last two variables are the analogous angles for the initial-state pair.

\begin{figure}[b]
    \centering
    \includegraphics[width=0.41\textwidth]{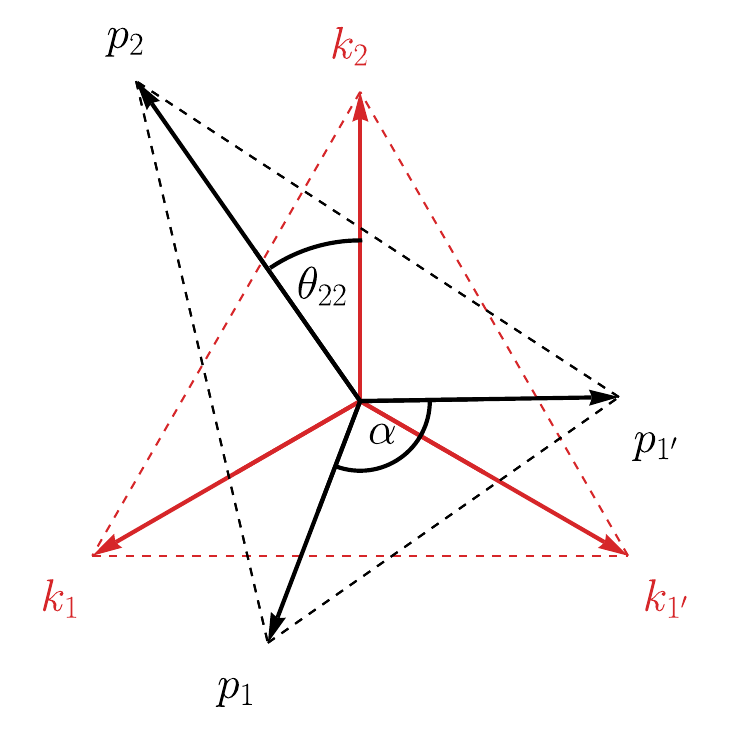}
    \caption{Kinematical configuration of incoming (red) and outgoing (black) momenta in the $2+1$ system in the three-body c.m. frame. In the initial state, we consider momenta forming bisectors of an equilateral triangle. In the final state, the momenta form bisectors of an isosceles triangle characterized by the angle $\alpha$. Two triangles are coplanar, but rotated with respect to each other by the angle $\theta_{22}$.}
    \label{fig:configs}
\end{figure}

Since the three-particle amplitudes can depend on many variables, it is difficult to meaningfully present the physical information they contain on a single two-dimensional graph. 
Many choices can be made for which variables to keep free and which to keep fixed. For instance, a Dalitz-like plot of the purely $s$-wave partial-wave-projected amplitude requires choosing values for three variables (e.g. two invariant masses of incoming pairs and the total invariant mass), while probing the dependence on two other variables (e.g. two invariant masses of final pairs). This is the choice made, for instance, in Ref.~\cite{Hansen:2020otl}.

To simplify this problem, we find it natural to consider certain symmetric, physical configurations of the incoming/outgoing momenta that effectively reduce the number of kinematical variables the amplitude depends on. For concreteness, in this work we consider two such special kinematic configurations.

We discuss the kinematical configurations using a $2+1$ outgoing state as a general example. Systems of indistinguishable particles are described by the following equations in the limit $M_1, M_{1'}, M_2 \to M$, where $M$ is the mass of identical mesons. Additionally, following the notation explained above, one should replace particle labels according to $1 \to 1$, $1' \to 2$, and $2 \to 3$.

For the momenta of the outgoing state we assume that,
    %%%%%
    \begin{align}
        \label{eq:momentum3}
    & \bm p_{2} = p_2 \, (0, \, 0, \, -1 ) \, , \\
        \label{eq:momentum1}
    & \bm p_1 = p_{\rm is} \, ( \sin \alpha/2, \, 0,\, \cos \alpha/2) \, ,\\
    \label{eq:momentum2}
    & \bm p_{1'} = p_{\rm is} \, ( -\sin \alpha/2, \, 0,\, \cos \alpha/2) \, . 
    \end{align}
    %%%%%
These momenta form bisectors of an isosceles triangle, with an angle $\alpha$ between $\bm p_1$ and $\bm p_{1'}$. Assuming total energy $E$ and vanishing total momentum, energy-momentum conservation yields,
    %%%%%
    \begin{multline}
    \label{eq:mom-iso-dis}
    p_{\rm is} = \frac{1}{2 s_\alpha^2} \Bigg[ \left(c_\alpha^2 + 1 \right) E^2 - s^2_\alpha \left(4 M_1^2-M_2^2\right) \\ 
    %%%
    \qquad - 2 E \sqrt{c_\alpha^2 \left(E^2 - 4 s_\alpha^2 M_1^2\right) + s_\alpha^2 M_2^2} \Bigg]^{1/2} \, ,
    \end{multline}
while
    \begin{equation}
    p_2 = 2 \, c_\alpha \, p_{\rm is} \,.
    \label{eq:mom-iso-dis2}
    \end{equation}
    %%%%%
where $c_\alpha = \cos( \alpha/2)$ and $s_\alpha = \sin( \alpha/2)$. The orientation of the pair's relative momentum $\bm q_{p_1}^\star$ is given by the angle $\vartheta^\star_{p_1}$, where,
    \begin{align}
    \label{eq:cos_iso}
    \cos \vartheta_{p_1}^\star = - \frac{p_{\rm is} \gamma_{p_1}}{q_{p_1}^\star} \Big( \cos\alpha + \frac{\omega_{p_{1'}}}{E-\omega_{p_1}} \Big)
    \, .
    \end{align}
    %%%%%% 
Lorentz factors,
     %%%%%
    \begin{align}
    \gamma_{p_1} = \frac{1}{\sqrt{1- \beta_{p_1}^2}} \, , \quad \bm \beta_{p_1} = \frac{\bm p_1}{E - \omega_{p_1}} \, ,
    \end{align}
    %%%%%
define the boost from the total c.m.~frame to the final pair's c.m. frame and,
    %%%%%
    \begin{align}
    q_{p_1}^\star = \frac{\lambda^{1/2}(\sigma_{p_1}, M_1^2, M_2^2)}{2 \sqrt{\sigma_{p_1}}} \, ,
    \end{align}
    %%%%%
where $\sigma_{p_1}$ is the invariant mass squared of the pair composed of particles $1'$ and $2$. The quantity $\cos \vartheta_{p_{1'}}^\star$ is given by the analogous equations, while $\cos\vartheta^\star_{p_2}=0$. 

For simplicity, we assume that the planes spanned by the final and initial momenta coincide. Although this assumption does not affect the $J=0$ amplitudes, it greatly simplifies symmetrization formulas presented in the next section for $J=1,2$. The initial state kinematic is described by \Cref{eq:momentum3,eq:momentum1,eq:momentum2,eq:mom-iso-dis} with three modifications. First, outgoing momenta $\bm p$ are replaced by incoming momenta $\bm k$. Second, the initial-state $z$ axis (aligned with the momentum $-\bm k_2$) is different than the final-state $z$ axis. We choose the angle between $\bm p_2$ and $\bm k_2$ to be $\theta_{22}$, as shown in \Cref{fig:configs}. Third, for the initial kinematics, we choose 
$\alpha = 2\pi/3$.

It is relatively straightforward to check that limits $\alpha \to 2\pi/3$ and $M_{1'}, M_{1}, M_2 \to M$ applied to the above formulas give the expected results. For instance, when $\alpha = 2\pi/3$,
    %%%%%
    \begin{align}
    \label{eq:cosine_p1}
    \cos \vartheta^\star_{p_1} = - \frac{\gamma_{p_1} \beta_{p_1}}{2 q_{p_1}^\star} \, (\omega_{p_{1'}} - \omega_{p_2}) \, ,
    \end{align}
    %%%%%
which reduces to the expected $\cos \vartheta^\star_{p_1} = 0$ when masses become identical. In the same scenario, momenta of particles reduce to the anticipated $p_{\rm is} = p_2 = \sqrt{ E^2/9 - M^2 }$.

In the following, we will refer to the kinematical configuration with $\alpha = 2\pi/3$ in the final state as the ``equilateral configuration.'' In this case all momenta have equal magnitudes which we call $p_{\rm eq}$ (final) and $k_{\rm eq}$ (initial). When the final angle $\alpha \neq 2\pi/3$ we use ``isosceles configuration'' instead. 
Results for the amplitudes for these two configurations are presented in the main text, \Cref{sec:amplitudes}.

\subsection{Partial-wave projection}
\label{sec:PW}

A significant simplification of the three-body integral equations presented in \Cref{sec:inteqs} can be achieved by partial-wave projection of the pair-spectator amplitudes~\cite{Jackura:2020bsk, Jackura:2023qtp}.
Here we follow the presentation of Ref.~\cite{Dawid:2024dgy}.
The projection is done in several steps. 
First, we decompose the amplitude $X = \cM^{(u,u)}_3$ or $\cM^{(u,u)}_{3,ij}$ into waves corresponding to definite spins and helicities of external pairs,
    %%%%%
    \begin{align}
    & X(\{\bm p\},\{\bm k\}) \\ \nonumber 
    & \quad = 4\pi 
    \sum_{ \substack{s'\lambda' \\ s \lambda} } 
    Y_{s'\lambda'}(\hat q_p^\star) \, X_{s'\lambda'; s \lambda}(p, k; E; \Theta) \,
    Y_{s \lambda}^*(\hat q_k^\star) \, .
    \end{align} 
    %%%%%
Here, the resulting amplitude $X_{s'\lambda'; s \lambda}$ depends on magnitudes of momenta $p, k$ of external spectators, 
total energy $E$, and the scattering angle $\Theta$.\footnote{%
One must also specify a phase convention corresponding to a choice of where the angles $\phi_p^\star$ and $\phi_k^\star$ vanish. A consistent set of choices is described in Appendix A of Ref.~\cite{Dawid:2024dgy}, and we use these choices in the results below.
}
It is labeled by the angular momentum (spin) of the initial pair, denoted $s$, and its component in the $-\hat{k}$ direction (helicity), denoted $\lambda$. The spin and helicity of the final pair is $s',\lambda'$, respectively. 

Secondly, we partial-wave decompose the amplitude $X_{s'\lambda'; s \lambda}$ into waves corresponding to definite total angular momentum, $J$. This is given by,
    %%%%%
    \begin{align}
    \label{eq:masterK3}
    & X^J_{s'\lambda'; s \lambda}(p,k;E) \\ \nonumber 
    & \quad = 
    \int\limits_{-1}^{1} \frac{dc_\Theta}2\,
    d_{-\lambda', -\lambda}^J(\Theta) \, 
    X_{s'\lambda'; s \lambda}(p,k; E; \Theta) \, ,
    \end{align}
    %%%%%
where $c_{\Theta} = \cos\Theta$ and $d^{j}_{m'm}$ is the Wigner $d$ function.

In the final step we transform the resulting amplitude to the $J\ell s$ (orbital) basis, where we recall that $\ell$ is the angular momentum of the pair relative to the spectator. This yields,
    %%%%%
    \begin{align}
    \label{eq:LS-basis}
    X^J_{\ell's';\ell s} = \sum_{\lambda'\lambda} 
    \mathcal U^{J}_{\ell' s' \lambda'} \, 
    X^J_{s' \lambda'; s \lambda} \, 
    \mathcal U^{J}_{\ell s \lambda} \, ,
    \end{align}
    %%%%%
where the recoupling matrix is given in terms of Clebsch-Gordon coefficients~\cite{Chung:1971ri},
    %%%%%
    \begin{align}
    \mathcal U^J_{\ell s \lambda} = \sqrt{\frac{2 \ell + 1}{2 J + 1} } \, \langle J,-\!\lambda | \ell, 0; s, -\!\lambda \rangle \, .
    \end{align}
    %%%%%

Following the sequence of steps described above, the integral equations provided in \Cref{sec:inteqs} are partial-wave projected onto definite $J$ and $(\ell's'; \ell s)$ quantum numbers. 
The equations in \Cref{sec:inteqs}, specifically \Cref{eq:ladder,eq:Mdf3,eq:Teq}, retain their form, but now involve matrices with fixed overall $J$ and with indices $\ell', s';\ell, s$, and involve integrals only over the magnitudes of spectator momenta.
A complete description is given in Ref.~\cite{Dawid:2024dgy} for the $2+1$ system;
for identical particles the flavor indices are absent and the form is as given in Refs.~\cite{Jackura:2023qtp, Briceno:2024ehy}. 

The main challenge in performing the decomposition into partial waves arises from the complicated forms of the one-particle exchange amplitude $G$ and the three-body K matrix. In \Cref{app:PWG,app:PWK3} below, we provide partial-wave-decomposed versions of these objects for the cases that are needed for the results presented in the main text. 

\begin{widetext}

\subsection{Symmetrization}
\label{app:symm}

The solutions of the one-dimensional integral equations are combined according to \Cref{eq:symmetrization,eq:symmetrization2} to obtain the full three-body amplitude. The $\cZ$ function is,
    %%%%%
    \begin{equation}
    \cZ^{J}_{\ell' s'; \ell s}(\theta_{ij}, \hat{\bm{q}}_{p_i}^\star, \hat{\bm{q}}_{k_j}^\star )
    = 4 \pi \,  (2J+1) \sum_{\lambda' \lambda}   
    \mathcal U_{\ell' s' \lambda'}^J \, Y_{s'\lambda'}(\hat{\bm{q}}_{p_i}^\star) \, 
    d^J_{-\lambda', -\lambda}(\theta_{ij}) \, 
    Y^*_{s\lambda}(\hat{\bm{q}}_{k_j}^\star) \, \mathcal U_{\ell s \lambda}^J \, .
    \label{eq:newZ}
    \end{equation}
    %%%%%
In the case of identical particles, \Cref{eq:symmetrization2} has no dependence on the flavors of the spectators, and summations over momentum permutations include in total nine possible configurations of external momenta. Moreover, the symmetry factors $\mathcal X_{ij}$ are absent (all equal to 1) in this case.

In the following, we present the explicit expressions for symmetrization that result from \Cref{eq:symmetrization2,eq:newZ} for the kinematics discussed above, namely the simple equilateral and isosceles triangle configurations. We recall that the unsymmetrized amplitudes for identical particles have indices and arguments
    \begin{equation}
    \cM^{(u,u)J}_{3;\ell' s';\ell s}(p,k;E) \,,
    \end{equation}
while those for the $2+1$ systems are given by
    \begin{equation}
    \cM^{(u,u)J}_{3;i\ell' s';j\ell s}(p,k;E) \,,
    \end{equation}
where $i$ and $j$ label the flavor of the spectator, with the label $1$ corresponding to the spectator being one of the identical pair, while label $2$ indicates that the spectator is the nonidentical member of the triplet.
Here we display the full set of arguments; in the following we drop $E$ for the sake of brevity. In the expressions below, and all calculations, we neglect contributions from $\ell' , \ell > 2$ and $s', s > 2$.

Although \Cref{eq:symmetrization,eq:symmetrization2,eq:newZ} provide a complete definition of the symmetrized amplitudes, we provide explicit expressions for the cases we use in the plots in the main text, so as to allow checking by the intrepid reader.

\subsubsection{The isosceles/equilateral triangle configuration for $J^P=0^-$}

For identical particles, the partial waves that contribute are $(\ell,s)=(0,0)$ and $(2,2)$. The result is,
    %%%%%
    \begin{align} 
    \begin{split}
    &\cM_3^{J=0}(\{\bm p\},\{\bm k\}) = 
    3 \left[ \cM_{00;00}^{(u,u)0}(p_3, k_{\rm eq}) - 
    \frac{\sqrt{5}}{2} \cM_{00;22}^{(u,u)J=0}(p_3, k_{\rm eq}) 
    -\frac{\sqrt{5}}{2}\cM_{22;00}^{(u,u)J=0}(p_3, k_{\rm eq}) 
    + \frac{5}{4} \cM_{22;22}^{(u,u)J=0}(p_3, k_{\rm eq}) 
    \right] 
    \\
    &+ 6 \left[ \cM_{00;00}^{(u,u)J=0}(p_{\rm is}, k_{\rm eq}) 
    - \frac{\sqrt{5}}{2}\cM_{00;22}^{(u,u)J=0}(p_{\rm is}, k_{\rm eq}) 
    + \frac{\sqrt{5}}{2}(3 c^2 \!-\! 1) \cM_{22;00}^{(u,u)J=0}(p_{\rm is}, k_{\rm eq})
    + \frac{5}{4} (1 \!-\! 3 c^2) \cM_{22;22}^{(u,u)J=0}(p_{\rm is}, k_{\rm eq}) \right] \,,
    \end{split}
    \end{align}
    %%%%%
with $c=\cos\vartheta^*_{p_1}$. Note that this result is independent of the relative orientation of the initial and final reaction planes.

For the $2+1$ case, for spectator flavor $i=1$ the contributing waves are $(0,0)$ and $(1,1)$ (since we do not allow $s=2$ for nondegenerate pairs), while for $i=2$ the waves are $(0,0)$ and $(2,2)$.
We find the following result,
    \begin{align}
    \begin{split}
    %% 11
    & \cM_3^{J=0}(\{ \bm p \}; \{ \bm k \}) =
    \\
    & 4 \left[
    \cM_{3, 100; 100}^{(u,u) J=0}(p_{\rm is}, k_{\rm eq}) 
    - \sqrt{3} \, c \, 
    \cM_{3, 100; 111}^{(u,u) J=0}(p_{\rm is}, k_{\rm eq})
    - \sqrt{3} \, c' \, 
    \cM_{3, 111; 100}^{(u,u) J=0}(p_{\rm is}, k_{\rm eq})  
    +3 \, c' c \, 
    \cM_{3, 111; 111}^{(u,u) J=0}(p_{\rm is}, k_{\rm eq})
    \right] \\
    %%% 12
    & + 2 \sqrt{2} \left[ 
    \cM_{3, 100; 200}^{(u,u) J=0}(p_{\rm is},k_{\rm eq}) 
    - \sqrt{\frac54} \, 
    \cM_{3, 100; 222}^{(u,u) J=0}(p_{\rm is},k_{\rm eq})
    - \sqrt{3} c' \,
    \cM_{3, 111; 200}^{(u,u) J=0}(p_{\rm is},k_{\rm eq})
    + \sqrt{\frac{15}4} c' \,
    \cM_{3, 111; 222}^{(u,u) J=0}(p_{\rm is},k_{\rm eq})
    \right] \\
    %%% 21
    & + 2 \sqrt{2} \left[
    \cM_{3, 200; 100}^{(u,u) J=0}(p_2,k_{\rm eq}) 
    - \sqrt{\frac54}\cM_{3, 222; 100}^{(u,u) J=0}(p_2,k_{\rm eq}) 
    - \sqrt{3} \, c \cM_{3, 200; 111}^{(u,u) J=0}(p_2,k_{\rm eq}) 
    + \sqrt{\frac{15}{4}}c \cM_{3, 222; 111}^{(u,u) J=0}(p_2,k_{\rm eq})
    \right] \\
    %%%
    & +
    2 \cM_{3, 200; 200}^{(u,u) J=0}(p_2,k_{\rm eq}) 
    - \sqrt5 \cM_{3, 200; 222}^{(u,u) J=0}(p_2,k_{\rm eq})
    - \sqrt5 \cM_{3, 222; 200}^{(u,u) J=0}(p_2,k_{\rm eq}) 
    +\frac52 \cM_{3, 222; 222}^{(u,u) J=0}(p_2,k_{\rm eq}) 
     \, .
    \end{split}
    \end{align}
    %%%%%
with $c = \cos \vartheta^\star_{k_1}$
and 
$c' = \cos \vartheta^\star_{p_1}$.
Again, the result does not depend on the relative orientations of the reaction planes.

\subsubsection{The isosceles/equilateral triangle configuration for $J^P=1^+$}

We find that the $J^P = 1^+$ amplitude for identical particles vanishes. This is a consequence of choosing the highly symmetrical configuration of momenta in the initial state.
For $2+1$ systems, however, the result is nonzero. For $J^P=1^+$ the contributing waves for $i=1$ are $(\ell,s)=(0,1)$, $(1,0)$, and $(2,1)$, while for $i=2$ they are $(1,0)$ and $(1,2)$. The result is given by,
\begin{align}
     \begin{split}
      &\cM_3^{J = 1}(\{\bm p\}; \{\bm k\})  =
    3 \cos(\alpha/2) \Bigg\{
    2\cos\theta_{22} \mathcal M_{3, 110; 110}^{(u,u)J=1}(p_{\rm is},k_{\rm eq})
    %%%
    + 2 \cos(\theta_{22}\!-\!\vartheta_{k_1}^\star \!+\! \vartheta_{p_1}^\star)
    \mathcal M_{3, 101; 101}^{(u,u)J=1}(p_{\rm is},k_{\rm eq}) \\
    %%%
    &\qquad\qquad +
    2 \cos(\theta_{22}\!-\!\vartheta_{k_1}^\star) \mathcal M_{3, 110; 101}^{(u,u)J=1}(p_{\rm is},k_{\rm eq})
    %%%
    + 2 \cos(\theta_{22}\!+\!\vartheta_{p_1}^\star) \mathcal M_{3, 101; 110}^{(u,u)J=1}(p_{\rm is},k_{\rm eq}) \\
    %%%
    &\qquad\qquad +
    \left[-\sqrt8 \cos\theta_{22}\cos\vartheta_{k_1}^\star+\sqrt2\sin\theta_{22}\sin\vartheta_{k_1}^\star \right] \mathcal M_{3, 110; 121}^{(u,u)J=1}(p_{\rm is},k_{\rm eq})
    \\
    %%%
    &\qquad\qquad +
    \left[-\sqrt8 \cos\theta_{22} \cos\vartheta_{p_1^\star} + \sqrt2 \sin\theta_{22} \sin\vartheta_{p_1}^\star \right] 
    \mathcal M_{3, 121; 110}^{(u,u)J=1}(p_{\rm is},k_{\rm eq})
    \\
    &\qquad\qquad 
    - \sqrt{\frac12}
    \left[
    \cos(\theta_{22} \!+\! \vartheta_{p_1}^\star \!-\! \vartheta_{k_1}^\star)
    +3\cos(\theta_{22} \!+\! \vartheta_{p_1}^\star \!+\! \vartheta_{k_1}^\star)
    \right] 
    \mathcal M_{3, 101; 121}^{(u,u)J=1}(p_{\rm is},k_{\rm eq})
    \\
    &\qquad\qquad 
    - \sqrt{\frac12}
    \left[
    \cos(\theta_{22} \!+\! \vartheta_{p_1}^\star \!-\! \vartheta_{k_1}^\star)
    +3\cos(\theta_{22} \!-\! \vartheta_{p_1}^\star \!-\! \vartheta_{k_1}^\star)
    \right] 
    \mathcal M_{3, 121; 110}^{(u,u)J=1}(p_{\rm is},k_{\rm eq})
    \\
    &\qquad\qquad +
    \left[
    2 \sin\theta_{22} \sin(\vartheta_{p_1}^\star \!-\! \vartheta_{k_1}^\star)
    + 4 \cos\theta_{22} \cos\vartheta_{k_1}^\star \cos\vartheta_{p_1}^\star
    + \cos\theta_{22} \sin\vartheta_{k_1}^\star \sin\vartheta_{p_1}^\star
    \right]
    \mathcal M_{3, 121; 121}^{(u,u)J=1}(p_{\rm is},k_{\rm eq}) \Bigg\}
    \\
     &-\frac{3}{\sqrt2} \cos(\alpha/2) \Bigg\{
     \cos\theta_{22} \left[4 \cM_{3, 110; 210}^{(u,u)J=1}(p_{\rm is},k_{\rm eq})+ \sqrt8 \cM_{3, 110; 212}^{(u,u)J=1}(p_{\rm is},k_{\rm eq})\right]
     \\
     &\qquad\qquad\qquad\qquad + \cos(\theta_{22} \!+\! \vartheta_{p_1}^\star) \left[4 \cM_{3, 101; 210}^{(u,u)J=1}(p_{\rm is},k_{\rm eq})+ \sqrt8 \cM_{3, 101; 212}^{(u,u)J=1}(p_{\rm is},k_{\rm eq})\right]    
     \\
     &\qquad\qquad\qquad\qquad - \left[2\cos\theta_{22}\cos\vartheta_{p_1}^\star + \sin\theta_{22}\sin\vartheta_{p_1}^\star \right]
     \left[\sqrt8 \cM_{3, 121; 210}^{(u,u)J=1}(p_{\rm is},k_{\rm eq})+ 2 \cM_{3, 121; 212}^{(u,u)J=1}(p_{\rm is},k_{\rm eq})\right] \Bigg\}
\\
    &-\frac{3}{\sqrt8} \Bigg\{
     \cos\theta_{22} \left[4 \cM_{3, 210; 110}^{(u,u)J=1}(p_2,k_{\rm eq})+
     \sqrt8 \cM_{3, 212; 110}^{(u,u)J=1}(p_2,k_{\rm eq})\right]
      \\
      & \qquad\qquad +
      \cos(\theta_{22} \!-\! \vartheta_{k_1}^\star) 
      \left[4 \cM_{3, 210; 101}^{(u,u)J=1}(p_2,k_{\rm eq})+ 
      \sqrt8 \cM_{3, 212; 101}^{(u,u)J=1}(p_2,k_{\rm eq})\right]
      \\
      &\qquad\qquad - \left[3\cos\theta_{22}\cos\vartheta_{k_1}^\star + \sin\theta_{22}\sin\vartheta_{k_1}^\star \right]
     \left[\sqrt2 \cM_{3, 210; 121}^{(u,u)J=1}(p_2,k_{\rm eq})+  \cM_{3, 212; 121}^{(u,u)J=1}(p_2,k_{\rm eq})\right] \Bigg\}
    \\
    & + 3 \cos\theta_{22} 
    \left[
    2 \cM_{3, 210; 210}^{(u,u)J=1}(p_2,k_{\rm eq})
    +\sqrt{2} \cM_{3, 210; 212}^{(u,u)J=1}(p_2,k_{\rm eq})
    +\sqrt{2} \cM_{3, 212; 210}^{(u,u)J=1}(p_2,k_{\rm eq})
    + \cM_{3, 212; 212}^{(u,u)J=1}(p_2,k_{\rm eq}) \right]
     \end{split}
\end{align}
Note that this result that depends on $\theta_{22}$, a dependence that is expected in general for nonzero angular momentum.

\subsubsection{The isosceles/equilateral triangle configuration for $J^P=2^-$}

For identical particles we consider the $(\ell,s) = (2,0), (0,2),$ and $(2,2)$ partial waves. We find,
    %%%%%
     \begin{align} 
     \begin{split}
     & \cM_3^{J = 2}(\{\bm p\}; \{\bm k\})  = 
     \frac{150}7 \,M_{3, 22; 22}^{(u,u)J=2}(p_3,k_{\rm eq}) ~
     \\
     &+ \frac{15}{4}\left[ 
     \cM_{3, 20; 20}^{(u,u) J=2}(p_3,k_{\rm eq})
     +  \cM_{3, 20; 02}^{(u,u) J=2}(p_3,k_{\rm eq})
     + \cM_{3, 02; 02}^{(u,u)J=2}(p_3,k_{\rm eq})
     +  \cM_{3, 02; 22}^{(u,u)J=2}(p_3,k_{\rm eq}) \right]
    \\
     & + 15 \sqrt{\frac{5}{14}} \left[
     \cM_{3, 20; 22}^{(u,u)J=2}(p_3,k_{\rm eq})  
    + \cM_{3, 02; 20}^{(u,u) J=2}(p_3,k_{\rm eq})
    + \cM_{3, 22; 20}^{(u,u)J=2}(p_3,k_{\rm eq}) 
    + \cM_{3, 22; 02}^{(u,u)J=2}(p_3,k_{\rm eq}) \right]
    \\
    & + \frac{30}{4}\left[
     \cM_{3, 20; 20}^{(u,u)J=2}(p_{\rm is},k_{\rm eq}) 
      +  \cM_{3, 20; 02}^{(u,u)J=2}(p_{\rm is},k_{\rm eq})
      + \cM_{3, 02; 20}^{(u,u)J=2}(p_{\rm is},k_{\rm eq}) 
      + \cM_{3, 02; 02}^{(u,u)J=2}(p_{\rm is},k_{\rm eq}) \right]
\\
      & + 30 \sqrt{\frac{5}{14}} \left[
     \cM_{3, 20; 22}^{(u,u)J=2}(p_{\rm is},k_{\rm eq}) 
     + \cM_{3, 02; 22}^{(u,u)J=2}(p_{\rm is},k_{\rm eq})  \right]
\\
     & + \frac{15}4  \sqrt{\frac{5}{14}} (1 - 3 \bar{c}) \left[
    \cM_{3, 22; 20}^{(u,u)J=2}(p_{\rm is},k_{\rm eq}) 
    + \cM_{3, 22; 02}^{(u,u)J=2}(p_{\rm is},k_{\rm eq}) \right]
\\
    & + \frac{75}{7} (1 - 3 \bar{c} )
    \cM_{3, 22; 22}^{(u,u)J=2}(p_{\rm is}, k_{\rm eq}) \,,
    \end{split}
    \end{align}
where $\bar{c} = \cos(2 \vartheta_{p_1}^\star)$. Note that it does not depend on the angle $\theta_{22}$.

For non-identical particles, we consider $(\ell, s) = (2,0)$ and $(1,1)$ partial waves for $i = 1$, and $(\ell, s) = (2,0), (0,2),$ and $(2,2)$ partial waves for $i = 2$. The corresponding result for $2+1$ systems is,
\begin{align}
     \begin{split}
      &\cM_3^{J^P = 2^-}(\{\bm p\}; \{\bm k\})  = 
% 11 case, no symm factor 
    \left[5 - \tfrac{15}2 \cos\alpha \cos(2 \theta_{22}) \right]
     \cM_{3, 120; 120}^{(u,u)J=2}(p_{\rm is},k_{\rm eq})
\\
    &+ \left[ 6 \cos \vartheta_{k_1}^\star \cos\vartheta_{p_1}^\star - 9 \cos\alpha \cos(2 \theta_{22} \!-\! \vartheta_{k_1}^\star \!+\! \vartheta_{p_1}^\star)\right]
     \cM_{3, 111; 111}^{(u,u)J=2}(p_{\rm is},k_{\rm eq})
\\
    & - \sqrt{\frac{15}8} \left[
    3 \cos(\alpha \!+\! 2 \theta_{22} \!-\! \vartheta_{k_1}^\star) - 4 \cos\vartheta_{k_1}^\star +  3 \cos(\alpha \!-\! 2 \theta_{22} \!+\! \vartheta_{k_1}^\star)
   \right]
     \cM_{3, 120; 111}^{(u,u)J=2}(p_{\rm is},k_{\rm eq})
\\
    & - \sqrt{\frac{15}8} \left[
    3 \cos(\alpha \!-\! 2 \theta_{22} \!-\! \vartheta_{p_1}^\star) - 4 \cos\vartheta_{p_1}^\star +  3 \cos(\alpha \!+\! 2 \theta_{22} \!+\! \vartheta_{p_1}^\star)
   \right]
     \cM_{3, 111; 120}^{(u,u)J=2}(p_{\rm is},k_{\rm eq})\\
% 12 case, includes symm factor of sqrt2
    &+ \frac5{\sqrt2} \left[1 + 3 \cos\alpha \cos(2\theta_{22})\right]
     \cM_{3, 120; 220}^{(u,u)J=2}(p_{\rm is},k_{\rm eq})
    + \frac5{\sqrt2} \left[1 - 3 \cos\alpha \cos(2\theta_{22})\right]
     \cM_{3, 120; 202}^{(u,u)J=2}(p_{\rm is},k_{\rm eq})
\\
     &+ \sqrt{\frac{500}{7}} 
     \cM_{3, 120; 222}^{(u,u)J=2}(p_{\rm is},k_{\rm eq})
     + 10 \sqrt{\frac{6}{7}} \cos \vartheta_{p_1}^\star
     \cM_{3, 111; 222}^{(u,u)J=2}(p_{\rm is},k_{\rm eq})
     \\
     &+ \frac{\sqrt{15}}2 \left[ 
     3 \cos(\alpha \!-\! 2 \theta_{22} \!-\! \vartheta_{p_1}^\star) + 2 \cos\vartheta_{p_1}^\star + 3 \cos(\alpha \!+\! 2 \theta_{22} + \vartheta_{p_1}^\star) \right]
     \cM_{3, 111; 220}^{(u,u)J=2}(p_{\rm is},k_{\rm eq})
     \\
     & - \frac{\sqrt{15}}2 \left[3 \cos (\alpha \!-\!2 \theta_{22} \!-\!\vartheta_{p_1}^\star ) - 2 \cos \vartheta_{p_1}^\star + 3 \cos (\alpha +2 \theta_{22} \!+\! \vartheta_{p_1}^\star ) \right]
     \cM_{3, 111; 202}^{(u,u)J=2}(p_{\rm is},k_{\rm eq})
% 21 case, includes symm factor of Sqrt2
\\
     &+ \frac5{\sqrt8} \left[2 - 3 \cos(2\theta_{22})\right]
     \cM_{3, 220; 120}^{(u,u)J=2}(p_2,k_{\rm eq})
     + \frac5{\sqrt8} \left[2 + 3 \cos(2\theta_{22})\right]
     \cM_{3, 202; 120}^{(u,u)J=2}(p_2,k_{\rm eq})
\\
     &+ \sqrt{\frac{500}{7}} 
     \cM_{3, 222; 120}^{(u,u)J=2}(p_2,k_{\rm eq})
     + \sqrt{\frac{600}7} \cos\vartheta_{k_1}^\star
     \cM_{3, 222; 111}^{(u,u)J=2}(p_2,k_{\rm eq})
     \\
     &+ \sqrt{\frac{15}4}
     \left[ 2 \cos \vartheta_{k_1}^\star \!-\! 3 \cos(2 \theta_{22} \!- \!\vartheta_{k_1}) \right]
     \cM_{3, 220; 111}^{(u,u)J=2}(p_2,k_{\rm eq})
     \\
     &+ \sqrt{\frac{15}4}
     \left[ 2 \cos \vartheta_{k_1}^\star \!+\! 3 \cos(2 \theta_{22} \!-\! \vartheta_{k_1}^\star) \right]
     \cM_{3, 202; 111}^{(u,u)J=2}(p_2,k_{\rm eq})
\\
% 22 case, includes symm factor of 2
& + \frac52 \left[ 1 + 3 \cos(2 \theta_{22}) \right]
\left[
 \cM_{3, 220; 220}^{(u,u)J=2}(p_2,k_{\rm eq})
+ \cM_{3, 202; 202}^{(u,u)J=2}(p_2,k_{\rm eq}) \right]
+ \frac{100}7 \cM_{3, 222; 222}^{(u,u)J=2}(p_2,k_{\rm eq})
\\
& + \frac52 \left[1 - 3 \cos(2 \theta_{22}) \right]
\left[
\cM_{3, 220; 202}^{(u,u)J=2}(p_2,k_{\rm eq})
+ \cM_{3, 202; 220}^{(u,u)J=2}(p_2,k_{\rm eq})
\right]
\\
& + \sqrt{\frac{500}{14}} \left[ 
\cM_{3, 220; 222}^{(u,u)J=2}(p_2,k_{\rm eq})
+ \cM_{3, 222; 220}^{(u,u)J=2}(p_2,k_{\rm eq})
+ \cM_{3, 202; 222}^{(u,u)J=2}(p_2,k_{\rm eq})
+ \cM_{3, 222; 202}^{(u,u)J=2}(p_2,k_{\rm eq})
\right] 
     \end{split}
\end{align}

\subsection{Numerical solution}
\label{app:numerical_solution}

\begin{figure}
    \centering
    \includegraphics[width=0.9\textwidth]{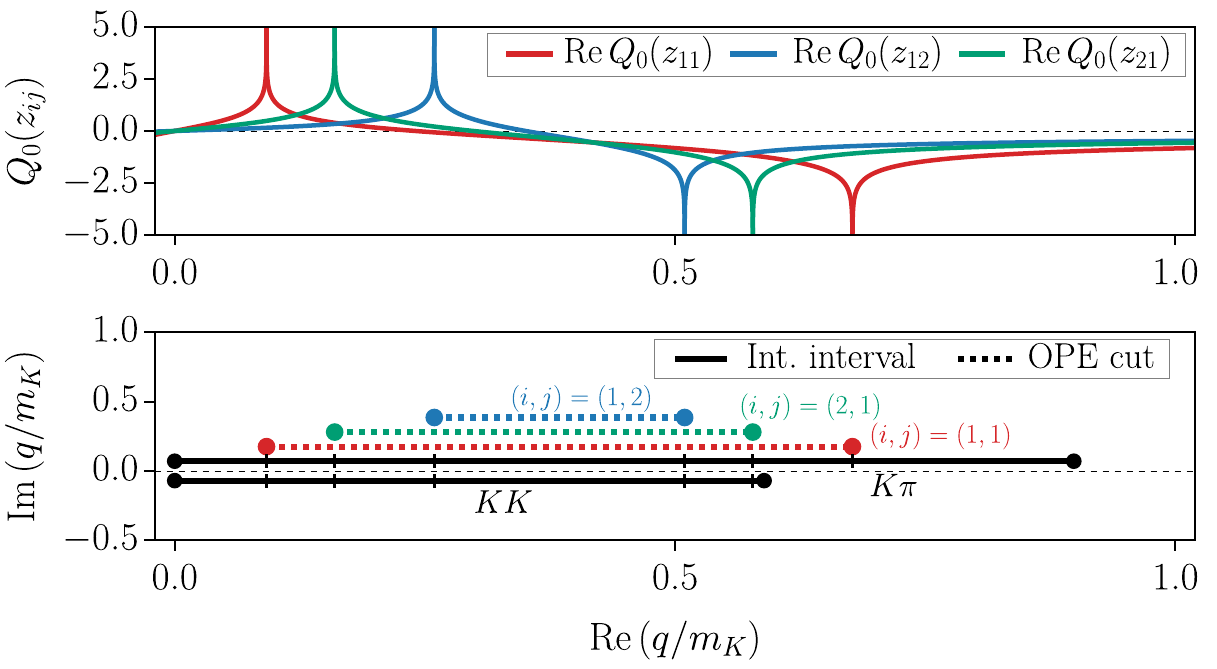}
    \caption{Example of singularities appearing in the integral equation kernel in the $K K \pi$ system in the equilateral configuration for $E = 2 m_K + 2.5m_\pi$ on the E250 ensemble. $q \equiv q_n$ is the magnitude of the integration variable, while the final spectator momentum $p_i = p_{\rm eq}$ for energy $E = 2 m_K + 2.5m_\pi$. The top panel shows real parts of the Legendre function $Q_0$, which leads to the singularities in the OPE kernel, for the three choices of external flavors. (Higher-order Legendre functions have singularities at the same positions as $Q_0$.) The OPE amplitude vanishes for $(i,j) = (2,2)$. 
    Branch-point singularities are placed at points where the function diverges. 
    The bottom panel compares the positions of the corresponding cuts (dashed lines) with the two integration intervals (solid black lines) corresponding to two possible choices of intermediate pairs. All lines are separated vertically to improve clarity. In applications, the OPE cuts are shifted away from the real axis by a distance $i\epsilon$. The integration intervals are divided into smaller parts determined by the positions of the branch points, as highlighted by small vertical ticks.}
    \label{fig:numerical}
\end{figure}

We solve the integral equations numerically, following similar methodology to that of Refs.~\cite{Hansen:2020otl, Jackura:2020bsk, Dawid:2023jrj, Dawid:2024dgy}, employing the standard Nystr\"om method~\cite{10.1007/BF02547521, delves1988computational} with Gauss-Legendre quadratures. 
Specifically, we discretize the momentum variables and reformulate the problem as algebraic equations for the amplitudes $\mathcal D^{(u,u)}$ and $\mathcal M^{(u,u)}_{\rm df, 3}$ rewritten as momentum-space matrices. The momentum-space matrices have an additional block structure with blocks corresponding to pair-spectator reactions of distinct angular momentum and spectator flavor~\cite{Dawid:2024dgy}. Finally, we interpolate the matrix solutions to continuous external momenta by applying the integral equation to the discrete amplitudes, as explained in Refs.~\cite{Jackura:2020bsk, Dawid:2023jrj}. This is done twice: once to interpolate the final momentum to the desired value, and once to interpolate the initial momentum.

The numerical solution for physical kinematics is hindered by the appearance of the one-particle exchange cuts in the kernel of the integral equation~\cite{Rubin:1966zz, Dawid:2023jrj, Sadasivan:2021emk}. These cuts originate from the Legendre functions of the second kind, $Q_n(z_{ij})$, contained in the partial-wave projections of the exchanged particle propagator, and lie in the range $z_{ij} \in [-1,+1]$. Here, the argument $z_{ij}$ is a function of the kinematical variables (external spectator momenta, total three-body energy), and also depends on the external spectator indices, $(i,j)$ (which are suppressed in the case of identical particles.) The explicit form of $z_{ij}$ and the partial-wave projections is discussed in the next subsection. These cuts can lie within the momentum integration region, as illustrated in \Cref{fig:numerical}. 
Explicitly, cuts of the partial-wave projection of $G_{in}(\bm p_i,\bm q_n)$ [appearing in the kernel of ~\Cref{eq:ladder}] in the $q_n = |\bm q_n|$ variable are placed at,
    %%%%%
    \begin{align}
    \label{eq:OPE_branch_point}
    q_n^{\pm}(p_i,E; x) = \frac{1}{2 \beta_x^i} 
    \left( p_i x \, (\beta_1^i - M_n^2 - M_{in}^2) \pm \sqrt{\beta_0^i} \sqrt{ (\beta_1^i + M_n^2 - M_{in}^2)^2 - 4 M_n^2 \beta^i_x  } \right) \, ,
    \end{align}
    %%%%%
where $\beta_{x}^i = (E - \omega_{p_i})^2 - x^2 p_i^2$ , $x \in [-1,1]$, and $M_{in}$ is the mass of the exchanged particle. 

In this work, we do not avoid these singularities via the contour deformation, as suggested in Refs.~\cite{Glockle:1978zz, Sadasivan:2022srs, Dawid:2023jrj}, but rather use the real integration contour while shifting the singularities away from the real integration interval by an imaginary distance $i \epsilon$.  This approach is required in the present formalism because the smooth real cutoff function $H$ does not possess an analytic continuation in the complex momentum plane. Note that for the scattering kinematics described in this paper, cuts with more complicated shapes, such as circular cuts, do not appear in the integration kernel. 

Since Gauss-Legendre quadratures are not suitable for approximating discontinuous functions, to improve {the stability} and the convergence of the solutions as the momentum-matrix sizes are taken to infinity, we divide the integration intervals into smaller pieces determined from positions of the relevant branch points, as highlighted in \Cref{fig:numerical} by vertical ticks on the integration intervals. The Gauss-Legendre points are introduced only in the smaller sub-intervals. {The number of these intervals varies depending on the exact set of the considered kinematical variables.} For instance, in the case of identical particles {in the equilateral configuration,} we use three sub-intervals, since only one OPE cut appears in the integration kernel in all angular momentum channels. In the case of non-identical particles {in the equilateral configuration,} since two integration intervals are used for different choices of the intermediate spectator, and three different cuts may appear on top of these intervals, we divide each interval into six or seven pieces. {The number of required intervals grows in the isosceles configuration case, since more possible external momenta enter \Cref{eq:OPE_branch_point} determining positions of the branch points.} Finally, the solution of the integral equations is obtained by taking the ordered limits $N\to\infty$ followed by $\epsilon \to 0$. In practice, a better than sub-percent level of convergence is achieved already at matrix sizes $N \approx 100$~\cite{Jackura:2020bsk, Dawid:2023jrj} and $\epsilon \approx 10^{-4}$.

The error bands of the integral equation solutions presented in \Cref{sec:amplitudes} are produced using stochastic propagation from the two- and three-body fit parameters given in \Cref{sec:fits}. For this purpose, 1000 random samples of the scattering parameters are generated, incorporating the full covariance matrix to respect the correlations between parameters. The integral equations are then solved for each parameter set in the sample, allowing an estimate of the mean and the 1-sigma (68.27\%) quantiles at each value of the kinematic parameters, which are used to define the error bands.

\subsection{Partial-wave projections of $G$}
\label{app:PWG}

In this section, we collect the results for the partial-wave projected form of the OPE amplitude $G_{ij}(\bm p_i, \bm k_j)$ [appearing in \Cref{eq:ladder}]
that are needed for the integral equations solved in the main text. The calculations needed to obtain these forms are explained in Refs.~\cite{Jackura:2023qtp, Dawid:2024dgy}, and we quote only the results. Some results are available in the literature, and we do not repeat those here but rather point to the appropriate references.

For $2+1$ systems, the results take the form,
    %%%%%
    \begin{align}
    G^{(ij)J}_{\ell's';\ell s}(p_i,k_j;E) = \frac{H_i(p_i) \, H_j(k_j)}{(q_{p_i}^\star)^{s'} (q_{k_j}^\star)^s} \sum_n a_n \, Q_n(z_{ij}) \, , 
    \end{align}
    %%%%%
where $Q_i(z)$ are Legendre functions of the second kind, and 
\begin{equation}
z_{ij} = \frac{1}{2 p_i k_j} \left[ (E-\omega_{p_i} - \omega_{k_j})^2 - p_i^2 - k_j^2 - M_{ij}^2 \right]\,,
\end{equation}
with $M_{ij}$ is the mass of the exchanged particle (i.e. the particle that is neither the initial nor the final spectator).
We are using an abbreviated notation in which the dependence on the masses of the spectator particles is implicit. For example, for the $\pi\pi K$ system, if the final state flavor is $i=1$, corresponding to the pion, then $\omega_p=\sqrt{p^2+M_\pi^2}$.
$H_i(p)$ and $H_j(k)$ are smooth cutoff functions, whose explicit form is given in \cite{Dawid:2024dgy}.
The coefficients $a_n$ depend on $i,j,J,\ell',s',\ell,s$, but we keep this dependence implicit. The sum over $n$ involves only a finite number of nonzero terms, starting at $n=0$.

The form for identical particles is essentially the same, except that the indices $i,j$ are dropped.
Our implicit notation for kinematic dependence means that that the coefficients $a_n$ have the same form for identical-particle and $2+1$ systems.

To simplify the results, we use the following abbreviations
    %%%%%%
    \begin{align}
    \label{eq:relativistic-factors}
     \beta_p = \frac{p}{E - \omega_p} \, , 
     %\quad \tilde{\beta}_p = \frac{p}{\omega_p} \,,
    %%%
    \quad \gamma_p = \frac{1}{\sqrt{1-\beta_p^2}} \, , \quad
    g_{pk} = \frac{\beta_p \gamma_p \omega_k}{k} \, . 
    \end{align}
    %%%%%%
Analogous formulas define $\beta_k$,
$\gamma_k$, and $g_{kp}$. We describe the projections using standard spectroscopic notation: $(J,\ell,s) \equiv {}^{2s+1} \ell_J$. Projections for $J^P = 1^+$ can be found in Appendix~B of Ref.~\cite{Dawid:2024dgy}, and are not repeated here.

\subsubsection{$J^P=0^-$}

The $^{1}S_0 \to {}^{1}S_0$, $^{1}S_0 \leftrightarrow {}^{3}P_0$, and $^{3}P_0 \to {}^{3}P_0$ results are given in Appendix~B of Ref.~\cite{Dawid:2024dgy}.
The remaining results that we need are as follows.
\begin{itemize}
%    \item The $^{1}S_0 \to {}^{1}S_0$ amplitude:
    %%%%%
%    \begin{align}
%    a_0 & = \frac{1}{2 p k} \, ,
%    \end{align}
    %%%%%
    \item 
    The ${}^{1}S_0 \to {}^{5}D_0$ amplitude $G^{J=0}_{22;00}$:
    %%%%%
    \begin{align}
    a_0 &=
    \frac{\sqrt{5} k}{6 p} \, 
    \left[ (\gamma_p^2 - 1) + 3 g_{pk}^2 \right] \, ,  \quad
    %%%
    a_1 =
    \frac{\sqrt{5} \, k }{p} \, 
    g_{pk} \gamma_p \, , \quad
    %%%
    a_2 =
    \frac{\sqrt{5} \, k}{6 p} \, (2\gamma_p^2+1) \, .
    \end{align}
    %%%%%
    The ${}^{5}D_0 \to {}^{1}S_0 $ amplitude $G^{J=0}_{00;22}$ is obtained by interchanging $p$ and $k$. 
    \item 
    The ${}^{5}D_0 \to {}^{5}D_0$ amplitude $G^{J=0}_{22;22}$:
    %%%%%
    \begin{align} 
    a_0 & = \frac{pk}{6} 
    \Big[ 20 g_{kp} g_{pk} \gamma_k \gamma_p 
     + 5 g_{pk}^2 ( \gamma_k^2 - 1 ) 
    + 5 g_{kp}^2 ( \gamma _p^2 - 1 )  
     + 15 g_{kp}^2 g_{pk}^2 
    -\gamma_k^2 - \gamma_p^2 + 3 \gamma_k^2 \gamma_p^2 + 2 \Big] \, ,
     \\ 
    a_1 &= pk 
    \Big[ 3 \gamma_k \gamma_p (g_{kp} \gamma_p + g_{pk} \gamma_k) 
     + 5 g_{kp} g_{pk} ( g_{pk} \gamma _k + g_{kp} \gamma_p) 
     - (g_{pk} \gamma _p + g_{kp} \gamma_k) \Big] \, , 
     \\ 
     a_2 &= \frac{5pk}{6}  \Big[ 8 g_{kp} g_{pk} \gamma_k \gamma_p 
    g_{pk}^2 \gamma_k^2 + 2 g_{kp}^2 \gamma_p^2 + g_{kp}^2 + g_{pk}^2
     - \tfrac{1}{7} \left( \gamma _k^2 - 12 \gamma_k^2 \gamma_p^2 +  \gamma _p^2 + 4 \right) \Big] \, , 
    \\
    a_3 &= pk \Big[ 
    g_{kp} \gamma_k (2 \gamma_p^2 + 1)
    + g_{pk} \gamma_p (2 \gamma_k^2 + 1) 
    \Big] \, , \\
    %%%
    a_4 &= \frac{pk}{7} (2 \gamma _k^2+1) (2 \gamma _p^2+1) \, ,
    \end{align}
    %%%%%
    \item
The ${}^{3}P_0 \to {}^{5}D_0$ amplitude $G^{J=0}_{22;11}$:
   \begin{align}
    a_0 &= \sqrt{\frac53}
    \frac{k}{2}
    \Big[ g_{kp} (\gamma_p^2 - 1) + 3 g_{kp} g_{pk}^2 +2 g_{pk}\gamma_p\gamma_k
    \Big] 
    \\
     a_1 &= \sqrt{\frac35}
    \frac{k}{2} \Big[ 10 g_{kp} g_{pk}\gamma_p + \gamma_k(3 \gamma_p^2-1) + 5 g_{pk}^2 \gamma_k
    \Big]  
    \\
    a_2 &= \sqrt{\frac53}
    \frac{k}{2} \Big[ g_{kp}(2\gamma_p^2+1) + 4 g_{pk} \gamma_p\gamma_k 
    \Big] 
    \\
    a_3 &= \sqrt{\frac35}
    \frac{k}{2} (2\gamma_p^2+1) \gamma_k \, .
    \end{align}
    %%%%%
    The ${}^{5}D_0 \to {}^{3}P_0 $ amplitude $G^{J=0}_{11;22}$ is obtained by interchanging $p$ and $k$. 
    \end{itemize}
    
\subsubsection{$J^P=2^-$}

Here we display only the results involving even values of $s$, since these are the only cases considered above. We find the nonzero coefficients to be as follows:
\begin{itemize}
    \item 
    The ${}^{1}D_2 \to {}^{1}D_2$ amplitude $G^{J=2}_{20;20}$,
    \begin{align}
    a_2 &= \frac1{2kp}
    \end{align}
    \item 
    The ${}^3P_2 \to {}^3 P_2$ amplitude $G_{11;11}^{J=2}$:
    \begin{align}
    a_0 &= \frac{1}{50} \left(2 \gamma _k+3\right) \left(2 \gamma _p+3\right) \, , \quad 
    a_1 = \frac{3}{25} \left(g_{pk} \left(2 \gamma _k+3\right)+g_{kp} \left(2 \gamma _p+3\right)\right) \, , \\
    %%%
    a_2 & = \frac{1}{70} \left( 42 g_{kp} g_{pk}+6 \gamma _k+22 \gamma _k \gamma _p+6 \gamma _p-27 \right)\, , \quad 
    a_3 = \frac{9}{25} \left(g_{pk} \left(\gamma _k-1\right)+g_{kp} \left(\gamma _p-1\right)\right) \, , \\
    %%%
    a_4 &= \frac{36}{175} \left(\gamma _k-1\right) \left(\gamma _p-1\right) \, .
    \end{align}
    \item The ${}^{3}P_2 \to {}^{1}D_2$ amplitude $G_{20;11}^{J=2}$:
    \begin{align}
    a_0 & = 0 \, , \quad 
    a_1 = - \sqrt{\frac{3}{10}} \frac{ \left(2 \gamma _k+3 \right)}{5 k} \, , \quad
    a_2 = - \sqrt{\frac{3}{10}} \frac{ g_{kp}}{k} \, , \quad
    a_3 = - \sqrt{\frac{3}{10}} \frac{3 \left(\gamma _k-1\right)}{5 k} \, .
    \end{align}
    The ${}^{1}D_2 \to {}^{3}P_2$ amplitude $G^{J=2}_{11;20}$ is obtained by interchanging $p$ and $k$.
    \item The ${}^{3}P_2 \to {}^{5}S_2$ amplitude $G^{J=2}_{02;11}$:
    \begin{align}
    a_0 & = -\frac{k}{5 \sqrt{30}} \bigg(g_{pk} \left(2 \gamma _k+3\right) \left(2 \gamma _p+3\right)+g_{kp} \left(2 \gamma _p \left(\gamma _p+3\right)+7\right)\bigg) \, , \\
    %%%
    a_1 &= -\frac{k}{35} \sqrt{\frac{3}{10}} \bigg( 7 g_{pk}^2 \left(2 \gamma _k+3\right)+14 g_{kp} g_{pk} \left(2 \gamma _p+3\right)+3 \left(4 \gamma _k+3\right) \gamma _p^2+3 \left(6 \gamma _k+1\right) \gamma _p +5 \gamma _k -12 \bigg) \, , \\
    %%%
    a_2 &= -\frac{k}{7 \sqrt{30}} \bigg( g_{pk} \big(\gamma _k \left(22 \gamma _p+6\right)+6 \gamma _p-27\big)+g_{kp} \big(21 g_{pk}^2+\left(\gamma _p-1\right) \left(11 \gamma _p+17 \right) \big) \bigg) \, , \\
    a_3 &= -\frac{k}{5 \sqrt{30}}  \bigg(9 g_{pk}^2 \left(\gamma _k-1\right)+18 g_{kp} g_{pk} \left(\gamma _p-1\right)+\left(\gamma _p-1\right) \left(5 \gamma _k+\left(7 \gamma _k-1\right) \gamma _p-8\right)\bigg)\, , \\
    a_4 &= -\frac{6 k }{35} \sqrt{\frac{6}{5}} \left(\gamma _p-1\right) \big(2 g_{pk} \left(\gamma _k-1\right)+g_{kp} \left(\gamma _p-1\right)\big) \, , \\
    %%%
    a_5 &= -\frac{2k}{7} \sqrt{\frac{2}{15}} \left(\gamma _k-1\right) \left(\gamma _p-1\right)^2 \, ,
    \end{align}
    The ${}^{5}S_2 \to {}^{3}P_2$ amplitude $G^{J=2}_{11;02}$ is obtained by interchanging $p$ and $k$.
    \item The ${}^{3}P_2 \to {}^{5}D_2$ amplitude $G^{J=2}_{22;11}$:
    \begin{align}
    a_0 & = \frac{k}{10 \sqrt{21}}  \bigg(g_{pk} \left(2 \gamma _k+3\right) \left(4 \gamma _p+3\right)+2 g_{kp} \left(\gamma _p-1\right) \left(2 \gamma _p+5\right)\bigg)\, , \\
    %%%
    a_1 & = \frac{k}{70} \sqrt{\frac{3}{7}} \bigg(14 g_{pk}^2 \left(2 \gamma _k+3\right)+14 g_{kp} g_{pk} \left(4 \gamma _p+3\right)-14 \gamma _k+6 \left(4 \gamma _k+3\right) \gamma _p^2+3 \left(6 \gamma _k+1\right) \gamma _p\bigg) \, , \\
    %%%
    a_2 & = \frac{k}{14 \sqrt{21}} \bigg(g_{pk} \left(\gamma _k \left(44 \gamma _p+6\right)+3 \left(4 \gamma _p-9\right)\right)+2 g_{kp} \left(21 g_{pk}^2+\gamma _p \left(11 \gamma _p+3\right)+7\right)\bigg)\, , \\
    %%%
    a_3 & = \frac{k}{10 \sqrt{21}} \bigg(18 g_{pk}^2 \left(\gamma _k-1\right)+18 g_{kp} g_{pk} \left(2 \gamma _p-1\right)+2 \left(7 \gamma _k-1\right) \gamma _p^2-\left(2 \gamma _k+7\right) \gamma _p +6 \gamma _k \bigg) \, , \\
    %%%
    a_4 & = \frac{12 k}{35} \sqrt{\frac{3}{7}} \left(g_{pk} \left(\gamma _k-1\right) \left(2 \gamma _p-1\right)+g_{kp} \left(\gamma _p-1\right) \gamma _p\right) \, , \\
    %%%
    a_5 & = \frac{4 k}{7 \sqrt{21}} \left(\gamma _k-1\right) \left(\gamma _p-1\right) \gamma _p \, .
    \end{align}
    The ${}^{5}D_2 \to {}^{3}P_2$ amplitude $G^{J=2}_{11;22}$ is obtained by interchanging $p$ and $k$.
    \item 
    The ${}^{5}S_2 \to {}^{5}S_2$ amplitude $G^{J=2}_{02;02}$:
    \begin{align}
    a_0 &= \frac{kp}{210} \Big\{ -16 + 6 \gamma_k \gamma_p + 12 \gamma_k^2\gamma_p^2
   + ( - 24 \gamma_k + 5 \gamma_k^2 + 18 \gamma_k^2 \gamma_p + k\leftrightarrow p)
    \\
    &\quad\quad\quad\quad + 126 g_{kp} g_{pk} + 56 g_{kp} g_{pk} \gamma_k\gamma_p +
    7 \left[ g_{kp}^2 (7 + 6 \gamma_p + 2 \gamma_p^2) +12 g_{kp} g_{pk}\gamma_k + k\leftrightarrow p \right] \Big \}
    \\
    a_1 &= \frac{kp}{35} \Big\{ g_{kp} \left[
    -12 + 5 \gamma_k + 3 \gamma_p + 18 \gamma_k \gamma_p + 9 \gamma_p^2 + 12 \gamma_k \gamma_p^2 + 7g_{kp}g_{pk} (3+ 2\gamma_p)\right]
   +  k\leftrightarrow p
    \Big\} \, ,
    \\
    \begin{split}
    a_2 &= \frac{kp}{42} \Big[8  - 6 \gamma_k \gamma_p + 
    9 \gamma_k^2 \gamma_p^2 + \left( 6 \gamma_k^2 \gamma_p - \gamma_k^2 + k\leftrightarrow p \right)
    \\
    &\qquad\qquad -54 g_{kp} g_{pk} + 44 \gamma_k \gamma_p g_{kp} g_{pk} + 21 g_{kp}^2 g_{pk}^2
     + \left(  6\gamma_p g_{kp}^2 - 17 g_{kp}^2 + 11 \gamma_p^2 g_{kp}^2 + 12 \gamma_k g_{kp} g_{pk} + k\leftrightarrow p
    \right)
    \Big]
    \end{split}
    \\
    \begin{split}
    a_3 &= \frac{k p}{15} \Big[
    (\gamma_k-1) (-8 + 5 \gamma_p - \gamma_k  + 7 \gamma_p\gamma_k) g_{pk} + 
   9 (\gamma_p-1) g_{kp}^2 g_{pk} 
   \\
   & \qquad\qquad + 9 (\gamma_k-1) g_{kp} g_{pk}^2 +
   g_{kp} (8 - 7 \gamma_p - \gamma_p^2 -5 \gamma_k  - 2 \gamma_p \gamma_k + 7 \gamma_p^2\gamma_k) \Big] 
   \end{split}
    \\
    \begin{split}
    a_4 &= \frac{2kp}{385} \Big[33 (\gamma_p -1 )^2 g_{kp}^2 + 33 (\gamma_k-1)^2 g_{pk}^2 \\ 
   & \qquad \qquad + 132 (\gamma_k-1) (\gamma_p-1) g_{kp} g_{pk} +
   2 (\gamma_k-1) (\gamma_p-1) (-16 + 5 \gamma_p + 5 \gamma_k + 17 \gamma_p\gamma_k)) 
    \Big] \,,
   \end{split}
    \\
    a_5 &= \frac{4 kp}{21} (\gamma_k-1) (\gamma_p-1) \left[(\gamma_p-1) g_{kp} + (\gamma_k-1) g_{pk} \right]\,,
    \\
    a_6 &= \frac{4kp}{77} (\gamma_p-1)^2(\gamma_k-1)^2\,,
    \end{align}
    \item 
    The ${}^{5}D_2 \to {}^{5}D_2$ amplitude $G^{J=2}_{22;22}$, 
    \begin{align}
        \begin{split}
        a_0 &= \frac{kp}{294} \Big\{ 112 + 3 \gamma_k\gamma_p + 24 \gamma_k^2\gamma_p^2 + \left(18 \gamma_k^2\gamma_p - 14 \gamma_k^2 + k\leftrightarrow p \right)
        \\
        &\quad\quad\quad\quad + 63 g_{kp} g_{pk} + 112 g_{kp} g_{pk} \gamma_k \gamma_p + 7\left[ g_{kp}^2 ( -10 + 6 \gamma_p + 4 \gamma_p^2) + 12 g_{kp}g_{pk}\gamma_k + k\leftrightarrow p \right]
        \Big\}\,,
        \end{split}
        \\
        a_1 &= \frac{kp}{14} \Big[ g_{kp} \left( 6 g_{kp} g_{pk} +8 g_{kp} g_{pk} \gamma_k - 4 \gamma_k - \gamma_p + 8 \gamma_k \gamma_p + 4 \gamma_p^2 + 4 \gamma_k \gamma_p^2 \right) + k\leftrightarrow p 
        \Big] \, ,
        \\
        \begin{split}
        a_2 &= \frac{5kp}{294} \Big\{ -32 -3 \gamma_k\gamma_p + 18 \gamma_k^2\gamma_p^2 
        + \left[\gamma_k^2(6\gamma_p-2) + k\leftrightarrow p\right]
        \\
        &\qquad\qquad+  \left[ g_{kp}^2 (14 + 6 \gamma_p + 22 \gamma_p^2 + 12 g_{kp}g_{pk} \gamma_p + k \leftrightarrow p \right] 
        + g_{kp}g_{pk} (-27 + 88 \gamma_k \gamma_p + 42 g_{kp}g_{pk})
        \Big\}\,,
        \end{split}
        \\
        a_3 &= \frac{kp}{42} \Big[
        g_{kp}(12 \gamma_k - 7 \gamma_p - 4 \gamma_k \gamma_p - 2 \gamma_p^2 + 28 \gamma_k\gamma_p^2 - 18 g_{kp}g_{pk} + 36 g_{kp}g_{pk}\gamma_p) + k\leftrightarrow p
        \Big]\,,
        \\
        \begin{split}
        a_4 &= \frac{2kp}{539} \Big\{
        44 - 9 \gamma_k \gamma_p + 68 \gamma_k^2 \gamma_p^2
        + (22 \gamma_k^2 - 24 \gamma_k^2\gamma_p + k\leftrightarrow p)
        \\
        &\qquad\qquad + 66 g_{kp} g_{pk} + 264 \gamma_k\gamma_p g_{kp}g_{pk}
        + \left[ 66 g_{kp}^2 \gamma_p (\gamma_p-1) - 132 \gamma_k g_{kp} g_{pk} 
        + k\leftrightarrow p \right]
         \Big\}\,,
        \end{split}
        \\
        a_5 &= \frac{20 kp }{147} \Big[
        g_{kp} \gamma_p (\gamma_p-1)(2\gamma_k-1) + k\leftrightarrow p 
        \Big]\,,
        \\
        a_6 &= \frac{40 kp}{539} \gamma_k (\gamma_k-1) \gamma_p (\gamma_p-1)\,.
        \end{align}
    \item 
    The ${}^{5}S_2 \to {}^{1}D_2$ amplitude $G^{J=2}_{20;02}$,
    \begin{align}
    a_0 &= \frac{p}{30 k} \left[7 + 2 \gamma_k (3 + \gamma_k) \right]\,,
    \qquad
    a_1 = \frac{p}{5 k} (3 + 2 \gamma_k) g_{kp}\,,
    \qquad
    a_2 = \frac{p}{42 k} \left[-17 + \gamma_k (6 + 11 \gamma_k) + 21 g_{kp}^2 \right]\,,
    \\
    & \qquad\qquad\qquad\qquad 
    a_3 = \frac{p}{5 k} 3 (-1 + \gamma_k) g_{kp}\,,
    \qquad
    a_4 = \frac{6 p}{35 k} (-1 + \gamma_k)^2 \,.
    \end{align}
    The ${}^{1}D_2 \to {}^{5}S_2$ amplitude $G^{J=2}_{02;20}$ is obtained by interchanging $p$ and $k$.
    \item 
    The ${}^{5}S_2 \to {}^{5}D_2$ amplitude $G^{J=2}_{22;02}$, 
    \begin{align}
    \begin{split}
        a_0 &= - \frac{k p}{21\sqrt{70}} \Big\{-28 + 3 \gamma_k \gamma_p + 12 \gamma_k^2\gamma_p^2 - 12 \gamma_p - 7 \gamma_k^2 + 5 \gamma_p^2 + 9 \gamma_k^2\gamma_p + 18 \gamma_k \gamma_p^2
        \\
        &\qquad\qquad\qquad 
        + 7 \left[ g_{kp}^2 ( -5 + 3 \gamma_p + 2 \gamma_p^2)
        +g_{kp} g_{pk} (9 + 6 \gamma_k + 12 \gamma_p + 8 \gamma_k \gamma_p)
        + g_{pk}^2 (7 + 6 \gamma_k + 2 \gamma_k^2)
        \right]
        \Big\}
    \end{split}
    \\
    \begin{split}
    a_1 &= - \frac{k p}{7\sqrt{70}} \Big\{
    g_{kp}( -14 \gamma_k + 3 \gamma_p + 18 \gamma_k \gamma_p + 18 \gamma_p^2 + 24 \gamma_k \gamma_p^2)
    + g_{pk} (-12 + 3 \gamma_k + 9 \gamma_k^2 + 10 \gamma_p + 36 \gamma_k \gamma_p + 24 \gamma_k^2 \gamma_p)
    \\
    &\qquad\qquad\qquad +7 g_{kp}^2 g_{pk} (3+4\gamma_p) + 14 g_{kp} g_{pk}^2(3 + 2 \gamma_k)
    \Big\}\,,
    \end{split}
    \\
    \begin{split}
    a_2 &= -  \frac{k p}{21}\sqrt{\frac5{14}}\Big\{
    8 - \gamma_k^2 - 3 \gamma_k \gamma_p + 3 \gamma_k^2 \gamma_p - \gamma_p^2 + 6 \gamma_k \gamma_p^2 +  9 \gamma_k^2 \gamma_p^2    
    + g_{kp}^2 (7 + 3\gamma_p + 11 \gamma_p^2)
    \\
    &\qquad\qquad\qquad
    + g_{kp}g_{pk} (-27 + 6\gamma_k + 12 \gamma_p + 44 \gamma_k \gamma_p)
    + g_{pk}^2 (-17 + 6 \gamma_k + 11 \gamma_k^2)
    + 21 g_{kp}^2 g_{pk}^2
    \Big\}\,,
    \end{split}
    \\
    \begin{split}
    a_3 &= \frac{k p}{3\sqrt{70}} \Big\{
    g_{kp} \left[-6 \gamma_k + 7 \gamma_p + 2  \gamma_k \gamma_p + 2 \gamma_p^2 - 14 \gamma_k\gamma_p^2 + 9 g_{kp} g_{pk} (1-2\gamma_p) \right]
    \\
    &\qquad\qquad\quad
    + g_{pk}\left[-8 + 7\gamma_k + \gamma_k^2 + 10 \gamma_p + 4 \gamma_k\gamma_p - 14 \gamma_k^2\gamma_p + 18 g_{kp}g_{pk}(1-\gamma_k) \right]
    \Big\}\,,
    \end{split}
    \\
    \begin{split}
    a_4 &= -\frac{2 k p}{77} \sqrt{\frac2{35}} \Big\{
    -11 + 11 \gamma_k^2 + 21 \gamma_p - 9 \gamma_k \gamma_p - 12 \gamma_k^2 \gamma_p - 
     10 \gamma_p^2 - 24 \gamma_k \gamma_p^2 + 34 \gamma_k^2 \gamma_p^2
     \\
    &\qquad\qquad\qquad\quad
    + 33 g_{kp}^2  \gamma_p (\gamma_p - 1) + 66 g_{kp} g_{pk} ( 1 - \gamma_k)(1 - 2 \gamma_p)
    + 33 g_{pk}^2(1 -   \gamma_k)^2
    \Big\}\,,
    \end{split}
    \\
    a_5 &= - \frac{2 k p}{21} \sqrt{\frac{10}7} 
    \Big\{ 2 g_{kp} (\gamma_k-1)(\gamma_p-1)\gamma_p + g_{pk} (\gamma_k-1)^2(2\gamma_p-1)
    \Big\}\,,
    \\
    a_6 &=  - \frac{4 k p}{77} \sqrt{\frac{10}7} (\gamma_k-1)^2 \gamma_p (\gamma_p-1)
    \,.
    \end{align}
    The ${}^{5}D_2 \to {}^{5}S_2$ amplitude $G^{J=2}_{02;22}$ is obtained by interchanging $p$ and $k$.
    \item 
    The ${}^{1}D_2 \to {}^{5}D_2$ amplitude $G^{J=2}_{22;20}$, 
    \begin{align}
        & \qquad\qquad\qquad
        a_0 = - \frac{k}{3\sqrt{70} p} (\gamma_p-1) (2\gamma_p + 5)\,,
        \qquad
        a_1 = - \frac{k}{\sqrt{70} p} (4 \gamma_p + 3) g_{pk}\,,
        \\
        a_2 &= - \frac{k}{21 p} \sqrt{\frac5{14}} 
        \left[ 7 + 3 \gamma_p + 11 \gamma_p^2 + 21 g_{pk}^2 \right]\,,
        \qquad
        a_3 = \frac{3 k}{\sqrt{70}p} (1 - 2 \gamma_p) g_{pk}\,,
        \qquad
        a_4 = - \frac{6 k}{7 p} \sqrt{\frac2{35}} \gamma_p (\gamma_p-1)\,.
    \end{align}
    The ${}^{5}D_2 \to {}^{1}D_2$ amplitude $G^{J=2}_{20;22}$ is obtained by interchanging $p$ and $k$.

\end{itemize}

\subsection{Partial-wave projections of $\Kdf$}
\label{app:PWK3}

In this section, we collect the results needed for the partial-wave-projected form of $\Kdf$.
These are obtained using \Cref{eq:masterK3,eq:LS-basis} applied to the forms for $\Kdf$ given in \Cref{eq:Kdfidentical,eq:Kdfidenticala} for identical particles and
\Cref{eq:KdfExpansion,eq:KdfExpansiona} for $2+1$ systems.
Further details on the methodology can be found in Appendix A of Ref.~\cite{Dawid:2024dgy}.

We use the kinematic notation of \Cref{app:PWG} and \Cref{eq:Kdfidenticala,eq:Kdfidenticalb}, augmented by the definition,
    %%%%%
    \begin{equation}
    \epsilon_p = \sigma_p + 3p^2 + 2 M^2= E^2 - 2 E \omega_p + 3 \omega_p^2\,,
    \end{equation}
    %%%%%
and similarly for $\epsilon_k$,
where $M=M_\pi$ ($M_K$) for the $3\pi$ ($3K$) system.
We also introduce the abbreviation
    %%%%%
    \begin{equation}
    \cK_3^{\rm iso} \equiv \cK_0 + \cK_1 \, \Delta  + \cK_2\,\Delta^2\,.
    \end{equation}
    %%%%%

 As discussed in Ref.~\cite{Dawid:2024dgy}, we simplify the solution of the integral equation for $\mathcal{M}_{\rm df, 3}^{(u,u)}$ by rewriting the partial-wave-projected $\Kdf$ in separable form. For identical particles this is,
    %%%%%
    \begin{align}
    M^2 \cK_{\rm df,3; \ell' s';\ell s}^{J}(p,k) &= \sum_a
    \cK_{L; \ell' s'}^{(a)}(p) \cK_{R; \ell s}^{(a)}(k)\,.
    \label{eq:separable_id}
    \end{align}
    %%%%%
The number of terms in the sum depends on $J$, as seen explicitly below. The corresponding form for $2+1$ systems is discussed below.

%%%%%%%%%%%%%%%%%%%%%%%%%%%%%%%%%%%
\subsubsection{Identical particles: $J^P=0^-$}
%%%%%%%%%%%%%%%%%%%%%%%%%%%%%%%%%%%

    \begin{itemize}
    \item The $^{1}S_0 \to {}^{1}S_0$ amplitude is,
    %%%%%
    \begin{align}
    \begin{split}
    \cK^{J=0}_{3; 00;00} &=  \cK_3^{\text{iso}} 
    \\
    %%% KA terms
    & + \cK_A \Bigg\{ - \Delta^2 + 
    \frac{16 q_{p}^{\star 4}}{E_{\rm th}^4}
    + \frac{16 q_{k}^{\star 4}}{E_{\rm th}^4} 
    %%%
    + \frac{2 (E \omega_p -3 E_{\rm th}^2)^2}{E_{\rm th}^4} 
    + \frac{2 (E \omega_k -3 E_{\rm th}^2)^2}{E_{\rm th}^4}
    %%%
    + \frac{8}{3 } \frac{q_{p}^{\star 2}}{E_{\rm th}^4} \frac{E^2 p^2}{ \sigma_p}
    + \frac{8 }{3} \frac{q_{k}^{\star 2}}{E_{\rm th}^4 } \frac{E^2 k^2}{\sigma_k} \Bigg\}
    \\
    %%% KB terms
    & + \cK_B \Bigg\{ - \Delta^2 -\frac{4}{9} (2\Delta + 1) + \frac{1}{E_{\rm th}^4} \left[ \epsilon_p \epsilon_k + 3 k^2 p^2 \right] 
    %%%
    + \frac{4}{3} \frac{q_p^{\star 2}}{E_{\rm th}^4} [\gamma_p^2 \beta_p^2 \epsilon_k + (2 + \gamma_p^2) k^2] 
    \\ & \qquad\qquad
    + \frac{4}{3} \frac{q_k^{\star 2}}{E_{\rm th}^4} [\gamma_k^2 \beta_k^2 \epsilon_p + (2 + \gamma_k^2) p^2] 
    %%%
    + \frac{q_p^{\star 2} q_k^{\star 2}}{E_{\rm th}^4} \frac{16}{27} \left[ (2+\gamma_k^2)(2+\gamma_p^2) + 3 (\gamma_p \gamma_k \beta_p \beta_k)^2 \right]
    \Bigg\} \, .
    \end{split}
    \end{align}
    %%%%%
    %%%%%
    \item The $^{1}S_0 \to {}^{5}D_0$ amplitude is,
    %%%%%
    \begin{align} 
    \begin{split}
    \cK^{J=0}_{3; 22;00} &= \cK_A \,  \frac{16}{3 \sqrt{5}}  \, \frac{q^{\star 2}_p}{E_{\rm th}^4} \,  \frac{E^2 p^2}{\sigma_p}  \\
    %%%
    & + \cK_B 
    \frac{8}{3\sqrt{5}} \, \frac{q^{\star 2}_p}{E_{\rm th}^4} \Bigg\{  \left[\gamma_p^2 \beta_p^2 \epsilon_k + (\gamma_p^2 - 1) k^2 \right]
    +
    \frac{4}{9} \, (q^\star_k)^2 \left[(2+\gamma_k^2)(\gamma_p^2-1) + 3 (\gamma_p \gamma_k \beta_p \beta_k)^2 \right] \Bigg\}
    \, ,
    \end{split}
    \end{align}
    with the $^{5}D_0 \to {}^{1}S_0$ amplitude $\cK_{3; 00;22}^{J=0}$ obtained by the interchange $k\leftrightarrow p$.
    %%%%%
    \item The $^{5}D_0 \to {}^{5}D_0$ amplitude is,
    %%%%%
    \begin{align} \nonumber
    \cK^{J=0}_{3; 22;22} = & \, \cK_B \,  \frac{64}{135}  \, \frac{q^{\star 2}_p q^{\star 2}_k} {E_{\rm th}^4}  \left[(1-\gamma_k^2)(1-\gamma_p^2) + 3 (\gamma_p \gamma_k \beta_p \beta_k)^2 \right] \, .
    \end{align}
    %%%%%
    \end{itemize}

The above results can be written in the separable form \Cref{eq:separable_id} where the sum over $a$ contains seven terms. For $\ell=s=0$ these are,
    \begin{align}
    \cK^{(1)}_{L; 00} &= \cK^{(1)}_{R; 00} = \Big( \cK_3^{\text{iso}} 
    -\cK_A \Delta^2
    - \cK_B \big[ \tfrac{4}{9} (2\Delta + 1) + \Delta^2 \big] \Big)^{1/2} \, ,  
    \\
    \cK^{(2)}_{L, 00} &=  \cK_A  \,, 
    \quad
    %%%
    \cK^{(2)}_{R; 00}(k) =  \frac{8}{3} \, \frac{q^{\star 2}_k}{E_{\rm th}^4} \,  \frac{E^2 k^2}{\sigma_k} \,,
    \\
    %%%
    \cK^{(3)}_{L; 00}(p) &= \frac{8}{3} \, \frac{q^{\star 2}_p}{E_{\rm th}^4} \,  \frac{E^2 p^2}{\sigma_p}  \, , 
    \quad\cK^{(3)}_{R; 00}   =\cK_A\,,
    \\
    %%%
    \cK^{(4)}_{L; 00}(p) &=
    \frac{16 q_{p}^{\star 4} + 2 (E \omega_p -3 E_{\rm th}^2)^2}{E_{\rm th}^4}  \, , 
    \quad 
    \cK^{(4)}_{R;00}  =\cK_A\,, 
    \\
    %%%
    \cK^{(5)}_{L;00} &=\cK_A\,, \quad
    \cK^{(5)}_{R; 00}(k) = \frac{16 q_{k}^{\star 4} +2 (E \omega_k -3 E_{\rm th}^2)^2}{E_{\rm th}^4} \,, 
    \\
    \cK^{(6)}_{L; 00}(p) &= - \frac{\sqrt{3\cK_B}}{E_{\rm th}^2} \, \left[ p^2 + \tfrac{4}{9} q_p^{\star 2} (2 + \gamma_p^2) \right] \, , 
    \quad
    \cK^{(6)}_{R; 00}(k) = - \frac{\sqrt{3\cK_B}}{E_{\rm th}^2} \, \left[ k^2 + \tfrac{4}{9} q_k^{\star 2} (2 + \gamma_k^2) \right] \, , 
    \\
    %%%
    \cK^{(7)}_{L; 00}(p) &= \frac{\sqrt{\cK_B}}{E_{\rm th}^2} \, 
    \left( \epsilon_p + \tfrac{4}{3} q_p^{\star 2} \gamma_p^2 \beta_p^2 \right) \,,
    \quad
    \cK^{(7)}_{R; 00}(k) = \frac{\sqrt{\cK_B}}{E_{\rm th}^2} \, 
    \left( \epsilon_k + \tfrac{4}{3} q_k^{\star 2} \gamma_k^2 \beta_k^2 \right)\,,
    \end{align}
while for $\ell=s=2$ the only nonzero terms are, 
    %%%%%
    \begin{align}
    \cK^{(2)}_{L; 22}(p) &= 0  \,,
    \qquad
    \cK^{(2)}_{R; 22}(k) = \frac{16}{3\sqrt{5}} \, 
    \frac{q^{\star 2}_k}{E_{\rm th}^4} \, \frac{E^2 k^2}{\sigma_k}\,,
    \\
    \cK^{(2)}_{L; 22}(p) &= \frac{16}{3\sqrt{5}} \, 
    \frac{q^{\star 2}_p}{E_{\rm th}^4} \, \frac{E^2 p^2}{\sigma_p}\,,
    \quad
    \cK^{(3)}_{R; 22}(k) = 0\,,
    \\
    \cK^{(6)}_{L; 22}(p) &=  \frac{8}{3} \sqrt{\frac{\cK_B}{15}} 
    \, \frac{q^{\star 2}_p} {E_{\rm th}^2} (1-\gamma_p^2) \, , 
    \qquad
    \cK^{(6)}_{R; 22}(k) =  \frac{8}{3} \sqrt{\frac{\cK_B}{15}} 
    \, \frac{q^{\star 2}_k} {E_{\rm th}^2} (1-\gamma_k^2) \, , 
    \\ 
    %%%
    \cK^{(7)}_{L;22}(p) &= \frac{8}{3} \sqrt{\frac{\cK_B}{5}} 
    \, \frac{q^{\star 2}_p} {E_{\rm th}^2}  \gamma_p^2 \beta_p^2 \, , 
    \qquad
    \cK^{(7)}_{R;22}(k) = \frac{8}{3} \sqrt{\frac{\cK_B}{5}} 
    \, \frac{q^{\star 2}_k} {E_{\rm th}^2}  \gamma_k^2 \beta_k^2 \, .   
    \end{align}
    %%%%%
%%%%%%%%%%%%%%%%%%%%%%%%%%%%%%%%%%%
\subsubsection{Identical particles: $J^P=1^+$}
%%%%%%%%%%%%%%%%%%%%%%%%%%%%%%%%%%%

For $J^P = 1^+$, the only contribution is from the $\cK_B$ term in \Cref{eq:Kdfidentical},
and involves only the $(\ell,s)=(1,0)$ and $(1,2)$ components. We quote the result directly in its separable form, which here requires only a single component,
    %%%%%
    \begin{align}
    \cK_{L; 10}^{(1)}(p) &= -\sqrt{ \frac{2 \cK_B}{27}} \frac{3p (E - 3 \omega_p) + 4 q_p^{\star 2} \beta_p \gamma_p^2}{E_{\rm th}^2} \,, 
    \qquad
    \cK_{R; 10}^{(1)}(k) = \sqrt{ \frac{2 \cK_B}{27}} \frac{3k (E - 3 \omega_k) + 4 q_k^{\star 2} \beta_k \gamma_k^2}{E_{\rm th}^2} \,, 
    \\
    %%%
    \cK^{(1)}_{L; 12}(p) &= \sqrt{ \frac{64 \cK_B}{675}} \frac{q_p^{\star 2}}{E_{\rm th}^2} \, \beta_p \gamma_p (3+2\gamma_p) \,,
    \qquad
    \cK^{(1)}_{R; 12}(k) = -\sqrt{ \frac{64 \cK_B}{675}} \frac{q_k^{\star 2}}{E_{\rm th}^2} \, \beta_k \gamma_k (3+2\gamma_k) \,.
    \end{align}
    %%%%%
Note that the relative signs between $\cK_L$ and $\cK_R$ components are necessary to produce the correct form of $\Kdf$.

%%%%%%%%%%%%%%%%%%%%%%%%%%%%%%%%%%%
\subsubsection{Identical particles: $J^P=2^-$}
%%%%%%%%%%%%%%%%%%%%%%%%%%%%%%%%%%%

For $J^P = 2^-$, again only $\cK_B$ contributes, with the $(\ell,s)$ values now being $(0,2)$, $(2,0)$, and $(2,2)$. We again only quote the separable form of the result, which requires a single component,
    %%%%%
    \begin{align}
    \cK^{(1)}_{L; 02}(p) &= \sqrt{ \frac{32 \cK_B}{3375} } \frac{ q_p^{\star 2} }{E_{\rm th}^2} 
    \left[ 7 + 6 \gamma_p + 2 \gamma_p^2 \right] \,, 
    \qquad
    \cK^{(1)}_{R; 02}(k) = \sqrt{ \frac{32 \cK_B}{3375} } \frac{ q_k^{\star 2} }{E_{\rm th}^2} 
    \left[ 7 + 6 \gamma_k + 2 \gamma_k^2 \right] \,,     
    \\ 
    %%%
    \cK^{(1)}_{L; 20}(p) &= \sqrt{\frac{2 \cK_B}{135}}  
    \frac{4 q_p^{\star 2} (\gamma_p^2-1) + 9 p^2}{E_{\rm th}^2} \,, 
    \qquad
    \cK^{(1)}_{R; 20}(k) = \sqrt{\frac{2 \cK_B}{135}}  
    \frac{4 q_k^{\star 2} (\gamma_k^2-1) + 9 k^2}{E_{\rm th}^2} \,,    
    \\
    %%%
    \cK^{(1)}_{L; 22}(p) &= - \sqrt{\frac{64 \cK_B}{4725}}  
    \frac{ q_p^{\star 2}}{E_{\rm th}^2} (\gamma_p-1)(5+2\gamma_p) \,,
    \qquad
    \cK^{(1)}_{R; 22}(k) = - \sqrt{\frac{64 \cK_B}{4725}}  
    \frac{ q_k^{\star 2}}{E_{\rm th}^2} (\gamma_k-1)(5+2\gamma_k) \,.
    \end{align}
    %%%%%
Note that in this case there are no relative signs between $\cK_L$ and $\cK_R$ components.

\subsubsection{$2+1$ systems}

We recall that $\Kdf$ for $2+1$ systems is given by \Cref{eq:KdfExpansion},
and contains, at the order we work, four dimensionless constants, $\cK_0$, $\cK_1$,
$\cK_B$ and $\cK_E$. Following Ref.~\cite{Dawid:2024dgy}, it can be written in separable form using a slight generalization of \Cref{eq:separable_id},
    \begin{align}
        M_1^2 \cK_{\rm df,3; i \ell' s';j\ell s}^{J}(p,k) &= \sum_a
        e_i \cK_{L; \ell' s'}^{(a),i}(p) \cK_{R; \ell s}^{(a),j}(k) e_j\,,
        \label{eq:separable_twoplusone}
    \end{align}
where we recall that $i$ and $j$ label the final and initial spectator flavors,
and $M_1$ is the mass of the particle that appears twice.
The quantities $e_1=1$ and $e_2=1/\sqrt{2}$ are symmetry factors.

Results for $\cK_L$ and $\cK_R$ for angular momenta $J^P=0^-$ and $J^P=1^+$ are given for $2+1$ systems in Appendix C of Ref.~\cite{Dawid:2024dgy}.
To the order we work, there are no contributions for higher angular momenta, and in particular for $J^P=2^-$.

\end{widetext}

\clearpage
\bibliography{cited_refs}   

\begin{thebibliography}{125}%
\makeatletter
\providecommand \@ifxundefined [1]{%
 \@ifx{#1\undefined}
}%
\providecommand \@ifnum [1]{%
 \ifnum #1\expandafter \@firstoftwo
 \else \expandafter \@secondoftwo
 \fi
}%
\providecommand \@ifx [1]{%
 \ifx #1\expandafter \@firstoftwo
 \else \expandafter \@secondoftwo
 \fi
}%
\providecommand \natexlab [1]{#1}%
\providecommand \enquote  [1]{``#1''}%
\providecommand \bibnamefont  [1]{#1}%
\providecommand \bibfnamefont [1]{#1}%
\providecommand \citenamefont [1]{#1}%
\providecommand \href@noop [0]{\@secondoftwo}%
\providecommand \href [0]{\begingroup \@sanitize@url \@href}%
\providecommand \@href[1]{\@@startlink{#1}\@@href}%
\providecommand \@@href[1]{\endgroup#1\@@endlink}%
\providecommand \@sanitize@url [0]{\catcode `\\12\catcode `\$12\catcode
  `\&12\catcode `\#12\catcode `\^12\catcode `\_12\catcode `\%12\relax}%
\providecommand \@@startlink[1]{}%
\providecommand \@@endlink[0]{}%
\providecommand \url  [0]{\begingroup\@sanitize@url \@url }%
\providecommand \@url [1]{\endgroup\@href {#1}{\urlprefix }}%
\providecommand \urlprefix  [0]{URL }%
\providecommand \Eprint [0]{\href }%
\providecommand \doibase [0]{https://doi.org/}%
\providecommand \selectlanguage [0]{\@gobble}%
\providecommand \bibinfo  [0]{\@secondoftwo}%
\providecommand \bibfield  [0]{\@secondoftwo}%
\providecommand \translation [1]{[#1]}%
\providecommand \BibitemOpen [0]{}%
\providecommand \bibitemStop [0]{}%
\providecommand \bibitemNoStop [0]{.\EOS\space}%
\providecommand \EOS [0]{\spacefactor3000\relax}%
\providecommand \BibitemShut  [1]{\csname bibitem#1\endcsname}%
\let\auto@bib@innerbib\@empty
%</preamble>
\bibitem [{\citenamefont {Detmold}\ and\ \citenamefont
  {Savage}(2008)}]{Detmold:2008gh}%
  \BibitemOpen
  \bibfield  {author} {\bibinfo {author} {\bibfnamefont {W.}~\bibnamefont
  {Detmold}}\ and\ \bibinfo {author} {\bibfnamefont {M.~J.}\ \bibnamefont
  {Savage}},\ }\bibfield  {title} {\bibinfo {title} {{The Energy of n Identical
  Bosons in a Finite Volume at $O(L^{-7})$}},\ }\href
  {https://doi.org/10.1103/PhysRevD.77.057502} {\bibfield  {journal} {\bibinfo
  {journal} {Phys. Rev. D}\ }\textbf {\bibinfo {volume} {77}},\ \bibinfo
  {pages} {057502} (\bibinfo {year} {2008})},\ \Eprint
  {https://arxiv.org/abs/0801.0763} {arXiv:0801.0763 [hep-lat]} \BibitemShut
  {NoStop}%
\bibitem [{\citenamefont {Beane}\ \emph {et~al.}(2007)\citenamefont {Beane},
  \citenamefont {Detmold},\ and\ \citenamefont {Savage}}]{Beane:2007qr}%
  \BibitemOpen
  \bibfield  {author} {\bibinfo {author} {\bibfnamefont {S.~R.}\ \bibnamefont
  {Beane}}, \bibinfo {author} {\bibfnamefont {W.}~\bibnamefont {Detmold}},\
  and\ \bibinfo {author} {\bibfnamefont {M.~J.}\ \bibnamefont {Savage}},\
  }\bibfield  {title} {\bibinfo {title} {{n-Boson Energies at Finite Volume and
  Three-Boson Interactions}},\ }\href
  {https://doi.org/10.1103/PhysRevD.76.074507} {\bibfield  {journal} {\bibinfo
  {journal} {Phys. Rev. D}\ }\textbf {\bibinfo {volume} {76}},\ \bibinfo
  {pages} {074507} (\bibinfo {year} {2007})},\ \Eprint
  {https://arxiv.org/abs/0707.1670} {arXiv:0707.1670 [hep-lat]} \BibitemShut
  {NoStop}%
\bibitem [{\citenamefont {Briceno}\ and\ \citenamefont
  {Davoudi}(2013)}]{Briceno:2012rv}%
  \BibitemOpen
  \bibfield  {author} {\bibinfo {author} {\bibfnamefont {R.~A.}\ \bibnamefont
  {Briceno}}\ and\ \bibinfo {author} {\bibfnamefont {Z.}~\bibnamefont
  {Davoudi}},\ }\bibfield  {title} {\bibinfo {title} {{Three-particle
  scattering amplitudes from a finite volume formalism}},\ }\href
  {https://doi.org/10.1103/PhysRevD.87.094507} {\bibfield  {journal} {\bibinfo
  {journal} {Phys. Rev.}\ }\textbf {\bibinfo {volume} {D87}},\ \bibinfo {pages}
  {094507} (\bibinfo {year} {2013})},\ \Eprint
  {https://arxiv.org/abs/1212.3398} {arXiv:1212.3398 [hep-lat]} \BibitemShut
  {NoStop}%
%%CITATION = ARXIV:1212.3398;%%
\bibitem [{\citenamefont {Polejaeva}\ and\ \citenamefont
  {Rusetsky}(2012)}]{Polejaeva:2012ut}%
  \BibitemOpen
  \bibfield  {author} {\bibinfo {author} {\bibfnamefont {K.}~\bibnamefont
  {Polejaeva}}\ and\ \bibinfo {author} {\bibfnamefont {A.}~\bibnamefont
  {Rusetsky}},\ }\bibfield  {title} {\bibinfo {title} {{Three particles in a
  finite volume}},\ }\href {https://doi.org/10.1140/epja/i2012-12067-8}
  {\bibfield  {journal} {\bibinfo  {journal} {Eur. Phys. J.}\ }\textbf
  {\bibinfo {volume} {A48}},\ \bibinfo {pages} {67} (\bibinfo {year} {2012})},\
  \Eprint {https://arxiv.org/abs/1203.1241} {arXiv:1203.1241 [hep-lat]}
  \BibitemShut {NoStop}%
%%CITATION = ARXIV:1203.1241;%%
\bibitem [{\citenamefont {Hansen}\ and\ \citenamefont
  {Sharpe}(2014)}]{Hansen:2014eka}%
  \BibitemOpen
  \bibfield  {author} {\bibinfo {author} {\bibfnamefont {M.~T.}\ \bibnamefont
  {Hansen}}\ and\ \bibinfo {author} {\bibfnamefont {S.~R.}\ \bibnamefont
  {Sharpe}},\ }\bibfield  {title} {\bibinfo {title} {{Relativistic,
  model-independent, three-particle quantization condition}},\ }\href
  {https://doi.org/10.1103/PhysRevD.90.116003} {\bibfield  {journal} {\bibinfo
  {journal} {Phys. Rev. D}\ }\textbf {\bibinfo {volume} {90}},\ \bibinfo
  {pages} {116003} (\bibinfo {year} {2014})},\ \Eprint
  {https://arxiv.org/abs/1408.5933} {arXiv:1408.5933 [hep-lat]} \BibitemShut
  {NoStop}%
\bibitem [{\citenamefont {Hansen}\ and\ \citenamefont
  {Sharpe}(2015)}]{Hansen:2015zga}%
  \BibitemOpen
  \bibfield  {author} {\bibinfo {author} {\bibfnamefont {M.~T.}\ \bibnamefont
  {Hansen}}\ and\ \bibinfo {author} {\bibfnamefont {S.~R.}\ \bibnamefont
  {Sharpe}},\ }\bibfield  {title} {\bibinfo {title} {{Expressing the
  three-particle finite-volume spectrum in terms of the three-to-three
  scattering amplitude}},\ }\href {https://doi.org/10.1103/PhysRevD.92.114509}
  {\bibfield  {journal} {\bibinfo  {journal} {Phys. Rev.}\ }\textbf {\bibinfo
  {volume} {D92}},\ \bibinfo {pages} {114509} (\bibinfo {year} {2015})},\
  \Eprint {https://arxiv.org/abs/1504.04248} {arXiv:1504.04248 [hep-lat]}
  \BibitemShut {NoStop}%
%%CITATION = ARXIV:1504.04248;%%
\bibitem [{\citenamefont {{Brice{\~n}o}}\ \emph {et~al.}(2017)\citenamefont
  {{Brice{\~n}o}}, \citenamefont {Hansen},\ and\ \citenamefont
  {Sharpe}}]{Briceno:2017tce}%
  \BibitemOpen
  \bibfield  {author} {\bibinfo {author} {\bibfnamefont {R.~A.}\ \bibnamefont
  {{Brice{\~n}o}}}, \bibinfo {author} {\bibfnamefont {M.~T.}\ \bibnamefont
  {Hansen}},\ and\ \bibinfo {author} {\bibfnamefont {S.~R.}\ \bibnamefont
  {Sharpe}},\ }\bibfield  {title} {\bibinfo {title} {{Relating the
  finite-volume spectrum and the two-and-three-particle $S$ matrix for
  relativistic systems of identical scalar particles}},\ }\href
  {https://doi.org/10.1103/PhysRevD.95.074510} {\bibfield  {journal} {\bibinfo
  {journal} {Phys. Rev.}\ }\textbf {\bibinfo {volume} {D95}},\ \bibinfo {pages}
  {074510} (\bibinfo {year} {2017})},\ \Eprint
  {https://arxiv.org/abs/1701.07465} {arXiv:1701.07465 [hep-lat]} \BibitemShut
  {NoStop}%
%%CITATION = ARXIV:1701.07465;%%
\bibitem [{\citenamefont {Hammer}\ \emph
  {et~al.}(2017{\natexlab{a}})\citenamefont {Hammer}, \citenamefont {Pang},\
  and\ \citenamefont {Rusetsky}}]{Hammer:2017uqm}%
  \BibitemOpen
  \bibfield  {author} {\bibinfo {author} {\bibfnamefont {H.-W.}\ \bibnamefont
  {Hammer}}, \bibinfo {author} {\bibfnamefont {J.-Y.}\ \bibnamefont {Pang}},\
  and\ \bibinfo {author} {\bibfnamefont {A.}~\bibnamefont {Rusetsky}},\
  }\bibfield  {title} {\bibinfo {title} {{Three-particle quantization condition
  in a finite volume: 1. The role of the three-particle force}},\ }\href
  {https://doi.org/10.1007/JHEP09(2017)109} {\bibfield  {journal} {\bibinfo
  {journal} {JHEP}\ }\textbf {\bibinfo {volume} {09}},\ \bibinfo {pages}
  {109}},\ \Eprint {https://arxiv.org/abs/1706.07700} {arXiv:1706.07700
  [hep-lat]} \BibitemShut {NoStop}%
%%CITATION = ARXIV:1706.07700;%%
\bibitem [{\citenamefont {Hammer}\ \emph
  {et~al.}(2017{\natexlab{b}})\citenamefont {Hammer}, \citenamefont {Pang},\
  and\ \citenamefont {Rusetsky}}]{Hammer:2017kms}%
  \BibitemOpen
  \bibfield  {author} {\bibinfo {author} {\bibfnamefont {H.~W.}\ \bibnamefont
  {Hammer}}, \bibinfo {author} {\bibfnamefont {J.~Y.}\ \bibnamefont {Pang}},\
  and\ \bibinfo {author} {\bibfnamefont {A.}~\bibnamefont {Rusetsky}},\
  }\bibfield  {title} {\bibinfo {title} {{Three particle quantization condition
  in a finite volume: 2. general formalism and the analysis of data}},\ }\href
  {https://doi.org/10.1007/JHEP10(2017)115} {\bibfield  {journal} {\bibinfo
  {journal} {JHEP}\ }\textbf {\bibinfo {volume} {10}},\ \bibinfo {pages}
  {115}},\ \Eprint {https://arxiv.org/abs/1707.02176} {arXiv:1707.02176
  [hep-lat]} \BibitemShut {NoStop}%
%%CITATION = ARXIV:1707.02176;%%
\bibitem [{\citenamefont {Mai}\ and\ \citenamefont
  {D\"oring}(2017)}]{Mai:2017bge}%
  \BibitemOpen
  \bibfield  {author} {\bibinfo {author} {\bibfnamefont {M.}~\bibnamefont
  {Mai}}\ and\ \bibinfo {author} {\bibfnamefont {M.}~\bibnamefont {D\"oring}},\
  }\bibfield  {title} {\bibinfo {title} {{Three-body Unitarity in the Finite
  Volume}},\ }\href {https://doi.org/10.1140/epja/i2017-12440-1} {\bibfield
  {journal} {\bibinfo  {journal} {Eur. Phys. J. A}\ }\textbf {\bibinfo {volume}
  {53}},\ \bibinfo {pages} {240} (\bibinfo {year} {2017})},\ \Eprint
  {https://arxiv.org/abs/1709.08222} {arXiv:1709.08222 [hep-lat]} \BibitemShut
  {NoStop}%
\bibitem [{\citenamefont {Brice\~no}\ \emph
  {et~al.}(2019{\natexlab{a}})\citenamefont {Brice\~no}, \citenamefont
  {Hansen},\ and\ \citenamefont {Sharpe}}]{Briceno:2018aml}%
  \BibitemOpen
  \bibfield  {author} {\bibinfo {author} {\bibfnamefont {R.~A.}\ \bibnamefont
  {Brice\~no}}, \bibinfo {author} {\bibfnamefont {M.~T.}\ \bibnamefont
  {Hansen}},\ and\ \bibinfo {author} {\bibfnamefont {S.~R.}\ \bibnamefont
  {Sharpe}},\ }\bibfield  {title} {\bibinfo {title} {{Three-particle systems
  with resonant subprocesses in a finite volume}},\ }\href
  {https://doi.org/10.1103/PhysRevD.99.014516} {\bibfield  {journal} {\bibinfo
  {journal} {Phys. Rev. D}\ }\textbf {\bibinfo {volume} {99}},\ \bibinfo
  {pages} {014516} (\bibinfo {year} {2019}{\natexlab{a}})},\ \Eprint
  {https://arxiv.org/abs/1810.01429} {arXiv:1810.01429 [hep-lat]} \BibitemShut
  {NoStop}%
\bibitem [{\citenamefont {Brice\~no}\ \emph {et~al.}(2018)\citenamefont
  {Brice\~no}, \citenamefont {Hansen},\ and\ \citenamefont
  {Sharpe}}]{Briceno:2018mlh}%
  \BibitemOpen
  \bibfield  {author} {\bibinfo {author} {\bibfnamefont {R.~A.}\ \bibnamefont
  {Brice\~no}}, \bibinfo {author} {\bibfnamefont {M.~T.}\ \bibnamefont
  {Hansen}},\ and\ \bibinfo {author} {\bibfnamefont {S.~R.}\ \bibnamefont
  {Sharpe}},\ }\bibfield  {title} {\bibinfo {title} {{Numerical study of the
  relativistic three-body quantization condition in the isotropic
  approximation}},\ }\href {https://doi.org/10.1103/PhysRevD.98.014506}
  {\bibfield  {journal} {\bibinfo  {journal} {Phys. Rev. D}\ }\textbf {\bibinfo
  {volume} {98}},\ \bibinfo {pages} {014506} (\bibinfo {year} {2018})},\
  \Eprint {https://arxiv.org/abs/1803.04169} {arXiv:1803.04169 [hep-lat]}
  \BibitemShut {NoStop}%
\bibitem [{\citenamefont {Pang}\ \emph {et~al.}(2019)\citenamefont {Pang},
  \citenamefont {Wu}, \citenamefont {Hammer}, \citenamefont {Mei\ss{}ner},\
  and\ \citenamefont {Rusetsky}}]{Pang:2019dfe}%
  \BibitemOpen
  \bibfield  {author} {\bibinfo {author} {\bibfnamefont {J.-Y.}\ \bibnamefont
  {Pang}}, \bibinfo {author} {\bibfnamefont {J.-J.}\ \bibnamefont {Wu}},
  \bibinfo {author} {\bibfnamefont {H.~W.}\ \bibnamefont {Hammer}}, \bibinfo
  {author} {\bibfnamefont {U.-G.}\ \bibnamefont {Mei\ss{}ner}},\ and\ \bibinfo
  {author} {\bibfnamefont {A.}~\bibnamefont {Rusetsky}},\ }\bibfield  {title}
  {\bibinfo {title} {{Energy shift of the three-particle system in a finite
  volume}},\ }\href {https://doi.org/10.1103/PhysRevD.99.074513} {\bibfield
  {journal} {\bibinfo  {journal} {Phys. Rev. D}\ }\textbf {\bibinfo {volume}
  {99}},\ \bibinfo {pages} {074513} (\bibinfo {year} {2019})},\ \Eprint
  {https://arxiv.org/abs/1902.01111} {arXiv:1902.01111 [hep-lat]} \BibitemShut
  {NoStop}%
\bibitem [{\citenamefont {Jackura}\ \emph
  {et~al.}(2019{\natexlab{a}})\citenamefont {Jackura}, \citenamefont {Dawid},
  \citenamefont {Fern\'andez-Ram\'\i{}rez}, \citenamefont {Mathieu},
  \citenamefont {Mikhasenko}, \citenamefont {Pilloni}, \citenamefont {Sharpe},\
  and\ \citenamefont {Szczepaniak}}]{Jackura:2019bmu}%
  \BibitemOpen
  \bibfield  {author} {\bibinfo {author} {\bibfnamefont {A.~W.}\ \bibnamefont
  {Jackura}}, \bibinfo {author} {\bibfnamefont {S.~M.}\ \bibnamefont {Dawid}},
  \bibinfo {author} {\bibfnamefont {C.}~\bibnamefont
  {Fern\'andez-Ram\'\i{}rez}}, \bibinfo {author} {\bibfnamefont
  {V.}~\bibnamefont {Mathieu}}, \bibinfo {author} {\bibfnamefont
  {M.}~\bibnamefont {Mikhasenko}}, \bibinfo {author} {\bibfnamefont
  {A.}~\bibnamefont {Pilloni}}, \bibinfo {author} {\bibfnamefont {S.~R.}\
  \bibnamefont {Sharpe}},\ and\ \bibinfo {author} {\bibfnamefont {A.~P.}\
  \bibnamefont {Szczepaniak}},\ }\bibfield  {title} {\bibinfo {title}
  {{Equivalence of three-particle scattering formalisms}},\ }\href
  {https://doi.org/10.1103/PhysRevD.100.034508} {\bibfield  {journal} {\bibinfo
   {journal} {Phys. Rev. D}\ }\textbf {\bibinfo {volume} {100}},\ \bibinfo
  {pages} {034508} (\bibinfo {year} {2019}{\natexlab{a}})},\ \Eprint
  {https://arxiv.org/abs/1905.12007} {arXiv:1905.12007 [hep-ph]} \BibitemShut
  {NoStop}%
\bibitem [{\citenamefont {Blanton}\ \emph {et~al.}(2019)\citenamefont
  {Blanton}, \citenamefont {Romero-L\'opez},\ and\ \citenamefont
  {Sharpe}}]{Blanton:2019igq}%
  \BibitemOpen
  \bibfield  {author} {\bibinfo {author} {\bibfnamefont {T.~D.}\ \bibnamefont
  {Blanton}}, \bibinfo {author} {\bibfnamefont {F.}~\bibnamefont
  {Romero-L\'opez}},\ and\ \bibinfo {author} {\bibfnamefont {S.~R.}\
  \bibnamefont {Sharpe}},\ }\bibfield  {title} {\bibinfo {title} {{Implementing
  the three-particle quantization condition including higher partial waves}},\
  }\href {https://doi.org/10.1007/JHEP03(2019)106} {\bibfield  {journal}
  {\bibinfo  {journal} {JHEP}\ }\textbf {\bibinfo {volume} {03}},\ \bibinfo
  {pages} {106}},\ \Eprint {https://arxiv.org/abs/1901.07095} {arXiv:1901.07095
  [hep-lat]} \BibitemShut {NoStop}%
\bibitem [{\citenamefont {Brice\~no}\ \emph
  {et~al.}(2019{\natexlab{b}})\citenamefont {Brice\~no}, \citenamefont
  {Hansen}, \citenamefont {Sharpe},\ and\ \citenamefont
  {Szczepaniak}}]{Briceno:2019muc}%
  \BibitemOpen
  \bibfield  {author} {\bibinfo {author} {\bibfnamefont {R.~A.}\ \bibnamefont
  {Brice\~no}}, \bibinfo {author} {\bibfnamefont {M.~T.}\ \bibnamefont
  {Hansen}}, \bibinfo {author} {\bibfnamefont {S.~R.}\ \bibnamefont {Sharpe}},\
  and\ \bibinfo {author} {\bibfnamefont {A.~P.}\ \bibnamefont {Szczepaniak}},\
  }\bibfield  {title} {\bibinfo {title} {{Unitarity of the infinite-volume
  three-particle scattering amplitude arising from a finite-volume
  formalism}},\ }\href {https://doi.org/10.1103/PhysRevD.100.054508} {\bibfield
   {journal} {\bibinfo  {journal} {Phys. Rev. D}\ }\textbf {\bibinfo {volume}
  {100}},\ \bibinfo {pages} {054508} (\bibinfo {year} {2019}{\natexlab{b}})},\
  \Eprint {https://arxiv.org/abs/1905.11188} {arXiv:1905.11188 [hep-lat]}
  \BibitemShut {NoStop}%
\bibitem [{\citenamefont {Romero-L\'opez}\ \emph {et~al.}(2019)\citenamefont
  {Romero-L\'opez}, \citenamefont {Sharpe}, \citenamefont {Blanton},
  \citenamefont {Brice\~no},\ and\ \citenamefont
  {Hansen}}]{Romero-Lopez:2019qrt}%
  \BibitemOpen
  \bibfield  {author} {\bibinfo {author} {\bibfnamefont {F.}~\bibnamefont
  {Romero-L\'opez}}, \bibinfo {author} {\bibfnamefont {S.~R.}\ \bibnamefont
  {Sharpe}}, \bibinfo {author} {\bibfnamefont {T.~D.}\ \bibnamefont {Blanton}},
  \bibinfo {author} {\bibfnamefont {R.~A.}\ \bibnamefont {Brice\~no}},\ and\
  \bibinfo {author} {\bibfnamefont {M.~T.}\ \bibnamefont {Hansen}},\ }\bibfield
   {title} {\bibinfo {title} {{Numerical exploration of three relativistic
  particles in a finite volume including two-particle resonances and bound
  states}},\ }\href {https://doi.org/10.1007/JHEP10(2019)007} {\bibfield
  {journal} {\bibinfo  {journal} {JHEP}\ }\textbf {\bibinfo {volume} {10}},\
  \bibinfo {pages} {007}},\ \Eprint {https://arxiv.org/abs/1908.02411}
  {arXiv:1908.02411 [hep-lat]} \BibitemShut {NoStop}%
\bibitem [{\citenamefont {Pang}\ \emph {et~al.}(2020)\citenamefont {Pang},
  \citenamefont {Wu},\ and\ \citenamefont {Geng}}]{Pang:2020pkl}%
  \BibitemOpen
  \bibfield  {author} {\bibinfo {author} {\bibfnamefont {J.-Y.}\ \bibnamefont
  {Pang}}, \bibinfo {author} {\bibfnamefont {J.-J.}\ \bibnamefont {Wu}},\ and\
  \bibinfo {author} {\bibfnamefont {L.-S.}\ \bibnamefont {Geng}},\ }\bibfield
  {title} {\bibinfo {title} {{$DDK$ system in finite volume}},\ }\href
  {https://doi.org/10.1103/PhysRevD.102.114515} {\bibfield  {journal} {\bibinfo
   {journal} {Phys. Rev. D}\ }\textbf {\bibinfo {volume} {102}},\ \bibinfo
  {pages} {114515} (\bibinfo {year} {2020})},\ \Eprint
  {https://arxiv.org/abs/2008.13014} {arXiv:2008.13014 [hep-lat]} \BibitemShut
  {NoStop}%
\bibitem [{\citenamefont {Blanton}\ and\ \citenamefont
  {Sharpe}(2020{\natexlab{a}})}]{Blanton:2020gha}%
  \BibitemOpen
  \bibfield  {author} {\bibinfo {author} {\bibfnamefont {T.~D.}\ \bibnamefont
  {Blanton}}\ and\ \bibinfo {author} {\bibfnamefont {S.~R.}\ \bibnamefont
  {Sharpe}},\ }\bibfield  {title} {\bibinfo {title} {{Alternative derivation of
  the relativistic three-particle quantization condition}},\ }\href
  {https://doi.org/10.1103/PhysRevD.102.054520} {\bibfield  {journal} {\bibinfo
   {journal} {Phys. Rev. D}\ }\textbf {\bibinfo {volume} {102}},\ \bibinfo
  {pages} {054520} (\bibinfo {year} {2020}{\natexlab{a}})},\ \Eprint
  {https://arxiv.org/abs/2007.16188} {arXiv:2007.16188 [hep-lat]} \BibitemShut
  {NoStop}%
\bibitem [{\citenamefont {Blanton}\ and\ \citenamefont
  {Sharpe}(2020{\natexlab{b}})}]{Blanton:2020jnm}%
  \BibitemOpen
  \bibfield  {author} {\bibinfo {author} {\bibfnamefont {T.~D.}\ \bibnamefont
  {Blanton}}\ and\ \bibinfo {author} {\bibfnamefont {S.~R.}\ \bibnamefont
  {Sharpe}},\ }\bibfield  {title} {\bibinfo {title} {{Equivalence of
  relativistic three-particle quantization conditions}},\ }\href
  {https://doi.org/10.1103/PhysRevD.102.054515} {\bibfield  {journal} {\bibinfo
   {journal} {Phys. Rev. D}\ }\textbf {\bibinfo {volume} {102}},\ \bibinfo
  {pages} {054515} (\bibinfo {year} {2020}{\natexlab{b}})},\ \Eprint
  {https://arxiv.org/abs/2007.16190} {arXiv:2007.16190 [hep-lat]} \BibitemShut
  {NoStop}%
\bibitem [{\citenamefont {Romero-L\'opez}\ \emph {et~al.}(2021)\citenamefont
  {Romero-L\'opez}, \citenamefont {Rusetsky}, \citenamefont {Schlage},\ and\
  \citenamefont {Urbach}}]{Romero-Lopez:2020rdq}%
  \BibitemOpen
  \bibfield  {author} {\bibinfo {author} {\bibfnamefont {F.}~\bibnamefont
  {Romero-L\'opez}}, \bibinfo {author} {\bibfnamefont {A.}~\bibnamefont
  {Rusetsky}}, \bibinfo {author} {\bibfnamefont {N.}~\bibnamefont {Schlage}},\
  and\ \bibinfo {author} {\bibfnamefont {C.}~\bibnamefont {Urbach}},\
  }\bibfield  {title} {\bibinfo {title} {{Relativistic $N$-particle energy
  shift in finite volume}},\ }\href {https://doi.org/10.1007/JHEP02(2021)060}
  {\bibfield  {journal} {\bibinfo  {journal} {JHEP}\ }\textbf {\bibinfo
  {volume} {02}},\ \bibinfo {pages} {060}},\ \Eprint
  {https://arxiv.org/abs/2010.11715} {arXiv:2010.11715 [hep-lat]} \BibitemShut
  {NoStop}%
\bibitem [{\citenamefont {Blanton}\ and\ \citenamefont
  {Sharpe}(2021{\natexlab{a}})}]{Blanton:2020gmf}%
  \BibitemOpen
  \bibfield  {author} {\bibinfo {author} {\bibfnamefont {T.~D.}\ \bibnamefont
  {Blanton}}\ and\ \bibinfo {author} {\bibfnamefont {S.~R.}\ \bibnamefont
  {Sharpe}},\ }\bibfield  {title} {\bibinfo {title} {{Relativistic
  three-particle quantization condition for nondegenerate scalars}},\ }\href
  {https://doi.org/10.1103/PhysRevD.103.054503} {\bibfield  {journal} {\bibinfo
   {journal} {Phys. Rev. D}\ }\textbf {\bibinfo {volume} {103}},\ \bibinfo
  {pages} {054503} (\bibinfo {year} {2021}{\natexlab{a}})},\ \Eprint
  {https://arxiv.org/abs/2011.05520} {arXiv:2011.05520 [hep-lat]} \BibitemShut
  {NoStop}%
\bibitem [{\citenamefont {M\"uller}\ \emph {et~al.}(2021)\citenamefont
  {M\"uller}, \citenamefont {Yu},\ and\ \citenamefont
  {Rusetsky}}]{Muller:2020vtt}%
  \BibitemOpen
  \bibfield  {author} {\bibinfo {author} {\bibfnamefont {F.}~\bibnamefont
  {M\"uller}}, \bibinfo {author} {\bibfnamefont {T.}~\bibnamefont {Yu}},\ and\
  \bibinfo {author} {\bibfnamefont {A.}~\bibnamefont {Rusetsky}},\ }\bibfield
  {title} {\bibinfo {title} {{Finite-volume energy shift of the three-pion
  ground state}},\ }\href {https://doi.org/10.1103/PhysRevD.103.054506}
  {\bibfield  {journal} {\bibinfo  {journal} {Phys. Rev. D}\ }\textbf {\bibinfo
  {volume} {103}},\ \bibinfo {pages} {054506} (\bibinfo {year} {2021})},\
  \Eprint {https://arxiv.org/abs/2011.14178} {arXiv:2011.14178 [hep-lat]}
  \BibitemShut {NoStop}%
\bibitem [{\citenamefont {Blanton}\ and\ \citenamefont
  {Sharpe}(2021{\natexlab{b}})}]{Blanton:2021mih}%
  \BibitemOpen
  \bibfield  {author} {\bibinfo {author} {\bibfnamefont {T.~D.}\ \bibnamefont
  {Blanton}}\ and\ \bibinfo {author} {\bibfnamefont {S.~R.}\ \bibnamefont
  {Sharpe}},\ }\bibfield  {title} {\bibinfo {title} {{Three-particle
  finite-volume formalism for $\pi^+\pi^+K^+$ and related systems}},\ }\href
  {https://doi.org/10.1103/PhysRevD.104.034509} {\bibfield  {journal} {\bibinfo
   {journal} {Phys. Rev. D}\ }\textbf {\bibinfo {volume} {104}},\ \bibinfo
  {pages} {034509} (\bibinfo {year} {2021}{\natexlab{b}})},\ \Eprint
  {https://arxiv.org/abs/2105.12094} {arXiv:2105.12094 [hep-lat]} \BibitemShut
  {NoStop}%
\bibitem [{\citenamefont {M\"uller}\ \emph {et~al.}(2022)\citenamefont
  {M\"uller}, \citenamefont {Pang}, \citenamefont {Rusetsky},\ and\
  \citenamefont {Wu}}]{Muller:2021uur}%
  \BibitemOpen
  \bibfield  {author} {\bibinfo {author} {\bibfnamefont {F.}~\bibnamefont
  {M\"uller}}, \bibinfo {author} {\bibfnamefont {J.-Y.}\ \bibnamefont {Pang}},
  \bibinfo {author} {\bibfnamefont {A.}~\bibnamefont {Rusetsky}},\ and\
  \bibinfo {author} {\bibfnamefont {J.-J.}\ \bibnamefont {Wu}},\ }\bibfield
  {title} {\bibinfo {title} {{Relativistic-invariant formulation of the NREFT
  three-particle quantization condition}},\ }\href
  {https://doi.org/10.1007/JHEP02(2022)158} {\bibfield  {journal} {\bibinfo
  {journal} {JHEP}\ }\textbf {\bibinfo {volume} {02}},\ \bibinfo {pages}
  {158}},\ \Eprint {https://arxiv.org/abs/2110.09351} {arXiv:2110.09351
  [hep-lat]} \BibitemShut {NoStop}%
\bibitem [{\citenamefont {M\"uller}\ \emph {et~al.}(2023)\citenamefont
  {M\"uller}, \citenamefont {Pang}, \citenamefont {Rusetsky},\ and\
  \citenamefont {Wu}}]{Muller:2022oyw}%
  \BibitemOpen
  \bibfield  {author} {\bibinfo {author} {\bibfnamefont {F.}~\bibnamefont
  {M\"uller}}, \bibinfo {author} {\bibfnamefont {J.-Y.}\ \bibnamefont {Pang}},
  \bibinfo {author} {\bibfnamefont {A.}~\bibnamefont {Rusetsky}},\ and\
  \bibinfo {author} {\bibfnamefont {J.-J.}\ \bibnamefont {Wu}},\ }\bibfield
  {title} {\bibinfo {title} {{Three-particle Lellouch-L\"uscher formalism in
  moving frames}},\ }\href {https://doi.org/10.1007/JHEP02(2023)214} {\bibfield
   {journal} {\bibinfo  {journal} {JHEP}\ }\textbf {\bibinfo {volume} {02}},\
  \bibinfo {pages} {214}},\ \Eprint {https://arxiv.org/abs/2211.10126}
  {arXiv:2211.10126 [hep-lat]} \BibitemShut {NoStop}%
\bibitem [{\citenamefont {Pang}\ \emph {et~al.}(2024)\citenamefont {Pang},
  \citenamefont {Bubna}, \citenamefont {M\"uller}, \citenamefont {Rusetsky},\
  and\ \citenamefont {Wu}}]{Pang:2023jri}%
  \BibitemOpen
  \bibfield  {author} {\bibinfo {author} {\bibfnamefont {J.-Y.}\ \bibnamefont
  {Pang}}, \bibinfo {author} {\bibfnamefont {R.}~\bibnamefont {Bubna}},
  \bibinfo {author} {\bibfnamefont {F.}~\bibnamefont {M\"uller}}, \bibinfo
  {author} {\bibfnamefont {A.}~\bibnamefont {Rusetsky}},\ and\ \bibinfo
  {author} {\bibfnamefont {J.-J.}\ \bibnamefont {Wu}},\ }\bibfield  {title}
  {\bibinfo {title} {{Lellouch-L\"uscher factor for the K \textrightarrow{}
  3\ensuremath{\pi} decays}},\ }\href {https://doi.org/10.1007/JHEP05(2024)269}
  {\bibfield  {journal} {\bibinfo  {journal} {JHEP}\ }\textbf {\bibinfo
  {volume} {05}},\ \bibinfo {pages} {269}},\ \Eprint
  {https://arxiv.org/abs/2312.04391} {arXiv:2312.04391 [hep-lat]} \BibitemShut
  {NoStop}%
\bibitem [{\citenamefont {Bubna}\ \emph {et~al.}(2023)\citenamefont {Bubna},
  \citenamefont {M\"uller},\ and\ \citenamefont {Rusetsky}}]{Bubna:2023oxo}%
  \BibitemOpen
  \bibfield  {author} {\bibinfo {author} {\bibfnamefont {R.}~\bibnamefont
  {Bubna}}, \bibinfo {author} {\bibfnamefont {F.}~\bibnamefont {M\"uller}},\
  and\ \bibinfo {author} {\bibfnamefont {A.}~\bibnamefont {Rusetsky}},\
  }\bibfield  {title} {\bibinfo {title} {{Finite-volume energy shift of the
  three-nucleon ground state}},\ }\href
  {https://doi.org/10.1103/PhysRevD.108.014518} {\bibfield  {journal} {\bibinfo
   {journal} {Phys. Rev. D}\ }\textbf {\bibinfo {volume} {108}},\ \bibinfo
  {pages} {014518} (\bibinfo {year} {2023})},\ \Eprint
  {https://arxiv.org/abs/2304.13635} {arXiv:2304.13635 [hep-lat]} \BibitemShut
  {NoStop}%
\bibitem [{\citenamefont {Brice\~no}\ \emph
  {et~al.}(2024{\natexlab{a}})\citenamefont {Brice\~no}, \citenamefont
  {Jackura}, \citenamefont {Pefkou},\ and\ \citenamefont
  {Romero-L\'opez}}]{Briceno:2024txg}%
  \BibitemOpen
  \bibfield  {author} {\bibinfo {author} {\bibfnamefont {R.~A.}\ \bibnamefont
  {Brice\~no}}, \bibinfo {author} {\bibfnamefont {A.~W.}\ \bibnamefont
  {Jackura}}, \bibinfo {author} {\bibfnamefont {D.~A.}\ \bibnamefont
  {Pefkou}},\ and\ \bibinfo {author} {\bibfnamefont {F.}~\bibnamefont
  {Romero-L\'opez}},\ }\bibfield  {title} {\bibinfo {title} {{Electroweak
  three-body decays in the presence of two- and three-body bound states}},\
  }\href {https://doi.org/10.1007/JHEP05(2024)279} {\bibfield  {journal}
  {\bibinfo  {journal} {JHEP}\ }\textbf {\bibinfo {volume} {05}},\ \bibinfo
  {pages} {279}},\ \Eprint {https://arxiv.org/abs/2402.12167} {arXiv:2402.12167
  [hep-lat]} \BibitemShut {NoStop}%
\bibitem [{\citenamefont {Xiao}\ \emph {et~al.}(2024)\citenamefont {Xiao},
  \citenamefont {Pang},\ and\ \citenamefont {Wu}}]{Xiao:2024dyw}%
  \BibitemOpen
  \bibfield  {author} {\bibinfo {author} {\bibfnamefont {Q.-C.}\ \bibnamefont
  {Xiao}}, \bibinfo {author} {\bibfnamefont {J.-Y.}\ \bibnamefont {Pang}},\
  and\ \bibinfo {author} {\bibfnamefont {J.-J.}\ \bibnamefont {Wu}},\
  }\bibfield  {title} {\bibinfo {title} {{Lattice spectra of the DDK three-body
  system with Lorentz covariant kinematic}},\ }\href
  {https://doi.org/10.1103/PhysRevD.110.094517} {\bibfield  {journal} {\bibinfo
   {journal} {Phys. Rev. D}\ }\textbf {\bibinfo {volume} {110}},\ \bibinfo
  {pages} {094517} (\bibinfo {year} {2024})},\ \Eprint
  {https://arxiv.org/abs/2408.16590} {arXiv:2408.16590 [hep-lat]} \BibitemShut
  {NoStop}%
\bibitem [{\citenamefont {Hansen}\ \emph {et~al.}(2024)\citenamefont {Hansen},
  \citenamefont {Romero-L\'opez},\ and\ \citenamefont
  {Sharpe}}]{Hansen:2024ffk}%
  \BibitemOpen
  \bibfield  {author} {\bibinfo {author} {\bibfnamefont {M.~T.}\ \bibnamefont
  {Hansen}}, \bibinfo {author} {\bibfnamefont {F.}~\bibnamefont
  {Romero-L\'opez}},\ and\ \bibinfo {author} {\bibfnamefont {S.~R.}\
  \bibnamefont {Sharpe}},\ }\bibfield  {title} {\bibinfo {title}
  {{Incorporating DD\ensuremath{\pi} effects and left-hand cuts in lattice QCD
  studies of the T$_{cc}$(3875)$^{+}$}},\ }\href
  {https://doi.org/10.1007/JHEP06(2024)051} {\bibfield  {journal} {\bibinfo
  {journal} {JHEP}\ }\textbf {\bibinfo {volume} {06}},\ \bibinfo {pages}
  {051}},\ \Eprint {https://arxiv.org/abs/2401.06609} {arXiv:2401.06609
  [hep-lat]} \BibitemShut {NoStop}%
\bibitem [{\citenamefont {Draper}\ \emph
  {et~al.}(2023{\natexlab{a}})\citenamefont {Draper}, \citenamefont {Hansen},
  \citenamefont {Romero-L\'opez},\ and\ \citenamefont
  {Sharpe}}]{Draper:2023xvu}%
  \BibitemOpen
  \bibfield  {author} {\bibinfo {author} {\bibfnamefont {Z.~T.}\ \bibnamefont
  {Draper}}, \bibinfo {author} {\bibfnamefont {M.~T.}\ \bibnamefont {Hansen}},
  \bibinfo {author} {\bibfnamefont {F.}~\bibnamefont {Romero-L\'opez}},\ and\
  \bibinfo {author} {\bibfnamefont {S.~R.}\ \bibnamefont {Sharpe}},\ }\bibfield
   {title} {\bibinfo {title} {{Three relativistic neutrons in a finite
  volume}},\ }\href {https://doi.org/10.1007/JHEP07(2023)226} {\bibfield
  {journal} {\bibinfo  {journal} {JHEP}\ }\textbf {\bibinfo {volume} {07}},\
  \bibinfo {pages} {226}},\ \Eprint {https://arxiv.org/abs/2303.10219}
  {arXiv:2303.10219 [hep-lat]} \BibitemShut {NoStop}%
\bibitem [{\citenamefont {Feng}\ \emph {et~al.}(2024)\citenamefont {Feng},
  \citenamefont {Gil}, \citenamefont {D\"oring}, \citenamefont {Molina},
  \citenamefont {Mai}, \citenamefont {Shastry},\ and\ \citenamefont
  {Szczepaniak}}]{Feng:2024wyg}%
  \BibitemOpen
  \bibfield  {author} {\bibinfo {author} {\bibfnamefont {Y.}~\bibnamefont
  {Feng}}, \bibinfo {author} {\bibfnamefont {F.}~\bibnamefont {Gil}}, \bibinfo
  {author} {\bibfnamefont {M.}~\bibnamefont {D\"oring}}, \bibinfo {author}
  {\bibfnamefont {R.}~\bibnamefont {Molina}}, \bibinfo {author} {\bibfnamefont
  {M.}~\bibnamefont {Mai}}, \bibinfo {author} {\bibfnamefont {V.}~\bibnamefont
  {Shastry}},\ and\ \bibinfo {author} {\bibfnamefont {A.}~\bibnamefont
  {Szczepaniak}},\ }\bibfield  {title} {\bibinfo {title} {{A unitary
  coupled-channel three-body amplitude with pions and kaons}}\ }\href
  {https://doi.org/10.1103/PhysRevD.110.094002} {10.1103/PhysRevD.110.094002}
  (\bibinfo {year} {2024}),\ \Eprint {https://arxiv.org/abs/2407.08721}
  {arXiv:2407.08721 [nucl-th]} \BibitemShut {NoStop}%
\bibitem [{\citenamefont {Jackura}\ and\ \citenamefont
  {Brice\~no}(2024)}]{Jackura:2023qtp}%
  \BibitemOpen
  \bibfield  {author} {\bibinfo {author} {\bibfnamefont {A.~W.}\ \bibnamefont
  {Jackura}}\ and\ \bibinfo {author} {\bibfnamefont {R.~A.}\ \bibnamefont
  {Brice\~no}},\ }\bibfield  {title} {\bibinfo {title} {{Partial-wave
  projection of the one-particle exchange in three-body scattering
  amplitudes}},\ }\href {https://doi.org/10.1103/PhysRevD.109.096030}
  {\bibfield  {journal} {\bibinfo  {journal} {Phys. Rev. D}\ }\textbf {\bibinfo
  {volume} {109}},\ \bibinfo {pages} {096030} (\bibinfo {year} {2024})},\
  \Eprint {https://arxiv.org/abs/2312.00625} {arXiv:2312.00625 [hep-ph]}
  \BibitemShut {NoStop}%
\bibitem [{\citenamefont {Hansen}\ and\ \citenamefont
  {Sharpe}(2019)}]{Hansen:2019nir}%
  \BibitemOpen
  \bibfield  {author} {\bibinfo {author} {\bibfnamefont {M.~T.}\ \bibnamefont
  {Hansen}}\ and\ \bibinfo {author} {\bibfnamefont {S.~R.}\ \bibnamefont
  {Sharpe}},\ }\bibfield  {title} {\bibinfo {title} {{Lattice QCD and
  Three-particle Decays of Resonances}}\ }\href
  {https://doi.org/10.1146/annurev-nucl-101918-023723}
  {10.1146/annurev-nucl-101918-023723} (\bibinfo {year} {2019}),\ \Eprint
  {https://arxiv.org/abs/1901.00483} {arXiv:1901.00483 [hep-lat]} \BibitemShut
  {NoStop}%
%%CITATION = ARXIV:1901.00483;%%
\bibitem [{\citenamefont {Rusetsky}(2019)}]{Rusetsky:2019gyk}%
  \BibitemOpen
  \bibfield  {author} {\bibinfo {author} {\bibfnamefont {A.}~\bibnamefont
  {Rusetsky}},\ }\bibfield  {title} {\bibinfo {title} {{Three particles on the
  lattice}},\ }\href {https://doi.org/10.22323/1.363.0281} {\bibfield
  {journal} {\bibinfo  {journal} {PoS}\ }\textbf {\bibinfo {volume}
  {LATTICE2019}},\ \bibinfo {pages} {281} (\bibinfo {year} {2019})},\ \Eprint
  {https://arxiv.org/abs/1911.01253} {arXiv:1911.01253 [hep-lat]} \BibitemShut
  {NoStop}%
\bibitem [{\citenamefont {Mai}\ \emph {et~al.}(2021{\natexlab{a}})\citenamefont
  {Mai}, \citenamefont {D\"oring},\ and\ \citenamefont
  {Rusetsky}}]{Mai:2021lwb}%
  \BibitemOpen
  \bibfield  {author} {\bibinfo {author} {\bibfnamefont {M.}~\bibnamefont
  {Mai}}, \bibinfo {author} {\bibfnamefont {M.}~\bibnamefont {D\"oring}},\ and\
  \bibinfo {author} {\bibfnamefont {A.}~\bibnamefont {Rusetsky}},\ }\bibfield
  {title} {\bibinfo {title} {{Multi-particle systems on the lattice and chiral
  extrapolations: a brief review}},\ }\href
  {https://doi.org/10.1140/epjs/s11734-021-00146-5} {\bibfield  {journal}
  {\bibinfo  {journal} {Eur. Phys. J. ST}\ }\textbf {\bibinfo {volume} {230}},\
  \bibinfo {pages} {1623} (\bibinfo {year} {2021}{\natexlab{a}})},\ \Eprint
  {https://arxiv.org/abs/2103.00577} {arXiv:2103.00577 [hep-lat]} \BibitemShut
  {NoStop}%
\bibitem [{\citenamefont {Romero-L\'opez}(2023)}]{Romero-Lopez:2022usb}%
  \BibitemOpen
  \bibfield  {author} {\bibinfo {author} {\bibfnamefont {F.}~\bibnamefont
  {Romero-L\'opez}},\ }\bibfield  {title} {\bibinfo {title} {{Multi-hadron
  interactions from lattice QCD}},\ }\href
  {https://doi.org/10.22323/1.430.0235} {\bibfield  {journal} {\bibinfo
  {journal} {PoS}\ }\textbf {\bibinfo {volume} {LATTICE2022}},\ \bibinfo
  {pages} {235} (\bibinfo {year} {2023})},\ \Eprint
  {https://arxiv.org/abs/2212.13793} {arXiv:2212.13793 [hep-lat]} \BibitemShut
  {NoStop}%
\bibitem [{\citenamefont {Mai}\ \emph {et~al.}(2021{\natexlab{b}})\citenamefont
  {Mai}, \citenamefont {Alexandru}, \citenamefont {Brett}, \citenamefont
  {Culver}, \citenamefont {D\"oring}, \citenamefont {Lee},\ and\ \citenamefont
  {Sadasivan}}]{Mai:2021nul}%
  \BibitemOpen
  \bibfield  {author} {\bibinfo {author} {\bibfnamefont {M.}~\bibnamefont
  {Mai}}, \bibinfo {author} {\bibfnamefont {A.}~\bibnamefont {Alexandru}},
  \bibinfo {author} {\bibfnamefont {R.}~\bibnamefont {Brett}}, \bibinfo
  {author} {\bibfnamefont {C.}~\bibnamefont {Culver}}, \bibinfo {author}
  {\bibfnamefont {M.}~\bibnamefont {D\"oring}}, \bibinfo {author}
  {\bibfnamefont {F.~X.}\ \bibnamefont {Lee}},\ and\ \bibinfo {author}
  {\bibfnamefont {D.}~\bibnamefont {Sadasivan}} (\bibinfo {collaboration}
  {GWQCD}),\ }\bibfield  {title} {\bibinfo {title} {{Three-Body Dynamics of the
  $a_1(1260)$ Resonance from Lattice QCD}},\ }\href
  {https://doi.org/10.1103/PhysRevLett.127.222001} {\bibfield  {journal}
  {\bibinfo  {journal} {Phys. Rev. Lett.}\ }\textbf {\bibinfo {volume} {127}},\
  \bibinfo {pages} {222001} (\bibinfo {year} {2021}{\natexlab{b}})},\ \Eprint
  {https://arxiv.org/abs/2107.03973} {arXiv:2107.03973 [hep-lat]} \BibitemShut
  {NoStop}%
\bibitem [{\citenamefont {Garofalo}\ \emph {et~al.}(2023)\citenamefont
  {Garofalo}, \citenamefont {Mai}, \citenamefont {Romero-L\'opez},
  \citenamefont {Rusetsky},\ and\ \citenamefont {Urbach}}]{Garofalo:2022pux}%
  \BibitemOpen
  \bibfield  {author} {\bibinfo {author} {\bibfnamefont {M.}~\bibnamefont
  {Garofalo}}, \bibinfo {author} {\bibfnamefont {M.}~\bibnamefont {Mai}},
  \bibinfo {author} {\bibfnamefont {F.}~\bibnamefont {Romero-L\'opez}},
  \bibinfo {author} {\bibfnamefont {A.}~\bibnamefont {Rusetsky}},\ and\
  \bibinfo {author} {\bibfnamefont {C.}~\bibnamefont {Urbach}},\ }\bibfield
  {title} {\bibinfo {title} {{Three-body resonances in the
  \ensuremath{\varphi}$^{4}$ theory}},\ }\href
  {https://doi.org/10.1007/JHEP02(2023)252} {\bibfield  {journal} {\bibinfo
  {journal} {JHEP}\ }\textbf {\bibinfo {volume} {02}},\ \bibinfo {pages}
  {252}},\ \Eprint {https://arxiv.org/abs/2211.05605} {arXiv:2211.05605
  [hep-lat]} \BibitemShut {NoStop}%
\bibitem [{\citenamefont {Yan}\ \emph {et~al.}(2024)\citenamefont {Yan},
  \citenamefont {Mai}, \citenamefont {Garofalo}, \citenamefont {Mei\ss{}ner},
  \citenamefont {Liu}, \citenamefont {Liu},\ and\ \citenamefont
  {Urbach}}]{Yan:2024gwp}%
  \BibitemOpen
  \bibfield  {author} {\bibinfo {author} {\bibfnamefont {H.}~\bibnamefont
  {Yan}}, \bibinfo {author} {\bibfnamefont {M.}~\bibnamefont {Mai}}, \bibinfo
  {author} {\bibfnamefont {M.}~\bibnamefont {Garofalo}}, \bibinfo {author}
  {\bibfnamefont {U.-G.}\ \bibnamefont {Mei\ss{}ner}}, \bibinfo {author}
  {\bibfnamefont {C.}~\bibnamefont {Liu}}, \bibinfo {author} {\bibfnamefont
  {L.}~\bibnamefont {Liu}},\ and\ \bibinfo {author} {\bibfnamefont
  {C.}~\bibnamefont {Urbach}},\ }\bibfield  {title} {\bibinfo {title}
  {{\ensuremath{\omega} Meson from Lattice QCD}},\ }\href
  {https://doi.org/10.1103/PhysRevLett.133.211906} {\bibfield  {journal}
  {\bibinfo  {journal} {Phys. Rev. Lett.}\ }\textbf {\bibinfo {volume} {133}},\
  \bibinfo {pages} {211906} (\bibinfo {year} {2024})},\ \Eprint
  {https://arxiv.org/abs/2407.16659} {arXiv:2407.16659 [hep-lat]} \BibitemShut
  {NoStop}%
\bibitem [{\citenamefont {Beane}\ \emph {et~al.}(2008)\citenamefont {Beane},
  \citenamefont {Detmold}, \citenamefont {Luu}, \citenamefont {Orginos},
  \citenamefont {Savage},\ and\ \citenamefont {Torok}}]{Beane:2007es}%
  \BibitemOpen
  \bibfield  {author} {\bibinfo {author} {\bibfnamefont {S.~R.}\ \bibnamefont
  {Beane}}, \bibinfo {author} {\bibfnamefont {W.}~\bibnamefont {Detmold}},
  \bibinfo {author} {\bibfnamefont {T.~C.}\ \bibnamefont {Luu}}, \bibinfo
  {author} {\bibfnamefont {K.}~\bibnamefont {Orginos}}, \bibinfo {author}
  {\bibfnamefont {M.~J.}\ \bibnamefont {Savage}},\ and\ \bibinfo {author}
  {\bibfnamefont {A.}~\bibnamefont {Torok}},\ }\bibfield  {title} {\bibinfo
  {title} {{Multi-Pion Systems in Lattice QCD and the Three-Pion
  Interaction}},\ }\href {https://doi.org/10.1103/PhysRevLett.100.082004}
  {\bibfield  {journal} {\bibinfo  {journal} {Phys. Rev. Lett.}\ }\textbf
  {\bibinfo {volume} {100}},\ \bibinfo {pages} {082004} (\bibinfo {year}
  {2008})},\ \Eprint {https://arxiv.org/abs/0710.1827} {arXiv:0710.1827
  [hep-lat]} \BibitemShut {NoStop}%
\bibitem [{\citenamefont {Detmold}\ \emph
  {et~al.}(2008{\natexlab{a}})\citenamefont {Detmold}, \citenamefont {Savage},
  \citenamefont {Torok}, \citenamefont {Beane}, \citenamefont {Luu},
  \citenamefont {Orginos},\ and\ \citenamefont {Parreno}}]{Detmold:2008fn}%
  \BibitemOpen
  \bibfield  {author} {\bibinfo {author} {\bibfnamefont {W.}~\bibnamefont
  {Detmold}}, \bibinfo {author} {\bibfnamefont {M.~J.}\ \bibnamefont {Savage}},
  \bibinfo {author} {\bibfnamefont {A.}~\bibnamefont {Torok}}, \bibinfo
  {author} {\bibfnamefont {S.~R.}\ \bibnamefont {Beane}}, \bibinfo {author}
  {\bibfnamefont {T.~C.}\ \bibnamefont {Luu}}, \bibinfo {author} {\bibfnamefont
  {K.}~\bibnamefont {Orginos}},\ and\ \bibinfo {author} {\bibfnamefont
  {A.}~\bibnamefont {Parreno}},\ }\bibfield  {title} {\bibinfo {title}
  {{Multi-Pion States in Lattice QCD and the Charged-Pion Condensate}},\ }\href
  {https://doi.org/10.1103/PhysRevD.78.014507} {\bibfield  {journal} {\bibinfo
  {journal} {Phys. Rev. D}\ }\textbf {\bibinfo {volume} {78}},\ \bibinfo
  {pages} {014507} (\bibinfo {year} {2008}{\natexlab{a}})},\ \Eprint
  {https://arxiv.org/abs/0803.2728} {arXiv:0803.2728 [hep-lat]} \BibitemShut
  {NoStop}%
\bibitem [{\citenamefont {Detmold}\ \emph
  {et~al.}(2008{\natexlab{b}})\citenamefont {Detmold}, \citenamefont {Orginos},
  \citenamefont {Savage},\ and\ \citenamefont {Walker-Loud}}]{Detmold:2008yn}%
  \BibitemOpen
  \bibfield  {author} {\bibinfo {author} {\bibfnamefont {W.}~\bibnamefont
  {Detmold}}, \bibinfo {author} {\bibfnamefont {K.}~\bibnamefont {Orginos}},
  \bibinfo {author} {\bibfnamefont {M.~J.}\ \bibnamefont {Savage}},\ and\
  \bibinfo {author} {\bibfnamefont {A.}~\bibnamefont {Walker-Loud}},\
  }\bibfield  {title} {\bibinfo {title} {{Kaon Condensation with Lattice
  QCD}},\ }\href {https://doi.org/10.1103/PhysRevD.78.054514} {\bibfield
  {journal} {\bibinfo  {journal} {Phys. Rev. D}\ }\textbf {\bibinfo {volume}
  {78}},\ \bibinfo {pages} {054514} (\bibinfo {year} {2008}{\natexlab{b}})},\
  \Eprint {https://arxiv.org/abs/0807.1856} {arXiv:0807.1856 [hep-lat]}
  \BibitemShut {NoStop}%
\bibitem [{\citenamefont {Detmold}\ and\ \citenamefont
  {Smigielski}(2011)}]{Detmold:2011kw}%
  \BibitemOpen
  \bibfield  {author} {\bibinfo {author} {\bibfnamefont {W.}~\bibnamefont
  {Detmold}}\ and\ \bibinfo {author} {\bibfnamefont {B.}~\bibnamefont
  {Smigielski}},\ }\bibfield  {title} {\bibinfo {title} {{Lattice QCD study of
  mixed systems of pions and kaons}},\ }\href
  {https://doi.org/10.1103/PhysRevD.84.014508} {\bibfield  {journal} {\bibinfo
  {journal} {Phys. Rev. D}\ }\textbf {\bibinfo {volume} {84}},\ \bibinfo
  {pages} {014508} (\bibinfo {year} {2011})},\ \Eprint
  {https://arxiv.org/abs/1103.4362} {arXiv:1103.4362 [hep-lat]} \BibitemShut
  {NoStop}%
\bibitem [{\citenamefont {Mai}\ and\ \citenamefont
  {Doring}(2019)}]{Mai:2018djl}%
  \BibitemOpen
  \bibfield  {author} {\bibinfo {author} {\bibfnamefont {M.}~\bibnamefont
  {Mai}}\ and\ \bibinfo {author} {\bibfnamefont {M.}~\bibnamefont {Doring}},\
  }\bibfield  {title} {\bibinfo {title} {{Finite-Volume Spectrum of
  $\pi^+\pi^+$ and $\pi^+\pi^+\pi^+$ Systems}},\ }\href
  {https://doi.org/10.1103/PhysRevLett.122.062503} {\bibfield  {journal}
  {\bibinfo  {journal} {Phys. Rev. Lett.}\ }\textbf {\bibinfo {volume} {122}},\
  \bibinfo {pages} {062503} (\bibinfo {year} {2019})},\ \Eprint
  {https://arxiv.org/abs/1807.04746} {arXiv:1807.04746 [hep-lat]} \BibitemShut
  {NoStop}%
%%CITATION = ARXIV:1807.04746;%%
\bibitem [{\citenamefont {H\"orz}\ and\ \citenamefont
  {Hanlon}(2019)}]{Horz:2019rrn}%
  \BibitemOpen
  \bibfield  {author} {\bibinfo {author} {\bibfnamefont {B.}~\bibnamefont
  {H\"orz}}\ and\ \bibinfo {author} {\bibfnamefont {A.}~\bibnamefont
  {Hanlon}},\ }\bibfield  {title} {\bibinfo {title} {{Two- and three-pion
  finite-volume spectra at maximal isospin from lattice QCD}},\ }\href
  {https://doi.org/10.1103/PhysRevLett.123.142002} {\bibfield  {journal}
  {\bibinfo  {journal} {Phys. Rev. Lett.}\ }\textbf {\bibinfo {volume} {123}},\
  \bibinfo {pages} {142002} (\bibinfo {year} {2019})},\ \Eprint
  {https://arxiv.org/abs/1905.04277} {arXiv:1905.04277 [hep-lat]} \BibitemShut
  {NoStop}%
\bibitem [{\citenamefont {Blanton}\ \emph {et~al.}(2020)\citenamefont
  {Blanton}, \citenamefont {Romero-L\'opez},\ and\ \citenamefont
  {Sharpe}}]{Blanton:2019vdk}%
  \BibitemOpen
  \bibfield  {author} {\bibinfo {author} {\bibfnamefont {T.~D.}\ \bibnamefont
  {Blanton}}, \bibinfo {author} {\bibfnamefont {F.}~\bibnamefont
  {Romero-L\'opez}},\ and\ \bibinfo {author} {\bibfnamefont {S.~R.}\
  \bibnamefont {Sharpe}},\ }\bibfield  {title} {\bibinfo {title} {{$I=3$
  Three-Pion Scattering Amplitude from Lattice QCD}},\ }\href
  {https://doi.org/10.1103/PhysRevLett.124.032001} {\bibfield  {journal}
  {\bibinfo  {journal} {Phys. Rev. Lett.}\ }\textbf {\bibinfo {volume} {124}},\
  \bibinfo {pages} {032001} (\bibinfo {year} {2020})},\ \Eprint
  {https://arxiv.org/abs/1909.02973} {arXiv:1909.02973 [hep-lat]} \BibitemShut
  {NoStop}%
\bibitem [{\citenamefont {Culver}\ \emph {et~al.}(2020)\citenamefont {Culver},
  \citenamefont {Mai}, \citenamefont {Brett}, \citenamefont {Alexandru},\ and\
  \citenamefont {D\"oring}}]{Culver:2019vvu}%
  \BibitemOpen
  \bibfield  {author} {\bibinfo {author} {\bibfnamefont {C.}~\bibnamefont
  {Culver}}, \bibinfo {author} {\bibfnamefont {M.}~\bibnamefont {Mai}},
  \bibinfo {author} {\bibfnamefont {R.}~\bibnamefont {Brett}}, \bibinfo
  {author} {\bibfnamefont {A.}~\bibnamefont {Alexandru}},\ and\ \bibinfo
  {author} {\bibfnamefont {M.}~\bibnamefont {D\"oring}},\ }\bibfield  {title}
  {\bibinfo {title} {{Three pion spectrum in the $I=3$ channel from lattice
  QCD}},\ }\href {https://doi.org/10.1103/PhysRevD.101.114507} {\bibfield
  {journal} {\bibinfo  {journal} {Phys. Rev. D}\ }\textbf {\bibinfo {volume}
  {101}},\ \bibinfo {pages} {114507} (\bibinfo {year} {2020})},\ \Eprint
  {https://arxiv.org/abs/1911.09047} {arXiv:1911.09047 [hep-lat]} \BibitemShut
  {NoStop}%
\bibitem [{\citenamefont {Mai}\ \emph {et~al.}(2020)\citenamefont {Mai},
  \citenamefont {D\"oring}, \citenamefont {Culver},\ and\ \citenamefont
  {Alexandru}}]{Mai:2019fba}%
  \BibitemOpen
  \bibfield  {author} {\bibinfo {author} {\bibfnamefont {M.}~\bibnamefont
  {Mai}}, \bibinfo {author} {\bibfnamefont {M.}~\bibnamefont {D\"oring}},
  \bibinfo {author} {\bibfnamefont {C.}~\bibnamefont {Culver}},\ and\ \bibinfo
  {author} {\bibfnamefont {A.}~\bibnamefont {Alexandru}},\ }\bibfield  {title}
  {\bibinfo {title} {{Three-body unitarity versus finite-volume
  $\pi^+\pi^+\pi^+$ spectrum from lattice QCD}},\ }\href
  {https://doi.org/10.1103/PhysRevD.101.054510} {\bibfield  {journal} {\bibinfo
   {journal} {Phys. Rev. D}\ }\textbf {\bibinfo {volume} {101}},\ \bibinfo
  {pages} {054510} (\bibinfo {year} {2020})},\ \Eprint
  {https://arxiv.org/abs/1909.05749} {arXiv:1909.05749 [hep-lat]} \BibitemShut
  {NoStop}%
\bibitem [{\citenamefont {Fischer}\ \emph {et~al.}(2021)\citenamefont
  {Fischer}, \citenamefont {Kostrzewa}, \citenamefont {Liu}, \citenamefont
  {Romero-L\'opez}, \citenamefont {Ueding},\ and\ \citenamefont
  {Urbach}}]{Fischer:2020jzp}%
  \BibitemOpen
  \bibfield  {author} {\bibinfo {author} {\bibfnamefont {M.}~\bibnamefont
  {Fischer}}, \bibinfo {author} {\bibfnamefont {B.}~\bibnamefont {Kostrzewa}},
  \bibinfo {author} {\bibfnamefont {L.}~\bibnamefont {Liu}}, \bibinfo {author}
  {\bibfnamefont {F.}~\bibnamefont {Romero-L\'opez}}, \bibinfo {author}
  {\bibfnamefont {M.}~\bibnamefont {Ueding}},\ and\ \bibinfo {author}
  {\bibfnamefont {C.}~\bibnamefont {Urbach}},\ }\bibfield  {title} {\bibinfo
  {title} {{Scattering of two and three physical pions at maximal isospin from
  lattice QCD}},\ }\href {https://doi.org/10.1140/epjc/s10052-021-09206-5}
  {\bibfield  {journal} {\bibinfo  {journal} {Eur. Phys. J. C}\ }\textbf
  {\bibinfo {volume} {81}},\ \bibinfo {pages} {436} (\bibinfo {year} {2021})},\
  \Eprint {https://arxiv.org/abs/2008.03035} {arXiv:2008.03035 [hep-lat]}
  \BibitemShut {NoStop}%
\bibitem [{\citenamefont {Hansen}\ \emph {et~al.}(2021)\citenamefont {Hansen},
  \citenamefont {Brice\~no}, \citenamefont {Edwards}, \citenamefont {Thomas},\
  and\ \citenamefont {Wilson}}]{Hansen:2020otl}%
  \BibitemOpen
  \bibfield  {author} {\bibinfo {author} {\bibfnamefont {M.~T.}\ \bibnamefont
  {Hansen}}, \bibinfo {author} {\bibfnamefont {R.~A.}\ \bibnamefont
  {Brice\~no}}, \bibinfo {author} {\bibfnamefont {R.~G.}\ \bibnamefont
  {Edwards}}, \bibinfo {author} {\bibfnamefont {C.~E.}\ \bibnamefont
  {Thomas}},\ and\ \bibinfo {author} {\bibfnamefont {D.~J.}\ \bibnamefont
  {Wilson}} (\bibinfo {collaboration} {Hadron Spectrum}),\ }\bibfield  {title}
  {\bibinfo {title} {{Energy-Dependent $\pi^+ \pi^+ \pi^+$ Scattering Amplitude
  from QCD}},\ }\href {https://doi.org/10.1103/PhysRevLett.126.012001}
  {\bibfield  {journal} {\bibinfo  {journal} {Phys. Rev. Lett.}\ }\textbf
  {\bibinfo {volume} {126}},\ \bibinfo {pages} {012001} (\bibinfo {year}
  {2021})},\ \Eprint {https://arxiv.org/abs/2009.04931} {arXiv:2009.04931
  [hep-lat]} \BibitemShut {NoStop}%
\bibitem [{\citenamefont {Alexandru}\ \emph {et~al.}(2020)\citenamefont
  {Alexandru}, \citenamefont {Brett}, \citenamefont {Culver}, \citenamefont
  {D\"oring}, \citenamefont {Guo}, \citenamefont {Lee},\ and\ \citenamefont
  {Mai}}]{Alexandru:2020xqf}%
  \BibitemOpen
  \bibfield  {author} {\bibinfo {author} {\bibfnamefont {A.}~\bibnamefont
  {Alexandru}}, \bibinfo {author} {\bibfnamefont {R.}~\bibnamefont {Brett}},
  \bibinfo {author} {\bibfnamefont {C.}~\bibnamefont {Culver}}, \bibinfo
  {author} {\bibfnamefont {M.}~\bibnamefont {D\"oring}}, \bibinfo {author}
  {\bibfnamefont {D.}~\bibnamefont {Guo}}, \bibinfo {author} {\bibfnamefont
  {F.~X.}\ \bibnamefont {Lee}},\ and\ \bibinfo {author} {\bibfnamefont
  {M.}~\bibnamefont {Mai}},\ }\bibfield  {title} {\bibinfo {title}
  {{Finite-volume energy spectrum of the $K^-K^-K^-$ system}},\ }\href
  {https://doi.org/10.1103/PhysRevD.102.114523} {\bibfield  {journal} {\bibinfo
   {journal} {Phys. Rev. D}\ }\textbf {\bibinfo {volume} {102}},\ \bibinfo
  {pages} {114523} (\bibinfo {year} {2020})},\ \Eprint
  {https://arxiv.org/abs/2009.12358} {arXiv:2009.12358 [hep-lat]} \BibitemShut
  {NoStop}%
\bibitem [{\citenamefont {Brett}\ \emph {et~al.}(2021)\citenamefont {Brett},
  \citenamefont {Culver}, \citenamefont {Mai}, \citenamefont {Alexandru},
  \citenamefont {D\"oring},\ and\ \citenamefont {Lee}}]{Brett:2021wyd}%
  \BibitemOpen
  \bibfield  {author} {\bibinfo {author} {\bibfnamefont {R.}~\bibnamefont
  {Brett}}, \bibinfo {author} {\bibfnamefont {C.}~\bibnamefont {Culver}},
  \bibinfo {author} {\bibfnamefont {M.}~\bibnamefont {Mai}}, \bibinfo {author}
  {\bibfnamefont {A.}~\bibnamefont {Alexandru}}, \bibinfo {author}
  {\bibfnamefont {M.}~\bibnamefont {D\"oring}},\ and\ \bibinfo {author}
  {\bibfnamefont {F.~X.}\ \bibnamefont {Lee}},\ }\bibfield  {title} {\bibinfo
  {title} {{Three-body interactions from the finite-volume QCD spectrum}},\
  }\href {https://doi.org/10.1103/PhysRevD.104.014501} {\bibfield  {journal}
  {\bibinfo  {journal} {Phys. Rev. D}\ }\textbf {\bibinfo {volume} {104}},\
  \bibinfo {pages} {014501} (\bibinfo {year} {2021})},\ \Eprint
  {https://arxiv.org/abs/2101.06144} {arXiv:2101.06144 [hep-lat]} \BibitemShut
  {NoStop}%
\bibitem [{\citenamefont {Blanton}\ \emph {et~al.}(2021)\citenamefont
  {Blanton}, \citenamefont {Hanlon}, \citenamefont {H\"orz}, \citenamefont
  {Morningstar}, \citenamefont {Romero-L\'opez},\ and\ \citenamefont
  {Sharpe}}]{Blanton:2021llb}%
  \BibitemOpen
  \bibfield  {author} {\bibinfo {author} {\bibfnamefont {T.~D.}\ \bibnamefont
  {Blanton}}, \bibinfo {author} {\bibfnamefont {A.~D.}\ \bibnamefont {Hanlon}},
  \bibinfo {author} {\bibfnamefont {B.}~\bibnamefont {H\"orz}}, \bibinfo
  {author} {\bibfnamefont {C.}~\bibnamefont {Morningstar}}, \bibinfo {author}
  {\bibfnamefont {F.}~\bibnamefont {Romero-L\'opez}},\ and\ \bibinfo {author}
  {\bibfnamefont {S.~R.}\ \bibnamefont {Sharpe}},\ }\bibfield  {title}
  {\bibinfo {title} {{Interactions of two and three mesons including higher
  partial waves from lattice QCD}},\ }\href
  {https://doi.org/10.1007/JHEP10(2021)023} {\bibfield  {journal} {\bibinfo
  {journal} {JHEP}\ }\textbf {\bibinfo {volume} {10}},\ \bibinfo {pages}
  {023}},\ \Eprint {https://arxiv.org/abs/2106.05590} {arXiv:2106.05590
  [hep-lat]} \BibitemShut {NoStop}%
\bibitem [{\citenamefont {Beane}\ \emph {et~al.}(2021)\citenamefont {Beane}
  \emph {et~al.}}]{NPLQCD:2020ozd}%
  \BibitemOpen
  \bibfield  {author} {\bibinfo {author} {\bibfnamefont {S.~R.}\ \bibnamefont
  {Beane}} \emph {et~al.} (\bibinfo {collaboration} {NPLQCD, QCDSF}),\
  }\bibfield  {title} {\bibinfo {title} {{Charged multihadron systems in
  lattice QCD+QED}},\ }\href {https://doi.org/10.1103/PhysRevD.103.054504}
  {\bibfield  {journal} {\bibinfo  {journal} {Phys. Rev. D}\ }\textbf {\bibinfo
  {volume} {103}},\ \bibinfo {pages} {054504} (\bibinfo {year} {2021})},\
  \Eprint {https://arxiv.org/abs/2003.12130} {arXiv:2003.12130 [hep-lat]}
  \BibitemShut {NoStop}%
\bibitem [{\citenamefont {Abbott}\ \emph {et~al.}(2023)\citenamefont {Abbott},
  \citenamefont {Detmold}, \citenamefont {Romero-L\'opez}, \citenamefont
  {Davoudi}, \citenamefont {Illa}, \citenamefont {Parre\~no}, \citenamefont
  {Perry}, \citenamefont {Shanahan},\ and\ \citenamefont
  {Wagman}}]{Abbott:2023coj}%
  \BibitemOpen
  \bibfield  {author} {\bibinfo {author} {\bibfnamefont {R.}~\bibnamefont
  {Abbott}}, \bibinfo {author} {\bibfnamefont {W.}~\bibnamefont {Detmold}},
  \bibinfo {author} {\bibfnamefont {F.}~\bibnamefont {Romero-L\'opez}},
  \bibinfo {author} {\bibfnamefont {Z.}~\bibnamefont {Davoudi}}, \bibinfo
  {author} {\bibfnamefont {M.}~\bibnamefont {Illa}}, \bibinfo {author}
  {\bibfnamefont {A.}~\bibnamefont {Parre\~no}}, \bibinfo {author}
  {\bibfnamefont {R.~J.}\ \bibnamefont {Perry}}, \bibinfo {author}
  {\bibfnamefont {P.~E.}\ \bibnamefont {Shanahan}},\ and\ \bibinfo {author}
  {\bibfnamefont {M.~L.}\ \bibnamefont {Wagman}} (\bibinfo {collaboration}
  {NPLQCD}),\ }\bibfield  {title} {\bibinfo {title} {{Lattice quantum
  chromodynamics at large isospin density}},\ }\href
  {https://doi.org/10.1103/PhysRevD.108.114506} {\bibfield  {journal} {\bibinfo
   {journal} {Phys. Rev. D}\ }\textbf {\bibinfo {volume} {108}},\ \bibinfo
  {pages} {114506} (\bibinfo {year} {2023})},\ \Eprint
  {https://arxiv.org/abs/2307.15014} {arXiv:2307.15014 [hep-lat]} \BibitemShut
  {NoStop}%
\bibitem [{\citenamefont {Abbott}\ \emph {et~al.}(2025)\citenamefont {Abbott},
  \citenamefont {Detmold}, \citenamefont {Illa}, \citenamefont {Parre\~no},
  \citenamefont {Perry}, \citenamefont {Romero-L\'opez}, \citenamefont
  {Shanahan},\ and\ \citenamefont {Wagman}}]{Abbott:2024vhj}%
  \BibitemOpen
  \bibfield  {author} {\bibinfo {author} {\bibfnamefont {R.}~\bibnamefont
  {Abbott}}, \bibinfo {author} {\bibfnamefont {W.}~\bibnamefont {Detmold}},
  \bibinfo {author} {\bibfnamefont {M.}~\bibnamefont {Illa}}, \bibinfo {author}
  {\bibfnamefont {A.}~\bibnamefont {Parre\~no}}, \bibinfo {author}
  {\bibfnamefont {R.~J.}\ \bibnamefont {Perry}}, \bibinfo {author}
  {\bibfnamefont {F.}~\bibnamefont {Romero-L\'opez}}, \bibinfo {author}
  {\bibfnamefont {P.~E.}\ \bibnamefont {Shanahan}},\ and\ \bibinfo {author}
  {\bibfnamefont {M.~L.}\ \bibnamefont {Wagman}} (\bibinfo {collaboration}
  {NPLQCD}),\ }\bibfield  {title} {\bibinfo {title} {{QCD Constraints on
  Isospin-Dense Matter and the Nuclear Equation of State}},\ }\href
  {https://doi.org/10.1103/PhysRevLett.134.011903} {\bibfield  {journal}
  {\bibinfo  {journal} {Phys. Rev. Lett.}\ }\textbf {\bibinfo {volume} {134}},\
  \bibinfo {pages} {011903} (\bibinfo {year} {2025})},\ \Eprint
  {https://arxiv.org/abs/2406.09273} {arXiv:2406.09273 [hep-lat]} \BibitemShut
  {NoStop}%
\bibitem [{\citenamefont {Kim}\ \emph {et~al.}(2005)\citenamefont {Kim},
  \citenamefont {Sachrajda},\ and\ \citenamefont {Sharpe}}]{Kim:2005gf}%
  \BibitemOpen
  \bibfield  {author} {\bibinfo {author} {\bibfnamefont {C.~H.}\ \bibnamefont
  {Kim}}, \bibinfo {author} {\bibfnamefont {C.~T.}\ \bibnamefont {Sachrajda}},\
  and\ \bibinfo {author} {\bibfnamefont {S.~R.}\ \bibnamefont {Sharpe}},\
  }\bibfield  {title} {\bibinfo {title} {{Finite-volume effects for two-hadron
  states in moving frames}},\ }\href
  {https://doi.org/10.1016/j.nuclphysb.2005.08.029} {\bibfield  {journal}
  {\bibinfo  {journal} {Nucl. Phys.}\ }\textbf {\bibinfo {volume} {B727}},\
  \bibinfo {pages} {218} (\bibinfo {year} {2005})},\ \Eprint
  {https://arxiv.org/abs/hep-lat/0507006} {arXiv:hep-lat/0507006 [hep-lat]}
  \BibitemShut {NoStop}%
%%CITATION = HEP-LAT/0507006;%%
\bibitem [{\citenamefont {Blanton}\ \emph {et~al.}(2022)\citenamefont
  {Blanton}, \citenamefont {Romero-L\'opez},\ and\ \citenamefont
  {Sharpe}}]{Blanton:2021eyf}%
  \BibitemOpen
  \bibfield  {author} {\bibinfo {author} {\bibfnamefont {T.~D.}\ \bibnamefont
  {Blanton}}, \bibinfo {author} {\bibfnamefont {F.}~\bibnamefont
  {Romero-L\'opez}},\ and\ \bibinfo {author} {\bibfnamefont {S.~R.}\
  \bibnamefont {Sharpe}},\ }\bibfield  {title} {\bibinfo {title} {{Implementing
  the three-particle quantization condition for
  \ensuremath{\pi}$^{+}$\ensuremath{\pi}$^{+}$K$^{+}$ and related systems}},\
  }\href {https://doi.org/10.1007/JHEP02(2022)098} {\bibfield  {journal}
  {\bibinfo  {journal} {JHEP}\ }\textbf {\bibinfo {volume} {02}},\ \bibinfo
  {pages} {098}},\ \Eprint {https://arxiv.org/abs/2111.12734} {arXiv:2111.12734
  [hep-lat]} \BibitemShut {NoStop}%
\bibitem [{\citenamefont {Dawid}\ \emph
  {et~al.}(2025{\natexlab{a}})\citenamefont {Dawid}, \citenamefont {Draper},
  \citenamefont {Hanlon}, \citenamefont {H\"orz}, \citenamefont {Morningstar},
  \citenamefont {Romero-L\'opez}, \citenamefont {Sharpe},\ and\ \citenamefont
  {Skinner}}]{Dawid:2025zxc}%
  \BibitemOpen
  \bibfield  {author} {\bibinfo {author} {\bibfnamefont {S.~M.}\ \bibnamefont
  {Dawid}}, \bibinfo {author} {\bibfnamefont {Z.~T.}\ \bibnamefont {Draper}},
  \bibinfo {author} {\bibfnamefont {A.~D.}\ \bibnamefont {Hanlon}}, \bibinfo
  {author} {\bibfnamefont {B.}~\bibnamefont {H\"orz}}, \bibinfo {author}
  {\bibfnamefont {C.}~\bibnamefont {Morningstar}}, \bibinfo {author}
  {\bibfnamefont {F.}~\bibnamefont {Romero-L\'opez}}, \bibinfo {author}
  {\bibfnamefont {S.~R.}\ \bibnamefont {Sharpe}},\ and\ \bibinfo {author}
  {\bibfnamefont {S.}~\bibnamefont {Skinner}},\ }\bibfield  {title} {\bibinfo
  {title} {{QCD predictions for physical multimeson scattering amplitudes}},\
  }\href@noop {} {\  (\bibinfo {year} {2025}{\natexlab{a}})},\ \Eprint
  {https://arxiv.org/abs/2502.14348} {arXiv:2502.14348 [hep-lat]} \BibitemShut
  {NoStop}%
\bibitem [{\citenamefont {Draper}\ \emph
  {et~al.}(2023{\natexlab{b}})\citenamefont {Draper}, \citenamefont {Hanlon},
  \citenamefont {H\"orz}, \citenamefont {Morningstar}, \citenamefont
  {Romero-L\'opez},\ and\ \citenamefont {Sharpe}}]{Draper:2023boj}%
  \BibitemOpen
  \bibfield  {author} {\bibinfo {author} {\bibfnamefont {Z.~T.}\ \bibnamefont
  {Draper}}, \bibinfo {author} {\bibfnamefont {A.~D.}\ \bibnamefont {Hanlon}},
  \bibinfo {author} {\bibfnamefont {B.}~\bibnamefont {H\"orz}}, \bibinfo
  {author} {\bibfnamefont {C.}~\bibnamefont {Morningstar}}, \bibinfo {author}
  {\bibfnamefont {F.}~\bibnamefont {Romero-L\'opez}},\ and\ \bibinfo {author}
  {\bibfnamefont {S.~R.}\ \bibnamefont {Sharpe}},\ }\bibfield  {title}
  {\bibinfo {title} {{Interactions of \ensuremath{\pi}K,
  \ensuremath{\pi}\ensuremath{\pi}K and KK\ensuremath{\pi} systems at maximal
  isospin from lattice QCD}},\ }\href {https://doi.org/10.1007/JHEP05(2023)137}
  {\bibfield  {journal} {\bibinfo  {journal} {JHEP}\ }\textbf {\bibinfo
  {volume} {05}},\ \bibinfo {pages} {137}},\ \Eprint
  {https://arxiv.org/abs/2302.13587} {arXiv:2302.13587 [hep-lat]} \BibitemShut
  {NoStop}%
\bibitem [{\citenamefont {Bruno}\ \emph {et~al.}(2015)\citenamefont {Bruno}
  \emph {et~al.}}]{Bruno:2014jqa}%
  \BibitemOpen
  \bibfield  {author} {\bibinfo {author} {\bibfnamefont {M.}~\bibnamefont
  {Bruno}} \emph {et~al.},\ }\bibfield  {title} {\bibinfo {title} {{Simulation
  of QCD with N$_{f} =$ 2 $+$ 1 flavors of non-perturbatively improved Wilson
  fermions}},\ }\href {https://doi.org/10.1007/JHEP02(2015)043} {\bibfield
  {journal} {\bibinfo  {journal} {JHEP}\ }\textbf {\bibinfo {volume} {02}},\
  \bibinfo {pages} {043}},\ \Eprint {https://arxiv.org/abs/1411.3982}
  {arXiv:1411.3982 [hep-lat]} \BibitemShut {NoStop}%
%%CITATION = ARXIV:1411.3982;%%
\bibitem [{\citenamefont {Bruno}\ \emph {et~al.}(2017)\citenamefont {Bruno},
  \citenamefont {Korzec},\ and\ \citenamefont {Schaefer}}]{Bruno:2016plf}%
  \BibitemOpen
  \bibfield  {author} {\bibinfo {author} {\bibfnamefont {M.}~\bibnamefont
  {Bruno}}, \bibinfo {author} {\bibfnamefont {T.}~\bibnamefont {Korzec}},\ and\
  \bibinfo {author} {\bibfnamefont {S.}~\bibnamefont {Schaefer}},\ }\bibfield
  {title} {\bibinfo {title} {{Setting the scale for the CLS $2 + 1$ flavor
  ensembles}},\ }\href {https://doi.org/10.1103/PhysRevD.95.074504} {\bibfield
  {journal} {\bibinfo  {journal} {Phys. Rev.}\ }\textbf {\bibinfo {volume}
  {D95}},\ \bibinfo {pages} {074504} (\bibinfo {year} {2017})},\ \Eprint
  {https://arxiv.org/abs/1608.08900} {arXiv:1608.08900 [hep-lat]} \BibitemShut
  {NoStop}%
%%CITATION = ARXIV:1608.08900;%%
\bibitem [{\citenamefont {Strassberger}\ \emph {et~al.}(2022)\citenamefont
  {Strassberger} \emph {et~al.}}]{Strassberger:2021tsu}%
  \BibitemOpen
  \bibfield  {author} {\bibinfo {author} {\bibfnamefont {B.}~\bibnamefont
  {Strassberger}} \emph {et~al.},\ }\bibfield  {title} {\bibinfo {title}
  {{Scale setting for CLS 2+1 simulations}},\ }\href
  {https://doi.org/10.22323/1.396.0135} {\bibfield  {journal} {\bibinfo
  {journal} {PoS}\ }\textbf {\bibinfo {volume} {LATTICE2021}},\ \bibinfo
  {pages} {135} (\bibinfo {year} {2022})},\ \Eprint
  {https://arxiv.org/abs/2112.06696} {arXiv:2112.06696 [hep-lat]} \BibitemShut
  {NoStop}%
\bibitem [{\citenamefont {Bali}\ \emph {et~al.}(2023)\citenamefont {Bali},
  \citenamefont {Collins}, \citenamefont {Georg}, \citenamefont {Jenkins},
  \citenamefont {Korcyl}, \citenamefont {Sch\"afer}, \citenamefont {Scholz},
  \citenamefont {Simeth}, \citenamefont {S\"oldner},\ and\ \citenamefont
  {Weish\"aupl}}]{RQCD:2022xux}%
  \BibitemOpen
  \bibfield  {author} {\bibinfo {author} {\bibfnamefont {G.~S.}\ \bibnamefont
  {Bali}}, \bibinfo {author} {\bibfnamefont {S.}~\bibnamefont {Collins}},
  \bibinfo {author} {\bibfnamefont {P.}~\bibnamefont {Georg}}, \bibinfo
  {author} {\bibfnamefont {D.}~\bibnamefont {Jenkins}}, \bibinfo {author}
  {\bibfnamefont {P.}~\bibnamefont {Korcyl}}, \bibinfo {author} {\bibfnamefont
  {A.}~\bibnamefont {Sch\"afer}}, \bibinfo {author} {\bibfnamefont {E.~E.}\
  \bibnamefont {Scholz}}, \bibinfo {author} {\bibfnamefont {J.}~\bibnamefont
  {Simeth}}, \bibinfo {author} {\bibfnamefont {W.}~\bibnamefont {S\"oldner}},\
  and\ \bibinfo {author} {\bibfnamefont {S.}~\bibnamefont {Weish\"aupl}}
  (\bibinfo {collaboration} {RQCD}),\ }\bibfield  {title} {\bibinfo {title}
  {{Scale setting and the light baryon spectrum in N$_{f}$ = 2 + 1 QCD with
  Wilson fermions}},\ }\href {https://doi.org/10.1007/JHEP05(2023)035}
  {\bibfield  {journal} {\bibinfo  {journal} {JHEP}\ }\textbf {\bibinfo
  {volume} {05}},\ \bibinfo {pages} {035}},\ \Eprint
  {https://arxiv.org/abs/2211.03744} {arXiv:2211.03744 [hep-lat]} \BibitemShut
  {NoStop}%
\bibitem [{\citenamefont {Luscher}\ and\ \citenamefont
  {Schaefer}(2013)}]{Luscher:2012av}%
  \BibitemOpen
  \bibfield  {author} {\bibinfo {author} {\bibfnamefont {M.}~\bibnamefont
  {Luscher}}\ and\ \bibinfo {author} {\bibfnamefont {S.}~\bibnamefont
  {Schaefer}},\ }\bibfield  {title} {\bibinfo {title} {{Lattice QCD with open
  boundary conditions and twisted-mass reweighting}},\ }\href
  {https://doi.org/10.1016/j.cpc.2012.10.003} {\bibfield  {journal} {\bibinfo
  {journal} {Comput. Phys. Commun.}\ }\textbf {\bibinfo {volume} {184}},\
  \bibinfo {pages} {519} (\bibinfo {year} {2013})},\ \Eprint
  {https://arxiv.org/abs/1206.2809} {arXiv:1206.2809 [hep-lat]} \BibitemShut
  {NoStop}%
\bibitem [{\citenamefont {Morningstar}\ \emph {et~al.}(2011)\citenamefont
  {Morningstar}, \citenamefont {Bulava}, \citenamefont {Foley}, \citenamefont
  {Juge}, \citenamefont {Lenkner}, \citenamefont {Peardon},\ and\ \citenamefont
  {Wong}}]{Morningstar:2011ka}%
  \BibitemOpen
  \bibfield  {author} {\bibinfo {author} {\bibfnamefont {C.}~\bibnamefont
  {Morningstar}}, \bibinfo {author} {\bibfnamefont {J.}~\bibnamefont {Bulava}},
  \bibinfo {author} {\bibfnamefont {J.}~\bibnamefont {Foley}}, \bibinfo
  {author} {\bibfnamefont {K.~J.}\ \bibnamefont {Juge}}, \bibinfo {author}
  {\bibfnamefont {D.}~\bibnamefont {Lenkner}}, \bibinfo {author} {\bibfnamefont
  {M.}~\bibnamefont {Peardon}},\ and\ \bibinfo {author} {\bibfnamefont {C.~H.}\
  \bibnamefont {Wong}},\ }\bibfield  {title} {\bibinfo {title} {{Improved
  stochastic estimation of quark propagation with Laplacian Heaviside smearing
  in lattice QCD}},\ }\href {https://doi.org/10.1103/PhysRevD.83.114505}
  {\bibfield  {journal} {\bibinfo  {journal} {Phys. Rev.}\ }\textbf {\bibinfo
  {volume} {D83}},\ \bibinfo {pages} {114505} (\bibinfo {year} {2011})},\
  \Eprint {https://arxiv.org/abs/1104.3870} {arXiv:1104.3870 [hep-lat]}
  \BibitemShut {NoStop}%
%%CITATION = ARXIV:1104.3870;%%
\bibitem [{\citenamefont {C\`e}\ \emph {et~al.}(2022)\citenamefont {C\`e},
  \citenamefont {G\'erardin}, \citenamefont {von Hippel}, \citenamefont
  {Meyer}, \citenamefont {Miura}, \citenamefont {Ottnad}, \citenamefont
  {Risch}, \citenamefont {San~Jos\'e}, \citenamefont {Wilhelm},\ and\
  \citenamefont {Wittig}}]{Ce:2022eix}%
  \BibitemOpen
  \bibfield  {author} {\bibinfo {author} {\bibfnamefont {M.}~\bibnamefont
  {C\`e}}, \bibinfo {author} {\bibfnamefont {A.}~\bibnamefont {G\'erardin}},
  \bibinfo {author} {\bibfnamefont {G.}~\bibnamefont {von Hippel}}, \bibinfo
  {author} {\bibfnamefont {H.~B.}\ \bibnamefont {Meyer}}, \bibinfo {author}
  {\bibfnamefont {K.}~\bibnamefont {Miura}}, \bibinfo {author} {\bibfnamefont
  {K.}~\bibnamefont {Ottnad}}, \bibinfo {author} {\bibfnamefont
  {A.}~\bibnamefont {Risch}}, \bibinfo {author} {\bibfnamefont
  {T.}~\bibnamefont {San~Jos\'e}}, \bibinfo {author} {\bibfnamefont
  {J.}~\bibnamefont {Wilhelm}},\ and\ \bibinfo {author} {\bibfnamefont
  {H.}~\bibnamefont {Wittig}},\ }\bibfield  {title} {\bibinfo {title} {{The
  hadronic running of the electromagnetic coupling and the electroweak mixing
  angle from lattice QCD}},\ }\href {https://doi.org/10.1007/JHEP08(2022)220}
  {\bibfield  {journal} {\bibinfo  {journal} {JHEP}\ }\textbf {\bibinfo
  {volume} {08}},\ \bibinfo {pages} {220}},\ \Eprint
  {https://arxiv.org/abs/2203.08676} {arXiv:2203.08676 [hep-lat]} \BibitemShut
  {NoStop}%
\bibitem [{\citenamefont {Luscher}\ and\ \citenamefont
  {Wolff}(1990)}]{Luscher:1990ck}%
  \BibitemOpen
  \bibfield  {author} {\bibinfo {author} {\bibfnamefont {M.}~\bibnamefont
  {Luscher}}\ and\ \bibinfo {author} {\bibfnamefont {U.}~\bibnamefont
  {Wolff}},\ }\bibfield  {title} {\bibinfo {title} {{How to Calculate the
  Elastic Scattering Matrix in Two-dimensional Quantum Field Theories by
  Numerical Simulation}},\ }\href
  {https://doi.org/10.1016/0550-3213(90)90540-T} {\bibfield  {journal}
  {\bibinfo  {journal} {Nucl. Phys.}\ }\textbf {\bibinfo {volume} {B339}},\
  \bibinfo {pages} {222} (\bibinfo {year} {1990})}\BibitemShut {NoStop}%
%%CITATION = NUPHA,B339,222;%%
\bibitem [{\citenamefont {Blossier}\ \emph {et~al.}(2009)\citenamefont
  {Blossier}, \citenamefont {Della~Morte}, \citenamefont {von Hippel},
  \citenamefont {Mendes},\ and\ \citenamefont {Sommer}}]{Blossier:2009kd}%
  \BibitemOpen
  \bibfield  {author} {\bibinfo {author} {\bibfnamefont {B.}~\bibnamefont
  {Blossier}}, \bibinfo {author} {\bibfnamefont {M.}~\bibnamefont
  {Della~Morte}}, \bibinfo {author} {\bibfnamefont {G.}~\bibnamefont {von
  Hippel}}, \bibinfo {author} {\bibfnamefont {T.}~\bibnamefont {Mendes}},\ and\
  \bibinfo {author} {\bibfnamefont {R.}~\bibnamefont {Sommer}},\ }\bibfield
  {title} {\bibinfo {title} {{On the generalized eigenvalue method for energies
  and matrix elements in lattice field theory}},\ }\href
  {https://doi.org/10.1088/1126-6708/2009/04/094} {\bibfield  {journal}
  {\bibinfo  {journal} {JHEP}\ }\textbf {\bibinfo {volume} {04}},\ \bibinfo
  {pages} {094}},\ \Eprint {https://arxiv.org/abs/0902.1265} {arXiv:0902.1265
  [hep-lat]} \BibitemShut {NoStop}%
%%CITATION = ARXIV:0902.1265;%%
\bibitem [{\citenamefont {Morningstar}\ \emph {et~al.}(2013)\citenamefont
  {Morningstar}, \citenamefont {Bulava}, \citenamefont {Fahy}, \citenamefont
  {Foley}, \citenamefont {Jhang} \emph {et~al.}}]{Morningstar:2013bda}%
  \BibitemOpen
  \bibfield  {author} {\bibinfo {author} {\bibfnamefont {C.}~\bibnamefont
  {Morningstar}}, \bibinfo {author} {\bibfnamefont {J.}~\bibnamefont {Bulava}},
  \bibinfo {author} {\bibfnamefont {B.}~\bibnamefont {Fahy}}, \bibinfo {author}
  {\bibfnamefont {J.}~\bibnamefont {Foley}}, \bibinfo {author} {\bibfnamefont
  {Y.}~\bibnamefont {Jhang}}, \emph {et~al.},\ }\bibfield  {title} {\bibinfo
  {title} {{Extended hadron and two-hadron operators of definite momentum for
  spectrum calculations in lattice QCD}},\ }\href
  {https://doi.org/10.1103/PhysRevD.88.014511} {\bibfield  {journal} {\bibinfo
  {journal} {Phys.Rev.}\ }\textbf {\bibinfo {volume} {D88}},\ \bibinfo {pages}
  {014511} (\bibinfo {year} {2013})},\ \Eprint
  {https://arxiv.org/abs/1303.6816} {arXiv:1303.6816 [hep-lat]} \BibitemShut
  {NoStop}%
%%CITATION = ARXIV:1303.6816;%%
\bibitem [{\citenamefont {Peardon}\ \emph {et~al.}(2009)\citenamefont
  {Peardon}, \citenamefont {Bulava}, \citenamefont {Foley}, \citenamefont
  {Morningstar}, \citenamefont {Dudek}, \citenamefont {Edwards}, \citenamefont
  {Joo}, \citenamefont {Lin}, \citenamefont {Richards},\ and\ \citenamefont
  {Juge}}]{Peardon:2009gh}%
  \BibitemOpen
  \bibfield  {author} {\bibinfo {author} {\bibfnamefont {M.}~\bibnamefont
  {Peardon}}, \bibinfo {author} {\bibfnamefont {J.}~\bibnamefont {Bulava}},
  \bibinfo {author} {\bibfnamefont {J.}~\bibnamefont {Foley}}, \bibinfo
  {author} {\bibfnamefont {C.}~\bibnamefont {Morningstar}}, \bibinfo {author}
  {\bibfnamefont {J.}~\bibnamefont {Dudek}}, \bibinfo {author} {\bibfnamefont
  {R.~G.}\ \bibnamefont {Edwards}}, \bibinfo {author} {\bibfnamefont
  {B.}~\bibnamefont {Joo}}, \bibinfo {author} {\bibfnamefont {H.-W.}\
  \bibnamefont {Lin}}, \bibinfo {author} {\bibfnamefont {D.~G.}\ \bibnamefont
  {Richards}},\ and\ \bibinfo {author} {\bibfnamefont {K.~J.}\ \bibnamefont
  {Juge}} (\bibinfo {collaboration} {Hadron Spectrum}),\ }\bibfield  {title}
  {\bibinfo {title} {{A Novel quark-field creation operator construction for
  hadronic physics in lattice QCD}},\ }\href
  {https://doi.org/10.1103/PhysRevD.80.054506} {\bibfield  {journal} {\bibinfo
  {journal} {Phys. Rev.}\ }\textbf {\bibinfo {volume} {D80}},\ \bibinfo {pages}
  {054506} (\bibinfo {year} {2009})},\ \Eprint
  {https://arxiv.org/abs/0905.2160} {arXiv:0905.2160 [hep-lat]} \BibitemShut
  {NoStop}%
%%CITATION = ARXIV:0905.2160;%%
\bibitem [{\citenamefont {Bulava}\ \emph {et~al.}(2023)\citenamefont {Bulava},
  \citenamefont {Hanlon}, \citenamefont {H\"orz}, \citenamefont {Morningstar},
  \citenamefont {Nicholson}, \citenamefont {Romero-L\'opez}, \citenamefont
  {Skinner}, \citenamefont {Vranas},\ and\ \citenamefont
  {Walker-Loud}}]{Bulava:2022vpq}%
  \BibitemOpen
  \bibfield  {author} {\bibinfo {author} {\bibfnamefont {J.}~\bibnamefont
  {Bulava}}, \bibinfo {author} {\bibfnamefont {A.~D.}\ \bibnamefont {Hanlon}},
  \bibinfo {author} {\bibfnamefont {B.}~\bibnamefont {H\"orz}}, \bibinfo
  {author} {\bibfnamefont {C.}~\bibnamefont {Morningstar}}, \bibinfo {author}
  {\bibfnamefont {A.}~\bibnamefont {Nicholson}}, \bibinfo {author}
  {\bibfnamefont {F.}~\bibnamefont {Romero-L\'opez}}, \bibinfo {author}
  {\bibfnamefont {S.}~\bibnamefont {Skinner}}, \bibinfo {author} {\bibfnamefont
  {P.}~\bibnamefont {Vranas}},\ and\ \bibinfo {author} {\bibfnamefont
  {A.}~\bibnamefont {Walker-Loud}},\ }\bibfield  {title} {\bibinfo {title}
  {{Elastic nucleon-pion scattering at $m_\pi=200$~MeV from lattice QCD}},\
  }\href {https://doi.org/10.1016/j.nuclphysb.2023.116105} {\bibfield
  {journal} {\bibinfo  {journal} {Nucl. Phys. B}\ }\textbf {\bibinfo {volume}
  {987}},\ \bibinfo {pages} {116105} (\bibinfo {year} {2023})},\ \Eprint
  {https://arxiv.org/abs/2208.03867} {arXiv:2208.03867 [hep-lat]} \BibitemShut
  {NoStop}%
\bibitem [{\citenamefont {Wagman}(2024)}]{Wagman:2024rid}%
  \BibitemOpen
  \bibfield  {author} {\bibinfo {author} {\bibfnamefont {M.~L.}\ \bibnamefont
  {Wagman}},\ }\bibfield  {title} {\bibinfo {title} {{Lanczos, the transfer
  matrix, and the signal-to-noise problem}},\ }\href@noop {} {\  (\bibinfo
  {year} {2024})},\ \Eprint {https://arxiv.org/abs/2406.20009}
  {arXiv:2406.20009 [hep-lat]} \BibitemShut {NoStop}%
\bibitem [{\citenamefont {Hackett}\ and\ \citenamefont
  {Wagman}(2024)}]{Hackett:2024xnx}%
  \BibitemOpen
  \bibfield  {author} {\bibinfo {author} {\bibfnamefont {D.~C.}\ \bibnamefont
  {Hackett}}\ and\ \bibinfo {author} {\bibfnamefont {M.~L.}\ \bibnamefont
  {Wagman}},\ }\bibfield  {title} {\bibinfo {title} {{Lanczos for lattice QCD
  matrix elements}},\ }\href@noop {} {\  (\bibinfo {year} {2024})},\ \Eprint
  {https://arxiv.org/abs/2407.21777} {arXiv:2407.21777 [hep-lat]} \BibitemShut
  {NoStop}%
\bibitem [{\citenamefont {Hansen}\ and\ \citenamefont
  {Peterken}(2024)}]{Hansen:2024cai}%
  \BibitemOpen
  \bibfield  {author} {\bibinfo {author} {\bibfnamefont {M.~T.}\ \bibnamefont
  {Hansen}}\ and\ \bibinfo {author} {\bibfnamefont {T.}~\bibnamefont
  {Peterken}},\ }\bibfield  {title} {\bibinfo {title} {{Discretization effects
  in finite-volume $2\to2$ scattering}},\ }\href@noop {} {\  (\bibinfo {year}
  {2024})},\ \Eprint {https://arxiv.org/abs/2408.07062} {arXiv:2408.07062
  [hep-lat]} \BibitemShut {NoStop}%
\bibitem [{\citenamefont {Luscher}(1986)}]{Luscher:1986pf}%
  \BibitemOpen
  \bibfield  {author} {\bibinfo {author} {\bibfnamefont {M.}~\bibnamefont
  {Luscher}},\ }\bibfield  {title} {\bibinfo {title} {{Volume Dependence of the
  Energy Spectrum in Massive Quantum Field Theories. 2. Scattering States}},\
  }\href {https://doi.org/10.1007/BF01211097} {\bibfield  {journal} {\bibinfo
  {journal} {Commun. Math. Phys.}\ }\textbf {\bibinfo {volume} {105}},\
  \bibinfo {pages} {153} (\bibinfo {year} {1986})}\BibitemShut {NoStop}%
\bibitem [{\citenamefont {Briceno}(2014)}]{Briceno:2014oea}%
  \BibitemOpen
  \bibfield  {author} {\bibinfo {author} {\bibfnamefont {R.~A.}\ \bibnamefont
  {Briceno}},\ }\bibfield  {title} {\bibinfo {title} {{Two-particle
  multichannel systems in a finite volume with arbitrary spin}},\ }\href
  {https://doi.org/10.1103/PhysRevD.89.074507} {\bibfield  {journal} {\bibinfo
  {journal} {Phys. Rev.}\ }\textbf {\bibinfo {volume} {D89}},\ \bibinfo {pages}
  {074507} (\bibinfo {year} {2014})},\ \Eprint
  {https://arxiv.org/abs/1401.3312} {arXiv:1401.3312 [hep-lat]} \BibitemShut
  {NoStop}%
%%CITATION = ARXIV:1401.3312;%%
\bibitem [{\citenamefont {Jackura}(2023)}]{Jackura:2022gib}%
  \BibitemOpen
  \bibfield  {author} {\bibinfo {author} {\bibfnamefont {A.~W.}\ \bibnamefont
  {Jackura}},\ }\bibfield  {title} {\bibinfo {title} {{Three-body scattering
  and quantization conditions from S-matrix unitarity}},\ }\href
  {https://doi.org/10.1103/PhysRevD.108.034505} {\bibfield  {journal} {\bibinfo
   {journal} {Phys. Rev. D}\ }\textbf {\bibinfo {volume} {108}},\ \bibinfo
  {pages} {034505} (\bibinfo {year} {2023})},\ \Eprint
  {https://arxiv.org/abs/2208.10587} {arXiv:2208.10587 [hep-lat]} \BibitemShut
  {NoStop}%
\bibitem [{\citenamefont {Blankenbecler}\ and\ \citenamefont
  {Sugar}(1966)}]{Blankenbecler:1965gx}%
  \BibitemOpen
  \bibfield  {author} {\bibinfo {author} {\bibfnamefont {R.}~\bibnamefont
  {Blankenbecler}}\ and\ \bibinfo {author} {\bibfnamefont {R.}~\bibnamefont
  {Sugar}},\ }\bibfield  {title} {\bibinfo {title} {{Linear integral equations
  for relativistic multichannel scattering}},\ }\href
  {https://doi.org/10.1103/PhysRev.142.1051} {\bibfield  {journal} {\bibinfo
  {journal} {Phys. Rev.}\ }\textbf {\bibinfo {volume} {142}},\ \bibinfo {pages}
  {1051} (\bibinfo {year} {1966})}\BibitemShut {NoStop}%
\bibitem [{\citenamefont {Taylor}(1966)}]{Taylor:1966zza}%
  \BibitemOpen
  \bibfield  {author} {\bibinfo {author} {\bibfnamefont {J.~G.}\ \bibnamefont
  {Taylor}},\ }\bibfield  {title} {\bibinfo {title} {{Relativistic
  Three-Particle Equations. I}},\ }\href
  {https://doi.org/10.1103/PhysRev.150.1321} {\bibfield  {journal} {\bibinfo
  {journal} {Phys. Rev.}\ }\textbf {\bibinfo {volume} {150}},\ \bibinfo {pages}
  {1321} (\bibinfo {year} {1966})}\BibitemShut {NoStop}%
\bibitem [{\citenamefont {Aaron}\ \emph {et~al.}(1968)\citenamefont {Aaron},
  \citenamefont {Amado},\ and\ \citenamefont {Young}}]{Aaron:1968aoz}%
  \BibitemOpen
  \bibfield  {author} {\bibinfo {author} {\bibfnamefont {R.}~\bibnamefont
  {Aaron}}, \bibinfo {author} {\bibfnamefont {R.~D.}\ \bibnamefont {Amado}},\
  and\ \bibinfo {author} {\bibfnamefont {J.~E.}\ \bibnamefont {Young}},\
  }\bibfield  {title} {\bibinfo {title} {{Relativistic three-body theory with
  applications to pi-minus n scattering}},\ }\href
  {https://doi.org/10.1103/PhysRev.174.2022} {\bibfield  {journal} {\bibinfo
  {journal} {Phys. Rev.}\ }\textbf {\bibinfo {volume} {174}},\ \bibinfo {pages}
  {2022} (\bibinfo {year} {1968})}\BibitemShut {NoStop}%
\bibitem [{\citenamefont {Brayshaw}(1978)}]{Brayshaw:1978tx}%
  \BibitemOpen
  \bibfield  {author} {\bibinfo {author} {\bibfnamefont {D.~D.}\ \bibnamefont
  {Brayshaw}},\ }\bibfield  {title} {\bibinfo {title} {{Diffractive Production
  and Rescattering of Three Particle Systems}},\ }\href
  {https://doi.org/10.1103/PhysRevD.18.2638} {\bibfield  {journal} {\bibinfo
  {journal} {Phys. Rev. D}\ }\textbf {\bibinfo {volume} {18}},\ \bibinfo
  {pages} {2638} (\bibinfo {year} {1978})}\BibitemShut {NoStop}%
\bibitem [{\citenamefont {Lindesay}\ and\ \citenamefont
  {Noyes}(1980)}]{Lindesay:1980ib}%
  \BibitemOpen
  \bibfield  {author} {\bibinfo {author} {\bibfnamefont {J.~V.}\ \bibnamefont
  {Lindesay}}\ and\ \bibinfo {author} {\bibfnamefont {H.~P.}\ \bibnamefont
  {Noyes}},\ }\bibfield  {title} {\bibinfo {title} {{Minimal relativistic three
  particle equations}},\ }in\ \href@noop {} {\emph {\bibinfo {booktitle} {{NATO
  Advanced Study Institute on Nonlinear Phenomena in Physics and Biology}}}}\
  (\bibinfo {year} {1980})\BibitemShut {NoStop}%
\bibitem [{\citenamefont {Mai}\ \emph {et~al.}(2017)\citenamefont {Mai},
  \citenamefont {Hu}, \citenamefont {Doring}, \citenamefont {Pilloni},\ and\
  \citenamefont {Szczepaniak}}]{Mai:2017vot}%
  \BibitemOpen
  \bibfield  {author} {\bibinfo {author} {\bibfnamefont {M.}~\bibnamefont
  {Mai}}, \bibinfo {author} {\bibfnamefont {B.}~\bibnamefont {Hu}}, \bibinfo
  {author} {\bibfnamefont {M.}~\bibnamefont {Doring}}, \bibinfo {author}
  {\bibfnamefont {A.}~\bibnamefont {Pilloni}},\ and\ \bibinfo {author}
  {\bibfnamefont {A.}~\bibnamefont {Szczepaniak}},\ }\bibfield  {title}
  {\bibinfo {title} {{Three-body Unitarity with Isobars Revisited}},\ }\href
  {https://doi.org/10.1140/epja/i2017-12368-4} {\bibfield  {journal} {\bibinfo
  {journal} {Eur. Phys. J. A}\ }\textbf {\bibinfo {volume} {53}},\ \bibinfo
  {pages} {177} (\bibinfo {year} {2017})},\ \Eprint
  {https://arxiv.org/abs/1706.06118} {arXiv:1706.06118 [nucl-th]} \BibitemShut
  {NoStop}%
\bibitem [{\citenamefont {Sadasivan}\ \emph {et~al.}(2020)\citenamefont
  {Sadasivan}, \citenamefont {Mai}, \citenamefont {Akdag},\ and\ \citenamefont
  {D\"oring}}]{Sadasivan:2020syi}%
  \BibitemOpen
  \bibfield  {author} {\bibinfo {author} {\bibfnamefont {D.}~\bibnamefont
  {Sadasivan}}, \bibinfo {author} {\bibfnamefont {M.}~\bibnamefont {Mai}},
  \bibinfo {author} {\bibfnamefont {H.}~\bibnamefont {Akdag}},\ and\ \bibinfo
  {author} {\bibfnamefont {M.}~\bibnamefont {D\"oring}},\ }\bibfield  {title}
  {\bibinfo {title} {{Dalitz plots and lineshape of $a_1(1260)$ from a
  relativistic three-body unitary approach}},\ }\href
  {https://doi.org/10.1103/PhysRevD.101.094018} {\bibfield  {journal} {\bibinfo
   {journal} {Phys. Rev. D}\ }\textbf {\bibinfo {volume} {101}},\ \bibinfo
  {pages} {094018} (\bibinfo {year} {2020})},\ \bibinfo {note} {[Erratum:
  Phys.Rev.D 103, 019901 (2021)]},\ \Eprint {https://arxiv.org/abs/2002.12431}
  {arXiv:2002.12431 [nucl-th]} \BibitemShut {NoStop}%
\bibitem [{\citenamefont {Sadasivan}\ \emph {et~al.}(2022)\citenamefont
  {Sadasivan}, \citenamefont {Alexandru}, \citenamefont {Akdag}, \citenamefont
  {Amorim}, \citenamefont {Brett}, \citenamefont {Culver}, \citenamefont
  {D\"oring}, \citenamefont {Lee},\ and\ \citenamefont
  {Mai}}]{Sadasivan:2021emk}%
  \BibitemOpen
  \bibfield  {author} {\bibinfo {author} {\bibfnamefont {D.}~\bibnamefont
  {Sadasivan}}, \bibinfo {author} {\bibfnamefont {A.}~\bibnamefont
  {Alexandru}}, \bibinfo {author} {\bibfnamefont {H.}~\bibnamefont {Akdag}},
  \bibinfo {author} {\bibfnamefont {F.}~\bibnamefont {Amorim}}, \bibinfo
  {author} {\bibfnamefont {R.}~\bibnamefont {Brett}}, \bibinfo {author}
  {\bibfnamefont {C.}~\bibnamefont {Culver}}, \bibinfo {author} {\bibfnamefont
  {M.}~\bibnamefont {D\"oring}}, \bibinfo {author} {\bibfnamefont {F.~X.}\
  \bibnamefont {Lee}},\ and\ \bibinfo {author} {\bibfnamefont {M.}~\bibnamefont
  {Mai}},\ }\bibfield  {title} {\bibinfo {title} {{Pole position of the
  $a_1(1260)$ resonance in a three-body unitary framework}},\ }\href
  {https://doi.org/10.1103/PhysRevD.105.054020} {\bibfield  {journal} {\bibinfo
   {journal} {Phys. Rev. D}\ }\textbf {\bibinfo {volume} {105}},\ \bibinfo
  {pages} {054020} (\bibinfo {year} {2022})},\ \Eprint
  {https://arxiv.org/abs/2112.03355} {arXiv:2112.03355 [hep-ph]} \BibitemShut
  {NoStop}%
\bibitem [{\citenamefont {Dawid}\ \emph
  {et~al.}(2025{\natexlab{b}})\citenamefont {Dawid}, \citenamefont
  {Romero-L\'opez},\ and\ \citenamefont {Sharpe}}]{Dawid:2024dgy}%
  \BibitemOpen
  \bibfield  {author} {\bibinfo {author} {\bibfnamefont {S.~M.}\ \bibnamefont
  {Dawid}}, \bibinfo {author} {\bibfnamefont {F.}~\bibnamefont
  {Romero-L\'opez}},\ and\ \bibinfo {author} {\bibfnamefont {S.~R.}\
  \bibnamefont {Sharpe}},\ }\bibfield  {title} {\bibinfo {title} {{Finite- and
  infinite-volume study of DD\ensuremath{\pi} scattering}},\ }\href
  {https://doi.org/10.1007/JHEP01(2025)060} {\bibfield  {journal} {\bibinfo
  {journal} {JHEP}\ }\textbf {\bibinfo {volume} {01}},\ \bibinfo {pages}
  {060}},\ \Eprint {https://arxiv.org/abs/2409.17059} {arXiv:2409.17059
  [hep-lat]} \BibitemShut {NoStop}%
\bibitem [{\citenamefont {Jackura}\ \emph {et~al.}(2021)\citenamefont
  {Jackura}, \citenamefont {Brice\~no}, \citenamefont {Dawid}, \citenamefont
  {Islam},\ and\ \citenamefont {McCarty}}]{Jackura:2020bsk}%
  \BibitemOpen
  \bibfield  {author} {\bibinfo {author} {\bibfnamefont {A.~W.}\ \bibnamefont
  {Jackura}}, \bibinfo {author} {\bibfnamefont {R.~A.}\ \bibnamefont
  {Brice\~no}}, \bibinfo {author} {\bibfnamefont {S.~M.}\ \bibnamefont
  {Dawid}}, \bibinfo {author} {\bibfnamefont {M.~H.~E.}\ \bibnamefont
  {Islam}},\ and\ \bibinfo {author} {\bibfnamefont {C.}~\bibnamefont
  {McCarty}},\ }\bibfield  {title} {\bibinfo {title} {{Solving relativistic
  three-body integral equations in the presence of bound states}},\ }\href
  {https://doi.org/10.1103/PhysRevD.104.014507} {\bibfield  {journal} {\bibinfo
   {journal} {Phys. Rev. D}\ }\textbf {\bibinfo {volume} {104}},\ \bibinfo
  {pages} {014507} (\bibinfo {year} {2021})},\ \Eprint
  {https://arxiv.org/abs/2010.09820} {arXiv:2010.09820 [hep-lat]} \BibitemShut
  {NoStop}%
\bibitem [{\citenamefont {Brice\~no}\ \emph
  {et~al.}(2024{\natexlab{b}})\citenamefont {Brice\~no}, \citenamefont
  {Costa},\ and\ \citenamefont {Jackura}}]{Briceno:2024ehy}%
  \BibitemOpen
  \bibfield  {author} {\bibinfo {author} {\bibfnamefont {R.~A.}\ \bibnamefont
  {Brice\~no}}, \bibinfo {author} {\bibfnamefont {C.~S.~R.}\ \bibnamefont
  {Costa}},\ and\ \bibinfo {author} {\bibfnamefont {A.~W.}\ \bibnamefont
  {Jackura}},\ }\bibfield  {title} {\bibinfo {title} {{Partial-wave projection
  of relativistic three-body amplitudes}},\ }\href@noop {} {\  (\bibinfo {year}
  {2024}{\natexlab{b}})},\ \Eprint {https://arxiv.org/abs/2409.15577}
  {arXiv:2409.15577 [hep-ph]} \BibitemShut {NoStop}%
\bibitem [{\citenamefont {Jackura}\ \emph
  {et~al.}(2019{\natexlab{b}})\citenamefont {Jackura}, \citenamefont
  {Fern\'andez-Ram\'\i{}rez}, \citenamefont {Mathieu}, \citenamefont
  {Mikhasenko}, \citenamefont {Nys}, \citenamefont {Pilloni}, \citenamefont
  {Salda\~na}, \citenamefont {Sherrill},\ and\ \citenamefont
  {Szczepaniak}}]{Jackura:2018xnx}%
  \BibitemOpen
  \bibfield  {author} {\bibinfo {author} {\bibfnamefont {A.}~\bibnamefont
  {Jackura}}, \bibinfo {author} {\bibfnamefont {C.}~\bibnamefont
  {Fern\'andez-Ram\'\i{}rez}}, \bibinfo {author} {\bibfnamefont
  {V.}~\bibnamefont {Mathieu}}, \bibinfo {author} {\bibfnamefont
  {M.}~\bibnamefont {Mikhasenko}}, \bibinfo {author} {\bibfnamefont
  {J.}~\bibnamefont {Nys}}, \bibinfo {author} {\bibfnamefont {A.}~\bibnamefont
  {Pilloni}}, \bibinfo {author} {\bibfnamefont {K.}~\bibnamefont {Salda\~na}},
  \bibinfo {author} {\bibfnamefont {N.}~\bibnamefont {Sherrill}},\ and\
  \bibinfo {author} {\bibfnamefont {A.~P.}\ \bibnamefont {Szczepaniak}}
  (\bibinfo {collaboration} {JPAC}),\ }\bibfield  {title} {\bibinfo {title}
  {{Phenomenology of Relativistic $\mathbf{3} \to \mathbf{3}$ Reaction
  Amplitudes within the Isobar Approximation}},\ }\href
  {https://doi.org/10.1140/epjc/s10052-019-6566-1} {\bibfield  {journal}
  {\bibinfo  {journal} {Eur. Phys. J. C}\ }\textbf {\bibinfo {volume} {79}},\
  \bibinfo {pages} {56} (\bibinfo {year} {2019}{\natexlab{b}})},\ \Eprint
  {https://arxiv.org/abs/1809.10523} {arXiv:1809.10523 [hep-ph]} \BibitemShut
  {NoStop}%
\bibitem [{\citenamefont {Virtanen}\ \emph {et~al.}(2020)\citenamefont
  {Virtanen}, \citenamefont {Gommers}, \citenamefont {Oliphant}, \citenamefont
  {Haberland}, \citenamefont {Reddy}, \citenamefont {Cournapeau}, \citenamefont
  {Burovski}, \citenamefont {Peterson}, \citenamefont {Weckesser},
  \citenamefont {Bright}, \citenamefont {{van der Walt}}, \citenamefont
  {Brett}, \citenamefont {Wilson}, \citenamefont {Millman}, \citenamefont
  {Mayorov}, \citenamefont {Nelson}, \citenamefont {Jones}, \citenamefont
  {Kern}, \citenamefont {Larson}, \citenamefont {Carey}, \citenamefont {Polat},
  \citenamefont {Feng}, \citenamefont {Moore}, \citenamefont {{VanderPlas}},
  \citenamefont {Laxalde}, \citenamefont {Perktold}, \citenamefont {Cimrman},
  \citenamefont {Henriksen}, \citenamefont {Quintero}, \citenamefont {Harris},
  \citenamefont {Archibald}, \citenamefont {Ribeiro}, \citenamefont
  {Pedregosa}, \citenamefont {{van Mulbregt}},\ and\ \citenamefont {{SciPy 1.0
  Contributors}}}]{2020SciPy-NMeth}%
  \BibitemOpen
  \bibfield  {author} {\bibinfo {author} {\bibfnamefont {P.}~\bibnamefont
  {Virtanen}}, \bibinfo {author} {\bibfnamefont {R.}~\bibnamefont {Gommers}},
  \bibinfo {author} {\bibfnamefont {T.~E.}\ \bibnamefont {Oliphant}}, \bibinfo
  {author} {\bibfnamefont {M.}~\bibnamefont {Haberland}}, \bibinfo {author}
  {\bibfnamefont {T.}~\bibnamefont {Reddy}}, \bibinfo {author} {\bibfnamefont
  {D.}~\bibnamefont {Cournapeau}}, \bibinfo {author} {\bibfnamefont
  {E.}~\bibnamefont {Burovski}}, \bibinfo {author} {\bibfnamefont
  {P.}~\bibnamefont {Peterson}}, \bibinfo {author} {\bibfnamefont
  {W.}~\bibnamefont {Weckesser}}, \bibinfo {author} {\bibfnamefont
  {J.}~\bibnamefont {Bright}}, \bibinfo {author} {\bibfnamefont {S.~J.}\
  \bibnamefont {{van der Walt}}}, \bibinfo {author} {\bibfnamefont
  {M.}~\bibnamefont {Brett}}, \bibinfo {author} {\bibfnamefont
  {J.}~\bibnamefont {Wilson}}, \bibinfo {author} {\bibfnamefont {K.~J.}\
  \bibnamefont {Millman}}, \bibinfo {author} {\bibfnamefont {N.}~\bibnamefont
  {Mayorov}}, \bibinfo {author} {\bibfnamefont {A.~R.~J.}\ \bibnamefont
  {Nelson}}, \bibinfo {author} {\bibfnamefont {E.}~\bibnamefont {Jones}},
  \bibinfo {author} {\bibfnamefont {R.}~\bibnamefont {Kern}}, \bibinfo {author}
  {\bibfnamefont {E.}~\bibnamefont {Larson}}, \bibinfo {author} {\bibfnamefont
  {C.~J.}\ \bibnamefont {Carey}}, \bibinfo {author} {\bibfnamefont
  {{\.I}.}~\bibnamefont {Polat}}, \bibinfo {author} {\bibfnamefont
  {Y.}~\bibnamefont {Feng}}, \bibinfo {author} {\bibfnamefont {E.~W.}\
  \bibnamefont {Moore}}, \bibinfo {author} {\bibfnamefont {J.}~\bibnamefont
  {{VanderPlas}}}, \bibinfo {author} {\bibfnamefont {D.}~\bibnamefont
  {Laxalde}}, \bibinfo {author} {\bibfnamefont {J.}~\bibnamefont {Perktold}},
  \bibinfo {author} {\bibfnamefont {R.}~\bibnamefont {Cimrman}}, \bibinfo
  {author} {\bibfnamefont {I.}~\bibnamefont {Henriksen}}, \bibinfo {author}
  {\bibfnamefont {E.~A.}\ \bibnamefont {Quintero}}, \bibinfo {author}
  {\bibfnamefont {C.~R.}\ \bibnamefont {Harris}}, \bibinfo {author}
  {\bibfnamefont {A.~M.}\ \bibnamefont {Archibald}}, \bibinfo {author}
  {\bibfnamefont {A.~H.}\ \bibnamefont {Ribeiro}}, \bibinfo {author}
  {\bibfnamefont {F.}~\bibnamefont {Pedregosa}}, \bibinfo {author}
  {\bibfnamefont {P.}~\bibnamefont {{van Mulbregt}}},\ and\ \bibinfo {author}
  {\bibnamefont {{SciPy 1.0 Contributors}}},\ }\bibfield  {title} {\bibinfo
  {title} {{{SciPy} 1.0: Fundamental Algorithms for Scientific Computing in
  Python}},\ }\href {https://doi.org/10.1038/s41592-019-0686-2} {\bibfield
  {journal} {\bibinfo  {journal} {Nature Methods}\ }\textbf {\bibinfo {volume}
  {17}},\ \bibinfo {pages} {261} (\bibinfo {year} {2020})}\BibitemShut
  {NoStop}%
\bibitem [{\citenamefont {Garcia-Martin}\ \emph {et~al.}(2011)\citenamefont
  {Garcia-Martin}, \citenamefont {Kaminski}, \citenamefont {Pelaez},
  \citenamefont {Ruiz~de Elvira},\ and\ \citenamefont
  {Yndurain}}]{Garcia-Martin:2011iqs}%
  \BibitemOpen
  \bibfield  {author} {\bibinfo {author} {\bibfnamefont {R.}~\bibnamefont
  {Garcia-Martin}}, \bibinfo {author} {\bibfnamefont {R.}~\bibnamefont
  {Kaminski}}, \bibinfo {author} {\bibfnamefont {J.~R.}\ \bibnamefont
  {Pelaez}}, \bibinfo {author} {\bibfnamefont {J.}~\bibnamefont {Ruiz~de
  Elvira}},\ and\ \bibinfo {author} {\bibfnamefont {F.~J.}\ \bibnamefont
  {Yndurain}},\ }\bibfield  {title} {\bibinfo {title} {{The Pion-pion
  scattering amplitude. IV: Improved analysis with once subtracted Roy-like
  equations up to 1100 MeV}},\ }\href
  {https://doi.org/10.1103/PhysRevD.83.074004} {\bibfield  {journal} {\bibinfo
  {journal} {Phys. Rev. D}\ }\textbf {\bibinfo {volume} {83}},\ \bibinfo
  {pages} {074004} (\bibinfo {year} {2011})},\ \Eprint
  {https://arxiv.org/abs/1102.2183} {arXiv:1102.2183 [hep-ph]} \BibitemShut
  {NoStop}%
\bibitem [{\citenamefont {Bijnens}\ and\ \citenamefont
  {Lu}(2011)}]{Bijnens:2011fm}%
  \BibitemOpen
  \bibfield  {author} {\bibinfo {author} {\bibfnamefont {J.}~\bibnamefont
  {Bijnens}}\ and\ \bibinfo {author} {\bibfnamefont {J.}~\bibnamefont {Lu}},\
  }\bibfield  {title} {\bibinfo {title} {{Meson-meson Scattering in QCD-like
  Theories}},\ }\href {https://doi.org/10.1007/JHEP03(2011)028} {\bibfield
  {journal} {\bibinfo  {journal} {JHEP}\ }\textbf {\bibinfo {volume} {03}},\
  \bibinfo {pages} {028}},\ \Eprint {https://arxiv.org/abs/1102.0172}
  {arXiv:1102.0172 [hep-ph]} \BibitemShut {NoStop}%
\bibitem [{\citenamefont {Pel\'aez}\ and\ \citenamefont
  {Rodas}(2022)}]{Pelaez:2020gnd}%
  \BibitemOpen
  \bibfield  {author} {\bibinfo {author} {\bibfnamefont {J.~R.}\ \bibnamefont
  {Pel\'aez}}\ and\ \bibinfo {author} {\bibfnamefont {A.}~\bibnamefont
  {Rodas}},\ }\bibfield  {title} {\bibinfo {title} {{Dispersive
  \ensuremath{\pi}K\textrightarrow{}\ensuremath{\pi}K and
  \ensuremath{\pi}\ensuremath{\pi}\textrightarrow{}\ensuremath{K\bar K}
  amplitudes from scattering data, threshold parameters, and the lightest
  strange resonance \ensuremath{\kappa} or $K_0^*(700)$}},\ }\href
  {https://doi.org/10.1016/j.physrep.2022.03.004} {\bibfield  {journal}
  {\bibinfo  {journal} {Phys. Rept.}\ }\textbf {\bibinfo {volume} {969}},\
  \bibinfo {pages} {1} (\bibinfo {year} {2022})},\ \Eprint
  {https://arxiv.org/abs/2010.11222} {arXiv:2010.11222 [hep-ph]} \BibitemShut
  {NoStop}%
\bibitem [{\citenamefont {Sharpe}(2017)}]{Sharpe:2017jej}%
  \BibitemOpen
  \bibfield  {author} {\bibinfo {author} {\bibfnamefont {S.~R.}\ \bibnamefont
  {Sharpe}},\ }\bibfield  {title} {\bibinfo {title} {{Testing the threshold
  expansion for three-particle energies at fourth order in $\phi^4$ theory}},\
  }\href {https://doi.org/10.1103/PhysRevD.96.054515} {\bibfield  {journal}
  {\bibinfo  {journal} {Phys. Rev. D}\ }\textbf {\bibinfo {volume} {96}},\
  \bibinfo {pages} {054515} (\bibinfo {year} {2017})},\ \bibinfo {note}
  {[Erratum: Phys.Rev.D 98, 099901 (2018)]},\ \Eprint
  {https://arxiv.org/abs/1707.04279} {arXiv:1707.04279 [hep-lat]} \BibitemShut
  {NoStop}%
\bibitem [{\citenamefont {Borsanyi}\ \emph {et~al.}(2021)\citenamefont
  {Borsanyi} \emph {et~al.}}]{Borsanyi:2020mff}%
  \BibitemOpen
  \bibfield  {author} {\bibinfo {author} {\bibfnamefont {S.}~\bibnamefont
  {Borsanyi}} \emph {et~al.},\ }\bibfield  {title} {\bibinfo {title} {{Leading
  hadronic contribution to the muon magnetic moment from lattice QCD}},\ }\href
  {https://doi.org/10.1038/s41586-021-03418-1} {\bibfield  {journal} {\bibinfo
  {journal} {Nature}\ }\textbf {\bibinfo {volume} {593}},\ \bibinfo {pages}
  {51} (\bibinfo {year} {2021})},\ \Eprint {https://arxiv.org/abs/2002.12347}
  {arXiv:2002.12347 [hep-lat]} \BibitemShut {NoStop}%
\bibitem [{\citenamefont {Jay}\ and\ \citenamefont {Neil}(2021)}]{Jay:2020jkz}%
  \BibitemOpen
  \bibfield  {author} {\bibinfo {author} {\bibfnamefont {W.~I.}\ \bibnamefont
  {Jay}}\ and\ \bibinfo {author} {\bibfnamefont {E.~T.}\ \bibnamefont {Neil}},\
  }\bibfield  {title} {\bibinfo {title} {{Bayesian model averaging for analysis
  of lattice field theory results}},\ }\href
  {https://doi.org/10.1103/PhysRevD.103.114502} {\bibfield  {journal} {\bibinfo
   {journal} {Phys. Rev. D}\ }\textbf {\bibinfo {volume} {103}},\ \bibinfo
  {pages} {114502} (\bibinfo {year} {2021})},\ \Eprint
  {https://arxiv.org/abs/2008.01069} {arXiv:2008.01069 [stat.ME]} \BibitemShut
  {NoStop}%
\bibitem [{\citenamefont {Neil}\ and\ \citenamefont
  {Sitison}(2023)}]{Neil:2023pgt}%
  \BibitemOpen
  \bibfield  {author} {\bibinfo {author} {\bibfnamefont {E.~T.}\ \bibnamefont
  {Neil}}\ and\ \bibinfo {author} {\bibfnamefont {J.~W.}\ \bibnamefont
  {Sitison}},\ }\bibfield  {title} {\bibinfo {title} {{Model averaging
  approaches to data subset selection}},\ }\href
  {https://doi.org/10.1103/PhysRevE.108.045308} {\bibfield  {journal} {\bibinfo
   {journal} {Phys. Rev. E}\ }\textbf {\bibinfo {volume} {108}},\ \bibinfo
  {pages} {045308} (\bibinfo {year} {2023})},\ \Eprint
  {https://arxiv.org/abs/2305.19417} {arXiv:2305.19417 [stat.ME]} \BibitemShut
  {NoStop}%
\bibitem [{\citenamefont {Pefkou}\ \emph {et~al.}(2022)\citenamefont {Pefkou},
  \citenamefont {Hackett},\ and\ \citenamefont {Shanahan}}]{Pefkou:2021fni}%
  \BibitemOpen
  \bibfield  {author} {\bibinfo {author} {\bibfnamefont {D.~A.}\ \bibnamefont
  {Pefkou}}, \bibinfo {author} {\bibfnamefont {D.~C.}\ \bibnamefont
  {Hackett}},\ and\ \bibinfo {author} {\bibfnamefont {P.~E.}\ \bibnamefont
  {Shanahan}},\ }\bibfield  {title} {\bibinfo {title} {{Gluon gravitational
  structure of hadrons of different spin}},\ }\href
  {https://doi.org/10.1103/PhysRevD.105.054509} {\bibfield  {journal} {\bibinfo
   {journal} {Phys. Rev. D}\ }\textbf {\bibinfo {volume} {105}},\ \bibinfo
  {pages} {054509} (\bibinfo {year} {2022})},\ \Eprint
  {https://arxiv.org/abs/2107.10368} {arXiv:2107.10368 [hep-lat]} \BibitemShut
  {NoStop}%
\bibitem [{\citenamefont {Baeza-Ballesteros}\ \emph {et~al.}(2023)\citenamefont
  {Baeza-Ballesteros}, \citenamefont {Bijnens}, \citenamefont {Husek},
  \citenamefont {Romero-L\'opez}, \citenamefont {Sharpe},\ and\ \citenamefont
  {Sj\"o}}]{Baeza-Ballesteros:2023ljl}%
  \BibitemOpen
  \bibfield  {author} {\bibinfo {author} {\bibfnamefont {J.}~\bibnamefont
  {Baeza-Ballesteros}}, \bibinfo {author} {\bibfnamefont {J.}~\bibnamefont
  {Bijnens}}, \bibinfo {author} {\bibfnamefont {T.}~\bibnamefont {Husek}},
  \bibinfo {author} {\bibfnamefont {F.}~\bibnamefont {Romero-L\'opez}},
  \bibinfo {author} {\bibfnamefont {S.~R.}\ \bibnamefont {Sharpe}},\ and\
  \bibinfo {author} {\bibfnamefont {M.}~\bibnamefont {Sj\"o}},\ }\bibfield
  {title} {\bibinfo {title} {{The isospin-3 three-particle K-matrix at NLO in
  ChPT}},\ }\href {https://doi.org/10.1007/JHEP05(2023)187} {\bibfield
  {journal} {\bibinfo  {journal} {JHEP}\ }\textbf {\bibinfo {volume} {05}},\
  \bibinfo {pages} {187}},\ \Eprint {https://arxiv.org/abs/2303.13206}
  {arXiv:2303.13206 [hep-ph]} \BibitemShut {NoStop}%
\bibitem [{\citenamefont {Aoki}\ \emph {et~al.}(2024)\citenamefont {Aoki} \emph
  {et~al.}}]{FlavourLatticeAveragingGroupFLAG:2024oxs}%
  \BibitemOpen
  \bibfield  {author} {\bibinfo {author} {\bibfnamefont {Y.}~\bibnamefont
  {Aoki}} \emph {et~al.} (\bibinfo {collaboration} {Flavour Lattice Averaging
  Group (FLAG)}),\ }\bibfield  {title} {\bibinfo {title} {{FLAG Review 2024}},\
  }\href@noop {} {\  (\bibinfo {year} {2024})},\ \Eprint
  {https://arxiv.org/abs/2411.04268} {arXiv:2411.04268 [hep-lat]} \BibitemShut
  {NoStop}%
\bibitem [{\citenamefont {Sharpe}\ and\ \citenamefont
  {Singleton}(1998)}]{Sharpe:1998xm}%
  \BibitemOpen
  \bibfield  {author} {\bibinfo {author} {\bibfnamefont {S.~R.}\ \bibnamefont
  {Sharpe}}\ and\ \bibinfo {author} {\bibfnamefont {R.~L.}\ \bibnamefont
  {Singleton}, \bibfnamefont {Jr}},\ }\bibfield  {title} {\bibinfo {title}
  {{Spontaneous flavor and parity breaking with Wilson fermions}},\ }\href
  {https://doi.org/10.1103/PhysRevD.58.074501} {\bibfield  {journal} {\bibinfo
  {journal} {Phys. Rev.}\ }\textbf {\bibinfo {volume} {D58}},\ \bibinfo {pages}
  {074501} (\bibinfo {year} {1998})},\ \Eprint
  {https://arxiv.org/abs/hep-lat/9804028} {arXiv:hep-lat/9804028 [hep-lat]}
  \BibitemShut {NoStop}%
%%CITATION = HEP-LAT/9804028;%%
\bibitem [{\citenamefont {Bar}\ \emph {et~al.}(2004)\citenamefont {Bar},
  \citenamefont {Rupak},\ and\ \citenamefont {Shoresh}}]{Bar:2003mh}%
  \BibitemOpen
  \bibfield  {author} {\bibinfo {author} {\bibfnamefont {O.}~\bibnamefont
  {Bar}}, \bibinfo {author} {\bibfnamefont {G.}~\bibnamefont {Rupak}},\ and\
  \bibinfo {author} {\bibfnamefont {N.}~\bibnamefont {Shoresh}},\ }\bibfield
  {title} {\bibinfo {title} {{Chiral perturbation theory at $O(a^2)$ for
  lattice QCD}},\ }\href {https://doi.org/10.1103/PhysRevD.70.034508}
  {\bibfield  {journal} {\bibinfo  {journal} {Phys. Rev. D}\ }\textbf {\bibinfo
  {volume} {70}},\ \bibinfo {pages} {034508} (\bibinfo {year} {2004})},\
  \Eprint {https://arxiv.org/abs/hep-lat/0306021} {arXiv:hep-lat/0306021}
  \BibitemShut {NoStop}%
\bibitem [{\citenamefont {Bazavov}\ \emph {et~al.}(2010)\citenamefont {Bazavov}
  \emph {et~al.}}]{MILC:2010hzw}%
  \BibitemOpen
  \bibfield  {author} {\bibinfo {author} {\bibfnamefont {A.}~\bibnamefont
  {Bazavov}} \emph {et~al.} (\bibinfo {collaboration} {MILC}),\ }\bibfield
  {title} {\bibinfo {title} {{Results for light pseudoscalar mesons}},\ }\href
  {https://doi.org/10.22323/1.105.0074} {\bibfield  {journal} {\bibinfo
  {journal} {PoS}\ }\textbf {\bibinfo {volume} {LATTICE2010}},\ \bibinfo
  {pages} {074} (\bibinfo {year} {2010})},\ \Eprint
  {https://arxiv.org/abs/1012.0868} {arXiv:1012.0868 [hep-lat]} \BibitemShut
  {NoStop}%
\bibitem [{\citenamefont {Dowdall}\ \emph {et~al.}(2013)\citenamefont
  {Dowdall}, \citenamefont {Davies}, \citenamefont {Lepage},\ and\
  \citenamefont {McNeile}}]{Dowdall:2013rya}%
  \BibitemOpen
  \bibfield  {author} {\bibinfo {author} {\bibfnamefont {R.~J.}\ \bibnamefont
  {Dowdall}}, \bibinfo {author} {\bibfnamefont {C.~T.~H.}\ \bibnamefont
  {Davies}}, \bibinfo {author} {\bibfnamefont {G.~P.}\ \bibnamefont {Lepage}},\
  and\ \bibinfo {author} {\bibfnamefont {C.}~\bibnamefont {McNeile}},\
  }\bibfield  {title} {\bibinfo {title} {{Vus from pi and K decay constants in
  full lattice QCD with physical u, d, s and c quarks}},\ }\href
  {https://doi.org/10.1103/PhysRevD.88.074504} {\bibfield  {journal} {\bibinfo
  {journal} {Phys. Rev. D}\ }\textbf {\bibinfo {volume} {88}},\ \bibinfo
  {pages} {074504} (\bibinfo {year} {2013})},\ \Eprint
  {https://arxiv.org/abs/1303.1670} {arXiv:1303.1670 [hep-lat]} \BibitemShut
  {NoStop}%
\bibitem [{\citenamefont {Helmes}\ \emph {et~al.}(2018)\citenamefont {Helmes},
  \citenamefont {Jost}, \citenamefont {Knippschild}, \citenamefont {Kostrzewa},
  \citenamefont {Liu}, \citenamefont {Pittler}, \citenamefont {Urbach},\ and\
  \citenamefont {Werner}}]{Helmes:2018nug}%
  \BibitemOpen
  \bibfield  {author} {\bibinfo {author} {\bibfnamefont {C.}~\bibnamefont
  {Helmes}}, \bibinfo {author} {\bibfnamefont {C.}~\bibnamefont {Jost}},
  \bibinfo {author} {\bibfnamefont {B.}~\bibnamefont {Knippschild}}, \bibinfo
  {author} {\bibfnamefont {B.}~\bibnamefont {Kostrzewa}}, \bibinfo {author}
  {\bibfnamefont {L.}~\bibnamefont {Liu}}, \bibinfo {author} {\bibfnamefont
  {F.}~\bibnamefont {Pittler}}, \bibinfo {author} {\bibfnamefont
  {C.}~\bibnamefont {Urbach}},\ and\ \bibinfo {author} {\bibfnamefont
  {M.}~\bibnamefont {Werner}} (\bibinfo {collaboration} {ETM}),\ }\bibfield
  {title} {\bibinfo {title} {{Hadron-Hadron Interactions from $N_f=2+1+1$
  Lattice QCD: $I=3/2$ $\pi K$ Scattering Length}},\ }\href
  {https://doi.org/10.1103/PhysRevD.98.114511} {\bibfield  {journal} {\bibinfo
  {journal} {Phys. Rev. D}\ }\textbf {\bibinfo {volume} {98}},\ \bibinfo
  {pages} {114511} (\bibinfo {year} {2018})},\ \Eprint
  {https://arxiv.org/abs/1809.08886} {arXiv:1809.08886 [hep-lat]} \BibitemShut
  {NoStop}%
\bibitem [{\citenamefont {Helmes}\ \emph {et~al.}(2015)\citenamefont {Helmes},
  \citenamefont {Jost}, \citenamefont {Knippschild}, \citenamefont {Liu},
  \citenamefont {Liu}, \citenamefont {Liu}, \citenamefont {Urbach},
  \citenamefont {Ueding}, \citenamefont {Wang},\ and\ \citenamefont
  {Werner}}]{Helmes:2015gla}%
  \BibitemOpen
  \bibfield  {author} {\bibinfo {author} {\bibfnamefont {C.}~\bibnamefont
  {Helmes}}, \bibinfo {author} {\bibfnamefont {C.}~\bibnamefont {Jost}},
  \bibinfo {author} {\bibfnamefont {B.}~\bibnamefont {Knippschild}}, \bibinfo
  {author} {\bibfnamefont {C.}~\bibnamefont {Liu}}, \bibinfo {author}
  {\bibfnamefont {J.}~\bibnamefont {Liu}}, \bibinfo {author} {\bibfnamefont
  {L.}~\bibnamefont {Liu}}, \bibinfo {author} {\bibfnamefont {C.}~\bibnamefont
  {Urbach}}, \bibinfo {author} {\bibfnamefont {M.}~\bibnamefont {Ueding}},
  \bibinfo {author} {\bibfnamefont {Z.}~\bibnamefont {Wang}},\ and\ \bibinfo
  {author} {\bibfnamefont {M.}~\bibnamefont {Werner}} (\bibinfo {collaboration}
  {ETM}),\ }\bibfield  {title} {\bibinfo {title} {{Hadron-hadron interactions
  from N$_{f}$ = 2 + 1 + 1 lattice QCD: isospin-2 $\pi\pi$ scattering
  length}},\ }\href {https://doi.org/10.1007/JHEP09(2015)109} {\bibfield
  {journal} {\bibinfo  {journal} {JHEP}\ }\textbf {\bibinfo {volume} {09}},\
  \bibinfo {pages} {109}},\ \Eprint {https://arxiv.org/abs/1506.00408}
  {arXiv:1506.00408 [hep-lat]} \BibitemShut {NoStop}%
%%CITATION = ARXIV:1506.00408;%%
\bibitem [{\citenamefont {Helmes}\ \emph {et~al.}(2017)\citenamefont {Helmes},
  \citenamefont {Jost}, \citenamefont {Knippschild}, \citenamefont {Kostrzewa},
  \citenamefont {Liu}, \citenamefont {Urbach},\ and\ \citenamefont
  {Werner}}]{Helmes:2017smr}%
  \BibitemOpen
  \bibfield  {author} {\bibinfo {author} {\bibfnamefont {C.}~\bibnamefont
  {Helmes}}, \bibinfo {author} {\bibfnamefont {C.}~\bibnamefont {Jost}},
  \bibinfo {author} {\bibfnamefont {B.}~\bibnamefont {Knippschild}}, \bibinfo
  {author} {\bibfnamefont {B.}~\bibnamefont {Kostrzewa}}, \bibinfo {author}
  {\bibfnamefont {L.}~\bibnamefont {Liu}}, \bibinfo {author} {\bibfnamefont
  {C.}~\bibnamefont {Urbach}},\ and\ \bibinfo {author} {\bibfnamefont
  {M.}~\bibnamefont {Werner}},\ }\bibfield  {title} {\bibinfo {title}
  {{Hadron-Hadron Interactions from $N_f=2+1+1$ lattice QCD: Isospin-1 $KK$
  scattering length}},\ }\href {https://doi.org/10.1103/PhysRevD.96.034510}
  {\bibfield  {journal} {\bibinfo  {journal} {Phys. Rev. D}\ }\textbf {\bibinfo
  {volume} {96}},\ \bibinfo {pages} {034510} (\bibinfo {year} {2017})},\
  \Eprint {https://arxiv.org/abs/1703.04737} {arXiv:1703.04737 [hep-lat]}
  \BibitemShut {NoStop}%
\bibitem [{\citenamefont {Beane}\ \emph {et~al.}(2012)\citenamefont {Beane},
  \citenamefont {Chang}, \citenamefont {Detmold}, \citenamefont {Lin},
  \citenamefont {Luu}, \citenamefont {Orginos}, \citenamefont {Parreno},
  \citenamefont {Savage}, \citenamefont {Torok},\ and\ \citenamefont
  {Walker-Loud}}]{NPLQCD:2011htk}%
  \BibitemOpen
  \bibfield  {author} {\bibinfo {author} {\bibfnamefont {S.~R.}\ \bibnamefont
  {Beane}}, \bibinfo {author} {\bibfnamefont {E.}~\bibnamefont {Chang}},
  \bibinfo {author} {\bibfnamefont {W.}~\bibnamefont {Detmold}}, \bibinfo
  {author} {\bibfnamefont {H.~W.}\ \bibnamefont {Lin}}, \bibinfo {author}
  {\bibfnamefont {T.~C.}\ \bibnamefont {Luu}}, \bibinfo {author} {\bibfnamefont
  {K.}~\bibnamefont {Orginos}}, \bibinfo {author} {\bibfnamefont
  {A.}~\bibnamefont {Parreno}}, \bibinfo {author} {\bibfnamefont {M.~J.}\
  \bibnamefont {Savage}}, \bibinfo {author} {\bibfnamefont {A.}~\bibnamefont
  {Torok}},\ and\ \bibinfo {author} {\bibfnamefont {A.}~\bibnamefont
  {Walker-Loud}} (\bibinfo {collaboration} {NPLQCD}),\ }\bibfield  {title}
  {\bibinfo {title} {{The I=2 $\pi\pi$ S-wave Scattering Phase Shift from
  Lattice QCD}},\ }\href {https://doi.org/10.1103/PhysRevD.85.034505}
  {\bibfield  {journal} {\bibinfo  {journal} {Phys. Rev. D}\ }\textbf {\bibinfo
  {volume} {85}},\ \bibinfo {pages} {034505} (\bibinfo {year} {2012})},\
  \Eprint {https://arxiv.org/abs/1107.5023} {arXiv:1107.5023 [hep-lat]}
  \BibitemShut {NoStop}%
\bibitem [{\citenamefont {Wilson}\ \emph {et~al.}(2015)\citenamefont {Wilson},
  \citenamefont {Dudek}, \citenamefont {Edwards},\ and\ \citenamefont
  {Thomas}}]{Wilson:2014cna}%
  \BibitemOpen
  \bibfield  {author} {\bibinfo {author} {\bibfnamefont {D.~J.}\ \bibnamefont
  {Wilson}}, \bibinfo {author} {\bibfnamefont {J.~J.}\ \bibnamefont {Dudek}},
  \bibinfo {author} {\bibfnamefont {R.~G.}\ \bibnamefont {Edwards}},\ and\
  \bibinfo {author} {\bibfnamefont {C.~E.}\ \bibnamefont {Thomas}},\ }\bibfield
   {title} {\bibinfo {title} {{Resonances in coupled $\pi K, \eta K$ scattering
  from lattice QCD}},\ }\href {https://doi.org/10.1103/PhysRevD.91.054008}
  {\bibfield  {journal} {\bibinfo  {journal} {Phys. Rev. D}\ }\textbf {\bibinfo
  {volume} {91}},\ \bibinfo {pages} {054008} (\bibinfo {year} {2015})},\
  \Eprint {https://arxiv.org/abs/1411.2004} {arXiv:1411.2004 [hep-ph]}
  \BibitemShut {NoStop}%
\bibitem [{\citenamefont {Baeza-Ballesteros}\ \emph {et~al.}(2024)\citenamefont
  {Baeza-Ballesteros}, \citenamefont {Bijnens}, \citenamefont {Husek},
  \citenamefont {Romero-L\'opez}, \citenamefont {Sharpe},\ and\ \citenamefont
  {Sj\"o}}]{Baeza-Ballesteros:2024mii}%
  \BibitemOpen
  \bibfield  {author} {\bibinfo {author} {\bibfnamefont {J.}~\bibnamefont
  {Baeza-Ballesteros}}, \bibinfo {author} {\bibfnamefont {J.}~\bibnamefont
  {Bijnens}}, \bibinfo {author} {\bibfnamefont {T.}~\bibnamefont {Husek}},
  \bibinfo {author} {\bibfnamefont {F.}~\bibnamefont {Romero-L\'opez}},
  \bibinfo {author} {\bibfnamefont {S.~R.}\ \bibnamefont {Sharpe}},\ and\
  \bibinfo {author} {\bibfnamefont {M.}~\bibnamefont {Sj\"o}},\ }\bibfield
  {title} {\bibinfo {title} {{The three-pion K-matrix at NLO in ChPT}},\ }\href
  {https://doi.org/10.1007/JHEP03(2024)048} {\bibfield  {journal} {\bibinfo
  {journal} {JHEP}\ }\textbf {\bibinfo {volume} {03}},\ \bibinfo {pages}
  {048}},\ \Eprint {https://arxiv.org/abs/2401.14293} {arXiv:2401.14293
  [hep-ph]} \BibitemShut {NoStop}%
\bibitem [{\citenamefont {Bijnens}\ \emph {et~al.}(2022)\citenamefont
  {Bijnens}, \citenamefont {Husek},\ and\ \citenamefont
  {Sj\"o}}]{Bijnens:2022zsq}%
  \BibitemOpen
  \bibfield  {author} {\bibinfo {author} {\bibfnamefont {J.}~\bibnamefont
  {Bijnens}}, \bibinfo {author} {\bibfnamefont {T.}~\bibnamefont {Husek}},\
  and\ \bibinfo {author} {\bibfnamefont {M.}~\bibnamefont {Sj\"o}},\ }\bibfield
   {title} {\bibinfo {title} {{Six-meson amplitude in QCD-like theories}},\
  }\href {https://doi.org/10.1103/PhysRevD.106.054021} {\bibfield  {journal}
  {\bibinfo  {journal} {Phys. Rev. D}\ }\textbf {\bibinfo {volume} {106}},\
  \bibinfo {pages} {054021} (\bibinfo {year} {2022})},\ \Eprint
  {https://arxiv.org/abs/2206.14212} {arXiv:2206.14212 [hep-ph]} \BibitemShut
  {NoStop}%
\bibitem [{\citenamefont {Bijnens}\ and\ \citenamefont
  {Husek}(2021)}]{Bijnens:2021hpq}%
  \BibitemOpen
  \bibfield  {author} {\bibinfo {author} {\bibfnamefont {J.}~\bibnamefont
  {Bijnens}}\ and\ \bibinfo {author} {\bibfnamefont {T.}~\bibnamefont
  {Husek}},\ }\bibfield  {title} {\bibinfo {title} {{Six-pion amplitude}},\
  }\href {https://doi.org/10.1103/PhysRevD.104.054046} {\bibfield  {journal}
  {\bibinfo  {journal} {Phys. Rev. D}\ }\textbf {\bibinfo {volume} {104}},\
  \bibinfo {pages} {054046} (\bibinfo {year} {2021})},\ \Eprint
  {https://arxiv.org/abs/2107.06291} {arXiv:2107.06291 [hep-ph]} \BibitemShut
  {NoStop}%
\bibitem [{\citenamefont {Rubin}\ \emph {et~al.}(1966)\citenamefont {Rubin},
  \citenamefont {Sugar},\ and\ \citenamefont {Tiktopoulos}}]{Rubin:1966zz}%
  \BibitemOpen
  \bibfield  {author} {\bibinfo {author} {\bibfnamefont {M.}~\bibnamefont
  {Rubin}}, \bibinfo {author} {\bibfnamefont {R.}~\bibnamefont {Sugar}},\ and\
  \bibinfo {author} {\bibfnamefont {G.}~\bibnamefont {Tiktopoulos}},\
  }\bibfield  {title} {\bibinfo {title} {{Dispersion Relations for
  Three-Particle Scattering Amplitudes. I}},\ }\href
  {https://doi.org/10.1103/PhysRev.146.1130} {\bibfield  {journal} {\bibinfo
  {journal} {Phys. Rev.}\ }\textbf {\bibinfo {volume} {146}},\ \bibinfo {pages}
  {1130} (\bibinfo {year} {1966})}\BibitemShut {NoStop}%
\bibitem [{\citenamefont {Potapov}\ and\ \citenamefont
  {Taylor}(1977{\natexlab{a}})}]{PhysRevA.16.2264}%
  \BibitemOpen
  \bibfield  {author} {\bibinfo {author} {\bibfnamefont {V.~S.}\ \bibnamefont
  {Potapov}}\ and\ \bibinfo {author} {\bibfnamefont {J.~R.}\ \bibnamefont
  {Taylor}},\ }\bibfield  {title} {\bibinfo {title} {Three-particle scattering
  rates and singularities of the $t$ matrix. i.},\ }\href
  {https://doi.org/10.1103/PhysRevA.16.2264} {\bibfield  {journal} {\bibinfo
  {journal} {Phys. Rev. A}\ }\textbf {\bibinfo {volume} {16}},\ \bibinfo
  {pages} {2264} (\bibinfo {year} {1977}{\natexlab{a}})}\BibitemShut {NoStop}%
\bibitem [{\citenamefont {Potapov}\ and\ \citenamefont
  {Taylor}(1977{\natexlab{b}})}]{PhysRevA.16.2276}%
  \BibitemOpen
  \bibfield  {author} {\bibinfo {author} {\bibfnamefont {V.~S.}\ \bibnamefont
  {Potapov}}\ and\ \bibinfo {author} {\bibfnamefont {J.~R.}\ \bibnamefont
  {Taylor}},\ }\bibfield  {title} {\bibinfo {title} {Three-particle scattering
  rates and singularities of the $t$ matrix. ii.},\ }\href
  {https://doi.org/10.1103/PhysRevA.16.2276} {\bibfield  {journal} {\bibinfo
  {journal} {Phys. Rev. A}\ }\textbf {\bibinfo {volume} {16}},\ \bibinfo
  {pages} {2276} (\bibinfo {year} {1977}{\natexlab{b}})}\BibitemShut {NoStop}%
\bibitem [{\citenamefont {Gomez~Nicola}\ and\ \citenamefont
  {Pelaez}(2002)}]{GomezNicola:2001as}%
  \BibitemOpen
  \bibfield  {author} {\bibinfo {author} {\bibfnamefont {A.}~\bibnamefont
  {Gomez~Nicola}}\ and\ \bibinfo {author} {\bibfnamefont {J.~R.}\ \bibnamefont
  {Pelaez}},\ }\bibfield  {title} {\bibinfo {title} {{Meson meson scattering
  within one loop chiral perturbation theory and its unitarization}},\ }\href
  {https://doi.org/10.1103/PhysRevD.65.054009} {\bibfield  {journal} {\bibinfo
  {journal} {Phys. Rev. D}\ }\textbf {\bibinfo {volume} {65}},\ \bibinfo
  {pages} {054009} (\bibinfo {year} {2002})},\ \Eprint
  {https://arxiv.org/abs/hep-ph/0109056} {arXiv:hep-ph/0109056} \BibitemShut
  {NoStop}%
\bibitem [{\citenamefont {Dawid}\ \emph {et~al.}(2023)\citenamefont {Dawid},
  \citenamefont {Islam},\ and\ \citenamefont {Brice\~no}}]{Dawid:2023jrj}%
  \BibitemOpen
  \bibfield  {author} {\bibinfo {author} {\bibfnamefont {S.~M.}\ \bibnamefont
  {Dawid}}, \bibinfo {author} {\bibfnamefont {M.~H.~E.}\ \bibnamefont
  {Islam}},\ and\ \bibinfo {author} {\bibfnamefont {R.~A.}\ \bibnamefont
  {Brice\~no}},\ }\bibfield  {title} {\bibinfo {title} {{Analytic continuation
  of the relativistic three-particle scattering amplitudes}},\ }\href
  {https://doi.org/10.1103/PhysRevD.108.034016} {\bibfield  {journal} {\bibinfo
   {journal} {Phys. Rev. D}\ }\textbf {\bibinfo {volume} {108}},\ \bibinfo
  {pages} {034016} (\bibinfo {year} {2023})},\ \Eprint
  {https://arxiv.org/abs/2303.04394} {arXiv:2303.04394 [nucl-th]} \BibitemShut
  {NoStop}%
\bibitem [{\citenamefont {Chung}()}]{Chung:1971ri}%
  \BibitemOpen
  \bibfield  {author} {\bibinfo {author} {\bibfnamefont {S.~U.}\ \bibnamefont
  {Chung}},\ }\bibfield  {title} {\bibinfo {title} {{Spin Formalisms}},\
  }\bibfield  {journal} {\bibinfo  {journal}
  {https://suchung.web.cern.ch/spinfm1.pdf}\ }\href
  {https://doi.org/10.5170/CERN-1971-008} {10.5170/CERN-1971-008}\BibitemShut
  {NoStop}%
\bibitem [{\citenamefont {Nystr{\"o}m}(1930)}]{10.1007/BF02547521}%
  \BibitemOpen
  \bibfield  {author} {\bibinfo {author} {\bibfnamefont {E.~J.}\ \bibnamefont
  {Nystr{\"o}m}},\ }\bibfield  {title} {\bibinfo {title} {{\"Uber Die
  Praktische Aufl\"osung von Integralgleichungen mit Anwendungen auf
  Randwertaufgaben}},\ }\href {https://doi.org/10.1007/BF02547521} {\bibfield
  {journal} {\bibinfo  {journal} {Acta Mathematica}\ }\textbf {\bibinfo
  {volume} {54}},\ \bibinfo {pages} {185 } (\bibinfo {year}
  {1930})}\BibitemShut {NoStop}%
\bibitem [{\citenamefont {Delves}\ and\ \citenamefont
  {Mohamed}(1988)}]{delves1988computational}%
  \BibitemOpen
  \bibfield  {author} {\bibinfo {author} {\bibfnamefont {L.}~\bibnamefont
  {Delves}}\ and\ \bibinfo {author} {\bibfnamefont {J.}~\bibnamefont
  {Mohamed}},\ }\href {https://books.google.com/books?id=n1c7AAAAIAAJ} {\emph
  {\bibinfo {title} {Computational Methods for Integral Equations}}}\ (\bibinfo
   {publisher} {Cambridge University Press},\ \bibinfo {year}
  {1988})\BibitemShut {NoStop}%
\bibitem [{\citenamefont {Glockle}(1978)}]{Glockle:1978zz}%
  \BibitemOpen
  \bibfield  {author} {\bibinfo {author} {\bibfnamefont {W.}~\bibnamefont
  {Glockle}},\ }\bibfield  {title} {\bibinfo {title} {{S-matrix pole trajectory
  in a three-neutron model}},\ }\href {https://doi.org/10.1103/PhysRevC.18.564}
  {\bibfield  {journal} {\bibinfo  {journal} {Phys. Rev. C}\ }\textbf {\bibinfo
  {volume} {18}},\ \bibinfo {pages} {564} (\bibinfo {year} {1978})}\BibitemShut
  {NoStop}%
\bibitem [{\citenamefont {Sadasivan}\ \emph {et~al.}(2023)\citenamefont
  {Sadasivan}, \citenamefont {Mai}, \citenamefont {D\"oring}, \citenamefont
  {Mei\ss{}ner}, \citenamefont {Amorim}, \citenamefont {Klucik}, \citenamefont
  {Lu},\ and\ \citenamefont {Gen}}]{Sadasivan:2022srs}%
  \BibitemOpen
  \bibfield  {author} {\bibinfo {author} {\bibfnamefont {D.}~\bibnamefont
  {Sadasivan}}, \bibinfo {author} {\bibfnamefont {M.}~\bibnamefont {Mai}},
  \bibinfo {author} {\bibfnamefont {M.}~\bibnamefont {D\"oring}}, \bibinfo
  {author} {\bibfnamefont {U.-G.}\ \bibnamefont {Mei\ss{}ner}}, \bibinfo
  {author} {\bibfnamefont {F.}~\bibnamefont {Amorim}}, \bibinfo {author}
  {\bibfnamefont {J.~P.}\ \bibnamefont {Klucik}}, \bibinfo {author}
  {\bibfnamefont {J.-X.}\ \bibnamefont {Lu}},\ and\ \bibinfo {author}
  {\bibfnamefont {L.-S.}\ \bibnamefont {Gen}},\ }\bibfield  {title} {\bibinfo
  {title} {{New insights into the pole parameters of the $\Lambda(1380)$, the
  $\Lambda(1405)$ and the $\Sigma(1385)$}},\ }\href
  {https://doi.org/10.3389/fphy.2023.1139236} {\bibfield  {journal} {\bibinfo
  {journal} {Front. Phys.}\ }\textbf {\bibinfo {volume} {11}},\ \bibinfo
  {pages} {1139236} (\bibinfo {year} {2023})},\ \Eprint
  {https://arxiv.org/abs/2212.10415} {arXiv:2212.10415 [nucl-th]} \BibitemShut
  {NoStop}%
\end{thebibliography}%

\end{document}